\documentclass[12pt,twoside]{book}
\pdfoutput=1
\usepackage{vuwthesis}
\usepackage{graphicx}
\usepackage{amssymb}
\usepackage{amsfonts} 
\usepackage{amsmath}
\usepackage{makeidx}
\usepackage{bm}
\usepackage{graphicx,epsfig}
\usepackage[dvips]{color}
\usepackage{vuw}
\usepackage{dayofweek}
\usepackage{daytime}
\usepackage{fancyhdr}
\usepackage{mathrsfs} 
\usepackage{amssymb,amsmath,amsthm}

\def\d{{\mathrm{d}}}

\newtheorem{theorem}{Theorem}[chapter]

\def\d{{\mathrm{d}}}

\def\lint{\hbox{\Large $\displaystyle\int$}} 

\def\d{{\mathrm{d}}}
\def\tr{{\mathrm{tr}}}
\def\sech{{\mathrm{sech}}}
\def\etc{\emph{etc}}

\numberwithin{equation}{section} 

\def\Re{{\mathrm{Re}}}
\def\Im{{\mathrm{Im}}}
\def\J{{\mathscr{J}}}
\def\sech{{\mathrm{sech}}}

\begin{document}
\pagestyle{fancy}
\fancyhead{ }
\bibliographystyle{plain}
\include{titlepage}
\thispagestyle{empty}
\frontmatter

\title{ \centerline{---\crest---}
\vskip 1 cm
Rigorous bounds \\ on \\ Transmission, Reflection, \\ and \\ Bogoliubov coefficients}

\author{Petarpa Boonserm}

\subject{Mathematics}

\abstract{This thesis describes the development of some basic mathematical tools of wide relevance to mathematical physics. Transmission and reflection coefficients are associated with quantum tunneling phenomena, while Bogoliubov coefficients are associated with the mathematically related problem of excitations of a parametric oscillator. While many approximation techniques for these quantities are known, very  little is known about rigorous upper and lower bounds.

In this thesis four separate problems relating to rigorous bounds on transmission, reflection and Bogoliubov coefficients are considered, divided into four separate themes: 
\begin{itemize}
\item 
Bounding the Bogoliubov coefficients;
\item
 Bounding the greybody factors for Schwarzschild black holes;
 \item
 Transformation probabilities and the Miller--Good transformation; 
 \item
 Analytic bounds on transmission probabilities. 
\end{itemize}

\vfill
\begin{center}
25 February 2009;  \LaTeX-ed \DayOfWeek, \today; \daytime\\
\copyright\ Petarpa Boonserm
\end{center}
}

\phd

\maketitle
\frontmatter
\chapter{Acknowledgements}
I would like to thank my supervisor, Professor Matt Visser for his guidance and 
support. I am lucky to have such an active and hard working supervisor. 
I really appreciate the effort he made for my thesis, making sure that 
things got done properly and on time.

I am grateful to the School of Mathematics, Statistics, and Operations Research
for providing me with an office and all the facilities, and the Thai 
Government Scholarship that provided me with funding.

I also would like to say thanks to my family who were also very supportive 
and listened to me.

Finally, I would like to extend my gratitude to my friends, Adisorn Juntrasook, Nattakarn Kajohnwongsatit, Trin Sunathvanichkul, Jirayu Chotimongkol, Narit Pidokrajt, Anon Khamwon, Celine Cattoen, and Garoon Pongsart for always being there to listen to me. I am not sure that this thesis would have been finished without the support they showed.

\chapter{Preface}

This thesis looks at a number of problems related to the derivation of rigorous bounds on transmission, reflection, and Bogoliubov coefficients: 
To set the stage, we shall first briefly describe the general ideas underlying the Schr\"odinger equation, and the concept of the WKB approximation for barrier penetration probability. In addition, we shall present a discussion of some general features of scattering theory in one space dimension. By considering one-dimensional problems involving an incident beam of particles, we shall derive an important connection between reflection and transmission amplitudes. Furthermore, we shall collect several known analytic results, and show how they relate to the general results presented in this thesis. We shall also review and concisely describe the concept of  \emph{quasinormal modes},  and see how most of the concepts introduced here are important tools for comparing the bounds  derived in the body of the thesis with known analytic results.

The technical heart of the thesis is this: We shall rewrite the second-order Schr\"odinger equation as a set of two coupled first-order linear differential equations (for which bounds can relatively easily be established). Systems of differential equations of this type are often referred to as Shabat--Zakharov systems or Zhakarov--Shabat systems. After this initial  investigation, we shall use this system of ODEs to derive our first bound, and then continue by finding several slightly different ways of recasting the Schr\"odinger equation as a 1st-order Shabat--Zakharov system, in this way deriving a number of slightly different bounds.

Regarding the chapter  ``Bounding the Bogoliubov coefficients'', we have developed a distinct method for deriving general bounds on the Bogoliubov coefficients, providing a largely independent derivation of the key results;  a seperate derivation that short-circuits much of the technical discussion. 

Proceeding further along this branch of our investigation, we shall consider the Regge--Wheeler equation for excitations of a scalar field defined on a Schwarzschild spacetime, and adapt the general analysis of the previous chapters to this specific case. We shall demonstrate that rigorous and explicit analytic bounds are indeed achievable. While these bounds may not answer all the physical questions one might legitimately wish to ask, they are definitely a solid step in the right direction. 

We shall then use the Miller--Good transformation (which maps an initial Schr\"odinger 
equation to a final Schr\"odinger equation for a different potential) to significantly generalize the previous bound. Moreover, we shall then use the Miller--Good transformation to generalize the bound to make it more efficient. 

Finally we shall consider analytic bounds on the transmission probabilities obtained by comparing a simple ``known'' potential with a more complicated ``unknown'' one. In this case we shall obtain yet another (distinct) Shabat--Zakharov system and use it to (partially and formally) ``solve'' the scattering problem. In this case we can derive both \emph{upper} and \emph{lower} bounds on the transmission coefficients and related Bogoliubov coefficients.

\section*{Chapter by chapter outline}
This thesis is divided into twelve main chapters. The first chapter is devoted to describing the general ideas of the Schr\"odinger equation and the concepts of WKB approximation for barrier penetration probability. Furthermore, we introduce the concept of the classical turning point, which is one of the key ideas in developing the WKB estimate. Moreover, these general concepts are important for understanding the bounds we will derive on transmission and reflection in Bogoliubov coefficients.

In chapter~\ref{C:consistency-2}, we shall introduce scattering theory in one space dimension. This is an elegant topic that is mathematically simple and physically transparent. We shall apply the Schr\"odinger equation to a generic system to identify the potential-energy function. Furthermore, we shall derive a significant relationship between reflection and transmission amplitudes by considering one-dimensional problems with an incident beam of particles. 

In chapter~\ref{C:consistency-3}, we shall concentrate our attention on collecting several known analytic results, and show how they relate to the general results presented in this thesis. We shall review and 
briefly describe the concept of \emph{quasinormal modes}, and see how most of the concepts introduced here are important for comparing the bounds we shall derive with known analytic results.
By taking specific cases of these bounds and related results it is possible to reproduce many analytically known results, such as those for the delta-function potential, double-delta-function potential, square potential barrier,  $\mathrm{tanh}$ potential, $\mathrm{sech}^2$ potential, asymmetric square-well potential, the Poeschl--Teller potential and its variants, and finally the general Eckart--Rosen--Morse--Poeschl--Teller potential.

The next two chapters, chapter~\ref{C:consistency-4} and chapter~\ref{C:consistency-5}, can be seen as two deeply interconnected chapters. The key idea in chapter~\ref{C:consistency-4} is to recast the Schr\"odinger equation as a 1st-order Shabat--Zakharov system.  In chapter~\ref{C:consistency-5}, we shall use the Shabat--Zakharov system of ODEs to derive our first bound on the transmission, reflection, and Bogoliubov coefficients.

In chapter~\ref{C:consistency-6}, we shall deal with some specific cases of these bounds and develop a number of interesting specializations.  We shall collect together a large number of results that otherwise appear quite unrelated, including reflection above and below the barrier. In addition, we have divided the special case bounds we consider into five special cases: special cases $1$---$4$, and ``future directions''. At the end of this chapter, we take further specific cases of these bounds and related result to reproduce many analytically known results.

In chapter~\ref{C:consistency-7},  we shall re-cast and represent these bounds in terms of the mathematical structure of parametric oscillations. This time-dependent problem is closely related to the spatial properties of the time-independent Schr\"odinger equation.

In chapter~\ref{C:consistency-8}, we shall re-assess the general bounds on the Bogoliubov coefficients developed in~\cite{bounds1}, providing a new and largely independent derivation of the key results, one that short-circuits much of the technical discussion in~\cite{bounds1}. 

In chapter~\ref{C:consistency-9}, we shall develop a complementary set of results--- we shall derive several rigorous analytic bounds that can be placed on the greybody factors. Furthermore, we shall consider the greybody factors in black hole physics, which modify the naive Planckian spectrum that is predicted for Hawking radiation when working in the limit of geometrical optics.

In chapter~\ref{C:consistency-10}, we shall use the Miller--Good transformation (which maps an initial 
Schr\"odinger equation to a final Schr\"odinger equation for a different potential) to significantly generalize the previous bound. At the end of this chapter, we shall discuss the possibility of using the Miller--Good transformation to derive generalized special-case bounds to make them more efficient.

In chapter~\ref{C:consistency-11}, we shall develop a new set of techniques that are more amenable to the development of \emph{both} upper and lower bounds. Moreover, we shall derive significantly different results (a number of rigorous bounds on transmission probabilities for one dimensional scattering problems), of both theoretical and practical interest. 

In chapter~\ref{C:consistency-12}, we finally conclude with a brief discussion of lessons learned from these rigorous bounds on transmission, reflection, and Bogoliubov coefficients.

\section*{Structure of the thesis}
This thesis has been written with the goal of being accessible to people with a basic background in non-relativistic quantum physics, especially in transmission, reflection, and Bogoliubov coefficients. Mathematically, the key feature is an analytic study of the properties of second-order linear differential equations, and the derivation of analytic bounds on the growth of solutions of these equations.

This thesis is made up of twelve chapters and five appendices. Four of the appendices are papers published or submitted on work relating to this thesis. All of them were produced in collaboration with my supervisor, Professor Matt Visser. At the time of writing three papers have been published~\cite{bounds2, greybody-factor, Miller-good-transformation-articles}, and the latest has been submitted for refereeing~\cite{analytic-bounds1}.

\section*{Use of references}

Regarding referencing --- For completely non controversial background information (and \emph{only} for completely noncontroversial items) we will often just reference {\sf Wikipedia} or similar reasonably definitive web resources. For more technical information, especially recent research,  we will always directly cite the appropriate scientific literature.

\thispagestyle{plain}
\clearpage
\fancyhead{}
\fancyhead[LE]{\textsl{\leftmark}}
\fancyhead[RO]{\textsl{\rightmark}}
\thispagestyle{plain}
\tableofcontents
\listoffigures
\listoftables
\mainmatter

\chapter{General introduction}
\label{C:consistency-1}
\section{Introduction}

This chapter is an introduction to the topic of developing rigorous bounds on transmission, reflection,  and Bogoliubov coefficients. We shall introduce the basic ideas underlying  the Schr\"odinger equation, and its application to the wave-function that describes the wavelike properties of a subatomic system.

We shall also review the concept of the WKB approximation, which is an important and  significant method to derive approximate solutions for the wave function. For instance, as we shall show, the WKB approach can be used as a ``basis'' for formally writing down the exact solutions. Most physicists, and many mathematicians, have seen how important the WKB approximation is for estimating barrier penetration probability. Unfortunately, the WKB approximation is an example of an uncontrolled approximation, and we do not know if the resulting estimate is high or low. As part of the main work reported in this thesis, we modify, improve, and extend the approach originally developed by Visser~\cite{bounds1}. 

We shall derive a number of rigourous bounds on transmission probabilities (and reflection probabilities, and Bogoliubov coefficients) for one-dimensional scattering problems. The derivation of these bounds generally proceeds by rewriting the Schr\"odinger equation in terms of some equivalent system of first-order equations, and then analytically bounding the growth of certain quantities related to the net flux of particles as one sweeps across the potential. 

While over the last century or more considerable effort has been put into the problem of finding \emph{approximate solutions} for wave equations in general, and quantum mechanical problems in particular, it appears that as yet relatively little work seems to have been put into the complementary problem of establishing rigorous \emph{bounds} on the exact solutions. We have in mind either bounds on parametric amplification and the related quantum phenomenon of particle production (as encoded in the Bogoliubov coefficients), or bounds on transmission and reflection coefficients. 

In this thesis, we introduce and prove several rigorous bounds on the Bogoliubov coefficients associated with a time-dependent potential, and also derive several rigorous analytic bounds that can be placed on barrier transmission probabilities. As a specific application, we shall then explore greybody factors in black hole physics, which modify the naive Planckian spectrum that is predicted for Hawking radiation when working in the limit of geometrical optics.

Additionally, we  will extend these ideas to address topics of considerable general interest in quantum physics, such as transmission through a potential barrier, and the related issue of particle production from a parametric resonance. This is an example of finding new physics (and new mathematics) in an old and apparently well-understood area.

To begin with, we need to briefly describe the concept of the \emph{Schr\"odinger equation},  and the rationale behind the \emph{WKB estimate} for barrier penetration probability, otherwise the rigorous bounds on transmission, reflection, and Bogoliubov coefficients will be difficult to understand.

\vfill

\section{The Schr\"odinger Equation}
The Schr\"odinger equation was discovered by the Austrian physicist Erwin Schr\"odinger in 1925,  it describes the space -- and time -- dependence of the quantum amplitude that characterizes quantum mechanical systems~\cite{Scho}.

Both Erwin Schr\"odinger and Werner Heisenberg independently developed different versions of the ``modern'' quantum theory. Schr\"odinger's method relates to partial differential equations, whereas Heisenberg's method uses infinite-dimensional matrices.  

However, both methods were soon shown to be mathematically equivalent. Furthermore, from the modern viewpoint it seems very clear that Schr\"odinger's equation has a clearer physical interpretation via the classical wave equation. Indeed, the Schr\"odinger equation can be shown to be a form of the wave equation applied to matter waves~\cite{Scho_1}.

It is apparent that this equation defines the behaviour of the wave function that describes the wavelike properties of a subatomic system. Furthermore, it deals with the kinetic energy and potential energy, both of which contribute to the total energy. It is solved to derive the different energy levels of the system. 
More generally, Schr\"odinger applied the equation to the hydrogen atom, and its properties can be predicted with remarkable precision. It should be remarked that the equation is applied widely in atomic, nuclear, and solid-state physics~\cite{Scho_online}.

Actually there are two slightly different equations which go by Schr\"o\-dinger's name as follows:

\subsection{The time-independent Schr\"odinger equation}
We start with the one-dimensional classical wave equation~\cite{Scho_1},
\begin{equation}
{\partial^2 u \over \partial x^2} = {1 \over v^2} \,  {\partial^2 u \over \partial t^2}.
\end{equation}
Let us consider the separation of variables 
\begin{equation}
u(x,t) = \psi (x) \; f(t),
\end{equation}
which then leads to
\begin{equation}
f(t) \, {\d^2 \psi(x) \over \d x^2} = {1 \over v^2} \, \psi(x) \, {\d^2 f(t) \over \d t^2}.
\end{equation}
When we introduce one of the standard wave equation solutions $f(t)$ such as $\exp(i \omega t)$, we easily obtain
\begin{equation}
\label{standard-wave-equation}
{\d^2 \psi(x) \over \d x^2} = {-\omega^2 \over v^2} \,  \psi(x).
\end{equation}
It is now easy to ``derive'' (in the sense of a physicist's plausibility argument)  an ordinary differential equation describing the spatial amplitude of the matter wave as a function of position. We note that the energy of a particle is the sum of kinetic and potential parts 
\begin{equation}
E = {p^2 \over m} + V(x),
\end{equation}
which can be solved for the momentum, $p$, to obtain
\begin{equation}
p = \{2m [E - V(x)] \}^{1/2}.
\end{equation}
We now see that it is convenient to use the \emph{de Broglie} formula to get an expression for the (position dependent) wavelength 
\begin{equation}
\lambda = {h \over p} = {h \over \{2m [E-V(x)]\}^{1/2}}.
\end{equation}
If we recall $\omega =  2 \pi \nu$  and $\nu \lambda = v $, then the term $\omega^2/v^2$ in equation (\ref{standard-wave-equation}) can be rewritten in terms of $\lambda$:
\begin{equation}
{\omega^2 \over v^2} = {4 \pi^2 \nu^2 \over v^2} = {4 \pi^2 \over \lambda^2} = {2m [E-V(x)] \over \hbar^2}.
\end{equation}
Additionally, when this result is substituted into equation (\ref{standard-wave-equation}), 
 we also ``derive'' the well-known time-independent Schr\"odinger equation, 
 \begin{equation}
{\d^2 \psi(x) \over \d x^2} + {2m \over \hbar^2} [E -V(x)] \psi(x) = 0.
\end{equation}
Let us now rewrite the above equation in a more standardized  form, we get 
\begin{equation}
- {\hbar^2 \over 2m} {\d^2 \psi(x) \over \d x^2} + V(x) \psi(x) = E \psi(x).
\end{equation}
We now have all the important information about our system. Moreover, this single-particle one-dimensional equation can clearly be extended to the case of three dimensions, where it becomes 
\begin{equation}
- {\hbar^2 \over 2m} \nabla^2 \psi(r) + V(r) \psi(r) = E \psi(r).
\end{equation}
A two-body problem can also be treated by this equation if the mass $m$ is replaced by the reduced mass $\mu$:
\begin{equation}
\mu = {1\over  \displaystyle {1\over m_1} + {1\over m_2} }.
\end{equation}

Nevertheless, it is important to point out that this analogy with the classical wave equation only goes so far. We cannot, for example, ``derive'' the time-dependent Schr\"odinger equation in an analogous fashion, at least not without several additional hypotheses.  (For instance, the  time-dependent Schr\"odinger equation involves the partial first derivative with respect to time instead of the partial second derivative.) Finally, we would like to comment that historically, Schr\"odinger presented his time-independent equation first, and then went back and postulated the more general time-dependent equation~\cite{Scho_1}.

\subsection{The time-dependent Schr\"odinger Equation}

In this section we now present the time-dependent version of the Schr\"odinger equation. Although we were able to ``derive''  the single-particle time-indepen\-dent Schr\"odinger equation starting from the classical wave equation and the \emph{de Broglie} relation, the time-dependent Schr\"odinger equation cannot be ``derived'' using elementary methods, and is generally given as a postulate of quantum mechanics~\cite{Scho_1}.

In other words, we shall postulate the single-particle three-dimensional time-dependent Schr\"odinger equation as
\begin{equation}
\label{three-dimensional-time-dependent}
i \hbar {\partial \psi(r,t) \over  \partial t} = - {\hbar^2 \over 2m} \nabla^2 \psi(r,t) + V(r) \psi(r,t).
\end{equation}
We now focus on the case where  $V$ is assumed to be a real function,  which represents the potential energy of the system. It is very easy to see that the time-dependent equation can be used to derive the time-independent equation. If we write the wavefunction as a product of spatial and temporal terms, $\psi(r,t) = \psi(r)\; f(t)$, then equation (\ref{three-dimensional-time-dependent}) becomes 
\begin{equation}
\psi(r) i \hbar \, {\d f(t) \over \d t} = f(t) \bigg[-{\hbar^2 \over 2m} \nabla^2 + V(r)\bigg] \psi(r),
\end{equation}
or
\begin{equation}
{i \hbar \over f(t)} {\d f(t) \over \d t} = {1 \over \psi(r)} \bigg[ - {\hbar^2 \over 2m} \nabla^2 + V(r)\bigg] \psi(r).
\end{equation}
It is easy to see that the left and right hand sides must each equal a constant $E$ when the left-hand side is a function of  $t$ only and the right hand side is a function of  $r$ only. (This is just the usual separation of variables technique.) 

Alternatively, if we appropriately denote this separation constant by $E$, (since the right-hand side clearly must have the dimensions of energy), then we extract two ordinary differential equations, specifically
\begin{equation}
\label{former-equation}
{1 \over f(t)} {\d f(t) \over \d t} = - {i E \over \hbar},
\end{equation} 
and
\begin{equation}
\label{extract-ode}
- {\hbar^2 \over 2m} \nabla^2 \psi(r) + V(r) \, \psi(r) = E \, \psi(r).
\end{equation}
The equation (\ref{extract-ode}) is once again the time-independent Schr\"odinger equation. Furthermore  equation (\ref{former-equation}) is easily solved to yield 
\begin{equation}
\label{previous-new-equation}
f(t) = \exp(-i E t / \hbar).
\end{equation}

Most generally, we can show that the Hamiltonian in equation (\ref{extract-ode}) is a Hermitian operator, and  that the eigenvalues of a Hermitian operator must be real, so $E$  is real.  This implies that the solutions $f(t)$ are purely oscillatory, since  $f(t)$ never changes in magnitude [recall Euler's formula  $\exp(\pm i \theta) = \mathrm{cos} \, \theta \pm i \,  \mathrm{sin} \, \theta$]. 
In the following if we set 
\begin{equation}
\label{euler-formula}
\psi(r,t) = \psi(r) \exp(-i E t/\hbar),
 \end{equation}
then the total wave function $\psi(r,t)$ differs from $\psi(r)$ only by a phase factor of constant magnitude. 
 
We can easily show that the quantity $|\psi(r,t)|^2$ is time independent as follows:
 \begin{equation}
 |\psi(r,t)|^2 = \psi^*(r,t) \, \psi(r,t) = \exp(i Et/ \hbar) \, \psi^*(r) \exp(-iEt/\hbar) \, \psi(r) = \psi^*(r) \psi(r). 
 \end{equation}
 Furthermore, if  $\psi(r,t)$ satisfies (\ref{euler-formula}), then the expectation value for any time-independent operator is also time-independent. Thus it is easy to see that
 \begin{equation}
 \langle A \rangle = \int \psi^*(r,t) \hat{A} \psi(r,t) = \int \psi^*(r) \hat{A} \psi(r).
 \end{equation}
Wave functions of the form (\ref{euler-formula}) are called stationary states. The state  $\psi(r,t)$ is ``stationary'', but the particle it describes is not.
It is now easy to see that equation (\ref{euler-formula}) represents a particular solution to equation (\ref{three-dimensional-time-dependent}). The general solution to equation (\ref{three-dimensional-time-dependent}) will be a linear combination of these particular solutions~\cite{Scho_1}
\begin{equation}
\psi(r,t) = \sum_{i} c_{i} \exp(-iE_{i}t/\hbar)  \psi_{i}(r).
\end{equation}
In the next section, we shall introduce an important technique, the \emph{WKB approximation}, which will be used several times in the body of this thesis.

\section{WKB approximation}
The \emph{WKB (Wentzel--Kramers--Brillouin) approximation} is also known as the \emph{WKBJ (Wentzel-Kramers-Brillouin-Jeffreys) approximation}, or sometimes the \emph{JWKB} approximation.
The basic idea is it estimates a real Schr\"odinger wave function by a sinusoidal vibration whose phase is presented by the space integral of the classical momentum, the phase integral, and whose amplitude varies inversely as the fourth root of the classical momentum. In fact, in its original 1800's incarnation as the \emph{Jeffreys approximation}, the \emph{WKB approximation} was already a meaningful expression for the physical waves of optics, acoustics, and hydrodynamics. After 1925, this approximation  was rapidly applied to the new Schr\"odinger \emph{probability} waves~\cite{wkb_history}.

The \emph{WKB approximation} is an important method to derive approximate solutions and estimates for many physical problems. For instance, it is mainly applicable to problems of wave propagation in which the frequency of the wave is very high, or equivalently, the wavelength of the wave is very short (compared to the typical distance over which the potential varies). Despite the fact that the \emph{WKB solutions} are approximate solutions,  sometimes they are amazingly accurate~\cite{wkb_intro}.

Let us begin with the one-dimensional time-independent Schr\"odinger equation~\cite{wkb_math}
\begin{equation}
\label{Schordinger}
- {\hbar^2 \over 2 m} {\d^2 \over \d x^2} \, \psi(x) + V(x) \, \psi(x) = E \, \psi(x),
\end{equation}
which can be rewritten as
\begin{equation}
\label{psi_1}
{\d^2 \over \d x^2} \, \psi(x) = {2 m \over \hbar^2} \, (V(x) -E) \, \psi(x),
\end{equation}
where \, $ \d^2 \psi(x) / \d x^2 =$  second derivative with respect to $x$,  \, $\psi(x) =$ Schr\"odinger wave function, $E =$  energy and $V =$ potential energy.

We can now write the wavefunction in terms of the exponential function by putting it in the form~\cite{Schwabl}
\begin{equation}
\label{psi_eq}
\psi(x) = A(x) \, \exp(i S(x)/\hbar) .
\end{equation}
Substituting $\psi(x)$ into equation (\ref{psi_1}), we derive
\begin{equation}
\label{eq1}
A(x) \, S'(x)^2 - i \hbar \, A(x) \, S''(x) - 2 i \hbar A'(x) \, S'(x) - \hbar^2 \, A''(x) = 2 m \, (E - V) \, A(x) .
\end{equation}
By comparing the first two terms, we expect that the quasi-classical region is given by
\begin{equation}
S'(x)^2 \gg \hbar \, S''(x) .
\end{equation}
We take the real and imaginary parts of equation (\ref{eq1}):
\begin{eqnarray}
\label{S_eq1}
S'(x)^2 &=& 2 m \, (E - V) + \hbar^2 \, A''(x) / A(x) \; ,
\\
\label{S_eq2}
- S''(x) &=& 2 S'(x) \, \big(\d \ln \, (A'(x)) / \d x \big) .
\end{eqnarray}
Now we are considering only one-dimensional problems, which of course also include radial motion in central potentials. One can then express equation (\ref{S_eq2}) in the form
\begin{equation}
{\d \over \d x} \, \bigg({1 \over 2} \, \log \, {\d \, S(x) \over \d x} + \log A \bigg) = 0 \, ,
\end{equation}
and one finds
\begin{equation}
\label{a_eq1}
A(x) = {C \over \sqrt{S'(x)}}  .
\end{equation}
In equation (\ref{S_eq1}), let us neglect the term $\hbar^2 \, A''(x) /A(x)$ compared to $S'(x)^2$. (This is where the approximation is made.) The resulting equation
\begin{equation}
\label{s_eq3}
S'(x)^2 = 2 m (E - V(x)),
\end{equation}
can then easily be integrated:
\begin{equation}
\label{s_eq4}
S(x) = \pm \int^x \, \d x' \, \sqrt{2 m (E - V(x'))} .
\end{equation}
Substituting (\ref{a_eq1}) and (\ref{s_eq4}) into (\ref{psi_eq}), one finds
\begin{equation}
\label{psieqn}
\psi(x) = \sum_{\pm} \, {C_{\pm} \over \sqrt{p(x)}} \, \exp \, \bigg\{\pm i \, \int \d x \, p(x)/ \hbar \bigg\}
\end{equation}
with momentum
\begin{equation}
p(x) = \sqrt{2 m (E - V(x))} \, ,
\end{equation}
Now we introduce the notation
\begin{equation}
\label{introduce-the-notation}
k(x)^2 = {2 m [E - V(x)] \over \hbar^2}.
\end{equation}
By the JWKB approximation, we derive
\begin{equation}
\psi \approx A \, {\exp [i \int{k(x)}] \over \sqrt{k(x)}} + B \, {\exp [-i \int{k(x)}] \over \sqrt{k(x)}}.
\end{equation}
This shows that the \emph{JWKB approximation} is a fruitful method of calculation, that can be used to develop a perturbation theory.

\section{Classical turning points}
We can start by considering one of the most interesting aspects regarding the \emph{WKB approximation};  that being what happens at the classical turning points where $V(x) = E$. In fact it is easy to realize that as long as we keep away from these points, the approximation works very well indeed. To get near or pass through 
a turning point one has to go beyond the WKB approximation. The most straightforward way to do so is by using a linear approximation to the Taylor series expansion of the potential in the vicinity of the classical turning point. The \emph{exact} solution to this \emph{approximate} problem is given in terms of  an \emph{Airy function}. (\emph{Bessel function} of order ${1\over3}$.) Using this, the standard approach is now to  derive a specific way of 
patching the wave functions on either side of the turning point --- this leads to the so-called ``connection conditions''. 
Finally, it is interesting to note that historically the WKB approach to barrier penetration application very quickly yielded significant achievements in terms of understanding alpha decay lifetimes~\cite{turning-points}.

\section{Discussion}

In this chapter, we introduced the Schr\"odinger equation, which is a specific partial differential equation used in the development of the ``new'' (1925) quantum theory. The Schr\"odinger equation was discovered by the Austrian physicist Erwin Schr\"odinger in 1925, and describes the space --and time-- dependence of quantum mechanical systems~\cite{Scho}.
In addition, physicists quickly applied the \emph{WKB approximation} to the new Schr\"odinger  \emph{probability}  waves.   The \emph{WKB approximation} is generally applicable to problems of wave propagation in which the frequency of the wave is very high, or equivalently, the wavelength of the wave is very short.

The problem of finding approximate solutions for wave equations in general, and quantum mechanical problems in particular,  has been extensively considered over the last century or two.
However, it appears that as yet relatively little work seems to have been put into the complementary problem of establishing rigourous bounds on the exact solutions. 

As the theory of the \emph{WKB approximation}, and the concept of the time-independent  Schr\"odinger equation, both underlie all our subsequent analyses, we have presented a very general introduction to these concepts first --- so that the bounds we will soon derive on transmission, reflection, and Bogoliubov coefficients will be easier to understand.

Finally we believe that this introduction has provided sufficient context for the reader  to appreciate the role played by the various topics to be 
discussed in this thesis, and to place them into a wider perspective. In brief, quantum mechanics is  a generic tool for addressing empirical reality, and in this thesis we are  
probing the complementary problem of establishing rigorous \emph{bounds} on the exact solutions.

\chapter{Scattering problems}
\label{C:consistency-2}

\section{Introduction}
In this chapter we shall present quantum scattering theory in one space dimension. It is a beautiful subject that is mathematically simple and physically transparent. Moreover, it still contains various important results~\cite{Matt01}. 

One-dimensional scattering problems appear in a vast variety of physical contexts.
For instance, in acoustics one might be interested in the propagation of sounds waves down a long pipe, while in electromagnetism one might be interested in the physics of wave-guides. Another important context which we want to stress in this chapter is that in quantum physics the canonical examples related to one-dimensional scattering theory are barrier penetration and reflection. In contrast, in classical physics an equivalent problem is the analysis of parametric resonances~\cite{bounds1}.

Furthermore, when considering  the basic ideas of  ``reflection and transmission probabilities'', we shall introduce a useful technique to derive a connection between reflection and transmission coefficients, showing that they are related via a conceptually simple formalism. This technique will be used several times in the main part of this thesis. 

In particular, at the end of this chapter we shall (purely as an example) illustrate how to derive either transmitted or reflected probability waves as a result of scattering of an object in the delta-potential well. More generally, we are specifically interested in the Schr\"odinger equation as shown below in equation~(\ref{E:SDE-scattering-renew}) in conditions where the potential $V(x)$ is zero outside of a finite interval---mathematically we are most interested in considering potentials of compact support. (Though much of what we will have to say will also apply to potentials with suitably rapid falloff properties as one moves to spatial infinity.)

\section{Reflection and Transmission Probabilities}
Let us consider the one-dimensional time-independent Schr\"odinger equation~\cite{Landau}--\cite{Messiah}
\begin{equation}
\label{E:SDE-scattering-renew}
-{\hbar^2\over2m} {\d^2\over \d x^2} \psi(x) + V(x) \; \psi(x) = E \; \psi(x).
\end{equation}
If the potential asymptotes to a constant,
\begin{equation}
V (x \rightarrow \pm \infty) \rightarrow V_{\pm \infty},
\end{equation}
then in each of the two asymptotic regions there are two independent solutions to the Schr\"odinger equation
\begin{equation}
\psi_\pm  (x\to\pm \infty) \approx {\exp (\pm i k_{\pm \infty} x) \over \sqrt{k_{\pm \infty}}}.
\end{equation}
Here the $\pm$ distinguishes right-moving modes $e^{+ ikx}$ from left-moving modes $e^{- ikx}$, while the $\pm \infty$ specifies which of the asymptotic regions we are in. Furthermore
\begin{equation}
k_{\pm \infty} = {\sqrt{2 m \, (E-V_{\pm \infty}) \over \hbar}}.
\end{equation}
To even begin to set up a scattering problem the minimum requirements are that potential asymptote to some constant, and this assumption will be made henceforth.
The so-called Jost solutions~\cite{Chadan} are exact solutions $\mathcal{J}_{\pm} (x)$ of the Schr\"odinger equation that satisfy
\begin{equation}
\mathcal{J}_{+} (x \rightarrow -\infty) \rightarrow {\exp (+ik_{-\infty} x) \over \sqrt{k_{-\infty}}},
\end{equation}
\begin{equation}
\mathcal{J}_{+} (x \rightarrow + \infty) \rightarrow \alpha_{+} {\exp (+ik_{+ \infty} x) \over \sqrt{k_{+ \infty}}} + \beta_{+} {\exp (-i k_{+\infty} x) \over \sqrt{k_{+ \infty}}},
\end{equation}
and
\begin{equation}
\mathcal{J}_{-}(x \rightarrow + \infty) \rightarrow {\exp \, (-ik_{+ \infty} x) \over \sqrt{k_{+ \infty}}},
\end{equation}
\begin{equation}
\mathcal{J}_{-}(x \rightarrow - \infty) \rightarrow \alpha_{-} {\exp (-ik_{- \infty} x) \over \sqrt{k_{- \infty}}} + \beta_{-} {\exp (+ik_{- \infty} x) \over \sqrt{k_{-\infty}}}.
\end{equation}
\begin{proof}[Identifying the reflection and transmission coefficients]\ \\
There are unfortunately at least four distinct sets of conventions in common use, depending on whether or not one absorbs factors of $\sqrt{k_{\pm\infty}}$ into $r$ and $t$ respectively, and on whether one chooses to focus on left-moving or right-moving waves as being primary. Let us, for the current section, adopt the convention of \emph{not} absorbing the factors of $\sqrt{k_{\pm\infty}}$ into $r$ and $t$. (We shall discuss the other convention a little later in this chapter). We start by introducing a minor variant of Messiah's notation~\cite{Messiah}
\begin{equation}
\mathcal{J}_{+} (x \rightarrow -\infty) \rightarrow  t_+ \exp (+ik_{-\infty} x),
\end{equation}
\begin{equation}
\mathcal{J}_{+} (x \rightarrow + \infty) \rightarrow  \exp (+ik_{+ \infty} x)  + r_{+} {\exp (-i k_{+\infty} x) \over \sqrt{k_{+ \infty}}},
\end{equation}
By comparing these two different forms for the asymptotic form of the Jost function we see that in this situation the ratios of the amplitudes are given by
\begin{equation}
{1\over\sqrt{k_{- \infty}}} : {\alpha_+ \over \sqrt{k_{+ \infty}}} : {\beta_+\over \sqrt{k_{+ \infty}}}
=
t_+: 1 : r_+.
\end{equation}
Thus we obtain
\begin{equation}
r_{+} = {\beta_{+} \over \sqrt{k_{+ \infty}}} \, {\sqrt{k_{+ \infty}} \over \alpha_{+} }  = {\beta_{+} \over \alpha_{+}}.
\end{equation}
We also derive (in this set of conventions)
\begin{equation}
t_{+} ={1 \over \sqrt{k_{-\infty}}} \, {\sqrt{k_{+ \infty}} \over \alpha_{+}} = 
\sqrt{k_{+\infty}\over k_{-\infty}} \; {1 \over \alpha_{+}}.
\end{equation}
\end{proof}
Thus we have demonstrated that  $\alpha_{+}$ and $\beta_{+}$, the (right-moving) Bogoliubov coefficients, are related to the (left-moving) reflection and transmission amplitudes by
\begin{equation}
r_{+} = {\beta_{+} \over \alpha_{+}}; \qquad t_{+} = \sqrt{k_{+\infty}\over k_{-\infty}} \; {1 \over \alpha_{+}}.
\end{equation}
Without further calculation we can also deduce 
\begin{equation}
r_{-} = {\beta_{-} \over \alpha_{-}}; \qquad t_{-} = \sqrt{k_{+\infty}\over k_{-\infty}} \; {1 \over \alpha_{-}}.
\end{equation}
The explicit occurrence of $k_{+\infty}$ and $k_{-\infty}$  is an annoyance, which is why many authors adopt the alternative normalization we shall discuss later on in this chapter.

In Bogoliubov language these conventions correspond to an incoming flux of right-moving particles (incident from the left) being amplified to amplitude $\alpha_{+}$ at a cost of a backflow of amplitude $\beta_{+}$. In scattering language one should consider the complex conjugate $\mathcal{J}_{+}^{*}$ ---   this is equivalent to an incoming flux of left-moving particles (incident from the right) of amplitude $\alpha_{+}^{*}$ being partially transmitted (amplitude unity) and partially scattered (amplitude $\beta_{+}^{*}$). If the potential has even parity, then the left-moving Bogoliubov coefficients are just the complex conjugates of the right-moving coefficients, however if the potential is asymmetric a more subtle analysis is called for.

The second interesting issue is that we can deal exclusively with $\alpha_{+}$ and $\beta_{+}$, dropping the suffix for brevity --- if information about $\alpha_{-}$ and $\beta_{-}$ is desired simply work with the reflected potential $V(-x)$. It should also be borne in mind that the phases of $\beta$ and $\beta^{*}$ are physically meaningless in that they can be arbitrarily changed simply by moving the origin of coordinates (or equivalently, physically moving the location of the potential). The phases of $\alpha$ and $\alpha^{*}$ on the other hand do contain real and significant physical information.

For completely arbitrary potentials, with no parity restriction (so the potential is neither even nor odd), a Wronskian analysis yields (see for example~\cite{Messiah}, noting that an overall minus sign between Messiah and the conventions above neatly cancels):
\begin{equation}
\label{k_minus1}
k_{- \infty} [1 - |r_{+}|^2] =  k_{+ \infty} |t_{+}|^2;
\end{equation}
\begin{equation}
\label{k_minus2}
k_{- \infty} \, |t_{-}|^2 = k_{+ \infty} \, [1 - |r_{-}|^2];
\end{equation}
\begin{equation}
\label{k_minus3}
k_{- \infty} \, t_{-} = k_{+ \infty} \, t_{+};
\end{equation}
\begin{equation}
\label{k_minus4}
k_{- \infty} \, r_{+} t_{+}^{*} = -k_{+ \infty} \, r_{-} t_{-}^{*};
\end{equation}
with equivalent relations for $\alpha$ and $\beta$. Then
\begin{equation}
T_{+} = {k_{+ \infty} \over {k_{- \infty}}} |t_{+}|^2 = {k_{- \infty} \over k_{+ \infty}} |t_{-}|^2 = T_{-}
\end{equation}
and barrier transmission is independent of direction. We also have
\begin{equation}
\mathrm{phase} \, (t_{+}) = \mathrm{phase} \, (t_{-})
\end{equation}
and
\begin{equation}
\mathrm{phase} \, (r_{+}/t_{+}) = \pi - \mathrm{phase} \, (r_{-}/ t_{-})
\end{equation}
with equivalent relations for $\alpha$ and $\beta$.

If we now adopt the (to our minds) more useful convention, by absorbing factors of $k_{+\infty}$ and $k_{-\infty}$ into the definitions of $r$ and $t$ then things simplify considerably:  We restart the calculation by now defining
\begin{equation}
\mathcal{J}_{+} (x \rightarrow -\infty) \rightarrow  t_+ {\exp (+ik_{-\infty} x)\over \sqrt{k_{- \infty}}},
\end{equation}
\begin{equation}
\mathcal{J}_{+} (x \rightarrow + \infty) \rightarrow  {\exp (+ik_{+ \infty} x)\over \sqrt{k_{+ \infty}}}  
+ r_{+} {\exp (-i k_{+\infty} x) \over \sqrt{k_{+ \infty}} },
\end{equation}
By comparing these two different forms for the asymptotic form of the Jost function we see that in this situation the ratios of the amplitudes are given by the much simpler formulae
\begin{equation}
1:  \alpha_+  : \beta_+ =  t_+: 1 : r_+.
\end{equation}
We now have
\begin{equation}
r_{+}  = {\beta_{+} \over \alpha_{+}},
\end{equation}
and 
\begin{equation}
t_{+} = {1 \over \alpha_{+}}.
\end{equation}
We see that by putting the factors of $\sqrt{k_{\pm\infty}}$ into the asymptotic form of the Jost functions, where they really belong, the formulae for $r$ and $t$ are suitably simplified.

For completely arbitrary potentials, with no parity restriction (so the potential is neither even nor odd), a modified Wronskian analysis now yields (in analogy with that reported by Messiah~\cite{Messiah}):
\begin{equation}
\label{k_minus1}
 |t_{+}|^2 = 1 - |r_{+}|^2 ;
\end{equation}
\begin{equation}
\label{k_minus2}
|t_{-}|^2 =1 - |r_{-}|^2;
\end{equation}
\begin{equation}
\label{k_minus3}
t_{-} = t_{+};
\end{equation}
\begin{equation}
\label{k_minus4}
r_{+} t_{+}^{*} = -r_{-} t_{-}^{*};
\end{equation}
with equivalent relations for $\alpha$ and $\beta$. Then
\begin{equation}
T_{+} =  |t_{+}|^2 =  |t_{-}|^2 = T_{-}
\end{equation}
and barrier transmission is independent of direction. Because they are independent of any overall scaling by a real number, also retain the previous results
\begin{equation}
\mathrm{phase} \, (t_{+}) = \mathrm{phase} \, (t_{-})
\end{equation}
and
\begin{equation}
\mathrm{phase} \, (r_{+}/t_{+}) = \pi - \mathrm{phase} \, (r_{-}/ t_{-})
\end{equation}
with equivalent relations for $\alpha$ and $\beta$.
It is this modified set of conventions, because they have much nicer normalization properties, that we shall prefer for the bulk of the thesis.

We shall now derive some very general bounds on $|\alpha|$ and $|\beta|$, which also lead to general bounds on the reflection and transmission probabilities
\begin{equation}
R = |r|^2; \qquad T = |t|^2.
\end{equation}

\section{Probability currents} 
The expressions for reflection and transmission coefficients were  based on the assumption that the intensity of a beam is the product of the speed of its particles and their linear number density. In classical physics, the assumption seems very natural, however, we should always be careful about carrying over classical concepts into quantum physics~\cite{Scattering-probability-currents}.


\begin{center}
  \setlength{\fboxsep}{0.45 cm} 
   \framebox{\parbox[t]{12.5cm}{
       {\bf  Definition (Unbound state)}: 
     Provided $V_{\pm\infty}=0$, the Schr\"odinger equation (\ref{Schordinger}) can be solved for any positive value of energy, when $E > 0$. In addition, the positive energies can be shown to define a  \emph{continuous} spectrum. Nevertheless, the corresponding eigenfunctions do not vanish at infinity; their asymptotic behavior is analogous to that of the plane wave $\exp (i k x)$. More accurately, the absolute value of wave functions ($|\psi(x)|$) approaches a non-zero constant when $x \rightarrow \infty$. Otherwise, the absolute value oscillates indefinitely between limits, one of which at least is not zero. It is clear that the particle does not remain localized in any finite region. This type of wave function is commonly applied to collision problems; the usual language is that one is dealing with an \emph{unbound state}, or \emph{stationary state of collision}~\cite{Messiah}.
}}
\end{center}

\section{Reflection and Transmission of Waves in unbound states}

The Schr\"odinger equation also can be analyzed in terms of the functions $u$ and $v$, as defined by Messiah~\cite{Messiah},  and their complex conjugates $u^*$ and $v^*$. Moreover, the \emph{Wronskian} of any two such solutions is independent of $x$; especially, it takes on the same value in the two asymptotic regions. 
Actually our approach can be seen as equating these two values;  we now derive a relation between the coefficients $r_{+}, t_{+}, r_{-}, t_{-}$, or their complex conjugates.  Six such relations can be formed with the four functions $u, v, u^*$ and $v^*$. From what we have seen earlier it is clear that they are very basic relations which must be maintained whatever the form of the potential function $V(x)$~\cite{Messiah}.

Specifically, we derive (in Messiah-like conventions)
\begin{eqnarray}
\label{uu}
{i \over 2} W(u, u^*)  &=& k_{+ \infty} (1 - |r_{+}|^2) = k_{- \infty} |t_{+}|^2;
\\
\label{vv}
{i \over 2} W(v, v^*) &=& k_{- \infty} (1 - |r_{-}|^2) = k_{+\infty} |t_{-}|^2;
\\
\label{uv}
{i \over 2} W(u, v) &=& k_{+ \infty} t_{-} =  k_{- \infty} t_{+};
\\
\label{uv*}
{i \over 2} W (u, v^*) &=& - k_{+ \infty} r_{+} t_{-}^* = k_{- \infty} r_{-}^* t_{+}.
\end{eqnarray}
The equations~(\ref{uu}) and~(\ref{vv}) are called the \emph{relations of conservation of flux}. They should always be true, and this should be verified in special cases. This name comes from the following statements regarding the wave function $\psi$ of an \emph{unbound state} in the asymptotic region. 
We let  $A \, \exp (i k x) + B \, \exp (- i k x)$ be the expression of the wave function $\psi$ in one of the asymptotic regions, for  $- \infty$ case. 

The \emph{total flux of particles} when passing a given point  is the difference between the flux $(\hbar k / m) |A|^2$ of particles traveling in the positive sense, and the flux $(\hbar k / m) |B|^2$  of particles traveling in the negative sense. This flux is equal, to within a constant, to the \emph{Wronskian} $W (\psi, \psi^*)$~\cite{Messiah}:
\begin{equation}
{\hbar k \over m} [|A|^2 - |B|^2] = {i \over 2} \, {\hbar k \over m} \, W(\psi, \psi^*)
\end{equation}

The equality of the \emph{Wronskian} $W (\psi, \psi^*)$ at both ends of the interval $(- \infty, + \infty)$, denotes that the number of particles entering the interaction region per unit time is equal to the number which leave it. In accordance with this interpretation, one or the other of equation~(\ref{uu}) and~(\ref{vv}) can be written as: 
 \begin{equation}
 \mathrm{incident \, flux} - \mathrm{reflected \, flux} = \mathrm{transmitted \, flux}.
 \end{equation}

Considering the same interpretation, we now can define the \emph{transmission coefficient} (transmission probability) $T$ as follows:
\begin{equation}
\label{Tu_tv}
T = {\mathrm{transmitted \, flux} \over {\mathrm{incident \, flux}}}.
\end{equation}
We have in particular
\begin{equation}
T_{+} = {k_{- \infty} \over k_{+ \infty}} |t_{+ \infty}|^2 , \qquad T_{-} = {k_{+ \infty} \over k_{- \infty}} |t_{- \infty}|^2.
\end{equation}
This result shows that the absolute values of the two sides of equation (\ref{uv}) are equal, and one obtains the equality
\begin{equation}
T_{-} = T_{+}.
\end{equation}
Thus the transmission coefficient of a wave at a given energy is independent of the direction of travel. This is the \emph{reciprocity property of the transmission coefficient}. It is just as hard to traverse a potential barrier in one direction as in the other.

The equality of the absolute values of the two sides of equation (\ref{uv*}), coupled with the conservation relations (\ref{uu}) and (\ref{vv}), again yields the reciprocity relation (\ref{Tu_tv}) we also obtain relations between the phases of the reflection and transmission amplitudes:
\begin{eqnarray}
\nonumber
\mathrm{phase} (t_{+}) &= & \mathrm{phase} (t_{-});
\\
\nonumber
\mathrm{phase} \bigg({r_{+} \over t_{+}} \bigg) &=& \pi - \mathrm{phase} \bigg({r_{-} \over t_{-}} \bigg). 
\end{eqnarray}
The most interesting point for these relations is the fact (not further investigated in this thesis) that the phases are related to ``retardation" effects in the propagation of the wave packets, with equivalent relations for $\alpha$ and $\beta$.

As previously, we can re-scale $r$ and $t$ by absorbing appropriate factors of $\sqrt{k_{\pm\infty}}$, and so simplify the discussion as in the previous section. (We will not repeat the details of the analysis, as it is straightforward.)
We shall now generalize some very general bounds on $|\alpha|$ and $|\beta|$, which also lead to general bounds on the reflection and transmission probabilities
\begin{equation}
R = |r|^2; \qquad T = |t|^2.
\end{equation}

\begin{center}
  \setlength{\fboxsep}{0.45 cm} 
   \framebox{\parbox[t]{12.5cm}{
       {\bf  Definition  (Bound states)}: 
     Provided $V_{\pm\infty}=0$, when $E<0$, the Schr\"odinger equation (\ref{Schordinger}) has solutions only for certain particular values of energy forming a discrete spectrum. The eigenfunction $\psi(x)$ corresponding to it --- or each of the eigenfunctions when several exist --- vanishes at infinity. More accurately, the integral $\int |\psi(r)|^2 \, \d r$ extended over the whole configuration space is convergent. There is a vanishing probability of finding the particle at infinity and the particle remains practically localized in a finite region. The particle can now be defined to be in \emph{a bound state}~\cite{Messiah}.
}}
\end{center}

\section{Bogoliubov transformation}
\begin{center}
  \setlength{\fboxsep}{0.45 cm} 
   \framebox{\parbox[t]{12.5cm}{
       {\bf  Definition (Bogoliubov transformation)}: 
     This is a unitary transformation from a unitary representation of some canonical commutation relation algebra or canonical anticommutation relation algebra into another unitary representation~\cite{Scho}. 
}}
\end{center}
\bigskip

\noindent
To see the import of this definition, let us consider the canonical commutation relation for bosonic creation and annihilation operators in the harmonic basis
\begin{equation}
[\hat{a}, \hat{a}^{\dagger}] = 1.
\end{equation}
Using this method we can derive a new pair of operators.
\begin{eqnarray}
\label{hat-b-first}
\hat{b} &=& u \hat{a} + v \hat{a}^{\dagger};
\\
\label{hat-b-second}
\hat{b}^{\dagger} &=& u^{*} \hat{a}^{\dagger} + v^* \hat{a} ;
\end{eqnarray}
where the equation (\ref{hat-b-first}) is the hermitian conjugate of the equation (\ref{hat-b-second}).  

This transformation is a canonical transformation of these operators.  It is easy to find the conditions on the constants $u$ and $v$. For instance, the transformation remains canonical by extending the commutator.
\begin{equation}
[\hat{b}, \hat{b}^{\dagger}] = [u \hat{a} + v \hat{a}^{\dagger}, u^{*} \hat{a}^{\dagger} + v^* \hat{a}] = \bigg( |u|^2 - |v|^2 \bigg) [\hat{a}, \hat{a}^{\dagger}].
\end{equation} 
It can be seen that 
\begin{equation}
|u|^2 - |v|^2 = 1
\end{equation}
is the condition for which the transformation is canonical.  (This normalization condition will occur and re-occur many times in the calculations which follow.)
Finally we note that since the form of this condition is reminiscent of the hyperbolic identity 
\begin{equation}
\cosh^2 r  - \sinh^2 r = 1
\end{equation}
between $\cosh$ and $\sinh$, the constants $u$ and $v$ are usually parameterized as
\begin{eqnarray}
u &=& \exp(i \theta) \, \mathrm{cosh}  \, r;
\\
v &=& \exp(i \theta) \, \mathrm{sinh}  \, r .
\end{eqnarray}

\section{Transfer matrix representation}
We can also investigate quantum mechanical tunneling by the so-called ``transfer matrix method'' or ``transfer matrix representation''. Ultimately, of course, this is still equivalent to extracting the transmission coefficient from the solution to the one-dimensional, time-independent Schr\"odinger equation. As before, the transmission coefficient is the ratio of the flux of particles that penetrate a potential barrier to the flux of particles incident on the barrier. It is related to the probability that tunneling will occur~\cite{transfer-matrix-method}. 
We again consider a one-dimensional problem which is characterized by an incident beam of particles that is either transmitted or reflected as a result of scattering from an object~\cite{Scattering-probability-currents}. For current purposes it is easiest to work with potentials of compact support, where $V(x)=0$ except in some finite region $[a,b]$. 

As long as the potential $V(x)$ is of compact support, it 
splits the space  in three parts ($x < a, x\in[a,b], x >b$). In both $(-\infty,a]$ and $[b,\infty)$ the potential energy is zero. Moreover, in each of these two regions the solution of the Schr\"odinger equation can be presented as a \emph{superposition} of exponentials by
\begin{eqnarray}
\label{delta-left-equation}
\psi_{L}(x) &=& A_{r} \exp(ikx) + A_{l} \exp(- ikx) \, , \qquad x <a, \quad \mathrm{and}
\\
\label{delta-right-equation}
\psi_{R}(x) &=& B_{r} \exp(ikx) + B_{l} \exp(- ikx) \, , \qquad x > b,
\end{eqnarray}
where $A_{l/r}$ and $B_{l/r}$ are at this stage unspecified, and $k = \sqrt{2mE}/\hbar$.
But because $\psi_L$ and $\psi_R$ are solutions to the Schr\"odinger equation that can be extended to the entire real line, and because the Schr\"odinger equation is a second-order differential equation so that its solution space is two-dimensional, there must be some linear relation between the coefficients appearing in   $\psi_L$ and $\psi_R$ --- specifically, there must be a $2\times2$ matrix $M$ such that
\begin{equation}
\left[\begin{array}{c} B_l\\  B_r \end{array}\right] = M \, \left[\begin{array}{c} A_l \\ A_r \end{array}\right].
\end{equation}
The  $2\times2$ matrix $M$ depends, in a complicated way, on the potential $V(x)$ in the region $[a,b]$.  In the transfer matrix approach we shall seek to extract as much information as possible without explicitly calculating $M$. 

\begin{figure}[ht] 
\hskip - 2.25 cm
\includegraphics[scale=1]{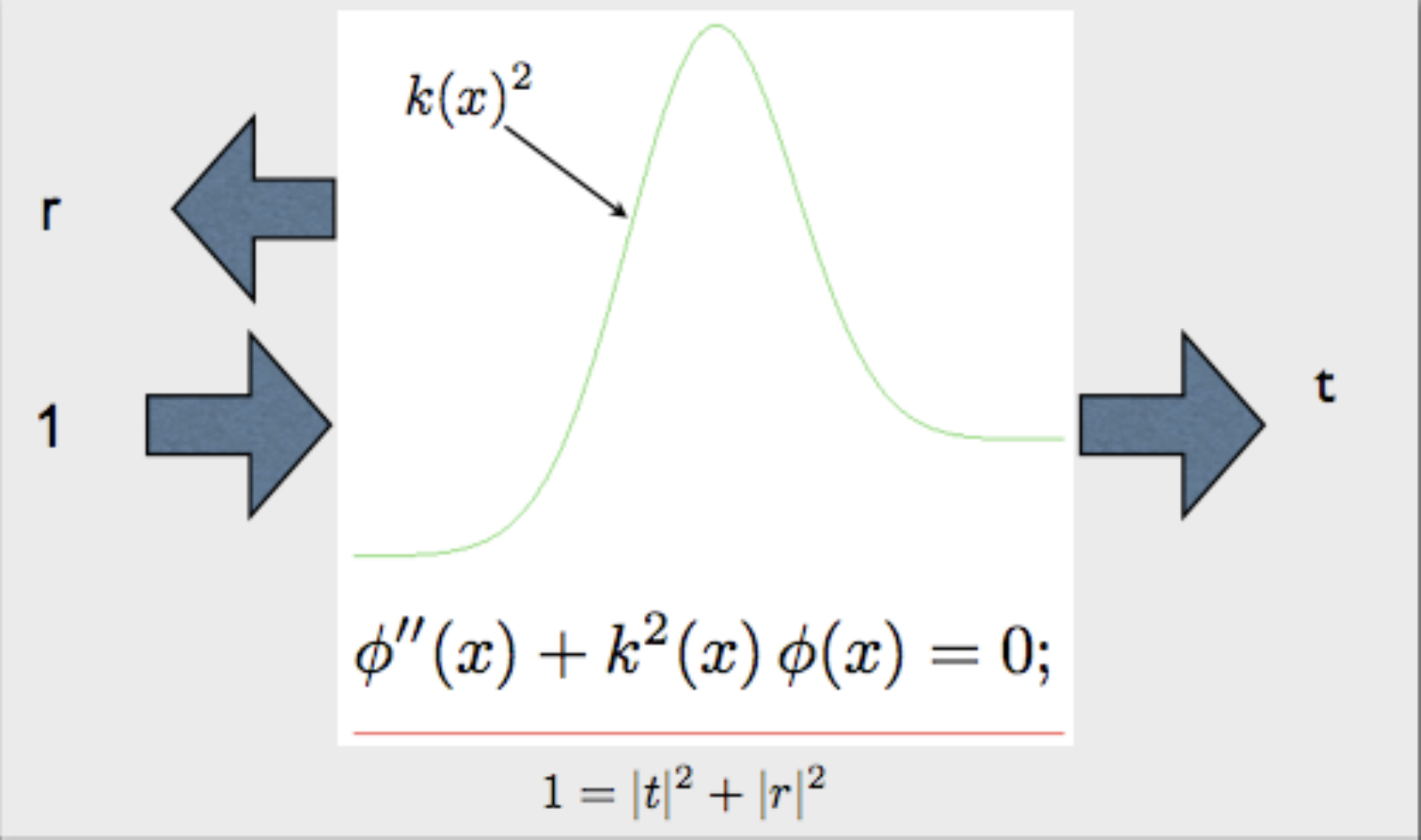}
\caption[Transmission and reflection amplitudes]
{\label{F:Transmission-and-reflection}
This shows an incoming flux of particles from the left, being partially transmitted to the right (with amplitude $t$),  and partially reflected back to the left (with amplitude $r$)~\cite{bounds-matt-seminar}.}
\end{figure}

To now derive amplitudes for reflection and transmission for incidence from the left, we put $A_{r} =1$ (incoming particles), $A_{l} =r$ (reflection), $B_{l} = 0$ (no incoming particle from the right) and $B_{r} = t$ (transmission) in equations (\ref{delta-left-equation}) and (\ref{delta-right-equation}). We now derive
\begin{equation}
\label{psiL}
\psi_{L}(x) = \exp (ikx) + r_{L} \exp (- ikx) \, ,
\end{equation}
where $r_{L}$ is the left-moving reflection amplitude and on the right of the potential
\begin{equation}
\label{psiR}
\psi_{R} (x) = t_{L} \exp(ikx) .
\end{equation}
where $t_{L}$ is the left-moving transmission amplitude. This tells us that
\begin{equation}
\left[\begin{array}{c} t_{L}\\ 0 \end{array}\right] = M \, \left[\begin{array}{c} 1\\ r_{L} \end{array}\right].
\end{equation}
But since the Schr\"odinger equation (\ref{euler-formula}) is real, the complex conjugate of any solution is also a solution. Therefore the solution which on the left has the form
\begin{equation}
\psi_{L} = \exp (- ikx) + r^*_{L} \exp (+ ikx) \, ,
\end{equation}
must on the right have the form
\begin{equation}
\psi_{R} (x) = t^*_{L} \exp (-ikx) \, ,
\end{equation}
and so we also have
\begin{equation}
\left[\begin{array}{c} 0\\ t^*_{L} \end{array}\right] = M \, \left[\begin{array}{c} r^*_{L}\\ 1 \end{array}\right].
\end{equation}
These two matrix equations now imply
\begin{equation}
M = {1 \over {1-r^*_{L} r_{L}}} \, \left[\begin{array}{cc} t_{L} & - t_{L} r^*_{L}\\ -t^*_{L} r_{L} & t^*_{L} \end{array}\right].
\end{equation}
But by conservation of flux we must have
\begin{equation}
\label{conservation-flux-part1}
|t_{L}|^2 + |r_{L}|^2 = 1.
\end{equation}

We just have seen an important connection between reflection and transmission amplitudes. In addition, it is interesting to show how to derive the above equation by following.

From the equation (\ref{psiL}), we can see that  this corresponds to a flux in the positive $x$ direction. For $x<a$ this is  of magnitude
\begin{eqnarray}
\nonumber
\mathscr{J} &=& {\hbar \over 2 m i} \, \bigg(\psi^* {\partial \psi \over \partial x} - {\partial \psi^* \over \partial x} \, \psi\bigg) \, ,
\\
\nonumber
&=& {\hbar \over 2 m i} \, \bigg(\big(\exp(- ikx) + r^*_{L} \, \exp(+ ikx)\big) \big(ik \exp(ikx)
\\
\nonumber
&& -   r_{L} ik \exp(- ikx) \big)\ - \mathrm{complex \; conjugate} \bigg) \, ,
\\
\nonumber
&=& {\hbar \over 2 m i} \, \bigg(2 i k - 2 ik|r_{L}|^2 \bigg) \, ,
\\
&=& {\hbar \,  k \over m} \, \bigg(1 - |r_{L}|^2\bigg),
\end{eqnarray}
and for $x > b$, we similarly derive from equation (\ref{psiR}) the fact that we can write the flux corresponding to this equation is
\begin{eqnarray}
\nonumber
\mathscr{J} &=& {\hbar \over 2 m i} \bigg(\big(t^*_{L} \exp(- ikx) \times ik(t_{L} \exp(ikx))\big)
\\
\nonumber
&& - \big(t_{L} \exp(ikx) \times - ik (t^*_{L} \exp(- ikx))\big) \bigg) \, ,
\\
\nonumber
&=& {\hbar \over 2 m i} \bigg( i k |t_{L}|^2 + i k |t_{L}|^2\bigg) \, ,
\\
&=& {\hbar \, k \over m} \bigg(|t_{L}|^2\bigg).
\end{eqnarray}
\begin{center}
  \setlength{\fboxsep}{0.45 cm} 
   \framebox{\parbox[t]{12.5cm}{
       {\bf  Definition}: 
   The \emph{probability current} $\mathscr{J}$ of the wave function $\psi(x)$ is defined as
   \begin{equation}
   \label{conservation-law}
   \mathscr{J} = {\hbar \over 2 m i} \, \bigg(\psi^* {\partial \psi \over \partial x} - {\partial \psi^* \over \partial x} \, \psi\bigg) \, ,
   \end{equation}
   in the position basis and satisfies the quantum mechanical continuity equation
   \begin{equation}
   {\partial \over \partial t} \rho  (x, t) + {\partial \over \partial x}  \mathscr{J}(x, t) = 0 \, ,
   \end{equation}
   where $\rho (x,t)$ is probability density~\cite{Probability-current}.
}}
\end{center}

Since there is no time dependence in the problem, the conservation law in equation (\ref{conservation-law}) implies that $\mathscr{J}(x)$ is independent of $x$. Hence the flux on the left must be equal to the flux on the right, that is, we expect that
\begin{eqnarray}
\nonumber
{\hbar \,  k \over m} \, \bigg(1 - |r_{L}|^2\bigg) &=& {\hbar \, k \over m} \bigg(|t_{L}|^2\bigg).
\\
\nonumber
1 - |r_{L}|^2 &=& |t_{L}|^2.
\end{eqnarray}
therefore,
\begin{equation}
|t_{L}|^2 + |r_{L}|^2 = 1\, ,
\end{equation}
so
\begin{equation}
{1 \over 1 - r^*_{L} r_{L}} = {1 \over 1 - |r_{L}|^2} = {1 \over |t_{L}|^2} \, ,
\end{equation}
whence 
\begin{equation}
M = {1 \over |t_{L}|^2} \left[\begin{array}{cc} t_{L} & - t_{L} r^*_{L}\\ - t^*_{L} r_{L} & t^*_{L} \end{array}\right] = \left[\begin{array}{cc} 1/t^*_{L} & - r^*_{L}/ t^*_{L} \\ - r_{L}/t_{L} & 1/t_{L} \end{array}\right].
\end{equation}
Similary, consider a wave moving in from the right
\begin{equation}
\exp(- ikx)
\end{equation}
which then hits the potential, is partially reflected and partially transmitted. In this case, on the \emph{right} of the potential we have
\begin{equation}
\psi_{R} (x) = \exp(- ikx) + r_{R} \exp(+ ikx) \, ,
\end{equation}
where $r_{R}$ is the right-moving reflection amplitude and on the \emph{left} of the potential
\begin{equation}
\psi_{L}(x) = t_{R} \exp (- ikx) \, ,
\end{equation}
where $t_{R}$ is the left-moving transmission amplitude. This tells us that
\begin{equation}
\left[\begin{array}{c} r_{R}\\ 1 \end{array}\right] = M \, \left[\begin{array}{c} 0\\ t_{R} \end{array}\right].
\end{equation}
Again, since the Schr\"odinger equation is real, the complex conjugate of any solution is also a solution. Therefore a related interesting solution which on the left can be cast  in the form
\begin{equation}
\psi_{L}(x) = t^*_{R} \exp(+ ikx) \, ,
\end{equation}
must on the right have the form
\begin{equation}
\psi_{R}(x) = \exp(+ ikx) + r^*_{R} \exp(- ikx) \, ,
\end{equation}
whence
\begin{equation}
\left[\begin{array}{c} 1\\ r^*_{R}  \end{array}\right] = M \, \left[\begin{array}{c} t^*_{R}\\ 0 \end{array}\right].
\end{equation}
But now these two matrix equations imply
\begin{equation}
M = \left[\begin{array}{cc} 1/t^*_{R} &  r_{R}/ t_{R} \\  r^*_{R}/t^*_{R} & 1/t_{R} \end{array}\right] .
\end{equation}
Combining the information from left moving and right moving cases we have first that
\begin{equation}
t_{L} = t_{R}.
\end{equation}
So we again derive the equality of the transmission amplitudes. 

Similarly we see that
\begin{equation}
{r_{R} \over t_{R}} = - {r_{L}^* \over t_{L}^*} \, ,
\end{equation}
implying
\begin{equation}
r_{R} = - r_{L}^* \, {t_{L} \over t_{L}^*} \, ; \qquad |r_{R}| = |r_{L}|.
\end{equation}
Note that we \emph{cannot} in general deduce $r_{L} = r_{R}$. Indeed,  in general this is false.

So for \emph{any} potential we have 
\begin{equation}
T = |t_{L}|^2 = |t_{R}|^2 ; \qquad R = |r_{L}|^2 = |r_{R}|^2 \, ,
\end{equation}
implying (in the same manner as the previous argument) that the transmission and reflection coefficients are independent on whether or not the particle is incident from the left or the right  --- and we have \emph{not} made any assumption here about any symmetry for the potential $V(x)$ itself. We conclude
\begin{equation}
M = \left[\begin{array}{cc} 1/t^* & - r_{L}^*/ t^* \\  - r_{L}/t & 1/t \end{array}\right] = \left[\begin{array}{cc} 1/t^* & r_{R}/ t \\  r^*_{R}/t^* & 1/t \end{array}\right] .
\end{equation}

Note the key step in this general derivation: In any region where the potential is zero we simply need to solve
\begin{equation}
- {\hbar^2 \over 2 m} \, {\d^2 \over \d x^2} \psi(x) = E \, \psi(x),
\end{equation}
for which the two independent solutions are
\begin{equation}
\exp (\pm i k x)  ; \qquad k = {\sqrt{2 m E} \over \hbar},
\end{equation}
or more explicitly
\begin{equation}
\exp \bigg(\pm i {\sqrt{2 m E} \over \hbar} \, x\bigg).
\end{equation}
To the \emph{left} of the potential we have
\begin{equation}
\psi_{L} (x) = a \, \exp(ikx) + b \, \exp(- ikx) \, ,
\end{equation}
while to the \emph{right} of the potential we have
\begin{equation}
\psi_{R} (x) = c \, \exp (ikx) + d \, \exp (-ikx).
\end{equation}
Even without knowing anything more about the potential $V(x)$, the linearity of the Schr\"odinger ODE guarantees that there will be some $2 \times 2$ transfer matrix $M$ such that
\begin{equation}
\left[\begin{array}{c} c\\ d \end{array}\right] = M \, \left[\begin{array}{c} a\\ b \end{array}\right].
\end{equation}
This transfer matrix relates the situation to the \emph{left} of the potential with the wave-function to the \emph{right} of the potential. For this reason we shall now use this formalism, for instance, to think about the propagation of electrons down a wire (approximately one-dimensional) with $V(x)$ used to describe various barriers placed in the path of the electron. Moreover, similar matrices also occur in optics, where they are referred to as ``Jones matrices''.

The components of the transfer matrix $M$ will be some horrible nonlinear function of the potential $V(x)$, but by linearity of the Schr\"odinger ODE these matrix components must be independent of the parameters $a, b, c,$ and $d$. In some particularly simple situations we may be able to calculate the matrix $M$ explicitly (see in particular the next chapter), but in general it will be a complicated mess. 

From the above discussion we now understand, from at least two different points of view,  the basic concepts of transmission and reflection. The probability that a given incident particle is reflected is called the ``reflection coefficient'', ${R}$. While the probability that it is transmitted is called the ``transmission coefficient'', ${T}$~\cite{Scattering-probability-currents}.

\section{Discussion}
In this chapter, we have presented basic aspects of scattering theory in one dimension. For a one-dimensional model, only one of the three coordinates of 3-dimensional physical space is explicitly involved. Specifically, we considered potentials of compact support, when the potential $V(x)$ is  mathematically zero outside of a finite interval. The situation where the potential is zero is referred to as ``the free particle''.  These one-dimensional models provide solid examples  exhibiting all the basic  features and ideas needed to derive the properties of quantum states of definite energy $E$.

A further step in our investigation is that we have just seen an important connection between reflection and transmission amplitudes. In particular, it is interesting to show how to derive  them directly by using scattering theory. Furthermore, we have now introduced the concept of transmission and reflection. We called the probability that a given incident particle is reflected as the ``reflection coefficient''. While the probability that it is transmitted is called the ``transmission coefficient''.

More importantly, we introduced the probability current to express the reflection and transmission coefficients.  The probability current is  based on the assumption that the intensity of a beam is the product of the speed of its particles and their linear number density. It is then a mathematical theorem that this probability current is conserved. We then introduced important ideas of reflection and transmission of waves in both unbound and bound states. By considering reflection and transmission of waves in unbound states, we  have seen that in principle they are completely specified by the potential function $V(x)$. 

For instance,  the linearity of the Schr\"odinger ODE guarantees that there will be some $2 \times 2$ transfer matrix.
Moreover, this transfer matrix can be represented by investigating quantum mechanical tunneling by extracting the transmission coefficient from the solution to the one-dimensional, time-independent Schr\"odinger equation.

\chapter{Known analytic results}
\label{C:consistency-3}
\section{Introduction}
In this chapter we shall collect a number of known analytic results in a form amenable to comparison with the   general results presented in subsequent chapters. We shall review and briefly describe the concept of  \emph{quasinormal modes},  and see how most of the concepts introduced here are important tools for comparing the bounds with known analytic results. Furthermore, we shall reproduce many analytically known results, such as the tunnelling probabilities and quasinormal modes [QNM] of the delta-function potential, double-delta-function potential,  square potential barrier, $\mathrm{tanh}$ potential, $\mathrm{sech}^2$ potential, asymmetric square-well potential, the Poeschl--Teller potential and its variants, and finally the Eckart--Rosen--Morse--Poeschl--Teller potential.

In the following, we shall first introduce the \emph{quasinormal modes}, which are the modes of energy dissipation of a perturbed object or field. In particular, the most outstanding and well-known example is the perturbation of a wine glass with a knife: the glass begins to ring, it rings with a set, or superposition, of its natural frequencies -- its modes of sonic energy dissipation. In the absence of any damping, when the glass goes on ringing forever, we can call these modes \emph{normal}. In the presence of damping, when the amplitude of oscillation decays in time, we call the modes \emph{quasi-normal}~\cite{quasinormal-mode}. 

To a very high degree of accuracy, quasinormal ringing can be approximated by 
\begin{equation}
\psi(t) \approx \exp(- \omega'' t) \, \mathrm{cos} (\omega' t),
\end{equation}
where $\psi(t)$ is the amplitude of oscillation, $\omega'$ is the frequency and $\omega''$  is the decay rate. We can express the \emph{quasinormal frequency} in two numbers,
\begin{equation}
\omega = (\omega', \omega''),
\end{equation}
or more compactly
\begin{equation}
\psi(t) \approx \exp(i \omega t),
\end{equation}
\begin{equation}
\omega = \omega' + i \omega'',
\end{equation}
where  for $\psi(t)$ we are to understand that we are only interested in  the real part. In our explanation here, $\omega$ is generally referred to as the \emph{quasinormal mode frequency}. The most interesting point is that it is a complex number with two pieces. One of them is a real part which describes the temporal oscillation, and the other part is an imaginary part which describes the temporal exponential decay. Formally,  quasinormal modes are most easily found by looking for complex frequencies where the transmission amplitude becomes infinite.

In theoretical physics, \emph{a quasinormal mode} is a formal solution of  some linear differential equations with a complex eigenvalue. In black hole physics these linear differential equations typically come from linearizing the full Einstein equations. 
It is important to note that black holes have many \emph{quasinormal modes} that express the exponential decrease of asymmetry of the black hole in time as it evolves towards the perfect spherical shape~\cite{quasinormal-mode}.
Experience obtained from black hole physics and related fields has shown that it is quite common for QNM to be approximately of the form
\begin{equation}
\omega_n =  a +  i n b + \mathcal{O}(1/n),
\end{equation}
where $a$ is a complex number called the ``offset'' and $b$ is a real number known as the ``gap''~\cite{qnmv}.

\section{Delta--function potential}
As a first approach to the comparison of the bounds with known analytic results we can start by studying in detail the concept of the delta--function potential. It is important to understand that the delta--function potential is one limiting case of a square well. It is a very narrow deep well, which can adequately be approximated by a mathematical delta function when the range of variation of the wave function is much greater than the range of the potential~\cite{Delta-function-definition}.

The time--independent Schr\"odinger equation for the wave function $\psi(x)$ is
\begin{equation}
H \, \psi(x) = \bigg[-{\hbar^2 \over 2m} \, {\d^2 \over \d x^2} + V(x) \, \bigg] \, \psi(x) = E \, \psi(x),
\end{equation}
where $H$ is the Hamiltonian, $\hbar$ is the (reduced) Planck constant, $m$ is the mass, $E$ is the energy of the particle and the potential $V(x)$ 
is the delta function well with strength $\alpha <0$ concentrated at the origin. Without changing the physical results, any other shifted position is also possible~\cite{delta-function-potential}, i.e.

For a delta function potential take
\begin{equation}
 V(x) = \alpha \, \delta(x).
\end{equation}
In this case the transmission coefficient is well known to be (see, for instance,~\cite{Baym, Gasiorowicz})
\begin{equation}
T = {1 \over 1 +\displaystyle {m \alpha^2 \over 2 E \hbar^2}}.
\end{equation}
\bigskip

\noindent
{\bf Quasinormal modes:} \quad $T = \infty$ when
\begin{equation}
m \alpha^2 = - 2 E \hbar^2,
\end{equation}
that is
\begin{equation}
E = - {m \alpha^2 \over 2 \hbar^2}; \qquad k = \pm {i m \alpha \over \hbar^2}.
\end{equation}
Note that there is only one pair of complex conjugate QNM. Because the width of the delta function is zero,  the ``gap'' is infinite, and the other QNM are driven off to imaginary infinity.

\subsection*{Deriving the amplitudes:}
For completeness we will explicitly provide the calculations required to deal with the delta potential barrier --- this is a textbook problem of quantum mechanics. Generally, the problem consists of solving the time-independent Schr\"odinger equation for a particle in a delta function potential in one dimension~\cite{delta-function-potential}.

We now consider particles entering from the left traveling to the right with $E > 0$ encountering a potential of the form~\cite{Field}
\begin{equation}
V(x) =  \alpha \, \delta(x).
\end{equation}
We look for solutions of the time-independent Schr\"odinger equation
\begin{equation} 
- {\hbar^2 \over 2 m} {\d^2 \psi (x) \over \d x^2} + \alpha \delta (x)  \psi(x) = E \psi(x),
\end{equation}
with $\Psi (x, t) = \psi(x)  \exp(-i E t / \hbar)$. For $E > 0$ in the region $x \neq 0$ we have
\begin{equation}
{\d^2 \psi (x) \over \d x^2} = - {2 m E \over \hbar^2} \,  \psi(x) = - k^2 \psi(x),
\end{equation}
with $\kappa =  \displaystyle{\sqrt{{2 m E \over \hbar^2}}}$ or $E = -  \displaystyle{{\hbar^2 \, \kappa^2 \over 2 m}}$ (where $\kappa^2 = -k^2$).
\bigskip

\noindent
The most general solution is
\begin{eqnarray}
\psi_{L} (x) &=& A_{L}  \exp(+ i k x) + B_{L}  \exp(- i k x),  \quad x<0, \quad \mathrm{and}
\\
\psi_{R} (x) &=& A_{R}  \exp(+ i k x) + B_{R}  \exp(- i k x), \quad  x>0.
\end{eqnarray}

\begin{figure}[!htbp] 
\hskip - 2.25 cm
\includegraphics[scale=0.6]{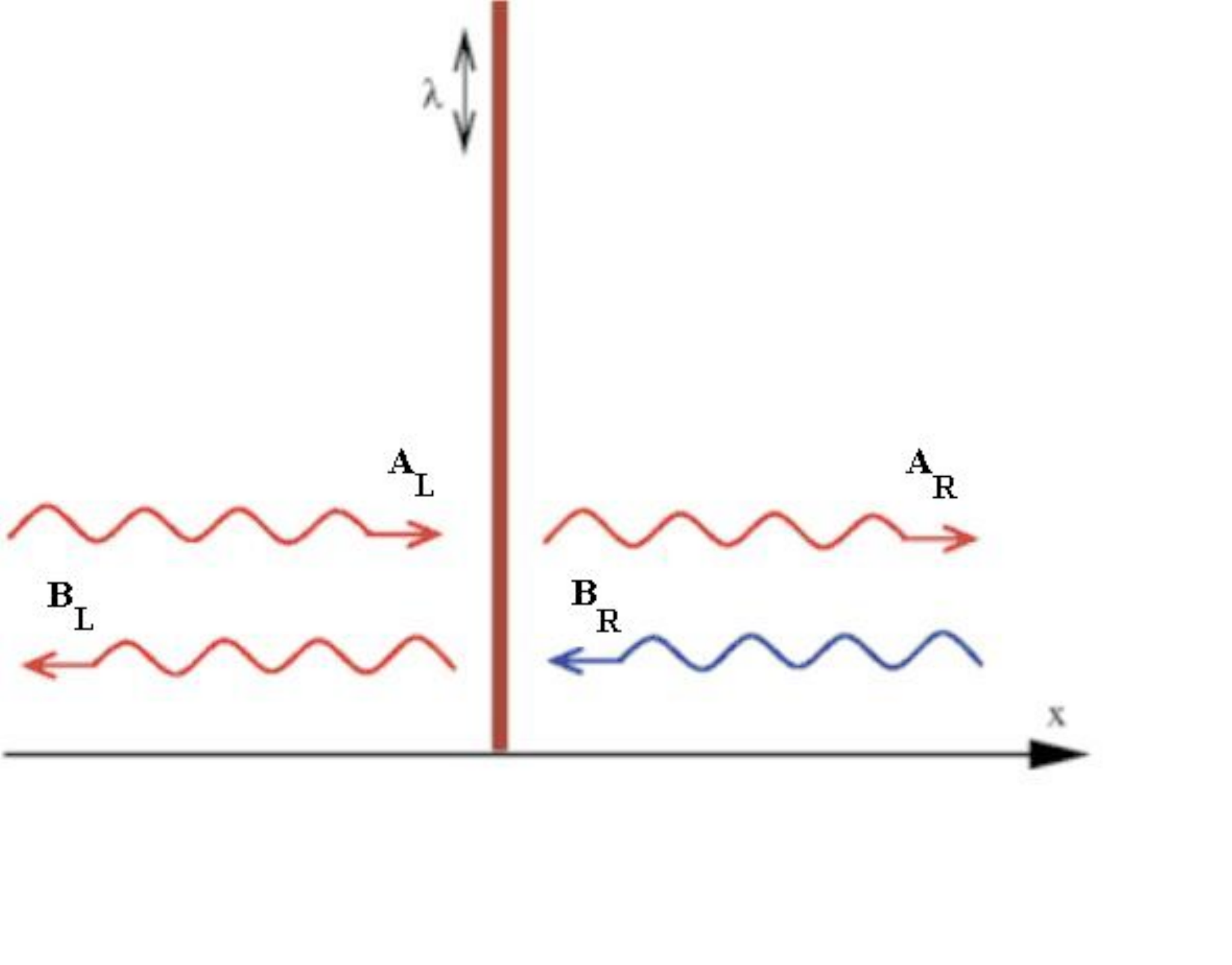}
\vskip - 2cm
\caption[Scattering for a $\delta$-function potential]{\label{F:delta-function}This diagram describes the scattering process at a delta-function potential of strength $\alpha$. The amplitudes and direction of left and right moving waves are indicated. We indicate, in red, the specific waves used for the derivation of the reflection and transmission amplitudes~\cite{Delta}.}
\end{figure}

\noindent
\subsubsection*{Boundary Conditions:}
We must require $\psi(x)$ to be continuous everywhere, and $\d \psi(x) / \d x$ to be  continuous (except possibly where $V$ is infinite). Thus, at $x = 0$, we have $\psi_{L} (0) = \psi_{R} (0)$ which implies that $A_{L} + B_{L} = A_{R} + B_{R}$. 

Also notice that for $\varepsilon \rightarrow 0$ we have
\begin{eqnarray}
\nonumber
\int_{- \varepsilon}^{+ \varepsilon} {\d^2 \psi(x) \over \d x^2} \, \d x &=& {2 m \over \hbar^2} \, \int_{- \varepsilon}^{+ \varepsilon} (V(x) - E) \psi(x) \, \d x ,
\nonumber
\\
&=& {2 m \over \hbar^2} \, \int_{- \varepsilon}^{+ \varepsilon} (\alpha \delta(x) - E) \psi(x) \, \d x = {2 m \alpha \over \hbar^2} \, \psi(0).
\nonumber
\\
&&
\end{eqnarray}
Thus,
\begin{equation}
{\d \psi_{R} (x) \over \d x} \bigg|_{x = + \varepsilon} - {\d \psi_{L} (x) \over \d x} \bigg|_{x = - \varepsilon} = {2 m \alpha \over \hbar^2} \, \psi_{R} (0) = {2 m \alpha \over \hbar^2} \psi_{L} (0),
\end{equation}
which implies that $[ik A_{R} - i k B_{R}] - [ik A_{L} - ik B_{L}] = \displaystyle{2 m \alpha \over \hbar^2} (A_{L}+B_{L})$.

Let us consider the second of these equations, which follows from integrating the Schr\"odinger equation with respect to $x$. The boundary conditions thus give the following restrictions on the coefficients~\cite{Delta}
\begin{equation}
\label{form_1}
A_{L} + B_{L} = A_{R} + B_{R};
\end{equation}
\begin{equation}
\label{form_2}
ik A_{R} - ik B_{R} - ik A_{L} + i k B_{L} = {2 m \alpha \over \hbar^2} \, (A_{L} + B_{L}).
\end{equation}

\subsubsection*{Reflection and Transmission:}
We shall see how to find the amplitudes for reflection and transmission for incidence from the left, by putting in the equations (\ref{form_1}) and (\ref{form_2}); $A_{L} = 1$ (incoming particle), $B_{L} = r$ (reflection), $B_{R} = 0$ (no incoming particle from the right) and $A_{R} = t$ (transmission). We obtain
\begin{equation}
1 + r = t,
\end{equation}
and
\begin{equation}
ik(1 - t - r) =  {2 m \alpha \over \hbar^2}  \, t.
\end{equation}
To find the amplitudes for reflection and transmission for incidence from the left, we now solve for $r$ and $t$:
\begin{equation}
t = {1 \over 1 - \displaystyle{im \alpha \over k \hbar^2}},
\end{equation}
and
\begin{equation}
r = {- 1 \over 1 + \displaystyle{ik \hbar^2 \over m \alpha}}.
\end{equation}
Now we can derive the probability for transmission and reflection (given by the transmission coefficient and the reflection coefficient)
\begin{equation}
T = |t|^2 =  {1 \over 1 + \displaystyle{m^2 \alpha^2 \over  k^2 \hbar^4}},
\end{equation}
and 
\begin{equation}
R = |r|^2 = {1 \over 1 + \displaystyle{k^2 \hbar^4 \over m^2 \alpha^2}}.
\end{equation}
This confirms the result we previously quoted, and by looking for poles of the transmission amplitudes, confirms the locations of the QNM.

\section{Double-delta-function potential}
For the double delta function
\begin{equation}
V(x) = \alpha \{\delta (x - L/2) + \delta(x + L/2)\} ,
\end{equation}
the transmission coefficient is known to be~\cite{Galindo}
\begin{equation}
T = {1 \over 1 + \left[\displaystyle{2 m \alpha \over \hbar^2 k} \, \mathrm{cos} \, (k L) + {1 \over 2} \displaystyle{\bigg({2 m \alpha \over \hbar^2 k}\bigg)^2 }\mathrm{sin} \, (k L) \right]^2}.
\end{equation}
It is an easy exercise to check that this satisfies the bounds (\ref{case1}) and (\ref{case1_2}) that we shall derive later on in this thesis.
\\
\\
{\bf Quasinormal modes:} \quad $T = \infty$ when
\begin{equation}
{2 m \alpha \over \hbar^2 k} \, \mathrm{cos} \, (k L) + {1 \over 2} \, \bigg({2 m \alpha \over \hbar^2 k} \bigg)^2 \, \mathrm{sin} \, (k L) = \pm i,
\end{equation}
which leads to quite horrible algebra, so that there is no \emph{explicit} formula for the QNM.
Implicitly, working with the transmission \emph{amplitude}:
 \begin{equation}
 t = \infty \, \Longleftrightarrow (k - i k_{0})^2 + k^2_0 \exp (2 ikL) = 0,
 \end{equation}
 so that
 \begin{equation}
 \exp(2 ikL) =  \bigg(1 + {i k \over k_0}\bigg)^2.
 \end{equation}

\subsection*{Deriving the amplitudes}
To derive the transmission and reflection amplitudes for the double-delta-function potential, we now start by considering the potential~\cite{Matt01}
\begin{equation}
V(x) = \alpha \{\delta (x - L/2) + \delta(x + L/2)\}.
\end{equation}
For a particle incident from the left we now have
\begin{equation}
\psi(x)  =  \left\{\begin{array}{r@{\quad \quad}l}
\exp(+ i k x) + r \exp(-i k x) & (\mathrm{at} \quad  x < - L/2); \\ 
A \exp(+ i k x) + B \exp(- i k  x)  & (\mathrm{at} 
\quad  - L/2 < x < L/2);  \\  
t \exp(+ i k x) & (\mathrm{at} \quad  x> L/2). \end{array} \right.
\end{equation}
Note that now we also have to explicitly consider the region between the two delta-functions. Applying the same sort of boundary conditions we now have four equations. From continuity at $x = -L/2$ we have
\begin{equation}
\exp(- i k L/2)  + r \exp(+ i k L/2) = A \exp(- i k L/2) + B \exp(+ i k L/2),
\end{equation}
while continuity at $x =  +L/2$ implies
\begin{equation}
A \exp(+ i k L/2) + B \exp(- i k L/2) = t \exp(i k L/2).
\end{equation}
Integrating across the delta functions leads to 
\begin{eqnarray}
&&\!\!\!\!\!\!\!\!
{\hbar^2 \over 2 m} \{ i k [\exp(-i k L/2) - r \exp(+ i k L/2)] - ik [A \exp(- i k L/2) - B \exp(+ i k L/2)]\}
\nonumber\\
&&
\qquad  + \alpha \, [A \exp(- i k L/2) + B \exp(+ i k L/2)] = 0,
\end{eqnarray}
and
\begin{eqnarray}
&&\!\!\!\!\!\!\!\!
{\hbar^2 \over 2 m} \{ i k [\exp(-i k L/2) - r \exp(+ i k L/2)] - ik [A \exp(- i k L/2) - B \exp(+ i k L/2)]\} 
\nonumber\\
&&
=  + \alpha \, [A \exp(- i k L/2) + B \exp(+ i k L/2)] .
\end{eqnarray}
We rearrange this to obtain
\begin{eqnarray}
&& \!\!\!\!\!\!\!\!
\{ i k [\exp(-i k L/2) - r \exp(+ i k L/2)] - ik [A \exp(- i k L/2) - B \exp(+ i k L/2)]\} 
\nonumber\\
&&=   {2 m \alpha \over \hbar^2} \, [A \exp(- i k L/2) + B \exp(+ i k L/2)] ,
\end{eqnarray}
and
\begin{eqnarray}
&& \!\!\!\!\!\!\!\!
 \{ i k [\exp(-i k L/2) - r \exp(+ i k L/2)] - ik [A \exp(- i k L/2) - B \exp(+ i k L/2)]\} 
\nonumber\\
&& =   2 k_0 \, [A \exp(- i k L/2) + B \exp(+ i k L/2)].
\end{eqnarray}
Some further rearrangements lead to 
\begin{eqnarray}
&& - {\hbar^2 \over 2m} \{ik [A \exp(+ i k L/2) - B \exp(- i k L/2)] - ik [t \exp(+ i k L/2)]\} 
\nonumber\\
&& \qquad\qquad
+ \alpha \, [t \exp(+i k L/2)] = 0,
\end{eqnarray}
and
\begin{eqnarray}
&&{\hbar^2 \over 2m} \{ik [A \exp(+ i k L/2) - B \exp(- i k L/2)] - ik [t \exp(+ i k L/2)]\} 
\nonumber
\\
&& \qquad \qquad  \qquad \qquad = \alpha \, [t \exp(+i k L/2)].
\end{eqnarray}
Finally we have
\begin{eqnarray}
&&\{ik [A \exp(+ i k L/2) - B \exp(- i k L/2)] - ik [t \exp(+ i k L/2)]\} 
\nonumber
\\
&& \qquad \qquad \qquad  \quad = { 2 m \alpha \over \hbar^2} \, [t \exp(+i k L/2)],
\end{eqnarray}
It is very useful to reduce clutter by defining
\begin{equation}
k_{0} = {m \alpha \over \hbar^2},
\end{equation}
then
\begin{eqnarray}
&&\{ik [A \exp(+ i k L/2) - B \exp(- i k L/2)] - ik [t \exp(+ i k L/2)]\} 
\nonumber
\\
&& \qquad \qquad \qquad \quad = 2 k_0 \, [t \exp(+i k L/2)] .
\end{eqnarray}
in which case our four boundary conditions become:
\begin{equation}
\exp(- i k L/2) + r \exp(+i k L/2) = A \exp(- i k L/2) + B \exp(i k L/2),
\end{equation}
\begin{equation}
A \exp(+ i k L/2) + B \exp(-i k L/2) = t \exp(+i k L/2),
\end{equation}
\begin{eqnarray}
\nonumber
&&  \{i k [\exp (- i k L/2) - r \exp(+ i k L/2)]  
\nonumber
\\
&& \quad - i k [A \exp(-i k L/2) - B \exp(+ i k L/2)] \} 
\nonumber
\\
&& \qquad = 2 k_{0} [A \exp(- i k L/2) + B \exp(+i kL/2)],
\end{eqnarray}
\begin{eqnarray}
\nonumber
&&\{i k [A \exp(+ i k L/2) - B \exp(-i k L/2) - i k [t \exp(+i k L/2)]]\} 
\nonumber
\\
&& \qquad \qquad \quad =  2 k_{0} [t \exp(+ i k L/2)].
\end{eqnarray}
These are four simultaneous linear equations for four unknowns: $r$, $A$, $B$, and $t$ (in terms of the known quantities $k$, $k_{0}$, and $a$). These can be solved, either by direct calculation or by {\sf Maple} or something similar. A little work then leads to
\begin{equation}
t = {k^2 \over (k - i k_{0})^2 + k_{0}^2 \, \exp(2 i k L)},
\end{equation}
and
\begin{equation}
r = {2 i k_{0} [k \, \mathrm{cos} (k L) - k_{0} \, \mathrm{sin} (k L)] \over (k - i k_{0})^2 + k_{0}^2 \exp(2 i k L)}.
\end{equation}
Note that $r/t$ is pure imaginary, in agreement with our general argument regarding definite parity potentials. Furthermore note that
\begin{equation}
R = |r|^2 \propto [k \, \mathrm{cos}(2ka) - k_{0} \, \mathrm{sin} (2ka)]^2,
\end{equation}
and so $R = 0$ whenever
\begin{equation}
\mathrm{tan}(2ka) = {k \over k_{0}}.
\end{equation}
That is, the system exhibits ``transmission resonances'' where $T \rightarrow 1$ and $R \rightarrow 0$. If we work at fixed energy then these resonances occur at equally spaced spatial separation for the two delta functions, namely:
\begin{equation}
a_{\mathrm{resonance}} = {1 \over 2 k} \{ \mathrm{tan}^{-1}(k/k_{0}) + n \pi\} ;  \qquad n \in Z.
\end{equation}
If we hold $a$ fixed and vary $k$ then the location of the resonances is determined by the transcendental equation
\begin{equation}
\mathrm{tan}(2ka) = {k \over k_{0}}.
\end{equation}
The existence of ``transmission resonances'' in one-dimensional scattering is in fact widespread, it is not specific to this particular example. A brief computation leads to the explicit transmission coefficient
\begin{equation}
T = {k^4 \over k^4 + 4 k_{0}^2 [k \, \mathrm{cos} (2ka) - k_{0} \mathrm{sin}(2ka)]^2},
\end{equation}
or
\begin{equation}
T = {1 \over 1 + \left[\displaystyle{2 m \alpha \over \hbar^2 k} \, \mathrm{cos} \, (k L) + \displaystyle{{1 \over 2} \bigg({2 m \alpha \over \hbar^2 k}\bigg)^2} \mathrm{sin} \, (k L) \right]^2},
\end{equation}
which agrees with the preceding argument on the location of the transmission  resonances. Finally note $T (k \rightarrow 0) \rightarrow 0$ and $T (k \rightarrow \infty) \rightarrow 1$. After we look at one more illustrative example, we will turn to the issue  of obtaining some general theorems governing one-dimensional scattering.

\section{Square barrier}
Let us now start by introducing another useful potential barrier --- the \emph{square barrier}~\cite{square}:
\begin{equation}
V(x)  =  \left\{\begin{array}{r@{\quad \quad}l}
V_{0} & (\mathrm{for} \quad  0 \leq x \leq L); \\ 0 & (\mathrm{otherwise}).
\end{array} \right.
\end{equation}
In our particular case, we set $V_{0} > 0$.
Tunneling \emph{over} a square barrier is an elementary problem which however is not always discussed in the textbooks. In contrast, tunneling \emph{under} a square barrier is much more popular. The exact transmission coefficient is known to be
\begin{equation}
T = {E (E - V_{e}) \over E(E - V_{e}) + {1 \over 4} V_{e}^{2} \; \mathrm{sin}^2 (\sqrt{2 m (E - V_{e})} L/\hbar)}.
\end{equation}
For more details see (for example) Landau and Lifshitz~\cite{Landau}, or Schiff~\cite{Schiff}. We can re-write this as
\begin{equation}
T = {1 \over 1 + \displaystyle{ {m V_{e}^2 L^2 \over 2 E \hbar^2} \, {{\mathrm{sin}^2}(\sqrt{2 m (E - V_{e})} L /\hbar) \over 2 m (E -V_{e}) L^2/ \hbar^2}} }\, .
\end{equation}
{\bf Quasinormal modes:} \quad $T = \infty$ when the numerator is nonzero and
\begin{equation}
\mathrm{sin} (\sqrt{2 m (E - V_{e})} L / \hbar) = \pm 2 \sqrt{E (E - V_{e})} / V_{e} \, ,
\end{equation}
which leads to hopeless algebra. Although $E = V_{e}$ is one solution, it corresponds to the numerator vanishing and is not a transmission pole. There is no simple explicit formula for the QNM.

\subsection*{Deriving the amplitudes:}
We shall now consider a particle of mass $m$ and energy $E > 0$ interacting with the simple square potential barrier.
 Let us consider $\psi(x)$ in the regions to the left and to the right of the barrier,  we have
\begin{equation}
\label{barrier_1}
{\d^2 \psi \over \d x^2} = -k^2 \psi,
\end{equation}
where $k^2 = 2m E / \hbar^2$. We choose the following solution of the above equation to the left of the barrier (i.e., $x<0$)
\begin{equation}
\psi(x) = \exp(ikx) + r \exp(-ikx).
\end{equation}
The composition of this solution is a plane-wave of unit amplitude traveling to the right [since the time dependent wave function is multiplied by a factor $\exp(-i E t/\hbar)$], and a plane wave of complex amplitude $r$ traveling to the left. 
Moreover it should be stressed that the first plane wave is incoming particle. Indeed, the second plane wave is a particle reflected by the potential barrier. Therefore  $|r|^2$ is the probability of reflection. This can also be seen by calculating the probability current in the region $(x<0)$, which takes the form
\begin{equation}
\mathscr{J}_{l} = {\hbar k \over m}(1-|r|^2).
\end{equation} 
We choose the following solution to equation~(\ref{barrier_1}) to the right of the barrier, that is, for  $x>L$:
\begin{equation}
\psi(x) = t  \exp(ikx).
\end{equation}
We have seen that this solution consists of a plane wave of complex amplitude $t$ traveling to the right. This implies that  this solution can be interpreted as a particle transmitted through the barrier. Consequently, $|t|^2$ is the probability of transmission. In fact we can write the probability current in the region $x>L$ as
\begin{equation}
\mathscr{J}_{r}= {\hbar k \over m} |t|^2.
\end{equation}
If we set $\mathscr{J}_{l} = \mathscr{J}_{r}$, then we derive
\begin{equation}
|r|^2 + |t|^2 = 1.
\end{equation}
At this point is easy to see that inside the barrier $(i.e., 0 \leq x \leq L)$, the wavefunction $\psi(x)$ satisfies
\begin{equation}
\label{barrier_2}
{\d^2 \psi \over \d x^2} = - q^2 \psi,
\end{equation}
where
\begin{equation}
q^2 = {2m (E - V_{0}) \over \hbar^2}.
\end{equation}
We consider the case where $E > V_{0}$. In addition, the general solution to equation~(\ref{barrier_2}) inside the barrier takes the form
\begin{equation}
\psi(x) = A \exp(+iqx) + B \exp(-iqx),
\end{equation}
where $q = \sqrt{2m (E - V_{0})/ \hbar^2}$.
From the continuity of $\psi$ and $\d \psi / \d x$ at the left edge of the barrier $(i.e., x = 0)$ we derive
\begin{eqnarray}
1 + r &=& A + B,
\\
k (1 - r) &=& q (A - B).
\end{eqnarray}
Moreover, continuity of $\psi$ and $\d \psi/ \d x$ at the right edge of the barrier, for $(x=L)$ gives
\begin{eqnarray}
A \exp(+ iqL) + B \exp(-iqL) &=& t \exp(+ ikL),
\\
q (A \exp(+ iqL) - B \exp(- iqL)) &=& k t \exp(+ ikL).
\end{eqnarray}
It is now relatively easy to see that, (after considerable algebra), the above four equations yield
\begin{eqnarray}
R &=&  |r|^2 = {(k^2 - q^2)^2  \, \mathrm{sin}^2 (qL) \over 4 k^2 q^2 + (k^2 - q^2)^2 \mathrm{sin}^2 (qL)},
\\
T &=& |t|^2 = {4 k^2 q^2 \over 4 k^2 q^2 + (k^2 - q^2)^2 \, \mathrm{sin}^2 (qL)}.
\end{eqnarray}

\section{Tanh potential}
For a smoothed step function of the form
\begin{equation}
V(x) =  {V_{-\infty} + V_{+\infty} \over 2} + {V_{+ \infty} - V_{- \infty} \over 2} \; \mathrm{tanh}\bigg({x \over L}\bigg),
\end{equation}
the reflection coefficient is known analytically to be (see, for instance, \cite{Landau}):
\begin{equation}
R = \bigg({\mathrm{sinh}[\pi (k_{- \infty} - k_{+ \infty}) L/2] \over \mathrm{sinh} [\pi (k_{- \infty} + k_{+ \infty}) L/2]}\bigg)^2 \, .
\end{equation}
This certainly satisfies the general bounds~(\ref{case2a1})--(\ref{case2a2}) that we shall derive later on in this thesis, see chapter~\ref{C:consistency-6}, and as $L \rightarrow 0$ approaches and saturates the bound.
\\
\\
{\bf Quasinormal modes: \quad $T = \infty$} when
\begin{equation}
\mathrm{sinh} [\pi (k_{- \infty} + k_{+ \infty}) L/2] = 0 \, ,
\end{equation}
that is
\begin{equation}
\mathrm{sin} [i \pi (k_{- \infty} + k_{+ \infty}) L/2] = 0 \, .
\end{equation}
This leads to
\begin{equation}
i \pi (k_{- \infty} + k_{+ \infty}) L/2 = n \pi \, ,
\end{equation}
that is
\begin{equation}
k_{+ \infty} + \sqrt{k_{+ \infty}^{2} + 4 m V_{+ \infty}/\hbar^2} = - i 2 n/ L \, ,
\end{equation}
so that
\begin{equation}
k_{+ \infty}(n) = i \bigg[{+ m V_{+ \infty} L \over \hbar^2 \, n} + {n \over L}\bigg]
\qquad n \neq 0 \, .
\end{equation}
Equivalently
\begin{equation}
k_{- \infty}(n) = i \,  \bigg[{- m V_{+ \infty} L \over \hbar^2 \, n} + {n \over L}\bigg] \qquad n \neq 0.
\end{equation}
Note the asympototic spacing as $n \rightarrow \infty$:
\begin{equation}
k_{+ \infty} \rightarrow i \, {n \over L}.
\end{equation}
Note that as $L \rightarrow 0$ all the QNM are driven to imaginary infinity --- this is compatible with the behaviour of the step potential for which there are no QNM.

\section{Sech$^2$ potential}
For a sech$^2$ potential of the form
\begin{equation}
V(x) = V_{e} \; \mathrm{sech}^2 (x/L) \, ,
\end{equation}
the transmission coefficient is known analytically to be (see for example~\cite{Landau}):
\begin{equation}
T = {\mathrm{sinh}^2 [\pi \sqrt{2 m E} L/\hbar] \over \mathrm{sinh}^2[\pi \sqrt{2 m E} L/\hbar] + \mathrm{cos}^2[{1 \over 2} \pi \sqrt{1 - 8 m V_{e} L^2/\hbar}]} \, ,
\end{equation}
provided $8 m V_{e} L^2 < \hbar^2$. This satisfies the general bounds derived later in this thesis, both 
the bound $T \geq \mathrm{sech}^2 \bigg( \int_{-\infty}^{+\infty} \vartheta \, \d x\bigg)$,  and the separate bound $T \geq (4 k_{+\infty} k_{-\infty}) /  (k_{+\infty} + k_{-\infty})^2$. (Though proving this is somewhat tedious.)  Start by noting that for this sech potential
\begin{equation}
 T \geq \mathrm{tanh}^2 [\pi \sqrt{2 m E} L/\hbar] \, ,
\end{equation}
and use the inequality $(x > 0)$
\begin{equation}
\mathrm{tanh}^2 \,  x > {x^2 \over 1 + x^2} > \mathrm{sech}^2 \, (1/x) \, .
\end{equation}
Then
\begin{eqnarray}
T &\geq& \mathrm{sech}^2[\hbar / (\pi \sqrt{2 m E} L)] \, ,
\\
&=& \mathrm{sech}^2 \bigg[{4 \over \pi}\sqrt{{m \over 2 E}} {2 L|V_{e}| \over \hbar} \, {\hbar^2 \over 8 m |V_{e}| L^2}\bigg] \, .
\end{eqnarray}
Provided that the extremum is a peak, $V_{\mathrm{peak}} > 0$ we can use the bound $8 m V_{\mathrm{peak}} L^2 < \hbar^2$ to deduce
\begin{equation}
T \geq \mathrm{sech}^2 \bigg[\sqrt{{m \over 2 E}} \, {2 L |V_{\mathrm{peak}}| \over \hbar} \bigg] \, .
\end{equation}
This is the particularization of the bound $T \geq \mathrm{tanh}^2 \bigg(\int_{-\infty}^{+\infty} \vartheta \, \d x\bigg)$ to the present case. If $V_{e} < 0$ we need a different analysis.
\\
\\
{\bf Quasinormal modes:} \quad $T = \infty$ when
\begin{equation}
\mathrm{sinh}^2 [\pi \sqrt{2 m E} L/h] + \mathrm{cos}^2 \bigg[{1 \over 2} \pi \sqrt{1 - 8 m V_{e} L^2/\hbar^2}\bigg] = 0 \, ,
\end{equation}
which leads to
\begin{equation}
- \mathrm{sin}^2 [i \pi \sqrt{2 m E} L/ \hbar] + \mathrm{cos}^2 \bigg[{1 \over 2} \pi \sqrt{1 - 8 m V_{e} L^2/ \hbar^2}\bigg] = 0.
\end{equation}
But then
\begin{equation}
\mathrm{sin}[i \pi \sqrt{2 m E} L/ \hbar] = \pm \mathrm{cos}[{1 \over 2} \pi \sqrt{1 - 8 m V_{e} L^2/\hbar^2}],
\end{equation}
and so
\begin{equation}
\mathrm{cos}[i \pi \sqrt{2 m E} L/\hbar - \pi/2] = \pm \mathrm{cos}\bigg[{1 \over 2} \pi \sqrt{1 - 8 m V_{e} L^2/ \hbar^2}\bigg].
\end{equation}
Therefore
\begin{equation}
i \pi \sqrt{2 m E}L/\hbar - \pi/2 = \pm {1 \over 2} \pi \sqrt{1 - 8 m V_{e} L^2/ \hbar^2} + \pi n,
\end{equation}
leading to
\begin{equation}
\sqrt{2 m E}/\hbar = \pm i {1 \over 2L}  \sqrt{1 - 8 m V_{e} L^2/ \hbar^2} + {i (2 n +1) \over 2 L}.
\end{equation}
Finally
\begin{equation}
k_{n} = \pm i {1 \over 2 L} \sqrt{1 - 8 m V_{e} L^2/ \hbar^2} + {i(2n +1) \over 2 L} \, .
\end{equation}
Again note the asymptotic spacing as $n \rightarrow \infty$
\begin{equation}
k_{n} \rightarrow i {n \over L}\, .
\end{equation}
Note that if $V_{e}$ is big, the offset term becomes real
\begin{equation}
k = \pm {1 \over 2 L} \sqrt{8 m V_{e} L^2/\hbar^2 - 1} + {i(2n + 1) \over 2 L} \, .
\end{equation}
Finally note what happens as $L$ becomes small,
\begin{equation}
k_{n} = \pm i {1 \over 2 L} \mp 2 i m V_{e} L/\hbar^2 + {i (2n +1) \over 2 L} + O(L^3) \, .
\end{equation}
In the limit only one pair of QNM survive
\begin{equation}
k = 2 i m \lim[V_{e} L]/ \hbar^2 \, ,
\end{equation}
the others being driven off to infinity. This agrees with the result for the single--delta--function potential.

\subsection*{Derivation of the amplitudes (sketch):}
Let us sketch how to determine the transmission coefficient for a potential barrier defined by the formula~\cite{Landau}
\begin{equation}
V(x) =  V_{e} \; \mathrm{sech}^2 (x/L).
\end{equation}
Comparing this with the analogous bound state computation, it is necessary merely to alter the sign of $V_{e}$ and to regard the energy $E$ now as positive. A calculation similar to that used for deriving the bound states when $E<0$~\cite{Landau}, now gives the solution
\begin{equation}
\psi(\xi) = (1- \xi^2)^{-ik/2L} F\left[-ik/L -s, -ik/L + s + 1, -ik/L + 1, {1\over 2} (1 - \xi)]\right],
\end{equation}
where
\begin{equation}
\xi = \mathrm{tanh} (x/L) , 
\end{equation}
\begin{equation}
 k = \sqrt{(2mE)/\hbar} \; ,
\end{equation}
\begin{equation}
s = {1 \over 2} \bigg(-1 + \sqrt{\bigg[ 1 - { 8 m V_{e} \over \alpha^2 \hbar^2}\bigg]}\bigg),
\end{equation}
and $F[\_\_,\_\_,\_\_,\_\_]$ is a hypergeometric function.
This solution satisfies the condition that, as $x \rightarrow \infty$  (i.e. as $\xi \rightarrow 1, (1 - \xi) \approx 2 \exp(-x/L)$), the wave function should include only the transmitted wave $(\sim \exp(ikx/L))$. The asymptotic form of the wave function as $x \rightarrow - \infty$, $(\xi \rightarrow -1)$ is found by transforming the hypergeometric function with the aid of formula 
\begin{equation}
\psi \sim \exp(-ik x) {\Gamma(ik L) \Gamma(1-ik L) \over \Gamma(-s) \Gamma(1 + s)} + \exp(ik x) {\Gamma(-ik L) \Gamma(1 - ik L) \over \Gamma(-i k L -s) \Gamma(-ik L + s +1)}.
\end{equation}
Taking the squared modulus of the ratio of coefficients in this function, we obtain the following expression for the transmission coefficient $T = 1 - R$:
\begin{equation}
T = {\mathrm{sinh}^2(\pi \sqrt{(2mE)} L/\hbar) \over \mathrm{sinh}^2(\pi \sqrt{(2mE)} L/\hbar) + \mathrm{cos}^2[{1 \over 2} \pi \sqrt{(1 - 8 m V_{e} L^2/ \hbar^2)}]} \, ,
\end{equation}
when $8 m V_{e} L^2/ \hbar^2 < 1$.
\section{Asymmetric Square-well potential}
For the asymmetric square well
\begin{equation}
V(x) = \left\{ \begin{array} {r@{, \qquad}l}
V_1 & x<a; \\ V_{2} & a <x <b; \\ V_3 & b<x.
\end{array} \right.
\end{equation}
We now define $k_{i} = \sqrt{2 m (E - V_{i})}/\hbar$. The transmission coefficient is (see for example~\cite{Messiah}):
\begin{equation}
T = {4 k_1 k_2^2 k_3 \over (k_1 + k_3)^2 k_2^2 + [k_1^2 k_3^2 + k_2^2(k_2^2 - k_1^2 -k_3^2)] \, \mathrm{sin}^2 (k_2 L)} \, .
\end{equation}
Then
\begin{equation}
T \geq {4 k_1 k_2^2 k_3 \over (k_2^2 + k_1 k_3)^2} \, .
\end{equation}
Similarly to the case for the symmetric square well, the transmission probability for the asymmetric square well oscillates between the bound (\ref{trans1}) that we shall subsequently derive, and the unitarity limit $T =1$. For certain values of the width of the well $[k_{2} L = (2 n + 1) \pi/2]$ the transmission coefficient saturates the bound thus showing that this bound cannot be improved \emph{unless additional hypotheses are made}. Because $V_{- \infty} \neq V_{+ \infty}$ the bound $T \geq \mathrm{sech}^2\bigg(\int_{-\infty}^{\infty} \vartheta \, \d x \bigg),$
is not applicable, at least not without modification from its original form.

\subsection*{Derivation of the amplitudes (sketch):}
As in the problem of the potential step, we build the eigenfunction of the form~\cite{Messiah}
\begin{equation}
\psi(x)  =  \left\{\begin{array}{r@{\quad \quad}l}
S \exp(-i k_{1} x)  &  x<a; \\ 
P \exp(-i k_{2} x) +Q  \exp(i k_{2} x) & a<x<b; \\ 
\exp(-i k_{3} x) + R \exp(i k_{3} x) & b<x.
\end{array} \right.
\end{equation}
The continuity conditions at points $a$ and $b$ give the values of $R$, $Q$, $P$, and $S$. Without entering into the specific details of the calculation, we simply list the results concerning the quantities $R$ and $S$. We use the following notation and conventions:
\begin{equation*}
a = -L, \qquad  \qquad b = 0, \qquad \qquad K = \sqrt{V_{3} - V_{1}},
\end{equation*}
\begin{equation*}
\xi = {k_{2} \over K}, \qquad \qquad \qquad \eta  = {k_{3} \over K}, \qquad \qquad \qquad \zeta = {k_{1} \over K}.
\end{equation*}
We now derive
\begin{eqnarray}
R &=& {k_{2}(k_{3} - k_{1}) \mathrm{cos}(k_{2}L) + i (k_{2}^2 - k_{3} k_{1}) \mathrm{sin}(k_{2}L) \over k_{2} (k_{3} + k_{1}) \mathrm{cos}(k_{2}L) - i (k_{2}^2 + k_{3} k_{1}) \mathrm{sin}(k_{2} L)},
\\
S &=& \exp(- i k_{1} L) {2 k_{2} k_{3} \over k_{2} (k_{3} + k_{1}) \mathrm{cos}(k_{2} L) - i(k_{2}^{2}+ k_{2} k_{3}) \mathrm{sin}(k_{2}L)}.
\end{eqnarray}
The wave is in general only partially transmitted, and we can define a transmission coefficient (note that we are now using Messiah-like conventions)
\begin{eqnarray}
T = {k_{1} \over k_{3}} |S|^2 &=& {4k_{1} k_{2}^2 k_{3} \over  k_{2}^2 (k_{3} + k_{1})^2 \mathrm{cos}^2(k_{2}L) + (k_{2}^2 + k_{3} k_{1})^2 \mathrm{sin}^2(k_{2} L)} \, ,
\\
\nonumber
&=& {4 k_1 k_2^2 k_3 \over (k_1 + k_3)^2 k_2^2 + [k_1^2 k_3^2 + k_2^2(k_2^2 - k_1^2 -k_3^2)] \, \mathrm{sin}^2 (k_2 L)} \, .
\end{eqnarray}
\section{Poeschl--Teller potential}
The Poeschl--Teller potential is most commonly written (see, \emph{e.g.},~\cite{Morse})
\begin{equation}
V(x) = V_{0} \; \mathrm{cosh}^2 \mu \{\mathrm{tanh}[(x - \mu L) /L] + \mathrm{tanh} \, \mu\}^2 \, ,
\end{equation}
we have 
\begin{equation}
V_{- \infty} = V_{0} e^{- 2 \mu}; \qquad V_{\mathrm{extremum}} = 0; \qquad V_{+ \infty} = V_{0} e^{+ 2 \mu} \, .
\end{equation}
The transmission coefficient is~\cite{Morse}
\begin{equation}
T = {2 \mathrm{sinh} [(\pi k_{- \infty} L) \mathrm{(\pi k_{+ \infty} L)}] \over \mathrm{cosh}[\pi (k_{- \infty} + k_{+ \infty})] + \mathrm{cos}\bigg[\pi \sqrt{1 + {8 m V_0 L^2 \over \hbar^2} \, \mathrm{cosh}^2 \, \mu}\bigg]}.
\end{equation}
It is now a straightforward if tedious exercise to check this analytic result against all the bounds derived in this thesis.

This Morse--Feshbach presentation of this potential~\cite{Morse}, as given above, is rather difficult to interpret --- it is much easier to first translate the potential ($x\to x+L$) to obtain
\begin{equation}
V(x) = V_{0} \, \mathrm{cosh}^2 \, \mu \{\mathrm{tanh} \, [x/L] + \mathrm{tanh} \, \mu\}^2 \, ,
\end{equation}
expand
\begin{equation}
V(x) = V_0 \, \mathrm{cosh}^2 \, \mu \left\{\mathrm{tanh}^2 [x/L] + 2 \,  \mathrm{tanh} \,\mu \, \mathrm{tanh} [x/L] + \mathrm{tanh}^2 \, \mu\right\} \, ,
\end{equation}
and then re-group terms as
\begin{equation}
V(x) = V_{0} \, \mathrm{cosh}^2 \, \mu \left\{- \mathrm{sech}^2[x/L] + 2 \, \mathrm{tanh} \, \mu \, \mathrm{tanh}[x/L] +2 - \mathrm{sech}^2 \, \mu\right\} \, ,
\end{equation}
to see that this is simply a linear combination of the ``sech$^2$'', ``tanh'',  and ``constant'' potentials. So without loss of generality we can re-write the Poeschl--Teller potential (with new definition for $V_{0} = - \, \mathrm{cosh}^2 \, \mu \, [V_{0}]_{\mathrm{old}}$) as:
\begin{equation}
V(x) = V_0 \, \mathrm{sech}^2 \, [x/L] + V_{\infty} \, \mathrm{tanh} \, [x/L] \, ,
\end{equation}
with
\begin{equation}
V(- \infty) = -V_{\infty}; \qquad V(0) = V_0; \qquad V(+ \infty) = + V_{\infty} \, ; \, ,
\end{equation}
in terms of which the analytically known transmission probability is
\begin{equation}
T = {2 \, \mathrm{sinh} \, (\pi k_{- \infty} L) \, \mathrm{sinh} \, (\pi k_{+ \infty} L) \over \mathrm{cosh} \, [\pi (k_{- \infty} + k_{+ \infty}) L] + \mathrm{cosh} \, \bigg[\pi \sqrt{1 - {8 m V_0 L^2 \over \hbar^2}} \bigg]} \, .
\end{equation}
Of course we have already seen that this has at least two much simpler limits: the $\sech^2$ and $\tanh$ potentials. In particular if we let $V_{\infty} \rightarrow 0$ and play with a few \emph{hyperbolic} identities we recover the results for the sech$^2$ potential, while if we let $V_0 \rightarrow 0$ and play with a few \emph{hyperbolic} identities we recover the results of the $\mathrm{tanh}$ potential.
\\
\\
{\bf Quasinormal modes:} \quad $T = \infty$ when
\begin{equation}
\mathrm{cos}[i \pi (k_{- \infty} + k_{+ \infty}) L] + \mathrm{cos}\bigg[\pi \sqrt{1 - {8 m V_0 L^2 \over \hbar^2}}\bigg] = 0.
\end{equation}
That is
\begin{equation}
i \pi (k_{- \infty} + k_{+ \infty}) L = \pm \pi \sqrt{1 - {8 m V_0 L^2 \over \hbar^2}} + (2 n + 1) \pi,
\end{equation}
whence
\begin{equation}
(k_{- \infty} + k_{+ \infty}) = \pm i {1 \over L} \sqrt{1 - {8 m V_0 L^2 \over \hbar^2}} + i {2 n + 1 \over L}.
\end{equation}
We now rearrange this as
\begin{eqnarray}
k_{+ \infty} + \sqrt{k_{+ \infty}^2 + 4 m V_{+ \infty}/\hbar^2} &=& \pm i {1 \over L} \sqrt{1 - {8 m V_0 L^2 \over \hbar^2}} + i {2 n + 1 \over L} ,
\nonumber
\\
&=& i \left({2 n +1 \pm \sqrt{1 - 8 m V_0 L^2/\hbar^2} \over L}\right),
\nonumber
\\
&&
\end{eqnarray}
and note that it has appropriate limits for the tanh and sech$^2$ potentials. Finally, because this is a simple quadratic equation, we solve for $k(n)$ to obtain
\begin{equation}
k(n) = i\left({2  m V_{\infty} L / \hbar^2 \over (2n + 1) \pm \sqrt{1 - 8 m V_0 L^2 /\hbar^2}} +  {2 n + 1 \pm \sqrt{1 - 8 m V_0 L^2/ \hbar^2} \over 2 L}\right).
\end{equation}
By shifting $n$ by $\pm 1$ as appropriate we can re-write this as
\begin{equation}
k(n) = i\left({2 m V_{\infty} L/\hbar^2 \over 2 n \pm \bigg[\sqrt{1 - 8 m V_0 L^2/ \hbar^2} -1 \bigg]} + {2 n \pm \bigg[\sqrt{1 - 8 m V_0 L^2/ \hbar^2} - 1\bigg] \over 2 L}\right).
\end{equation}
This now has the appropriate limits to reproduce both tanh and sech$^2$ quasinormal modes.
Note that asymptotically
\begin{equation}
k(n) \rightarrow i \, {n \over L} \pm i \, {\sqrt{1 - 8 m V_0 L^2/ \hbar^2} -1 \over 2 L} + \mathcal{O}(1/n),
\end{equation}
in accordance with the general suspicions based on black hole QNMs~\cite{qnmv}.

\section[Eckart--Rosen--Morse--Poeschl--Teller potential]
{Eckart--Rosen--Morse--Poeschl--Teller\\ potential} 

Many of the potentials commonly encountered in the literature are actually the same quantity in disguise. To start with, consider the following three potentials:
\\
{\bf Eckart (1930):}
\begin{equation}
V(x) = - {A \xi \over 1 - \xi} - {B \xi \over (1 - \xi)^2}; \qquad \xi = - \exp(2x/a) .
\end{equation}
{\bf Rosen--Morse (1932):}
\begin{equation}
V(x) = B \, \mathrm{tanh}(x/d) - C \, \mathrm{sech}^2(x/d).
\end{equation}
{\bf Poeschl--Teller (1933):}
\begin{equation}
V(x) = V_0 \, \mathrm{cosh}^2 \mu \{\mathrm{tanh}([x - \mu L]/L) + \mathrm{tanh} \, \mu\}^2.
\end{equation}
To see that all three of these potentials are actually the same, note that:
\begin{eqnarray}
{4 \xi \over (1 - \xi)^2} &=& {4 \over (\xi^{-1/2}+ \xi^{+1/2})^2} = {4 \over [\exp(-x/a) + \exp(x/a)]^2}\, ,
\nonumber
\\
&=& {1 \over \mathrm{cosh}^2(x/a)} = \mathrm{sech}^2(x/a).
\end{eqnarray}
Furthermore
\begin{eqnarray}
1 + {2 \xi \over 1 - \xi} &=& {1 + \xi \over 1 - \xi} = {1 - \exp(2x/a) \over 1 + \exp(2x/a)}\, ,
\nonumber
\\
&=& {\exp(-x/a) - \exp(x/a) \over \exp(-x/a) + \exp(x/a)} = - \mathrm{tanh}(x/a).
\end{eqnarray}
This is enough to show
\[
\hbox{(Eckart)} \Longleftrightarrow \hbox{(Rosen--Morse)}
\]
In fact in the article of Rosen and Morse~\cite{Rosen}, they cite Eckart~\cite{Eckart}, and describe Eckart's potential as begining ``somewhat like'' their own, but without noticing that the two potentials are in fact \emph{identical} up to trivial redefinitions of the parameters.

Now, for the Poeschl--Teller potential, note that by a trivial shift $x \rightarrow x + \mu L$ we have 
\begin{equation}
V(x)  \rightarrow V_0 \,  \mathrm{cosh}^2 \mu \{ \mathrm{tanh} (x/L) + \mathrm{tanh} \, \mu \}^2,
\end{equation}
which we can without loss of generality relabel as
\begin{eqnarray}
\nonumber
V(x) &\rightarrow& V_1 \, \{ \mathrm{tanh}(x/L) + D\}^2,
\nonumber
\\
&=& V_1 \{\mathrm{tanh}^2(x/L) + 2 D \, \mathrm{tanh}(x/L) + D^2\},
\nonumber
\\
&=& V_1 \{- \mathrm{sech}^2(x/L) + 2 D \mathrm{tanh}(x/L) + D^2 +1\},
\nonumber
\\
&=& V_2 \,  \mathrm{sech}^2 (x/L) + V_{3} \, \mathrm{tanh} (x/L) + V_4.
\end{eqnarray}
This is enough to show
\[
\hbox{(Poeschl--Teller)} \Longleftrightarrow \hbox{(Rosen--Morse)}
\]
and so all three potentials are completely identical up to trivial relabeling of the parameters and a shift in the zero of energy. 
In fact, including the offset, all three of these can be written in any one of the four general forms below:
\begin{eqnarray}
V(x) &=& A + B \, \mathrm{tanh} (x/a + \theta) + C \, \mathrm{tanh}^2 (x/a + \theta),
\nonumber
\\
&=& A_0 + [B_0 + C_0 \, \mathrm{tanh} (x/a + \theta)]^2,
\nonumber
\\
&=& A_0 + \bigg[{B_1 + C_1 \, \mathrm{tanh}(x/a) \over B_2 + C_2 \, \mathrm{tanh}(x/a)} \bigg]^2,
\nonumber
\\
&=& A_0 + \bigg[{E_1 + F_1 \, \exp(-2x/a) \over E_2 + F_2 \, \exp(-2 x/a)}\bigg]^2 .
\end{eqnarray}
Note that there is some redundancy here, but it is a useful redundancy. It makes it clear that the Eckart--Rosen--Morse--Poeschl--Teller potential is generally a Mobius function of the variable  $\exp(-2 x/a)$. Thus implies that without loss of generality we can set $E_1 F_2 - E_2 F_1$ as some convenient constant.

The ``best'' of these equivalent forms is arguably the Mobius form:
\begin{equation}
V(x) = V_0 + V_1 \bigg[{A+ B \, \exp(-2x/a) \over C + D \, \exp(-2x/a)}\bigg]^2.
\end{equation}
\\
{\bf Comment:}
Many authors seem to use the phrase ``Poeschl--Teller potential'' only to refer to the special case
\begin{equation}
V(x) = V_0/ \mathrm{cosh}^2 (x/a) = V_0 \, \mathrm{sech}^2 (x/a).
\end{equation}
This is historically inaccurate~\cite{Poeschl}, and we will have more to say on this later.

\section{Mobius potential}
Overall, the ``best'' general version of the potentials considered above is probably the Mobius form
\begin{equation}
V(x) = V_0 + V_1 \bigg[{A + B \, \exp(-2x/a) \over C + D \, \exp(-2x/a)}\bigg]^2.
\end{equation}
If you want to solve the Schr\"odinger equation
\begin{equation}
- {\hbar^2 \, \psi'' \over 2 m} + V \psi = E \psi,
\end{equation}
or 
\begin{equation}
\psi'' + {2m(E - V) \over \hbar^2} \, \psi = 0,
\end{equation}
then, absorbing the $\hbar^2/2m$ into a redefinition of $E, V_0$ and $V_1$, what you need to do is to solve
\begin{equation}
\psi'' + \bigg\{E - V_0 - V_1 \bigg[{A + B \, \exp(-2x/a) \over C + D \, \exp(-2x/a)}\bigg]^2 \bigg\} \, \psi = 0.
\end{equation}
This can either be solved ``by hand'', or with the aid of symbolic manipulation packages. For instance, 
{\sf Maple} still needs a little help. Without loss of generality, rescale $x \rightarrow xa/2$, and rescale $V_0 \rightarrow 0, V_1 \rightarrow V_0$, and demand $AD - BC =1$. This makes the new $E$ and $V_0$ dimensionless and we have:
\begin{equation}
\psi'' + \bigg\{E - V_0 \bigg[{A + B \, \exp(-x) \over C + D \, \exp(-x)} \bigg]^2 \bigg\} \, \psi = 0.
\end{equation}
This has an explicit solution in terms of hypergeometric functions. {\sf Maple} (after a little bit of convincing) gives
\begin{eqnarray}
\nonumber
\psi(x) &=& (C \, \exp(x) + D) 
\nonumber
\\
& \times& \left\{A_0 \, \exp \bigg(\sqrt{V_0 {B^2 \over D^2} - E} \, x \bigg)\; {_{2}F_{1}} \bigg(A_1, A_2; A_3; - {C \over D} \, \exp(x)\bigg)\right.
\nonumber
\\
&+& \left. B_0 \, \exp \bigg(- \sqrt{V_0 \, {B^2 \over D^2} - E} \, x \bigg)\;  {_{2}F_{1}} \bigg(B_1, B_2; B_3; - {C \over D \, \exp(x)} \bigg) \right\},
\nonumber
\\
&&
\end{eqnarray}
where the suppressed arguments on the hypergeometric functions are:
\begin{eqnarray}
\nonumber
A_1 &=& {1 \over 2} + \sqrt{V_0 \, {B^2 \over D^2} - E} - {1 \over 2} \sqrt{1 + {4 V_0 \over C^2 D^2}} - \sqrt{V_0 {(1 + BC)^2 \over C^2 D^2} - E};
\nonumber
\\
&&
\\
A_2 &=& {1 \over 2} + \sqrt{V_0 \, {B^2 \over D^2} - E} - {1 \over 2} \sqrt{1 + {4 V_0 \over C^2 D^2}} + \sqrt{V_0 {(1 + BC)^2 \over C^2 D^2} - E };
\nonumber
\\
&&
\\
A_3 &=& 1 + 2 \sqrt{V_0 \, {B^2 \over D^2} - E};
\nonumber
\\
&&
\end{eqnarray}
and
\begin{eqnarray}
\nonumber
B_1 &=& {1 \over 2} - \sqrt{V_0 \, {B^2 \over D^2} - E} - {1 \over 2} \sqrt{1 + {4 V_0 \over C^2 D^2}} + \sqrt{V_0 \, {A^2 \over C^2} - E};
\nonumber
\\
&&
\\
B_2 &=& {1 \over 2} - \sqrt{V_0 \, {B^2 \over D^2} - E} - {1 \over 2} \sqrt{1 + {4 V_0 \over C^2 D^2}} - \sqrt{V_0 \, {A^2 \over C^2} - E};
\nonumber
\\
&&
\\
B_3 &=& 1 - 2 \sqrt{V_0 \, {B^2 \over D^2} - E}.
\nonumber
\\
&&
\end{eqnarray}
By looking up tabulated properties of the hypergeometric function one can now determine the bound state eigenvalues, reflection and transmission coefficients, etc.
\\
\\
In general, the Mobius potential exhibits both bound and free states. (As it \emph{must}, since after all the way we have derived it is by showing that it is equivalent to any of the  Manning--Rosen, Poeschl--Teller, or Eckart potentials.) When the potential energy has a minimum but goes asymptotically to some higher  finite value at $x = + \infty$ and $- \infty$,  then some of the allowed energies will be discrete values, corresponding to states for which the particle is bound in the potential valley. For other ranges of energy, higher than the minimum of the two asymptotic values, all energies will be allowed, the particle being free to travel to infinity. 

For instance, consider the wave functions and allowed energies for a particle of mass $m$ in a Mobius potential that is written in Poeschl--Teller form~\cite{Morse}
\begin{equation}
V(x) = V_{0} \, \mathrm{cosh}^2 \mu \{ \mathrm{tanh}[(x - \mu L)/ L] + \mathrm{tanh}(\mu)\}^2.
\end{equation}
For $V_{0}$ positive, this potential field has its minimum value $(V=0)$ at $x=0$. As $x$ is increased positive, the potential increases to an asymptotic value $V_{0} \exp(2 \mu)$ for $x \rightarrow + \infty$; as $x$ is made negative, $V$ also rises to an asymptotic value $V_{0} \exp(-2 \mu)$, for $x \rightarrow - \infty$.
Classically, since in this form the potential is constrained to be positive semidefinite, the particle could not have a negative energy; for energies between zero and $V_{0} \exp(-2 \mu)$,  $(\mu >0)$, the particle would oscillate back and forth in the potential valley; for energies between $V_{0} \exp(-2 \mu)$ and $V_{0} \exp(2 \mu)$, the particle could come from $- \infty$, be reflected by the potential rise to the right of the minimum, and go back to $- \infty$ and for energies greater than $V_{0} \exp(2 \mu)$, the particle could move from $- \infty$ to $+ \infty$ or from $+ \infty$ to $- \infty$~\cite{Morse}.

\section{Other potentials:}
Furthermore, we would like to at least mention some other potentials:
\\
\\
{\bf Morse (1929)} 
\begin{equation}
V(x) = V_0 \, (1 - \exp (- [x - x_0]/a))^2.
\end{equation}
The Morse potential is actually a somewhat odd limit of the Mobius potential as various parameters go to unity or zero. In terms of the Mobius potential above we need $V_0 \rightarrow 0, V_1 \rightarrow V_0, A = B \rightarrow 1, C \rightarrow 1, D \rightarrow 0, x \rightarrow x - x_0$.
\\
\\
{\bf Manning--Rosen (1933)}
\begin{equation}
V(x) = B \, \mathrm{coth}(x/d) - C \, \mathrm{cosech}^2 (x/d).
\end{equation}
Warning: The relevant citation~\cite{Manning} is only an \emph{abstract} in a report of a conference. To find it with online tools such as {\sf PROLA} look up Phys.~Rev.~{\bf 44}\, (1933) \,951, and then manually scan for abstract $\# \, 10$.  The form actually given in the abstract is
\begin{equation}
V(x) = A {\exp(-2 x/b) \over [1 - \exp(-x/b)]^2} + B \, {\exp(-x/b) \over 1 - \exp(-x/b)},
\end{equation}
which you can manipulate into the form above by noting
\begin{eqnarray}
\nonumber
1 + 2 \, {\exp(-x/b) \over 1 - \exp(-x/b)} &=& {1 + \exp(-x/b) \over 1 - \exp(-x/b)},
\nonumber
\\
&=& {\exp(x/2b) + \exp(-x/2b) \over \exp(x/2b) - \exp(-x/2b)} = \mathrm{coth}(x/2b),
\nonumber
\\
&&
\end{eqnarray}
and 
\begin{equation}
\mathrm{coth}^2 z = 1 + \mathrm{cosech}^2 z.
\end{equation}
Note that the Manning--Rosen potential can be obtained from the Eckart potential by the substitution
\begin{equation}
\xi = - \exp(2x/a) \rightarrow + \exp(-2x/a).
\end{equation}
In particular, Manning--Rosen can be written in the form
\begin{eqnarray}
V(x) &=& A + B \, \mathrm{coth} (x/a) + C \, \mathrm{coth}^2 (x/a)\, ,
\nonumber
\\
&=& A_0 + [B_0 + C_0 \, \mathrm{coth}(x/a)]^2.
\end{eqnarray}
We can get this from the general Mobius form of the Eckart potential by appropriately choosing the parameters.
\\
\\
{\bf Hulthen $(1942)$}
\begin{equation}
V(x) = V_0 \, {\exp(-x/a) \over 1 - \exp(-x/a)}.
\end{equation}
The Hulthen potential~\cite{Hulthen} is actually a special case of the Manning--Rosen potential $(A = 0)$. We can also get this from the general Mobius form of the Eckart potential by appropriately choosing the parameters.
\\
\\
{\bf Tietz $(1963)$}
One version of the Tietz potential~\cite{Tietz} is:
\begin{equation}
V(x) = V_0 \bigg({\mathrm{sinh}([x - x_0]/a) \over \{\mathrm{sinh, cosh, \exp}\} (x/a)}\bigg)^2.
\end{equation}
We can get this from the general Mobius form of the Eckart potential by appropriately choosing the parameters.
\\
\\
{\bf Hua $(1990)$}
\begin{equation}
V(x) = V_0 \bigg({1 - \exp(-2x/a) \over 1 - q \, \exp(-2 x/a)}\bigg)^2.
\end{equation}
We can get this~\cite{exact} from the general Mobius form of the Eckart potential by appropriately choosing the parameters. We note
\begin{eqnarray}
V(x) &=& V_0 \bigg({\exp(x/a) - \exp(-x/a) \over \exp(x/a) - q \, \exp(-x/a)}\bigg)^2,
\nonumber
\\
&=& V_1 \bigg({\mathrm{sinh} (x/a) \over (1+q) \mathrm{sinh}(x/a) + (1-q) \mathrm{cosh}(x/a)}\bigg)^2.
\end{eqnarray}
If $q > 0$ define $1+ q = \mathrm{cosh} \theta$ and $1 - q = \mathrm{sinh} \theta$.\\
If $q < 0$ define $1- q = \mathrm{cosh} \theta$ and $1 + q = \mathrm{sinh} \theta$.\\
Then we see
\begin{eqnarray}
V(x) &=& V_1 \bigg({\mathrm{sinh}(x/a) \over \{\mathrm{sinh, cosh}\}(x/a + \theta)}\bigg)^2 
\qquad (q \neq 0) ,
\nonumber
\\
&=& V_1 \bigg({\mathrm{sinh}(\bar{x}/a - \theta) \over \{\mathrm{sinh, cosh, \exp}\} (\bar{x}/a)}\bigg) ^2,
\nonumber
\\
&=& \hbox{(Tietz potential)},
\nonumber
\\
&=& V_1 \bigg({A \, \mathrm{sinh}(\bar{x}/a) + B \, \mathrm{cosh} (\bar{x}/a) \over \{\mathrm{sinh, cosh}\}(\bar{x}/a)}\bigg)^2,
\nonumber
\\
&=& (\hbox{Eckart potential or Manning--Rosen potential as appropriate}).
\nonumber\\
&&
\end{eqnarray}

So all of these potentials are either identical to the Mobius potential, or special cases of the Mobius potential. To be historically accurate we should really just call this whole collection of potentials the Eckart potential, as Eckhart seems to have been the first author to have given the general form. Unfortunately other names are now in such common use that historical accuracy is difficult (if not impossible) to recover.

\section{Discussion}
Let us summarize the results that we have obtained from this chapter.

In this chapter, we collected many known analytic results in a form amenable to comparison with the general results we shall soon derive. In addition, we introduced the concept of \emph{quasinormal modes}. We shall use these tools for comparing the bounds with known analytic results. Moreover, we reproduced  some of the analytically known results, and showed (or at least sketched) how to derive their scattering amplitudes, and so calculate quantities such as the tunnelling probabilities and quasinormal modes [QNM]. We did this explicitly for the delta--function potential, double--delta--function potential, square potential barrier, tanh potential, sech$^2$ potential, asymmetric square-well potential, the Poeschl--Teller potential and its variants, and finally the Eckart--Rosen--Morse--Poeschl--Teller potential. 

In addition, we are able to gain some deeper understanding by realizing that
the Eckart--Rosen--Morse--Poeschl--Teller potential is generally a Mobius function of the variable  $\exp(-2 x/a)$. Furthermore, the Morse potential is actually a specific limit of the Mobius potential as various parameters go to unity or zero. We also demonstrate that the Hulthen potential is actually a special case of the Manning--Rosen potential $(A = 0)$.

As previously discussed, we have seen that a many of the ``exactly solvable'' potentials commonly encountered in the literature are actually the same quantity in disguise. For instance, we devoted that all of these potentials are either identical to the Mobius potential, or special cases of the Mobius potential. Moreover, we should really just call this the Eckart potential, as he seems to have been the first author to have given the general form. Unfortunately other names are now in such common use that historical accuracy is difficult (if not impossible) to recover.

\label{P:compare-3}
\begin{table}[!ht]
\centerline{
Some potentials for which the transmission probability is explicitly known}
\bigskip

\hskip-1cm
{\small
\begin{tabular}{|  l  |  l  |  l |}
\hline
Name & Potential $V(x)$ & Transmission coefficient $T = |t|^2$\\
\hline
\hline
Delta--function potential  & $ \alpha \delta(x)$ &
 $\vphantom{\Bigg|} \displaystyle{1 \over 1 + (m \alpha^2 / 2 E \hbar^2)}$ \\ 
\hline
Double--delta--function & $  \alpha \{\delta (x - L/2) + \delta(x + L/2)\}$ &
  $ \displaystyle{1 \over 1 + \left[{2 m \alpha \over \hbar^2 k} \, \mathrm{cos} \, (k L) + {1 \over 2} ({2 m \alpha \over \hbar^2 k})^2 \mathrm{sin} \, (k L) \right]^2}$ \\
\hline
Square potential barrier & $\left\{\begin{array}{r@{\quad \quad}l}
V_{0} & (\mathrm{for} \quad  0 \leq x \leq L) \\ 0 & (\mathrm{otherwise})
\end{array} \right.
$ & $\displaystyle{4 k^2 q^2 \over 4 k^2 q^2 + (k^2 - q^2)^2 \, \mathrm{sin}^2 (qL)}$ \\
\hline
Tanh potential & $V_{+ \infty} \, \mathrm{tanh} (x / L)$ & $1-  \displaystyle{\bigg({\mathrm{sinh}[\pi (k_{- \infty} - k_{+ \infty}) L/2] \over \mathrm{sinh} [\pi (k_{- \infty} + k_{+ \infty}) L/2]}\bigg)^2}$  \\
\hline
Sech$^2$ potential   & $V_{e} \, \mathrm{sech}^2 (x/L)$ & $ \geq \mathrm{sech}^2 \bigg[\sqrt{{m \over 2 E}} \, {2 L |V_{\mathrm{peak}}| \over \hbar} \bigg]$ \\
\hline
Asymmetric Square-well& $\left\{ \begin{array} {r@{, \qquad}l}
V_1 & x<a; \\ V_{2} & a <x <b; \\ V_3 & b<x.
\end{array} \right.$ & $ \geq \displaystyle{4 k_1 k_2^2 k_3 \over (k_2^2 + k_1 k_3)^2}$ \\
\hline
\hline
\end{tabular}
} 
\caption[Some exactly solvable potentials for which the transmission probability is explicitly known]{\label{T:compare-3} This table shows some of the potentials for which exact analytic results are known, and summarizes key properties of the transmission probabilities.}
\bigskip
\end{table}

\label{P:compare4}
\begin{table}[!ht]
\centerline{Inter-relationships between various ``exactly solvable'' potentials}
\bigskip
\hskip-1cm
{\small
\begin{tabular}{|  l  |  l  |  l |}
\hline
Name & Potential $V(x)$ & Properties \\
\hline
\hline
Eckart $(1930)$ & $- {A \xi \over 1 - \xi} - {B \xi \over (1 - \xi)^2};$ & $\mathrm{Eckart} \Leftrightarrow \hbox{Rosen--Morse}$ \\
& where $\quad \xi = - \exp(2x/a)$ & \\
\hline
Rosen--Morse $(1932)$ & $B \, \mathrm{tanh}(x/d) - C \, \mathrm{sech}^2(x/d)$ & $\hbox{Poeschl--Teller} \Leftrightarrow \hbox{Rosen--Morse}$ \\
\hline
Poeschl--Teller $(1933)$ & $V_0 \, \mathrm{cosh}^2 \mu$ & $\hbox{Rosen--Morse} \; 
\Leftrightarrow \hbox{Poeschl--Teller}$ \\
    &  $\times  \{\mathrm{tanh}([x - \mu L]/L) + \mathrm{tanh} \, \mu\}^2$ &  \\
\hline
Eckart--Rosen--Morse-& $A + B \, \mathrm{tanh} (x/a + \theta)$ &  This is generally a Mobius function  \\
-Poeschl--Teller  &  $+ C \, \mathrm{tanh}^2 (x/a + \theta)$  & of the variable  $\exp(-2 x/a)$  \\
\hline
Mobius & $  V_1 \bigg[{A + B \, \exp(-2x/a) \over C + D \, \exp(-2x/a)}\bigg]^2 $ & The ``best'' of these equivalent forms \\
\hline
Morse $(1929)$ & $V_0 \, (1 - \exp (- [x - x_0]/a))^2$   & Specific limit of the Mobius potential \\
\hline
Manning--Rosen $(1933)$ & $B \, \mathrm{coth}(x/d) - C \, \mathrm{cosech}^2 (x/d)$ & Obtained from Eckart by specific \\
& &
substitution  \\
\hline
Hulthen $(1942)$ & $V_0 \, {\exp(-x/a) \over 1 - \exp(-x/a)}$ & A special case of Manning--Rosen \\
&  &
$(A = 0)$  \\
\hline
Tietz $(1963)$ &  $V_0 \bigg({\mathrm{sinh}([x - x_0]/a) \over \{\mathrm{sinh, cosh, \exp}\} (x/a)}\bigg)^2$ & Obtained from the general Mobius form \\
&   & of Eckart by appropriately choosing \\
&      & the parameters\\
\hline
Hua $(1990)$ & $V_0 \bigg({1 - \exp(-2x/a) \over 1 - q \, \exp(-2 x/a)}\bigg)^2$  & $\hbox{Eckart  or  Manning--Rosen}$ \\
\hline
\hline
\end{tabular}
} 
\caption[Inter-relationships between selected ``exactly solvable'' potentials]{\label{T:compare-4} This table shows the inter-connections between many ``exactly solvable'' potentials. Many of these potentials are \emph{identical} to each other, though this may not always be obvious at first glance.} 

\bigskip
\end{table}


\chapter{Shabat--Zakharov systems}
\label{C:consistency-4}
\section{Introduction}
In this chapter we shall present the general concept of  the so-called ``Shabat--Zakharov systems'', (sometimes called ``Zakharov--Shabat'' systems). We shall re-write the second-order Schr\"odinger equation as a particular set of two coupled first order differential equations for which bounds  can be (relatively) easily established. 
In addition, we shall introduce the idea of the probability current  and demonstrate how to obtain the probability current density.  We shall then use the probability current (or probability flux) to describe the flow of probability density. 

Moreover, we shall present an ``auxiliary condition'' or  ``gauge condition'' that is used to relate two complex amplitudes $a(x)$ and $b(x)$ that we shall soon introduce, and to eliminate $\d a/\d x$ in favour of $\d b/\d x$. This allows us to write $d^2 \psi/\d x^2$ in either of two equivalent forms, which is key to developing a $2\times2$ matrix formalism. We shall represent the wave function in an inner product form which is the explicit general (but formal) solution to the Schr\"odinger equation. The general solution depends on three arbitrarily chosen functions $\varphi(x)$, $\Delta(x)$, and $\chi(x)$,  and a path-ordered exponential matrix.  

We shall consider path ordering as an ``elementary'' process to derive the holy grail of ODE theory (complete quadrature, albeit formal, of the second-order linear ODE). We shall then use the freedom to independently choose  $\varphi(x)$, $\Delta(x)$, and $\chi(x)$ to simplify the Bogoliubov coefficients (both the relevant ODEs and the general bounds) as much as possible.

\section{Ansatz}

Consider the one-dimensional time-independent Schr\"odinger equation~\cite{Landau}--\cite{Messiah}
\begin{equation}
\label{E:SDE}
-{\hbar^2\over2m} {\d^2\over \d x^2} \psi(x) + V(x) \; \psi(x) = E \; \psi(x).
\end{equation}
Introduce the notation
\begin{equation}
\label{E:k(x)}
k(x)^2 = {2m[E-V(x)]\over\hbar^2}.
\end{equation}
So we are really just trying to solve
\begin{equation}
 {\d^2\over \d x^2} \psi(x) + k(x)^2 \; \psi(x) =0 ,
\end{equation}
or equivalently in the time domain
\begin{equation}
 {\d^2\over \d t^2} \psi(t) + \omega(t)^2 \; \psi(t) =0 .
\end{equation}
Motivated by the JWKB approximation,
\begin{equation}
\label{E:JWKB}
\psi \approx A \; {\exp[i\int k(x)]\over \sqrt{k(x)}} 
+ B \; {\exp[-i\int k(x)]\over \sqrt{k(x)}},
\end{equation}
the key idea is to re-write the second-order Schr\"odinger
equation as a set of two coupled first-order linear differential equations (for which bounds can
relatively easily be established).
\begin{table}[!htb]
\begin{center}
 \setlength{\fboxsep}{0.45 cm} 
 \framebox{\parbox[t]{12.5cm}{
{\bf WKB (or JWKB) approximation}: In general, WKB (or JWKB) theory is a method for estimating the solution of a differential equation whose highest derivative is multiplied by a small parameter $\epsilon$. The method of approximation is as follows~\cite{wkb_math}:

For a differential equation
\begin{equation}
\epsilon \, {\d^n y \over \d x^n} + a(x) \, {\d^{n-1} \over \d x^{n-1}} + ... +k(x) \, {\d y \over \d x} + m(x) y = 0, 
\end{equation}
assume a solution of the form of an asymptotic series expansion as
\begin{equation}
y(x) \approx \exp\bigg[{1 \over \epsilon} \, \sum_{n=0}^{\infty} \epsilon^n \, S_{n}(x)\bigg].
\end{equation}
In the limit $\epsilon \rightarrow 0$, substitution of the above ansatz into the differential equation and canceling out the exponential terms allows one to solve for an arbitrary number of terms $S_{n}(x)$ in the expansion. 

The most common application of this general formalism is to the second-order differential equation presented in the standard form: 
\begin{equation}
\epsilon \, {\d^2 y \over \d x^2} + m(x) y = 0.
\end{equation}
Keeping only the first two terms in the WKB approximation yields
\begin{equation}
y(x) \approx {\exp\left\{  \pm i \int \sqrt{m(x)} \; \d x\right\} \over \sqrt[4]{m(x)}}.
\end{equation}
It is this ``standard'' form of the WKB approximation that we will be using extensively in this thesis.

}}
\end{center}
\end{table}
Systems of differential equations of this type are often referred to as 
Shabat--Zakharov~\cite{Eckhaus} systems. 
A similar representation of the Schr\"odinger
equation is briefly discussed by Peierls~\cite{Peierls} and related
representations are well-known, often being used without giving
an explicit reference (see \emph{e.g.}~\cite{Bordag}).  However an
exhaustive search has not uncovered prior use of the particular
representation presented here, (apart, of course, from~\cite{bounds1}),
nor the idea of using the representation to place bounds on
one-dimensional scattering.

We will start by introducing two arbitrary auxiliary functions $\varphi(x)$ and $\Delta(x)$
which may be either real or complex, though we do demand that
$\varphi'(x)\neq 0$, and then defining
\begin{equation}
\label{E:representation}
\psi(x) = 
a(x) \;{\exp(+i \varphi+i\Delta)\over\sqrt{\varphi'}} + 
b(x) \;{\exp(-i \varphi-i\Delta)\over\sqrt{\varphi'}}.
\end{equation}
This representation effectively seeks to use quantities resembling the ``phase integral'' wavefunctions as a basis for the true wavefunction~\cite{Froman}. We will ultimately want to interpret $a(x)$ and $b(x)$ as  ``position-dependent WKB-like coefficients''; in a scattering problem they can be thought of as
``position-dependent Bogoliubov coefficients''. This representation is of course extremely highly redundant, since one complex number $\psi(x)$ has been traded for two complex numbers $a(x)$ and $b(x)$, plus two essentially arbitrary auxiliary functions $\varphi(x)$ and $\Delta(x)$.  

To reduce this freedom, we introduce an ``auxiliary condition'' (or ``auxiliary constraint'', or ``gauge condition''):
\begin{equation}
\label{E:gauge-shabat}
{\d\over \d x}\left({a \exp(i \Delta) \over\sqrt{\varphi'}}\right) e^{+i \varphi} + 
{\d\over \d x}\left({b \exp(-i\Delta) \over\sqrt{\varphi'}}\right) e^{-i \varphi} 
= \chi(x) \; \psi(x).
\end{equation}
Here $\chi(x)$ is yet a third arbitrary function of position. It is allowed
to be complex, and may be zero. The original analysis, published in~\cite{bounds1} corresponds to the
special case $\Delta(x)=0$ and $\chi(x)=0$. Subject to this gauge condition,
\begin{equation}
\label{E:gradient}
{\d\psi\over \d x} = i \sqrt{\varphi'} 
\left\{ a(x) \exp(+i \varphi+i\Delta) - b(x) \exp(-i \varphi-i\Delta ) \right\} + \chi\; \psi.
\end{equation}
Repeated differentiation of this equation will soon lead to our desired result.

\begin{table}[!htb]
\begin{center}
 \setlength{\fboxsep}{0.45 cm} 
 \framebox{\parbox[t]{12.5cm}{
{\bf Definition (Zakharov--Shabat)}: We shall now turn to the study of the non self adjoint Zakharov--Shabat problem. In a rather general context, this may be viewed  as the first order system of differential equations in $\mathcal{R}$ defined by~\cite{Zakharov-Shabat}:
\begin{eqnarray}
\label{Zakharov-Shabat}
&& h \left(\begin{array}{c}
u'_{1}(x,z)\\
u'_{2}(x,z)
\end{array} \right) =
\nonumber
\\
&&\left(
\begin{array}{cc}
-iz& A(x) \exp(i S(x)/h)\\
-A(x) \exp(-iS(x)/h)&iz
\end{array}
\right)
\left(\begin{array}{c}
u_{1}(x,z)\\
u_{2}(x,z)
\end{array}
\right),
\nonumber
\\
&&
\end{eqnarray}
where $z$ is a complex eigenvalue parameter, and the prime denotes differentiation with respect to $x$.
This system appears (for instance) when one wants to solve the non linear initial value problem given by
\begin{equation}
\label{non-linear-eqn}
i h \partial_{t} \psi + {h^2 \over 2} \, \partial_{x}^{2} \psi + |\psi|^2 \, \psi = 0,
\end{equation}
\begin{equation}
\psi_{|t=0} = A(x) \, \exp(i S(x)/h) \, ,
\end{equation} 
with the inverse scattering method~\cite{Zakharov-Shabat}.

The same sort of system of ODEs will also arise in our particular problem: We are of course interested in the somewhat simpler problem of solving the \emph{linear} Schr\"odinger equation. 
}}
\end{center}
\end{table}
\section{Probability current}
We use the probability current (or probability flux) to describe the flow of probability density.
The following equation is use to describe the flow of a fluid, a process which occurs all the way through physics and applied mathematics~\cite{Probability-current-flux-density}
\begin{equation}
{\partial \over \partial t} \rho (x, t) = - {\partial \mathscr{J}\over \partial x}.
\end{equation}
The above equation tells us that the rate of change in density is equal to the negative of the difference between the amount of ``stuff''  (be it water, air, or ``probability'') flowing into the point and the amount flowing out.
Now in the context of the Schr\"odinger equation, we can write the probability density in the following form
\begin{equation}
\rho(x,t) = |\psi(x,t)|^2 = \psi^* \psi.
\end{equation}
In quantum mechanics we can certainly describe the movement of particles, however we also have an additional difficulty because our particles are not classical. Therefore we can only talk about the probability of the particle being at a certain place in time, now we can talk about a probability current or flux.
Consider the time--dependent Schr\"odinger Equation and its complex conjugate:
\begin{eqnarray}
\label{prob1}
i \hbar \, {\partial \psi \over \partial t} &=& H \psi(x, t) = - {\hbar^2 \over 2 m} \, {\partial^2 \psi \over \partial x^2} + V \psi;
\\
\label{prob2}
- i \hbar \, {\partial \psi^* \over \partial t} &=& H \psi^*(x, t) = - {\hbar^2 \over 2 m} \, {\partial^2 \psi^* \over \partial x^2} + V \psi^*.
\end{eqnarray}

Now let us differentiate the probability density with respect to time
\begin{eqnarray}
\nonumber
{\partial \rho \over \partial t} &=& {\partial \over \partial t} \, \bigg(\psi^* \psi\bigg) = \bigg({\partial \psi^* \over \partial t}\bigg) \, \psi + \psi^* \bigg({\partial \psi \over \partial t} \bigg),
\nonumber
\\
&=& {\hbar \over 2 m \, i} \,
\bigg({\partial^2 \psi^* \over \partial x^2} \, \psi - \psi^* \, {\partial^2 \psi \over \partial x^2}\bigg)
- {V \psi^* \psi \over i \hbar} +  {V \psi^* \psi \over i \hbar},
\nonumber
\\
&=& 
-{\partial \over \partial x} \bigg(\, {\hbar \over 2 m \, i}  
\bigg(\psi^* \, {\partial \psi \over \partial x} - {\partial  \psi^* \over \partial x} \, \psi \bigg) \bigg).
\end{eqnarray}
Apply
\begin{equation}
\label{conservation}
{\partial \over \partial t} \rho  (x, t) + {\partial \over \partial x}  \mathscr{J}(x, t) = 0.
\end{equation}
So we obtain the probability current density
\begin{equation}
\label{newdensity}
\mathscr{J} (x,t) = {\hbar \over 2 m  \, i} \, \bigg(\psi^* {\partial \psi \over \partial x} - {\partial \psi^* \over \partial x} \, \psi\bigg).
\end{equation}

We can re-write this as
\begin{equation}
\mathscr{J} (x,t) = {\hbar \over  m} \, \mathrm{Im} \bigg(\psi^* {\partial \psi \over \partial x}\bigg).
\end{equation}
Here $(\hbar / m)$ is just a normalization (that is often set $\rightarrow 1$ for 
convenience). There is nothing really important in this normalization (unless we want to calculate 
experimental numbers), so we might as well set
\begin{equation}
\mathscr{J} (x,t) =  \mathrm{Im} \bigg(\psi^* {\partial \psi \over \partial x}\bigg).
\end{equation}
Now at this stage $\varphi(x)$ and $\chi(x)$ are completely arbitrary possibly complex functions subject only to the constraint $\varphi' \neq 0$. For future use, compute the probability current using our WKB-based ansatz in terms of $a(x)$ and $b(x)$:
\begin{eqnarray}
\nonumber
\mathscr{J} &=& \mathrm{Im} \, \bigg\{ \psi^* {\d \psi \over \d x}\bigg\} ,
\\
&=& \mathrm{Im} \, \bigg\{ \psi^*\left[i \sqrt{\varphi'}\{a(x) \exp (+i \varphi) - b(x) \exp (-i \varphi)\} + \chi \, \psi \right] \bigg\} ,
\nonumber
\\
&&
\\
&=& \mathrm{Re} \, \Bigg\{\sqrt{{\varphi' \over \varphi'^*}}[a(x) \exp (+i \varphi) - b(x) \exp (-i \varphi)]
\nonumber
\\
&&
[a(x)^* \exp (-i \varphi^*) + b(x)^* \exp (+ i \varphi^*)] \Bigg\} 
+ 
\mathrm{Im} \{\chi\} \, \psi^* \psi,
\\
&=& \mathrm{Re} \, \Bigg\{\sqrt{{\varphi' \over \varphi'^*}}\Bigg\} \left[|a|^2 \mathrm{Re}\{e^{+i(\varphi-\varphi^*)}\} - |b|^2 \mathrm{Re}\{e^{-i(\varphi-\varphi^*)}\}\right] 
\nonumber\\ &&
+ \mathrm{Im} \, \Bigg\{\sqrt{\varphi' \over \varphi'^*} \Bigg\} \mathrm{Im}\left\{ab^* \, e^{i(\varphi + \varphi^*)}\right\} + \mathrm{Im} \{ \chi\} \, \psi^* \psi  ,
\\
&=& \mathrm{Re} \, \Bigg\{\sqrt{{\varphi' \over \varphi'^*}}\Bigg\} \left[|a|^2\mathrm{Re} \{e^{+2 \mathrm{Im}(\varphi)}\} - |b|^2 \mathrm{Re}\{e^{-2 \mathrm{Im}(\varphi)}\}\right] 
\nonumber\\ &&
+ \mathrm{Im} \Bigg\{\sqrt{\varphi' \over \varphi'^*}\Bigg\} \, \mathrm{Im} \{a b ^* \, e^{2 i \, \mathrm{Re}(\varphi)}\} + \mathrm{Im} \{\chi\} \, \psi^* \psi.
\end{eqnarray}
\section{SDE as a first order system}
%
We now re-write the Schr\"odinger equation in terms of two coupled
first-order differential equations for these position-dependent
WKB/Bogoliubov coefficients  $a(x)$ and $b(x)$. To do this we evaluate $\d^2\psi/ \d x^2$ in two different ways, making
repeated use of the gauge condition. First
\begin{eqnarray}
\label{E:double-gradient}
{\d^2\psi\over \d x^2} 
&=& 
{\d\over \d x} 
\left( i{\varphi'\over \sqrt{\varphi'}} 
\left\{ a e^{+i \varphi+i\Delta} - b e^{-i \varphi-i\Delta} \right\}  + \chi\;\psi
\right),
\\
&=& {(i\varphi')^2\over\sqrt{\varphi'}} 
\left\{ a e^{+i \varphi+i\Delta} + b e^{-i \varphi-i\Delta} \right\}
\nonumber\\
&&+ i \varphi'
\left\{ 
{\d\over \d x}\left({a\,e^{i\Delta}\over\sqrt{\varphi'}}\right) e^{+i \varphi} - 
{\d\over \d x}\left({b\, e^{-i\Delta}\over\sqrt{\varphi'}}\right) e^{-i \varphi} 
\right\} 
\nonumber\\
&&
+i{\varphi''\over\sqrt{\varphi'}} 
\left\{ a e^{+i \varphi+i\Delta} - b e^{-i \varphi-i\Delta} \right\}
+ \chi' \; \psi + \chi\; \psi' ,
\\
&=& -{\varphi'^2\over\sqrt{\varphi'}}  
\left\{ a e^{+i \varphi+i\Delta} + b e^{-i \varphi-i\Delta} \right\}
\nonumber\\
&&+i\varphi' \left\{ 2
{\d\over \d x}\left({a\,e^{i\Delta}\over\sqrt{\varphi'}}\right) e^{+i \varphi} - \chi \psi 
\right\}
\nonumber\\
&&
+i{\varphi''\over\sqrt{\varphi'}} 
\left\{ a e^{+i \varphi+i\Delta} - b e^{-i \varphi-i\Delta} \right\} + \chi' \; \psi + \chi\; \psi'.
\end{eqnarray}
But then
\begin{eqnarray}
\label{E:double-gradient-b}
{\d^2\psi\over \d x^2} 
&=& -\varphi'^2 \; \psi
+{2 i\varphi'\over\sqrt{\varphi'}}  {\d a\over \d x} e^{+i \varphi+i\Delta} 
-2\sqrt{\varphi'} \Delta' e^{i\varphi+i\Delta} a 
\nonumber\\
&& - i{\varphi''\over\sqrt{\varphi'}} b e^{-i \varphi-i\Delta} 
-i \varphi' \chi \psi
+ \chi' \psi 
\nonumber\\
&& + \chi\; \left[ i \sqrt{\varphi'} 
\left\{ a(x) e^{+i \varphi+i\Delta} - b(x) e^{-i \varphi-i\Delta} \right\} + \chi\; \psi \right].
\end{eqnarray}
So finally
\begin{eqnarray}
\label{E:double-gradient-c}
{\d^2\psi\over \d x^2} 
&=& \left[\chi^2+\chi'-(\varphi')^2\right] \psi
+{2 i\varphi'\over\sqrt{\varphi'}}  {\d a\over \d x} e^{+i \varphi+i\Delta} 
\nonumber\\
&&-2\sqrt{\varphi'} \Delta' e^{i\varphi+i\Delta} a 
-i{[\varphi''+2\chi\varphi']\over\sqrt{\varphi'}} b e^{-i \varphi-i\Delta}.
\nonumber\\
&&
\end{eqnarray}
Now use the gauge condition to eliminate $\d a/\d x$ in favour of $\d
b/\d x$.  This permits us to write $\d^2\psi/ \d x^2$ in
either of the two equivalent forms
\begin{eqnarray}
\label{E:double-gradient2}
{\d^2\psi\over \d x^2} 
&=& 
\left[\chi^2+\chi'-(\varphi')^2\right] \psi 
- 2 i \varphi' {\d b\over \d x} {e^{-i\varphi-i\Delta}\over\sqrt{\varphi'}} 
\nonumber \\
&&
-2\sqrt{\varphi'} \Delta' e^{-i\varphi-i\Delta} b 
+ i \left[ \varphi''+2\chi \varphi' \right] a {e^{+i\varphi+i\Delta}\over\sqrt{\varphi'}},
\end{eqnarray}
and
\begin{eqnarray}
\label{E:double-gradient2b}
{\d^2\psi\over \d x^2} 
&=& 
\left[\chi^2+\chi'-(\varphi')^2\right] \psi 
+ 2 i \varphi' {\d a\over \d x} {e^{+i\varphi+i\Delta}\over\sqrt{\varphi'}}
\nonumber
\\
&&
-2\sqrt{\varphi'} \Delta' e^{i\varphi+i\Delta} a 
- i \left[ \varphi''+2\chi \varphi' \right] b {e^{-i\varphi-i\Delta}\over\sqrt{\varphi'}}. 
\end{eqnarray}
Now insert these formulae into the Schr\"odinger equation written in the
form
\begin{equation}
{\d^2\psi\over \d x^2} + k(x)^2 \; \psi  = 0,
\end{equation}
to deduce the first-order system:
\begin{eqnarray}
\label{E:system-a}
\!\!\!
{\d a\over \d x} &=& + {1\over 2\varphi'} \Bigg\{ 
i\left[k^2(x)+\chi^2+\chi'-(\varphi')^2-2\varphi'\Delta'\right]  \; a
\nonumber \\
&&
+\left([\varphi''+2\chi\varphi']+ i\left[k^2(x)+\chi^2+\chi'-(\varphi')^2\right]\right) \; e^{-2i\varphi-2i\Delta} \; b
\Bigg\},\;
\\
\label{E:system-b}
\!\!\!
{\d b\over \d x} &=&  +{1\over 2\varphi'} \Bigg\{ 
\left( [\varphi''+2\chi\varphi'] - i\left[k^2(x)+\chi^2+\chi'-(\varphi')^2\right]  \right)\; 
e^{+2i\varphi+2i\Delta}  \; a
\nonumber \\
&&
- i\left[k^2(x)+\chi^2+\chi'-(\varphi')^2-2\varphi'\Delta'\right]  \; b
\Bigg\}.
\end{eqnarray}
It is easy to verify that this first-order system is compatible with
the ``gauge condition'' (\ref{E:gauge-shabat}), and that by iterating the
system twice (subject to this gauge condition) one recovers exactly
the original Schr\"odinger equation. These equations hold for arbitrary
$\varphi(x)$, $\Delta(x)$, and $\chi(x)$, real or complex. 

When written in $2\times2$ matrix
form, these equations exhibit a deep connection with the transfer matrix
formalism~\cite{Transfer}.  Let us define quantities
$\rho_1(x)$ and $\rho_2(x)$, not necessarily real, as
\begin{equation}
\rho_1 = \varphi''+2\chi \varphi';
\qquad
\rho_2 = k^2(x)+\chi^2+\chi'-(\varphi')^2.
\end{equation}
We can then re-write the Shabat--Zakharov system in matrix form as
\begin{equation}
{\d\over \d x} \left[\begin{matrix} a \\ b\end{matrix}\right] = 
{1\over2\varphi'} 
\left[
\begin{matrix}
i [\rho_2-2\phi'\Delta']   & 
\left\{ \rho_1+i\rho_2\right\} \exp(-2i\varphi-2i\Delta) \\
\left\{ \rho_1-i\rho_2\right\}  \exp(+2i\varphi+2i\Delta)  & 
-i[\rho_2-2\phi'\Delta]
\end{matrix}
\right]
\left[ \begin{matrix} a \cr b\end{matrix}\right].
\end{equation}
This has the formal solution
\begin{equation}
\left[\begin{matrix} a(x_f) \cr b(x_f)\end{matrix}\right] = 
E(x_f,x_i)  \;\left[\begin{matrix} a(x_i) \cr b(x_i)\end{matrix}\right],
\end{equation}
in terms of a generalized position-dependent ``transfer
matrix''~\cite{Transfer}
\begin{eqnarray}
&&
E(x_f,x_i) = 
\nonumber
\\
&&
{\cal P} \exp\left( 
\int_{x_i}^{x_f}
{1\over2\varphi'} 
\left[
\begin{matrix}
i [\rho_2-2\varphi'\Delta']   & 
\left\{ \rho_1+i\rho_2 \right\} \; e^{-2i\varphi-2i\Delta} \cr 
\left\{ \rho_1-i\rho_2 \right\}  \; e^{+2i\varphi+2i\Delta}  & 
-i[\rho_2-2\varphi'\Delta']
\end{matrix}
\right] 
 \d x \right),
\nonumber
\\
&&
\end{eqnarray}
where the symbol ${\cal P}$ denotes ``path ordering''. 

\begin{center}
 \setlength{\fboxsep}{0.45 cm} 
 \framebox{\parbox[t]{12.5cm}{
{\bf Path ordering:}
In theoretical physics, path--ordering is the procedure (or meta-operator ${\cal P}$) of ordering a product of many operators according to the value of one chosen parameter~\cite{path-ordering}:
\begin{equation}
{\cal P} [O_1(\sigma_1)O_2(\sigma_2)...O_n(\sigma_n)]=  O_{p(1)} (\sigma_{p(1)})O_{p(2)} (\sigma_{p(2)})...O_{p(n)} (\sigma_{p(n)}).
\end{equation}
Here $p: \{1,2,...,n\} \rightarrow \{1,2, ...,n\}$ is a permutation that orders the parameters: $\sigma_{p(1)} \leq \sigma_{p(2)} \leq...\leq \sigma_{p(n)}$.
For instance
\begin{equation}
{\cal P}[O_1(4) O_2(2) O_3 (3) O_4(1)]: = O_4(1) O_3 (3) O_2(2) O_1(4).
\end{equation}
}}
\end{center}

Equivalently if we were to be working in the time domain we would have 
\begin{eqnarray}
&&
E(t_f,t_i) = 
\nonumber
\\
&&
{\cal T} \exp\left( 
\int_{t_i}^{t_f}
{1\over2\varphi'} 
\left[
\begin{matrix}
i [\rho_2-2\varphi'\Delta']   & 
\left\{ \rho_1+i\rho_2 \right\} \; e^{-2i\varphi-2i\Delta} \cr 
\left\{ \rho_1-i\rho_2 \right\}  \; e^{+2i\varphi+2i\Delta}  & 
-i[\rho_2-2\varphi'\Delta']
\end{matrix}
\right] 
 \d t \right),
\nonumber
\\
&&
\end{eqnarray}
where ${\cal T}$ would now be the well-known ``time ordering'' operator (more usually encountered in quantum field theory) and we would now define
\begin{equation}
\rho_1 = \varphi''+2\chi \varphi';
\qquad
\rho_2 = \omega^2(t)+\chi^2-\chi'-(\varphi')^2,
\end{equation}
with $\varphi$, $\Delta$, and $\chi$ now being arbitrary functions of $t$ rather than $x$.

We can now write the wave function in
inner product form
\begin{equation}
\psi(x) =   
{1\over\sqrt{\varphi'}}
\left[\begin{matrix}
{\exp(+i \varphi+i\Delta)}; & 
{\exp(-i \varphi-i\Delta)}
\end{matrix} \right] \;
\left[\begin{matrix} a(x_i) \cr b(x_i)\end{matrix}\right],
\end{equation}
to yield a \emph{formal but completely general} solution for the Schr\"odinger
equation
\begin{equation}
\psi(x) = {1\over\sqrt{\varphi'}} 
\left[\begin{matrix}
{\exp(+i \varphi+i\Delta)}; & 
{\exp(-i \varphi-i\Delta)}
\end{matrix} \right] 
E(x,x_0)  
\left[\begin{matrix} a(x_0) \cr b(x_0)\end{matrix}\right].
\end{equation}
Explicitly
\begin{eqnarray}
&&
\psi(x) = 
{1\over\sqrt{\varphi'}}
\left[\begin{matrix}
{\exp(+i \varphi+i\Delta)}; & 
{\exp(-i \varphi+i\Delta)}
\end{matrix} \right]
\\
&&
{\cal P} \exp\left( 
\int_{x_0}^{x}
{1\over2\varphi'} 
\left[
\begin{matrix}
i [\rho_2-2\varphi'\Delta']   & 
\left\{ \rho_1+i\rho_2 \right\} \; e^{-2i\varphi-2i\Delta} \cr 
\left\{ \rho_1-i\rho_2 \right\}  \; e^{+2i\varphi+2i\Delta}  & 
-i[\rho_2-2\varphi'\Delta']
\end{matrix}
\right] 
\d \bar x \right)
\left[\begin{matrix} a(x_0) \cr b(x_0)\end{matrix}\right].
\nonumber
\end{eqnarray}
This is the explicit general solution to the Schr\"odinger equation. It
depends on the three arbitrarily chosen functions $\varphi(x)$, $\Delta(x)$, and
$\chi(x)$,  and a path-ordered exponential matrix.  If you consider path
ordering to be an ``elementary'' process, then this is indeed the holy grail
of ODE theory (complete quadrature, albeit formal, of the
second-order linear ODE).

\section{Bounding the coefficients $a(x)$ and $b(x)$}
From
\begin{eqnarray}
\label{E:system-a3}
{\d a\over \d x} &=& +
{1\over 2\varphi'} 
\Bigg\{ 
i\left[k^2(x)+\chi^2+\chi'-(\varphi')^2-2\varphi'\Delta'\right]  \; a
\nonumber\\
&&
\qquad
+
\left([\varphi''+2\chi\varphi']+ i\left[k^2(x)+\chi^2+\chi'-(\varphi')^2\right]\right) \; e^{-2i\varphi-2i\Delta} \; b
\Bigg\},
\nonumber
\\
&&
\\
\label{E:system-b3}
{\d b\over \d x} &=& +
{1\over 2\varphi'} 
\Bigg\{ 
\left( [\varphi''+2\chi\varphi'] - i\left[k^2(x)+\chi^2+\chi'-(\varphi')^2\right]  \right)\; 
e^{+2i\varphi+2i\Delta}  \; a
\nonumber\\
&&
\qquad
- i\left[k^2(x)+\chi^2+\chi'-(\varphi')^2-2\varphi'\Delta'\right]  \; b
\Bigg\},
\end{eqnarray}
we see
\begin{eqnarray}
a^* {\d a\over \d x} &=& +
{1\over 2\varphi'} 
\Bigg\{ 
i\left[k^2(x)+\chi^2+\chi'-(\varphi')^2-2\varphi'\Delta'\right]  \; a^* a
\nonumber\\
&&
\quad
+
\left([\varphi''+2\chi\varphi']+ i\left[k^2(x)+\chi^2+\chi'-(\varphi')^2\right]\right) \; e^{-2i\varphi-2i\Delta} \; a^* \; b
\Bigg\}.
\nonumber
\\
&&
\end{eqnarray}
Therefore (now assuming for the rest of this section that $\varphi$, $\Delta$, $\chi$ are all real and $\varphi' >0$)
\begin{equation}
a^* {\d a\over \d x} + a {\d a^*\over \d x} = {\Re\{ \left([\varphi''+2\chi\varphi']+ i\left[k^2(x)+\chi^2+\chi'-(\varphi')^2\right]\right) \; e^{-2i\varphi-2i\Delta} \; a^* \; b \}\over\varphi'}.
\end{equation}
This implies
\begin{equation}
{\d |a|^2\over \d x} =  {\Re\{ \left([\varphi''+2\chi\varphi']+ i\left[k^2(x)+\chi^2+\chi'-(\varphi')^2\right]\right) \; e^{-2i\varphi-2i\Delta} \; a^* \; b \}\over\varphi'}.
\end{equation}
But $\Re(A)\leq |A|$, so we have
\begin{equation}
{\d |a|^2\over \d x} \leq  {\left| \left([\varphi''+2\chi\varphi']+ i\left[k^2(x)+\chi^2+\chi'-(\varphi')^2\right]\right) \; e^{-2i\varphi-2i\Delta} \; a^* \; b \right|\over\varphi'},
\end{equation}
implying
\begin{equation}
2 |a| {\d |a|\over \d x} \leq  {\left| \; [\varphi''+2\chi\varphi']+ i\left[k^2(x)+\chi^2+\chi'-(\varphi')^2\right] \; \right| \;  |a| \; |b| \over\varphi'}.
\end{equation}
Thus
\begin{equation}
{\d |a|\over \d x} \leq  {\left| \; [\varphi''+2\chi\varphi']+ i\left[k^2(x)+\chi^2+\chi'-(\varphi')^2\right] \; \right| \; |b| \over2\varphi'},
\end{equation}
and so
\begin{equation}
{\d |a|\over \d x} \leq  {\sqrt{ [\varphi''+2\chi\varphi']^2 + \left[k^2(x)+\chi^2+\chi'-(\varphi')^2\right]^2} \over2\varphi'}  \; |b|.
\end{equation}
Now if  (as per our current assumption) $\varphi$, $\Delta$, $\chi$ are all real, then it is easy to check that
\begin{equation}
\mathscr{J} = \Im(\psi^* \psi') = |a|^2- |b|^2,
\end{equation}
so current conservation implies
\begin{equation}
\label{the-current-conservation}
|a|^2- |b|^2 = 1.
\end{equation}
Ultimately, it is this equation that allows us to interpret $a(x)$ and $b(x)$ as ``position dependent Bogoliubov coefficients''.

In view of the relation between $a(x)$ and $b(x)$  we have $|b| = \sqrt{|a|^2-1}$, so that we can deduce
\begin{equation}
{\d |a|\over \d x} \leq  {\sqrt{ [\varphi''+2\chi\varphi']^2 + \left[k^2(x)+\chi^2+\chi'-(\varphi')^2\right]^2} \over2\varphi'} \; \sqrt{|a|^2-1}.
\end{equation}
But this inequality can now be integrated. For convenience let us define
\begin{equation}
\label{E:theta}
\vartheta =  {\sqrt{ [\varphi''+2\chi\varphi']^2 + \left[k^2(x)+\chi^2+\chi'-(\varphi')^2\right]^2} \over2\varphi'}.
\end{equation}
Then 
\begin{equation}
{\d |a|\over \d x} \leq \vartheta \; \sqrt{|a|^2-1}.
\end{equation}
But now
\begin{equation}
\int {1\over  \sqrt{|a|^2-1} } \; {\d |a|\over \d x} \; \d x \leq \int \vartheta \; \d x,
\end{equation}
so that
\begin{equation}
\left\{ \cosh^{-1} |a| \right\}_{x_i}^{x_f}  \leq \int_{x_i}^{x_f}  \vartheta \; \d x.
\end{equation}
Now apply the boundary conditions: as $x_i\to-\infty$ we have chosen to set things up so that we have a pure transmitted wave, so $|b(-\infty)|=0$ and $|a(-\infty)|=1$.  On the other hand as  $x_f\to+\infty$ we have chosen to set things up so that $a(x)$ and $b(x)$ tend to $\alpha$ and $\beta$, the Bogoliubov coefficients we are interested in calculating. Thus taking the double limit  $x_i\to-\infty$  and  $x_f\to+\infty$  we see:
\begin{equation}
\cosh^{-1} |\alpha|  \leq \int_{-\infty}^{+\infty}  \vartheta \; \d x.
\end{equation}
That is
\begin{equation}
\label{E:alpha}
|\alpha|  \leq \cosh \left\{ \int_{-\infty}^{+\infty}  \vartheta \; \d x\right\} .
\end{equation}
This is the central result of this thesis --- it can be modified and rearranged in a number of ways, and related inequalities can be derived under slightly different hypotheses, but all the applications we are interested in will reduce in one way or another to an application of this inequality or one of its close variants.

For notational convenience, we often find it is useful to adopt the shorthand
\begin{equation}
\oint   =  \int_{-\infty}^{+\infty} ,
\end{equation}
since then
\begin{equation}
|\alpha|  \leq \cosh \left\{ \oint  \vartheta \; \d x\right\} .
\end{equation}
From the normalization condition (\ref{the-current-conservation}) we immediately deduce
\begin{equation}
|\beta|  \leq \sinh \left\{ \oint  \vartheta \; \d x\right\} .
\end{equation}
When translated into equivalent statements about transmission an reflection probabilities, we find
\begin{equation}
T  \geq \sech^2 \left\{ \oint  \vartheta \; \d x\right\} ,
\end{equation}
and
\begin{equation}
R  \leq \tanh^2 \left\{ \oint  \vartheta \; \d x\right\} .
\end{equation}

Now one of the the points of the exercise (and of this thesis) is to use the freedom to independently choose  $\varphi(x)$, $\Delta(x)$, and $\chi(x)$ to simplify life as much as possible. Here are a few special cases, chosen for their simplicity and the lessons they teach us.

\subsection{Case: $\Delta' = \rho_2/(2\varphi')$}

No one can prevent us from choosing
\begin{equation}
\Delta' = {\rho_2\over2\varphi'},
\end{equation}
that is
\begin{equation}
\Delta' = {k^2(x)+\chi^2+\chi'-(\varphi')^2\over2\varphi'},
\end{equation}
which implies
\begin{equation}
\Delta = \int {k^2(x)+\chi^2+\chi'-(\varphi')^2\over2\varphi'} \; \d x.
\end{equation}
Doing that greatly simplifies life since now the system of ODEs becomes
\begin{eqnarray}
\label{E:system-a1}
{\d a\over \d x} &=& +
{1\over 2\varphi'} 
\Bigg\{ 
\left([\varphi''+2\chi\varphi']+ i\left[k^2(x)+\chi^2+\chi'-(\varphi')^2\right]\right) \; e^{-2i\varphi-2i\Delta} \; b
\Bigg\},
\nonumber
\\
&&
\\
\label{E:system-b1}
{\d b\over \d x} &=& +
{1\over 2\varphi'} 
\Bigg\{ 
\left( [\varphi''+2\chi\varphi'] - i\left[k^2(x)+\chi^2+\chi'-(\varphi')^2\right]  \right)\; 
e^{+2i\varphi+2i\Delta}  \; a
\Bigg\}.
\nonumber
\\
&&
\end{eqnarray}
That is
\begin{equation}
{\d\over \d x} \left[\begin{matrix} a \\ b\end{matrix}\right] = 
{1\over2\varphi'} 
\left[
\begin{matrix}
0 & 
\left\{ \rho_1+i\rho_2\right\} \exp(-2i\varphi-2i\Delta) \\
\left\{ \rho_1-i\rho_2\right\}  \exp(+2i\varphi+2i\Delta)  & 
0
\end{matrix}
\right]
\left[ \begin{matrix} a \cr b\end{matrix}\right].
\end{equation}
If you now let $\varphi$ and $\chi$ be real, then the matrix above is Hermitian; unfortunately this does not tell us all that much about the evolution operator. (If we had $i$ times a Hermitian operator appearing above, then the evolution operator would have been unitary.) 
Using a specialization of the previous argument,
we can easily deduce
\begin{equation}
{\d |a| \over \d t} = {\sqrt{  [\varphi''+2\chi\varphi']^2 + \left[k^2(x)+\chi^2+\chi'-(\varphi')^2\right] ^2 }\over 2 |\varphi'| } \; |b|.
\end{equation}
We also have the same  constraint
\begin{equation}
|a|^2 - |b|^2 = 1,
\end{equation}
and so deduce
\begin{equation}
{\d |a| \over \d t} = {\sqrt{  [\varphi''+2\chi\varphi']^2 + \left[k^2(x)+\chi^2+\chi'-(\varphi')^2\right] ^2 }\over 2 |\varphi'| } \; \sqrt{ |a|^2-1}.
\end{equation}
Defining
\begin{equation}
\vartheta =  {\sqrt{[\varphi''+2\chi\varphi']^2 + \left[k^2(x)+\chi^2+\chi'-(\varphi')^2\right] ^2}\over 2
|\varphi'| },
\end{equation}
we have
\begin{equation}
{\d |a| \over \d t} = {\vartheta} \; \sqrt{ |a|^2-1}.
\end{equation}
This inequality can now be integrated in the manner discussed above, though doing so gives us no additional information.

\subsection{Case: $\Delta=-\varphi$}

No one can prevent us from choosing
\begin{equation}
\Delta(x) = -\varphi(x),
\end{equation}
in which case
\begin{eqnarray}
\label{E:system-a2}
{\d a\over \d x} &=& +
{1\over 2\varphi'} 
\Bigg\{ 
i\left[k^2(x)+\chi^2+\chi'+(\varphi')^2\right]  \; a
\nonumber\\
&&
+
\left([\varphi''+2\chi\varphi']+ i\left[k^2(x)+\chi^2+\chi'-(\varphi')^2\right]\right)  \; b
\Bigg\},
\\
\label{E:system-b2}
{\d b\over \d x} &=& +
{1\over 2\varphi'} 
\Bigg\{ 
\left( [\varphi''+2\chi\varphi'] - i\left[k^2(x)+\chi^2+\chi'-(\varphi')^2\right]  \right)\;   \; a
\nonumber\\
&&
- i\left[k^2(x)+\chi^2+\chi'+(\varphi')^2\right]  \; b
\Bigg\}.
\end{eqnarray}
The complicated phase structure has gone away, and we have 
\begin{eqnarray}
\!\!\!
a^* {\d a\over \d x} &=& +
{1\over 2\varphi'} 
\Bigg\{ 
i\left[k^2(x)+\chi^2+\chi'+(\varphi')^2\right]  \; a^*\; a
\nonumber\\
&&
+
\left([\varphi''+2\chi\varphi']+ i\left[k^2(x)+\chi^2+\chi'-(\varphi')^2\right]\right) \: a^* \; b
\Bigg\},\; 
\end{eqnarray}
whence
\begin{equation}
a^* {\d a\over \d x}  + a {\d a^*\over \d x} = 
{\Re\{ \left([\varphi''+2\chi\varphi']+ i\left[k^2(x)+\chi^2+\chi'-(\varphi')^2\right]\right) \: a^* \; b\} \over2\varphi'}.
\end{equation}
Perhaps more to the point, we can derive ODEs for the sums and differences:
\begin{eqnarray}
\label{E:system-a2a}
{\d (a+b)\over \d x} &=& +
{1\over \varphi'} 
\Bigg\{
 [\varphi''+2\chi\varphi'] (a+b) + 2 i (\varphi')^2 (a-b)
\Bigg\},
\\
\label{E:system-b2a}
{\d (a-b)\over \d x} &=& +
{1\over \varphi'} 
\Bigg\{ 
 - i\left[k^2(x)+\chi^2+\chi'\right] \;  (a+b)- [\varphi''+2\chi\varphi'] (a-b)
\Bigg\}.
\nonumber
\\
&&
\end{eqnarray}
that is
\begin{eqnarray}
\label{E:system-a2b}
{\d [i(a+b)]\over \d x} &=& +
{1\over \varphi'} 
\Bigg\{
 [\varphi''+2\chi\varphi'] (a+b) + 2  (\varphi')^2 [i(a-b)]
\Bigg\},
\\
\label{E:system-b2b}
{\d [i(a-b)]\over \d x} &=& +
{1\over \varphi'} 
\Bigg\{ 
 \left[k^2(x)+\chi^2+\chi'\right] \;  (a+b)- [\varphi''+2\chi\varphi'] \; [i(a-b)]
\Bigg\}.
\nonumber
\\
\end{eqnarray}
While this system of ODEs is somewhat simpler than those derived above, we have not been able to extract any significant improvement on or previous results in equations (\ref{E:theta}) and (\ref{E:alpha}).
\subsection{Case: $\Delta=0$}

We include this case for historical reasons, as it was  the first generalization we obtained of the original result published in~\cite{bounds1}. The derivation is somewhat simpler than the full $\Delta\neq0$ discussion presented above, and the ultimate bound we extract is no weaker.

We introduce an arbitrary auxiliary function $\varphi(x)$ which may be either real or complex, however we demand that $\varphi'(x) \neq 0$, and then define
\begin{equation}
\psi(x) = a(x) \, {\exp (+i \varphi) \over \sqrt{\varphi'}} + b(x) \, {\exp (-i \varphi) \over \sqrt{\varphi'}}.
\end{equation}
Again, to trim down the number of degrees of freedom it is useful to impose what can be thought of as a ``gauge condition'' (``auxiliary condition'')
\begin{equation}
\label{gauge_1}
{\d \over \d x} \left({a \over \sqrt{\varphi'}} \right) \, e^{+ i\varphi} + {\d \over \d x} \left({b \over \sqrt{\varphi'}}\right) e^{-i \varphi} = \chi(x) \, \psi(x).
\end{equation}
Here $\chi$ is some arbitrary function of position. The original analysis in~\cite{bounds1} corresponds to the special case $\chi(x) = 0$. Subject to this gauge condition,
\begin{equation}
{\d \psi \over \d x} = i\sqrt{\varphi'}\{a(x) \exp(+i\varphi) - b(x) \exp(-i\varphi)\} + \chi \psi.
\end{equation}
Now the Schr\"odinger equation can be rewritten in terms of two coupled first-order differential equations for these position-dependent Bogoliubov coefficients. We have to calculate $\d^2 \psi/ \d x^2$
making repeated use of the gauge condition:
\begin{eqnarray}
{\d^2 \psi \over \d x^2} &=& {\d \over \d x} \left(i {\varphi' \over \sqrt{\varphi'}} \{ae^{+i\varphi} - be^{-i\varphi}\} + \chi \psi \right),
\\
\nonumber
&=& {(i\varphi')^2 \over \sqrt{\varphi'}}\{ae^{+i \varphi} + b e^{-i\varphi}\}
+ i\varphi' \left\{{\d \over \d x} \left({a \over \sqrt{\varphi'}}\right)e^{+i \varphi} - {\d \over \d x} \left({b \over \sqrt{\varphi'}} \right) e^{-i \varphi} \right\}
\\
&& + i{\varphi'' \over \sqrt{\varphi'}}\{ ae^{+i \varphi} - b e^{-i \varphi}\} + \chi' \psi + \chi \psi',
\\
&=& -{\varphi'^2 \over \sqrt{\varphi'}} \{ae^{+i \varphi} + b e^{-i\varphi}\} 
+ i \varphi' \left\{2 {\d \over \d x} \left({a \over \sqrt{\varphi'}} \right)e^{+i \varphi}  - \chi \psi\right\} 
\nonumber\\
&&+ i {\varphi'' \over \sqrt{\varphi'}}
\{ae^{+i \varphi} - b e^{-i \varphi}\} + \chi' \, \psi + \chi \, \psi',
\\
&=& -\varphi'^2 \psi + {2i \varphi' \over \sqrt{\varphi'}} {\d a \over \d x}
e^{+i \varphi} - i {\varphi'' \over \sqrt{\varphi'}} be^{-i \varphi} - i \varphi' \chi \psi + \chi' \psi 
\nonumber\\
&&+ \chi \left[i \sqrt{\varphi'} \, \{a(x) e^{+i \varphi} - b(x)e^{-i\varphi}\} + \chi \,  \psi \right],
\\
&=&[\chi^2 + \chi' - (\varphi')^2] \psi + {2 i \varphi' \over \sqrt{\varphi'}} {\d a \over \d x} e^{+i \varphi} - i {[\varphi'' + 2 \chi \varphi'] \over \sqrt{\varphi'}} be^{-i \varphi}.
\end{eqnarray}
Now use the gauge condition to eliminate $\d a /\d x$ in favour of $\d b / \d x$. Finally, this permits us to write $\d^2 \psi/ \d x^2$ in either of the two equivalent forms
\begin{eqnarray}
{\d ^2 \psi \over \d x^2} &=& [\chi^2 + \chi' - (\varphi')^2] \, \psi -2i \varphi' {\d b \over \d x} {e^{-i \varphi} \over \sqrt{\varphi'}} + i[\varphi'' + 2 \chi \varphi'] \, a {e^{+i \varphi} \over \sqrt{\varphi'}};
\nonumber
\\ 
&&
\\
&=& [\chi^2 + \chi' - (\varphi')^2] \, \psi + 2i \varphi' {\d a  \over \d x} {e^{+i \varphi} \over \sqrt{\varphi'}} - i[\varphi'' + 2 \chi \varphi'] \, b {e^{-i \varphi} \over \sqrt{\varphi'}}.
\nonumber
\\
&&
\end{eqnarray}
Now insert these formulae into the Schr\"odinger equation written in the form
\begin{equation}
{\d^2 \psi \over \d x^2} = -k(x)^2 \psi \equiv - {2 m[E-V(x)] \over \hbar^2} \psi,
\end{equation}
to deduce
\begin{eqnarray}
\label{sch_eq}
{\d a \over \d x} &=&
+  {1 \over 2 \varphi'} \Bigg\{[\varphi'' + 2\chi \varphi'] \, b \, \exp(-2 i \varphi) 
\nonumber
\\
&&+ i \, [k^2(x) + \chi^2 + \chi' - (\varphi')^2] \, (a+ b \exp(-2i\varphi))\Bigg\},
\nonumber\\
&&\\
\label{sch_eq1}
{\d b \over \d x} &=&
+{1 \over 2 \varphi'} \Bigg\{[\varphi'' + 2\chi \varphi'] \, a \, \exp(+2 i \varphi) 
\nonumber
\\
&&- i \, [k^2(x) + \chi^2 + \chi' - (\varphi')^2] \, (b+ a \exp(+2i\varphi))\Bigg\}.
\nonumber\\
&& 
\end{eqnarray}
It is (again) easy to verify that this first-order system is compatible with the ``gauge condition'' (\ref{gauge_1}), and that by iterating the system twice (subject to this gauge condition) one recovers exactly the orginal Schr\"odinger equation. These equations hold for arbitrary $\varphi$ and $\chi$, real or complex, and when written in matrix form, exhibit a deep connection with the transfer matrix formalism~\cite{Transfer}. Let us define quantities $\rho_1(x)$ and $\rho_2(x)$, not necessarily real, as
\begin{equation}
\rho_1 = \varphi'' + 2 \chi \varphi'; \qquad \rho_2 = k^2(x) + \chi^2 + \chi' - (\varphi')^2.
\end{equation}
We can then re-write the Shabat--Zakharov system in matrix form as
\begin{equation}
{\d \over \d x} \left[\begin{array}{c} a\\b \end{array}\right] = {1 \over 2 \varphi'} \left[\begin{array}{cc} i \rho_2 & \{\rho_1 + i \rho_2\} \exp(-2 i \varphi) \\ \{\rho_1 - i \rho_2\} \exp(+2 i \varphi) & -i \rho_2 \end{array}\right] \left[\begin{array}{c} a\\b \end{array}\right].
\end{equation}
This has the formal solution
\begin{equation}
\left[\begin{array}{c} a(x_f)\\ b(x_f) \end{array}\right] = E(x_f, x_i) \left[\begin{array}{c} a(x_i)\\ b(x_i) \end{array}\right],
\end{equation}
in terms of a generalized position-dependent ``transfer matrix''~\cite{Transfer}
\begin{eqnarray}
&&E(x_f, x_i) = 
\nonumber
\\
&&\mathcal{P} \exp \left(\int_{x_i}^{x_f} {1 \over 2 \varphi'}\left[\begin{array}{cc} i \rho_2 & \{\rho_1 + i \rho_2\} \exp(-2 i \varphi) \\ \{\rho_1 - i \rho_2\} \exp(+2 i \varphi) & -i \rho_2 \end{array}\right] \, \d x \right),
\nonumber
\\
&&
\end{eqnarray}
where the symbol $\mathcal{P}$ denotes ``path ordering''.
(See the boxed text earlier in this chapter  for details.) Now we can write the wave function in inner product form
\begin{equation}
\psi(x) = {1 \over \sqrt{\varphi'}}
\left[\begin{array}{cc} \exp(+i \varphi); & \exp(-i \varphi) \end{array}\right] \left[\begin{array}{cc} a(x_i) \\ b(x_i) \end{array}\right],
\end{equation}
and where
\begin{equation}
\left[\begin{array}{c} a(x)\\ b(x) \end{array}\right] = E(x, x_0) \left[\begin{array}{c} a(x_0)\\ b(x_0) \end{array}\right].
\end{equation}
Therefore
\begin{equation}
\psi(x) = {1 \over \sqrt{\varphi'}}
\left[\begin{array}{cc} \exp(+i \varphi); & \exp(-i \varphi) \end{array}\right] E(x,x_0) \left[\begin{array}{cc} a(x_0) \\ b(x_0) \end{array}\right],
\end{equation}
to yield a formal but completely general solution for the Schr\"odinger equation
\begin{equation}
\sqrt{\varphi'} \psi(x) = \left[\begin{array}{cc} \exp(+i \varphi); & \exp(-i \varphi) \end{array}\right] E(x,x_0) \left[\begin{array}{cc} a(x_0) \\ b(x_0) \end{array}\right].
\end{equation}
Explicitly
\begin{eqnarray}
&&\psi(x) = {1 \over \sqrt{\varphi'}}
\left[\begin{array}{cc} \exp(+i \varphi); & \exp(-i \varphi) \end{array}\right] 
\nonumber
\\
&&\times
\mathcal{P} \exp \left(\int_{x_0}^{x} {1 \over 2 \varphi'}\left[\begin{array}{cc} i \rho_2 & \{\rho_1 + i \rho_2\} \exp(-2 i \varphi) \\ \{\rho_1 - i \rho_2\} \exp(+2 i \varphi) & -i \rho_2 \end{array}\right] \, \d \bar{x} \right)
\nonumber
\\
&&\times
\left[\begin{array}{cc} a(x_0) \\ b(x_0) \end{array}\right].
\end{eqnarray}
This is the explicit general solution to the Schr\"odinger equation. It depends on the two arbitrarily chosen functions $\varphi(x)$ and $\chi(x)$ and a path-ordered exponential matrix. If you consider path ordering to be an ``elementary'' process, then this is (again) the holy grail of ODE theory (complete quadrature, albeit formal, of the second-order linear ODE).

The development of bounds automatically follows as in the previous discussion, and we can without further calculation assert that equations (\ref{E:theta}) and (\ref{E:alpha}) hold in this situation as well.
\section{Discussion}
There are several ways to derive a number of rigourous bounds on transmission probabilities (and reflection probabilities and Bogoliubov coefficients) for one-dimensional scattering problems. The derivation of these bounds generally proceeds by rewriting the Schr\"odinger equation in terms of some equivalent system of first-order equations, and then analytically bounding the growth of certain 
quantities related to the net flux of particles as one sweeps across the potential. In this chapter we obtained a number of significant bounds, considerably stronger than those in~\cite{bounds1}, of both theoretical and practical interest. 

Even though the calculations we have presented are sometimes somewhat tedious, we feel however, they are more than worth the effort --- since there is a fundamental lesson to be learnt from them.
Technically, we demonstrated that the Schr\"odinger equation can be written as a Shabat--Zakharov system, which  can then be re-written in $2\times2$ matrix form. We rearranged this formation in terms of a generalized position-dependent ``transfer matrix''  involving  the symbol $\mathcal{P}$ which denotes ``path ordering''. Therefore the wavefunction $\psi(x)$ can be written in inner product form. This is the explicit general solution to the Schr\"odinger equation. It depends on the three arbitrarily chosen functions $\varphi(x)$, $\Delta(x)$,  and $\chi(x)$ and a path-ordered exponential matrix. If one considers path ordering to be an ``elementary'' process, then this is the holy grail of ODE theory (complete quadrature, albeit formal, of the second-order linear ODE).
We have seen that it is often convenient  to use the freedom to independently choose  $\varphi(x)$, $\Delta(x)$, and $\chi(x)$ to simplify life as much as possible. Furthermore, we have considered a few special cases. For instance, case  $\Delta' = \rho_2/(2\varphi')$, case $\Delta=\varphi$, and case $\Delta=0$.  The bounds that we have derived on the Bogoliubov coefficients $\alpha$ and $\beta$, and on the transmission and reflection probabilities $T$ and $R$, are the key results of this thesis --- the next few chapters will be devoted to developing several variants of these bounds, developing independent proofs that might ultimately lead to new bounds, and developing various applications of these bounds.

\chapter{First derivation of the bounds}
\label{C:consistency-5}
\section{Introduction}
In this chapter we shall review the analysis of~\cite{bounds1}, developing various techniques for estimating the scattering properties.  We shall review and briefly describe some very general bounds for reflection and transmission coefficients for one-dimensional potential scattering, and then indicate how the results of this thesis extend and expand on the earlier results. Equivalently, these results may be phrased as general bounds on the Bogoliubov coefficients, or statements about the transfer matrix~\cite{bounds1}. 

Finally, we shall re-demonstrate the use of Shabat--Zakharov system of ODEs (now in a greatly simplified context) to derive a first elementary bound on the transmission, reflection, and Bogoliubov coefficients. 
\section{Shabat--Zakharov systems}
 Consider the one-dimensional time-independent Schr\"odinger equation~\cite{Landau}--\cite{Messiah}
\begin{equation}
\label{E:SDE}
-{\hbar^2\over2m} {d^2\over dx^2} \psi(x) + V(x) \; \psi(x) = E \; \psi(x).
\end{equation}
If the potential asymptotes to a constant,
\begin{equation}
V(x\to\pm\infty) \to
V_{\pm\infty},
\end{equation}
then in each of the two asymptotic regions there are two
independent solutions to the Schr\"odinger equation
\begin{equation}
\psi(\pm i;\pm\infty;x) \approx 
{\exp(\pm i k_{\pm\infty} x) \over \sqrt{k_{\pm\infty}}}.
\end{equation}
Here the $\pm i$ distinguishes right-moving modes $e^{+ikx}$ from
left-moving modes $e^{-ikx}$, while the $\pm \infty$ specifies which
of the asymptotic regions we are in. Furthermore
\begin{equation}
k_{\pm\infty} = {\sqrt{2m(E-V_{\pm\infty})}\over\hbar}.
\end{equation}
To even begin to set up a scattering problem the minimum requirements
are that the potential asymptote to some constant, and this assumption
will be made henceforth.
\begin{table}[!htb]
\begin{center}
 \setlength{\fboxsep}{0.45 cm} 
 \framebox{\parbox[t]{12.5cm}{
{\bf Definition (Jost solutions)}: If we were working in three dimensions with radial symmetry then we would  study the solution of the radial Schr\"odinger equation~\cite{Parametric-Jost}
\begin{equation}
{\d^2 f(r,k) \over \d r^2} + (k^2 - V(r)) \, f(r,k) = 0 \, ,
\end{equation}
with the asymptotic conditions
\begin{equation}
\lim_{r \rightarrow + \infty} f(r \pm k) \, \exp(\pm ikr) =1,
\end{equation}
the so-called ``Jost solutions''.  In one dimension we would consider a formally identical equation for the full wave-function, now with boundary conditions at both $x\to+\infty$ and $x\to-\infty$. 
In addition, we now introduce the function
\begin{equation}
E(r,z) = \exp(\pm ikr) f(r, \pm k).
\end{equation}
which is also called a ``Jost solution'', and which satisfies the differential equation
\begin{equation}
{\d^2 E(r,z)\over \d r^2} - z \, {\d E(r,z)\over \d r} -V(r) \, E(r,z) = 0,
\end{equation}
with the following condition at infinity
\begin{equation}
\lim_{r \rightarrow + \infty} E(r,z) = 1,
\end{equation}
where
\begin{equation}
z = \pm 2ik .
\end{equation}
}}
\end{center}
\end{table}

The so-called Jost solutions~\cite{Chadan} are exact
solutions ${\cal J}_\pm(x)$ of the Schr\"odinger equation that satisfy
\begin{equation}
{\cal J}_+(x\to +\infty) \to {\exp( +i k_{+\infty} x) \over \sqrt{k_{+\infty}}},
\end{equation}
\begin{equation}
{\cal J}_-(x\to -\infty) \to {\exp( -i k_{-\infty} x) \over \sqrt{k_{-\infty}}},
\end{equation}
and
\begin{equation}
{\cal J}_+(x\to -\infty) \to 
\alpha {\exp( +i k_{-\infty} x) \over \sqrt{k_{-\infty}}} +
\beta  {\exp( -i k_{-\infty} x) \over \sqrt{k_{-\infty}}},
\end{equation}
\begin{equation}
{\cal J}_-(x\to +\infty) \to 
\alpha^* {\exp( -i k_{+\infty} x) \over \sqrt{k_{+\infty}}} +
\beta^*  {\exp( +i k_{+\infty} x) \over \sqrt{k_{+\infty}}}.
\end{equation}
Here $\alpha$ and $\beta$ are the (right-moving) Bogoliubov
coefficients, which are related to the (right-moving)
reflection and transmission amplitudes by
\begin{equation}
r = {\beta\over\alpha}; \qquad t = {1\over\alpha}.
\end{equation}
These conventions correspond to an incoming flux of right-moving
particles (incident from the left) being partially transmitted and
partially scattered.  The left-moving Bogoliubov coefficients are just
the complex conjugates of the right-moving coefficients, however it
should be borne in mind that the phases of $\beta$ and $\beta^*$ are
physically meaningless in that they can be arbitrarily changed simply
by moving the origin of coordinates. The phases of $\alpha$ and
$\alpha^*$ on the other hand do contain real physical information.
\begin{table}[!htbp]
\begin{center}
  \setlength{\fboxsep}{0.45 cm} 
   \framebox{\parbox[t]{12.5cm}{
 {\bf  Phase integral technique}:  We shall be concerned with the solution of the differential equation~\cite{Andersson, Froman}
 \begin{equation}
 {\d^2 \psi \over \d z^2} + R(z) \, \psi = 0 ,
 \end{equation}
 where $R(z)$ is an analytic function of $z$. This differential equation which could possibly result from separation of variables, describes large classes of important problems in various fields of physics, not only in quantum mechanics.
 
 The \emph{phase-integral} functions, in terms of which the solution will be expressed, are of the general form
 \begin{eqnarray}
 f_{1}(z) &=& q^{-1/2} (z) \, \exp\bigg(+i \int^{z} q(z) \, \d z\bigg) .
 \\
 f_{2}(z) &=& q^{-1/2} (z) \, \exp\bigg(-i \int^{z} q(z) \, \d z \bigg) .
 \end{eqnarray}

These phase integral functions are very closely related to the WKB approximation, and the lowest order approximation to the phase integral is essentially just the second-order WKB approximation.
}}

\end{center}
\end{table}
In this chapter we will derive some very general bounds on $|\alpha|$
and $|\beta|$, which also lead to general bounds on the reflection and
transmission probabilities
\begin{equation}
R = |r|^2; \qquad  T=|t|^2.
\end{equation}
The key idea is to re-write the second-order Schr\"odinger equation as a
particular type of Shabat--Zakharov~\cite{Eckhaus} system: a
particular set of two coupled first-order differential equations for
which bounds can be easily established. A similar representation of
the Schr\"odinger equation is briefly discussed by Peierls~\cite{Peierls}
and related representations are well-known, often being used
without giving an explicit reference (see {\em e.g.}~\cite{Bordag}).
However an exhaustive search has not uncovered prior use of the
particular representation of this chapter, nor the idea of using the
representation to place bounds on one-dimensional scattering.
\begin{table}[!htb]
\begin{center}
  \setlength{\fboxsep}{0.45 cm} 
   \framebox{\parbox[t]{12.5cm}{
       {\bf  Scattering matrix}: 
      In physics, the \emph{scattering matrix (or S-matrix)} relates the initial state and the final state for an interaction of particles. It is used in quantum mechanics, scattering theory and quantum field theory. In addition, it can also denote an infinite-dimensional matrix (or operator) that expresses the state of a scattering system consisting of waves or particles or both in the far future in terms of its state in the remote past; also called the \emph{S-matrix}. In the case of electromagnetic (or acoustic) waves, it connects the intensity, phase, and polarization of the outgoing waves in the far field at various angles to the direction and polarization of the beam pointed toward an obstacle~\cite{s-matrix, scattering-matrix-definition}.       
}}
\end{center}
\end{table}

\begin{table}[!htb]
\begin{center}
  \setlength{\fboxsep}{0.45 cm} 
   \framebox{\parbox[t]{12.5cm}{
       {\bf  Definition (Auxiliary functions)}: 
       They are not a rigorously defined type of function. In contrast, these functions are either explicitly constructed, or at least shown to exist, and are used to provide a formal solution to some assumed hypothesis, or otherwise prove the result in question. Devising an auxiliary function during the course of a proof in order to prove the result is not a technique exclusive to any particular branch of mathematics~\cite{auxiliary-functions}.
}}
\end{center}
\end{table}

\begin{table}[!htb]
\begin{center}
  \setlength{\fboxsep}{0.45 cm} 
   \framebox{\parbox[t]{12.5cm}{
       {\bf  Definition (Gauge conditions or Gauge fixing)}: 
       Indicates a mathematical scheme for coping with redundant degrees of freedom in field variables, most commonly in the physics of gauge theories. In Abelian or non-Abelian gauge theories one represents each physically distinct configuration of the system as an equivalence class of detailed local field configurations. Any two detailed configurations in the same equivalence group are related by a gauge transformation, equivalent to a shear along unphysical axes in configuration space~\cite{gauge-fixing}. 
       
       In the context of the current thesis things are somewhat simpler. There is a redundancy, which is eliminated by our ``gauge condition'', but one does not have to deal with the full complexity of an actual ``gauge theory'' in the sense of particle physics.
}}
\end{center}
\end{table}

We start by introducing an arbitrary auxiliary function $\varphi(x)$
which may be either real or complex, though we do demand that
$\varphi'(x)\neq 0$, and then defining
\begin{equation}
\label{E:representation}
\psi(x) = 
a(x) {\exp(+i \varphi)\over\sqrt{\varphi'}} + 
b(x) {\exp(-i \varphi)\over\sqrt{\varphi'}}.
\end{equation}
This representation effectively seeks to use quantities resembling the
``phase integral'' wavefunctions as a basis for the true
wavefunction~\cite{Froman-Froman}. This representation is of course
highly redundant, since one complex number $\psi(x)$ has been traded
for two complex numbers $a(x)$ and $b(x)$ plus an essentially
arbitrary auxiliary function $\varphi(x)$. In order for this
representation to be most useful it is best to arrange things so that
$a(x)$ and $b(x)$ asymptote to constants at spatial infinity, which we
shall soon see implies that we should pick the auxiliary function to
satisfy 
\begin{equation}
\varphi'(x) \to k_{\pm\infty} \qquad \hbox{as} \qquad x\to\pm\infty. 
\end{equation}
To trim down the number of degrees of freedom it is useful to impose a
``gauge condition''
\begin{equation}
\label{E:gauge-condition}
{\d\over \d x}\left({a\over\sqrt{\varphi'}}\right) e^{+i \varphi} + 
{\d\over \d x}\left({b\over\sqrt{\varphi'}}\right) e^{-i \varphi} = 0.
\end{equation}
Subject to this gauge condition,
\begin{equation}
\label{E:gradient}
{\d\psi\over \d x} = i \sqrt{\varphi'} 
\left\{ a(x) \exp(+i \varphi) - b(x) \exp(-i \varphi ) \right\}.
\end{equation}
We now re-write the Schr\"odinger equation in terms of two coupled
first-order differential equations for these position-dependent
Bogoliubov coefficients. To do this note that
\begin{eqnarray}
\label{E:double-gradient}
{\d^2\psi\over \d x^2} 
&=& 
{\d\over \d x} 
\left( i{\varphi'\over \sqrt{\varphi'}} 
\left\{ a e^{+i \varphi} - b e^{-i \varphi} \right\} 
\right)\, ,
\\
&=& {(i\varphi')^2\over\sqrt{\varphi'}} 
\left\{ a e^{+i \varphi} + b e^{-i \varphi} \right\}
\nonumber\\
&+&  i \varphi'
\left\{ 
{\d\over \d x}\left({a\over\sqrt{\varphi'}}\right) e^{+i \varphi} - 
{\d\over \d x}\left({b\over\sqrt{\varphi'}}\right) e^{-i \varphi} 
\right\}
\nonumber\\
&+&  i{\varphi''\over\sqrt{\varphi'}} 
\left\{ a e^{+i \varphi} - b e^{-i \varphi} \right\}\, ,
\\
&=& -{\varphi'^2\over\sqrt{\varphi'}}  
\left\{ a e^{+i \varphi} + b e^{-i \varphi} \right\}
\nonumber\\
&+&  {2 i\varphi'\over\sqrt{\varphi'}}  {\d a\over \d x} e^{+i \varphi} -
i{\varphi''\over\sqrt{\varphi'}} b e^{-i \varphi} \, ,
\\
&=& -{\varphi'^2\over\sqrt{\varphi'}}
  \left\{ a e^{+i \varphi} + b e^{-i \varphi} \right\}
\nonumber\\
&-&  {2 i\varphi'\over\sqrt{\varphi'}}  {\d b\over \d x} e^{-i \varphi} + 
i{\varphi''\over\sqrt{\varphi'}} a e^{+i \varphi}.
\end{eqnarray}
(The last two relations use the ``gauge condition''.) Now insert
these formulae into the Schr\"odinger equation written in the form
\begin{equation}
{\d^2\psi\over \d x^2} = - k(x)^2 \; \psi \equiv
- {2m(E-V(x))\over\hbar^2} \; \psi,
\end{equation}
to deduce
\begin{eqnarray}
\label{E:system-a-condition-1}
{\d a\over \d x} &=& +
{1\over 2\varphi'} 
\Bigg\{ 
\varphi'' \; b \; \exp(-2i\varphi) 
\nonumber\\
&&\qquad
+ 
i\left[k^2(x)-(\varphi')^2\right] \left( a + b \exp(-2i\varphi) \right)
\Bigg\},
\\
\label{E:system-b-condition-1}
{\d b\over \d x} &=& +
{1\over 2\varphi'} 
\Bigg\{ 
\varphi'' \; a \; \exp(+2i\varphi) 
\nonumber\\
&&\qquad
- i\left[k^2(x)-(\varphi')^2\right] \left( b + a \exp(+2i\varphi) \right)
\Bigg\}.
\end{eqnarray}
It is easy to verify that this first-order system is compatible with
the ``gauge condition'' (\ref{E:gauge-condition}), and that by iterating the
system twice (subject to this gauge condition) one recovers exactly
the original Schr\"odinger equation. These equations hold for arbitrary
$\varphi$, real or complex, and when written in matrix form, exhibit a
deep connection with the transfer matrix
formalism~\cite{Transfer-matrix}.

\section{Bounds} 

To obtain our bounds on the Bogoliubov coefficients we start by
restricting attention to the case that $\varphi(x)$ is a {\em real}\,
function of $x$.  (Since $\varphi$ is an essentially arbitrary
auxiliary function this is not a particularly restrictive
condition). Under this assumption the probability current is
\begin{equation}
\mathscr{J}
= \Im\left\{ \psi^* {\d\psi \over \d x} \right\} 
= \left\{ |a|^2 - |b|^2 \right\}.
\end{equation}
Now at $x\sim+\infty$ the wavefunction is purely right-moving and
normalized to 1, because we are considering one-dimensional Jost
solutions~\cite{Chadan}. Then for all $x$ we have a conserved
quantity
\begin{equation}
\label{E:conservation}
|a|^2 - |b|^2  = 1.
\end{equation}
{\em It is this result that makes it useful to interpret $a(x)$ and
$b(x)$ as position-dependent Bogoliubov coefficients relative to the
auxiliary function $\varphi(x)$}.  Now use the fact that
\begin{equation}
{d|a|\over dx} = 
{1\over2|a|} \left( a^* {\d a\over \d x} + a {\d a^* \over \d x} \right),
\end{equation}
and use equation (\ref{E:system-a-condition-1}) to obtain
\begin{eqnarray}
{\d|a|\over \d x} &=& 
{1\over2|a|} {1\over2\varphi'}
\Big( 
\varphi'' 
\left[a^* b \exp(-2i\varphi) + a b^* \exp(+2i\varphi) \right]
\nonumber\\
&&
+ i[k^2 -(\varphi')^2] 
\left[a^* b \exp(-2i\varphi) - a b^* \exp(+2i\varphi) \right]
\Big).
\nonumber\\
&&
\end{eqnarray}
That is
\begin{eqnarray}
{\d|a|\over \d x} &=& 
{1\over2|a|} {1\over2\varphi'}
\Re\Bigg( 
\left[\varphi'' + i[k^2 -(\varphi')^2] \right] \;
\left[a^* b \exp(-2i\varphi) \right]
\Bigg).
\nonumber\\
&&
\end{eqnarray}
The right hand side can now be bounded from above, by systematically
using $\Re(A\;B) \leq |A| \; |B|$. This leads to
\begin{equation}
{\d |a|\over \d x} \leq  
{\sqrt{(\varphi'')^2+ \left[k^2-(\varphi')^2\right]^2} \over 2 |\varphi'| } 
\; |b|.
\end{equation}
It is essential that $\varphi$ be real to have $|\exp(-2i\varphi)|=1$
which is the other key ingredient above.  Now define the non-negative
quantity
\begin{equation}
\label{E:vartheta}
\vartheta =\vartheta[\varphi(x),k(x)]
\equiv
{\sqrt{(\varphi'')^2+ \left[k^2(x)-(\varphi')^2\right]^2} 
\over 2 |\varphi'| }, 
\end{equation}
and use the conservation law (\ref{E:conservation}) to write
\begin{equation}
{\d |a|\over \d x} \leq  \vartheta \sqrt{|a|^2 -1}.
\end{equation}
Integrate this inequality
\begin{equation}
\left. \left\{\cosh^{-1} |a| \right\} \right|_{x_i}^{x_f} \leq  
\int_{x_i}^{x_f}  \vartheta \; \d x.
\end{equation}
Taking limits as $x_i\to-\infty$ and $x_f\to+\infty$
\begin{equation}
\cosh^{-1} |\alpha| \leq  
\int_{-\infty}^{+\infty}  \vartheta \; \d x.
\end{equation}
That is
\begin{equation}
\label{B:alpha0}
|\alpha| \leq  
\cosh\left(  \int_{-\infty}^{+\infty}  \vartheta \; \d x \right).
\end{equation}
Which automatically implies
\begin{equation}
\label{B:beta0}
|\beta| \leq  
\sinh\left(   \int_{-\infty}^{+\infty}  \vartheta \; \d x \right).
\end{equation}
Since this result holds for {\em all real} choices of the auxiliary
function $\varphi(x)$, (subject only to $\varphi' \neq 0$ and
$\varphi' \to k_{\pm\infty}$ as $x \to \pm\infty$), it encodes an
enormously wide class of bounds on the Bogoliubov coefficients. When
translated to reflection and transmission coefficients the equivalent
statements are
\begin{equation}
\label{B:T0}
T \geq 
\sech^2
\left(
\int_{-\infty}^{+\infty} \vartheta \; \d x
\right),
\end{equation}
and
\begin{equation}
\label{B:R0}
R \leq 
\tanh^2
\left(
\int_{-\infty}^{+\infty} \vartheta \; \d x
\right).
\end{equation}
We shall soon turn this general result into more specific theorems in chapter~\ref{C:consistency-6}.
\section{Transfer matrix representation} 
\begin{table}[!htbp]
\begin{center}
 \setlength{\fboxsep}{0.45 cm} 
 \framebox{\parbox[t]{12.5cm}{
{\bf Transfer matrix formalism}:
This is most commonly used in optics and acoustics to analyze the propagation of electromagnetic or acoustic waves through a stratified (layered) medium. In those situations, the stack layers are normal to the $z$ axis and the field within one layer can be represented as the superposition of a left and right traveling wave with wave number $k$~\cite{transfer-matrix-formalism},
\begin{equation}
E(z) = E_{r} \, \exp(ikz) + E_{l} \, \exp(-ikz), 
\end{equation}
We represent the field as the vector $(E(z),F(z))$, where
 \begin{equation}
F(z) = i k E_{r} \, \exp(ikz) - ik E_{l} \, \exp(-ikz).
\end{equation}
The propagation over a distance $L$ into the positive $z$ direction is described by the matrix
\begin{equation}
M = \left[\begin{array}{cc} \cos (kL) & {1\over k} \sin(kL)
\\ -k \sin(kL)  & \cos(kL)  \end{array}\right] ,
\end{equation}
and
\begin{equation}
\left[\begin{array}{cc} E(z+L) \\ F(z+L) \end{array}\right]  = M \times \left[\begin{array}{cc} E(z)   \\ F(z) \end{array}\right].
\end{equation}
The system transfer matrix is then built up by repeated multiplication
\begin{equation}
M_{s} = M_{N} \times \dots \times M_{2} \times M_{1}.
\end{equation}

A very similar formalism applies to the one-dimensional Schr\"odinger  equation, and the ``path ordered'' exponentials discussed in the body of the thesis are effectively transfer matrices built out of an infinite number of infinitesimally small steps.
}}
\end{center}
\end{table}
The system of equations (\ref{E:system-a-condition-1})--(\ref{E:system-b-condition-1}) can
also be written in matrix form. It is convenient to define
\begin{equation}
\rho 
\equiv
\varphi'' +i[k^2(x)-(\varphi')^2].
\end{equation}
Then
\begin{equation}
{\d \over \d x}\left[\begin{array}{c} a\\b \end{array}\right] ={1\over2\varphi'} \, \left[\begin{array}{cc} i \Im[\rho] & 
\rho\exp(-2i\varphi)  \\ \rho^*\exp(+2i\varphi)  & 
-i \Im[\rho] \end{array}\right] \left[\begin{array}{c} a\\b \end{array}\right].
\end{equation}
This has the formal solution
\begin{equation}
\left[\begin{array}{c}a(x_f)\\b(x_f) \end{array}\right] =E(x_f,x_i)  \, \left[\begin{array}{c}a(x_i)\\ 
b(x_i) \end{array}\right]  ,
\end{equation}
in terms of a generalized position-dependent ``transfer
matrix''~\cite{Transfer-matrix}
\begin{eqnarray}
&&
E(x_f,x_i) = 
\nonumber\\
&&
{\cal P} \exp\left( 
\int_{x_i}^{x_f}
{1\over2\varphi'} \left[\begin{array}{cc} i \Im[\rho] & 
\rho\exp(-2i\varphi)  \\ \rho^*\exp(+2i\varphi)  & 
-i \Im[\rho] \end{array}\right] \left[\begin{array}{c} a\\b \end{array}\right] \d x \right),
\nonumber\\
&&
\end{eqnarray}
where the symbol ${\cal P}$ denotes ``path ordering''. In particular,
if we take $x_i\to-\infty$ and $x_f\to+\infty$ we obtain a formal but
exact expression for the Bogoliubov coefficients
\begin{eqnarray}
&&
 \left[\begin{array}{cc} \alpha & \beta^*  \\ \beta & \alpha^*\end{array}\right]  =
E(\infty,-\infty) = 
\nonumber\\
&&
{\cal P} \exp\left( 
\int_{-\infty}^\infty
{1\over2\varphi'} 
\left[\begin{array}{cc} i \Im[\rho] & 
\rho\exp(-2i\varphi)  \\ \rho^*\exp(+2i\varphi)  & 
-i \Im[\rho] \end{array}\right]  \d x \right).
\nonumber\\
&& 
\end{eqnarray}
The matrix $E$ is {\em not} unitary, though it does have determinant
1. It is in fact an element of the group $SU(1,1)$. Taking
\begin{equation}
\sigma_z =  
\left[\begin{array}{cc} +1 &0  \\ 0 & -1\end{array}\right],
\end{equation}
then $(\sigma_z)^2 = +I$, and defining $E^\dagger = (E^*)^T$, it is
easy to see
\begin{equation}
E^\dagger \sigma_z E = \sigma_z.
\end{equation}
This is the analog of the invariance of the Minkowski metric for
Lorentz transformations in $SO(3,1)$. 

\begin{table}[!htb]
\begin{center}
 \setlength{\fboxsep}{0.45 cm} 
 \framebox{\parbox[t]{12.5cm}{
{\bf Complex structure}:  There is a minor (non-propagating) error in the analysis of~\cite{bounds1}.
 If we define the
``complex structure'' $J$ by
\begin{equation}
J = \left[\begin{array}{cc} 0 & 1  \\ -1 & 0\end{array}\right],
\end{equation}
then $J^2 = -I$ and it is claimed in~\cite{bounds1} that
\begin{equation}
E^\dagger =  J E J.
\end{equation}
This assertion is simply an error. Fortunately this is a side issue, and the error does not affect the rest of the argument.   
}}
\end{center}
\end{table}
\section{Discussion}
In this chapter we have re-cast, re-analyzed, and described the first derivation of scattering bounds as presented in~\cite{bounds1}. The formalism as developed here works in terms of one free function $\varphi(x)$. In other parts of this thesis we have established generalized bounds; some in terms of \emph{two} arbitrary functions $\varphi(x)$ and $\chi(x)$, and some in terms of \emph{three} arbitrary functions  $\varphi(x)$, $\Delta(x)$,  and $\chi(x)$. The derivation of the present chapter is noteworthy because of its brevity and simplicity --- and this chapter has acted as a ``seed'', suggesting and hinting at generalizations that have ultimately become the content of the previous chapter.  Of course, the ``simple'' calculation reported in this chapter is also the seed for the various published journal articles that have already arisen from this thesis, articles which are displayed in the appendices to the thesis.

\chapter{Bounds: Special cases}
\label{C:consistency-6}
In this chapter we shall deal with some specific cases of these general bounds and develop a number of interesting specializations. We shall collect together a large number of results that otherwise appear quite unrelated, including reflection above and below the barrier. We have divided the special case bounds we consider into five sub-cases: Special cases 1--4, and ``future directions''.
\section{Bounds: Special case 1}
We now reproduce the bounds of special case 1 in~\cite{bounds1}. 

Suppose now that the potential satisfies $V_{+ \infty} = V_{- \infty}$. Also, choose the phase function $\varphi(x)$ to be $\varphi = k_{\infty} x$ and take $\chi = 0$. We also require $k_{\infty} \neq 0$, that is $E > V_{\pm \infty}$. This is the special case discussed in a different context by Peierls~\cite{Peierls}. Then the evolution equations simplify tremendously.

We consider the quantity
\begin{equation}
\label{var_eq}
\vartheta [k(x), \varphi(x), \chi] = {\sqrt{(\varphi'' + 2 \chi \varphi')^2 + [k^2(x) + \chi^2 + \chi' - (\varphi')^2]^2} \over 2 |\varphi'|},
\end{equation}
which represents the generalization that we derived  earlier in this thesis, in equation (\ref{E:theta}),  of the bound reported in~\cite{bounds1}.
When  $\varphi = k_{\infty} x$,  then $\varphi'' = 0$, and if we additionally choose $\chi = 0$, then this simplifies to
\begin{equation}
\vartheta [k(x), \varphi(x), \chi] = {\sqrt{[k^2(x) - ( k_{\infty} )^2]^2} \over 2 | k_{\infty} |} =  {|k^2 - k^2_{\infty}| \over 2 |k_{\infty}|} .
\end{equation}
Inserting this into our general bound now reproduces case 1 of~\cite{bounds1} as desired.

Furthermore, we can calculate
\begin{eqnarray}
\nonumber
\vartheta [k(x), \varphi(x), \chi]  &=&  {|k^2 - k^2_{\infty}| \over 2 |k_{\infty}|} = {|2 m (E - V)|\over 2 \hbar^2  |k_{\infty}|} - {|2 m (E - V_{\infty})|\over 2 \hbar^2  |k_{\infty}|},
\\
\nonumber
&=&  {m |V(x) - V_{\infty}| \over \hbar^2 k_{\infty}}.
\end{eqnarray}
From the previous chapters \ref{C:consistency-4} and \ref{C:consistency-5}, we derive 
\begin{equation}
T  \geq \mathrm{sech^2} \bigg(\int^{+ \infty}_{- \infty} \vartheta[k, \varphi, \chi] \, \d x \bigg),
\end{equation}
when $\hbar k_{\infty} = \sqrt{2 m (E - V_{\infty})}$, then $k_{\infty} = \displaystyle{{\sqrt{2 m (E - V_{\infty})}\over \hbar}}$. Now we can find
\begin{eqnarray}
\nonumber
\vartheta [k(x), \varphi(x), \chi] &=& \bigg({m |V(x) - V_{\infty}| \over \hbar^2}\bigg) \bigg({\hbar \over \sqrt{2 m (E - V_{\infty})}}\bigg),
\nonumber
\\
&=& {m |V(x) - V_{\infty}| \over \hbar \sqrt{2 m (E - V_{\infty})}} = {1 \over \hbar} \, \sqrt{m \over 2 (E - V_{\infty})} \;\; |V(x) - V_{\infty}|.
\nonumber
\\
&&
\end{eqnarray}
Therefore,
\begin{eqnarray}
\nonumber
T  & \geq&  \mathrm{sech^2} \bigg(\int^{+ \infty}_{- \infty} \vartheta[k, \varphi, \chi] \, \d x \bigg)\, ,
\\
\nonumber
&= &  \mathrm{sech^2} \bigg(\int^{+ \infty}_{- \infty}  {1 \over \hbar} \, \sqrt{m \over 2 (E - V_{\infty})} \;\; |V(x) - V_{\infty}|\, \d x \bigg)\, ,
\\
\nonumber
&=& \mathrm{sech}^2 \bigg({1 \over \hbar} \sqrt{{m \over 2 (E - V_{\infty})}} \;\; \int_{- \infty}^{+ \infty} |V - V_{\infty}| \, \d x \bigg) ,
\end{eqnarray}
as desired.

Similarly one can find $R$ in terms of $[k(x), \varphi(x), \chi]$. Note
\begin{equation}
\vartheta  \rightarrow {|k^2 - k^2_{\infty}| \over 2 k_{\infty}} = {m |V(x) - V_{\infty}| \over \hbar^2 k_{\infty}}.
\end{equation}
Using $(\hbar k_{\infty})^2 = 2 m (E - V_{\infty})$, the bounds become
\begin{eqnarray}
\label{case1}
\nonumber
T &\geq& \mathrm{sech}^2 \bigg({1 \over \hbar} \sqrt{{m \over 2 (E - V_{\infty})}} \;\;
\int_{- \infty}^{+ \infty} |V - V_{\infty}| \, \d x \bigg) ,
\\
&=& \mathrm{sech}^2 \bigg(
{m \over \hbar^2 \, k_{\infty}} \;\;
\int_{- \infty}^{+ \infty} |V - V_{\infty}| \, \d x \bigg),
\end{eqnarray}
and
\begin{eqnarray}
\label{case1_1}
\nonumber
R &\leq& \mathrm{tanh^2} \bigg({1 \over \hbar} \sqrt{{m \over 2 (E - V_{\infty})}} \;\;
\int_{- \infty}^{+ \infty} |V - V_{\infty}| \, \d x \bigg) ,
\\
&=& \mathrm{tanh}^2 \bigg(
{m \over \hbar^2 \, k_{\infty}} \;\;
\int_{- \infty}^{+ \infty} |V - V_{\infty}| \, \d x \bigg).
\end{eqnarray}
These bounds are exact non-perturbative results, however for high energies it may be convenient to use the slightly less restrictive (but analytically much more tractable) bounds obtained by simply taking the first non-trivial term in the Taylor series.

When $E > V_{\pm \infty}$, using $(\hbar k_{\infty})^2 = 2 m (E-V_{\infty})$, the bounds become
\begin{eqnarray}
\label{case1_2}
\nonumber
T &\geq& 1 - {m \bigg(\int_{- \infty}^{+ \infty} |V - V_{\infty}| \d x \bigg)^2 \over 2 (E - V_{\infty}) \hbar^2} \, ,
\nonumber
\\
&=& 1 - {m^2 \bigg( \int_{- \infty}^{+ \infty} |V - V_{\infty}| \d x \bigg)^2 \over \hbar^4 k_{\infty}^2}.
\end{eqnarray}
and
\begin{equation}
\label{case1_3}
R \leq {m \bigg(\int_{- \infty}^{+ \infty} |V - V_{\infty}| \d x \bigg)^2 \over 2 (E - V_{\infty}) \hbar^2} 
= {m^2 \bigg( \int_{- \infty}^{+ \infty} |V - V_{\infty}| \d x \bigg)^2 \over k_\infty^2 \hbar^4}.
\end{equation}
This version of the bounds also holds for all energies, but is not very restrictive for low energy. (Somewhat better low energy bounds are developed in the chapter dealing with the Miller--Good transformation. See also the chapter \ref{C:consistency-9} on black hole greybody factors.) Note that the bounds of this subsection make perfectly good sense for both scattering ``over the barrier'' or ``under the barrier'', there is no requirement on the presence or absence of classical turning points.

The transfer matrices can be analyzed by checking that the evolution equations simplify to
\begin{equation}
\label{da}
{\d a \over \d x} = {- im (V - V_{\infty}) \over \hbar^2 k_{\infty}} \, \{a + b \, \exp(-2 i k_{\infty} x)\},
\end{equation}
\begin{equation}
\label{db}
{\d b \over \d x} = {+ im (V - V_{\infty}) \over \hbar^2 k_{\infty}} \, \{a \, \exp(+2 i k_{\infty} x) + b\}.
\end{equation}
This can be written in matrix form as
\begin{equation}
\label{ab_metric}
{\d \over \d x}\left[\begin{array}{c} a\\b \end{array}\right] = {- im (V - V_{\infty}) \over \hbar^2 k_{\infty}} \, \left[\begin{array}{cc} 1 & \exp(-2 ik_{\infty} x) \\ - \exp(+2i k_{\infty} x) & -1 \end{array}\right] \left[\begin{array}{c} a\\b \end{array}\right].
\end{equation}
This version of the Shabat--Zakharov system~\cite{Eckhaus} has a formal solution in terms of the relatively simple transfer matrix
\begin{equation}
E(x_f, x_i) = \mathcal{P} \, \exp \, \bigg({- im \over \hbar^2 k_{\infty}} \int_{x_i}^{x_f} (V(x) - V_{\infty}) \left[\begin{array}{cc} 1 & e^{-2 ik_{\infty} x} \\ - e^{+2i k_{\infty} x} & -1 \end{array}\right] \, \d x \bigg),
\end{equation}
The formal but exact expression for the Bogoliubov coefficients is now
\begin{eqnarray}
\nonumber
\left[\begin{array}{cc} \alpha & \beta^* \\ \beta & \alpha^* \end{array}\right] &=& E(\infty, - \infty) ,
\nonumber
\\
&=& \mathcal{P} \, \exp \, \bigg({- im \over \hbar^2 k_{\infty}} \int_{- \infty}^{\infty} (V(x) - V_{\infty}) \left[\begin{array}{cc} 1 & e^{-2 ik_{\infty} x} \\ - e^{+2i k_{\infty} x} & -1 \end{array}\right] \, \d x \bigg).
\nonumber
\\
&&
\end{eqnarray}
Furthermore, the form of the system (\ref{da})--(\ref{db}) suggests that it might be useful to define
\begin{equation}
\label{a_tilde}
a = \tilde{a} \, \exp \bigg[ + {im \over \hbar^2 k_{\infty}} \int_{- \infty}^{x} (V(y) - V_{\infty}) \, \d y \bigg],
\end{equation}
\begin{equation}
\label{b_tilde}
b = \tilde{b} \, \exp \bigg[ - {im \over \hbar^2 k_{\infty}} \int_{- \infty}^{x} (V(y) - V_{\infty}) \, \d y  \bigg].
\end{equation}
Then
\begin{figure}[!htb] 
\centering
\includegraphics[scale=1]{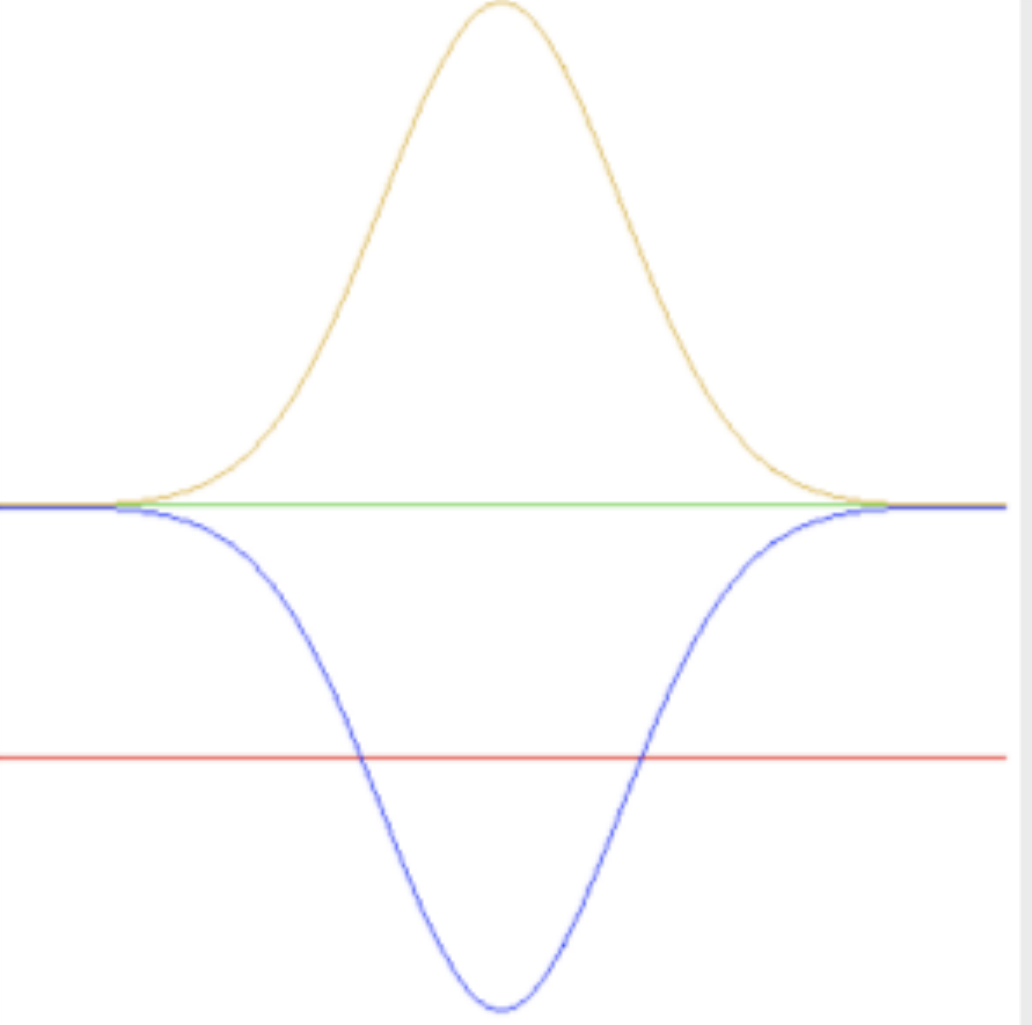}
\caption[Potential and wave-number for special case 1]{\label{F:special-case-1}This diagram (qualitatively) shows the potential $V(x)$, and associated function $k^2(x)$, for the scattering problem considered in special case~$1$~\cite{bounds-matt-seminar}.}
\end{figure}
we can substitute equations (\ref{a_tilde}) and (\ref{b_tilde})  into equation (\ref{ab_metric}). We derive
\begin{equation}
{\d \tilde{a} \over \d x} = {- im (V(x)-V_{\infty}) \over \hbar^2 k_{\infty}} \, \tilde{b} \, \exp(-2 i k_{\infty}x),
\end{equation}
and
\begin{equation}
{\d \tilde{b} \over \d x} = {+ im (V(x)-V_{\infty}) \over \hbar^2 k_{\infty}} \, \tilde{a} \, \exp(+2 i k_{\infty}x),
\end{equation}
respectively.
\noindent
This representation simplifies some of the results, for instance
\begin{eqnarray}
\nonumber
\left[\begin{array}{cc} \tilde{\alpha} & \tilde{\beta^*} \\ \tilde{\beta} & \tilde{\alpha^*} \end{array}\right] &=& \tilde{E}(\infty, - \infty) ,
\nonumber
\\
&=& \mathcal{P} \, \exp \, \bigg({- im \over \hbar^2 k_{\infty}} \int_{- \infty}^{\infty} (V(x) - V_{\infty}) \left[\begin{array}{cc} 0 & e^{-2 ik_{\infty} x} \\ - e^{+2i k_{\infty} x} & 0 \end{array}\right] \, \d x \bigg).
\nonumber
\\
&&
\end{eqnarray}
Also note that
\begin{equation}
\alpha = \tilde{\alpha} \, \exp \bigg[ + {im \over \hbar^2 k_{\infty}} \int_{- \infty}^{+\infty} (V(y) - V_{\infty}) \, \d y \bigg],
\end{equation}
\begin{equation}
\beta = \tilde{\beta} \, \exp \bigg[ - {im \over \hbar^2 k_{\infty}} \int_{- \infty}^{+\infty} (V(y) - V_{\infty}) \, \d y \bigg].
\end{equation}
This can be used as the basis of an approximation scheme for $\tilde{\beta}$. Suppose that for all $x$ we have $|\tilde{b}(x)| \ll 1$, so that $|\tilde{a}(x)| \approx 1$. Then
\begin{equation}
{\d \tilde{b} \over \d x} \approx {+ im (V(x)-V_{\infty}) \over \hbar^2 k_{\infty}} \, \exp(+2 i k_{\infty}x).
\end{equation}
This may be immediately integrated to yield
\begin{equation}
\tilde{\beta} \approx \int {+ im (V(x)-V_{\infty}) \over \hbar^2 k_{\infty}} \, \exp(+2 i k_{\infty}x) \; \d x.
\end{equation}
This is immediately recognizable as the (first) Born approximation. If we instead work in terms of the original definition $\beta$ we obtain a slightly different approximation
\begin{eqnarray}
\nonumber
\beta &\approx& {+ im \over \hbar^2 k_{\infty}} \, \exp \left[+ {im \over \hbar^2 k_{\infty}} \int^{+ \infty}_{- \infty} (V(x) - V_{\infty}) \, \d x \right]
\nonumber
\\
& \times &\int^{+ \infty}_{- \infty} (V(x) - V_{\infty})  \exp (+ 2 i k_{\infty} x)  \exp \left[- {im \over \hbar^2 k_{\infty}} \int^{x}_{- \infty} (V(x) - V_{\infty}) \, \d y \right] \, \d x.
\nonumber
\\
&&
\end{eqnarray}
This is one form of the distorted Born wave approximation.

\begin{table}[!thb]
\begin{center}
  \setlength{\fboxsep}{0.45 cm} 
   \framebox{\parbox[t]{12.5cm}{
       {\bf  Definition (The Born approximation in scattering theory)}: 
      The Born approximation consists of  taking the unperturbed incident field, in place of the total field, as the driving field at each point in the scatterer. It is the perturbation technique most commonly applied to scattering by an extended body. It is a good approximation if the scattered field is small, compared to the incident field, everywhere inside the scatterer~\cite{born-approximation}.
      
      For our purposes, the Born approximation is a good approximation whenever the ``reflected wave'', characterized in our notation by $b(x)$, is small everywhere,  $|b(x)|\ll1$. (This is a stronger statement than merely saying $|\beta|\ll 1$.)  Similarly, one requires the reflection amplitude to satisfy $|r| \ll 1$.
}}
\end{center}
\end{table}

In brief, the analysis of this chapter has so far collected together a large number of results that otherwise appear quite unrelated. By taking further specific cases of these bounds, and related results, it is possible to reproduce many analytically known results, such as those for delta-function potentials, double-delta-function potentials, square wells, and $\sech^2$ potentials, as discussed later in this chapter. 

\section{Bounds: special case 2}
Having developed a good understanding of the bounds for special case~1, now we reconsider the bounds for special case~2 in~\cite{bounds1}:
Suppose now we take $k(x) = \varphi'(x) \neq 0$ and again set $\chi = 0$. This implies that we are choosing our auxiliary function so that we use the $\mathrm{WKB}$ approximation for the true wavefunction as a ``basis'' for calculating the Bogoliubov coefficients.

This choice is perfectly capable of dealing with the case $V_{+ \infty} \neq V_{- \infty}$, but by reason of the assumed reality of $\varphi$ is limited to considering scattering $\emph{over}$ the potential barrier. (This is the special case implicit in a different context in~\cite{Bordag}). The evolution equations again simplify tremendously.

When we substitute $k(x) = \varphi'(x)$ and  $\chi = 0$ into equation (\ref{sch_eq})
and equation (\ref{sch_eq1}),  we  obtain: 
\begin{equation}
\label{system1}
{\d a \over \d x} = +{1 \over 2 \varphi'} \bigg\{\varphi'' \, b \exp(-2 i \varphi) \bigg\},
\end{equation}
\begin{equation}
\label{system2}
{\d b \over \d x} = +{1 \over 2 \varphi'} \bigg\{\varphi'' \, a \exp(+2 i \varphi) \bigg\}.
\end{equation}
This form of the evolution equations can be related to the qualitative discussion of scattering ``over a potential barrier''  presented by Migdal~\cite{Migdal0, Migdal1}.
For this choice of auxiliary functions we consider the equation (\ref{var_eq}) to derive
\begin{equation}
\vartheta[k(x), \varphi(x), \chi] = {\sqrt{(\varphi'')^2} \over 2 |\varphi'|} = {|\varphi''| \over 2 |\varphi'|} = {|k'| \over 2 |k|},
\end{equation}
\begin{equation}
\vartheta \rightarrow {|\varphi''| \over 2 |\varphi'|} = {|k'| \over 2 |k|},
\end{equation}
and the bounds become
\begin{equation}
\label{bound1}
T \geq \mathrm{sech^2} \bigg({1 \over 2} \int^{+ \infty}_{- \infty} {|k'| \over |k|} \, \d x \bigg),
\end{equation}
and
\begin{equation}
\label{bound2}
R \leq \mathrm{tanh^2} \bigg( {1 \over 2} \int^{+ \infty}_{- \infty} {|k'| \over |k|} \, \d x \bigg).
\end{equation}
The relevant transfer matrix is now
\begin{equation}
\label{bound3}
E(x_f, x_i) = \mathcal{P} \, \exp \bigg({1 \over 2} \int_{x_i}^{x_f} {\varphi'' \over \varphi'} \left[\begin{array}{cc} 0 & e^{-2 i\varphi} \\ - e^{+2i\varphi} & 0 \end{array}\right] \, \d x \bigg).
\end{equation}
The Bogoliubov coefficients are now
\begin{equation}
\label{bound4}
\left[\begin{array}{cc} \alpha & \beta^* \\ \beta & \alpha^* \end{array}\right] = E(\infty, - \infty) = \mathcal{P} \, \exp \bigg(\int_{-\infty}^{\infty} {\varphi'' \over \varphi'} \left[\begin{array}{cc} 0 & e^{-2 i\varphi} \\ - e^{+2i\varphi} & 0 \end{array}\right] \, \d x \bigg).
\end{equation}
\section{Reflection above the barrier}
The system (\ref{system1})-(\ref{system2}) can also be used as the basis of an approximation scheme for $\beta$. Suppose that for all $x$ we have $|b(x)| \ll 1$, so that $|a(x)|\approx 1$. Then
\begin{equation}
{\d b \over \d x} \approx {\varphi'' \over 2 \varphi'} \, \exp(+ 2 i \varphi).
\end{equation}
This may be immediately integrated to yield
\begin{equation}
\beta \approx {1 \over 2} \, \int_{- \infty}^{+ \infty} {\varphi''(x) \over \varphi'(x)} \, \exp (+2 i \varphi) \, \d x.
\end{equation}
Or the equivalent
\begin{equation}
\beta \approx {1 \over 2} \, \int_{- \infty}^{+ \infty} {k'(x) \over k(x)} \, \exp \bigg(+ 2 i \, \int_{-\infty}^{x} k(y) \, \d y \bigg) \, \d x.
\end{equation}
This result serves to clarify the otherwise quite mysterious discussion of so-called ``reflection above the barrier'' given by Migdal~\cite{Migdal0, Migdal1}. Even though the WKB wavefunctions are buried in the representation of the wavefunction underlying the analysis leading to this approximation, the validity of this result for $|\beta|$ does not require validity of the WKB approximation.

If the shifted potential, $V- V_{\infty}$, is ``small'' then we can recover the Born approximation in the usual manner. In that case $k' \equiv m V'/(\hbar^2 k) \approx m V' / (\hbar^2 k_{\infty})$, while $\exp (2 i \int k) \approx \exp(2 i k_{\infty} x)$. A single integration by parts then yields
\begin{equation}
\beta \approx -i {m \over \hbar^2 k_{\infty}} \int_{-\infty}^{+ \infty} (V(x) - V_{\infty}) \, \exp (+ 2 i k_{\infty} x) \, \d x.
\end{equation}
This is now equivalent to the first Born approximation, in this particular context.
\begin{table}[!htb]
\begin{center}
  \setlength{\fboxsep}{0.45 cm} 
   \framebox{\parbox[t]{12.5cm}{
{\bf  Quantum effect}:
The reflection probability $R = |r|^2$ is a function of the energy of the incoming particles. It can always in principle be evaluated by numerically integrating the Schr\"odinger equation for different values of kinetic energy, $E$ and potential energy, $V(x)$. (However, the central idea of this thesis is to see just how much we can do analytically, without resorting to numerical integration.) Furthermore, ``quantum effects'', such as \emph{over-barrier reflection} and \emph{under-barrier transmission}, are dominated by regions of the potential where the naive semiclassical treatment fails.  Due to tunneling, the reflection probabilities at energies \emph{above the barrier} are larger than zero, and the \emph{under-barrier reflection probabilities} are smaller than one~\cite{Segve}.
}}
\end{center}
\end{table}
\section{Under the barrier?}
What goes wrong when we try to extend this analysis into the classically forbidden region? Analytically continuing the system (\ref{system1})--(\ref{system2}) is trivial, replace
\begin{equation}
k^2 = {2 m [E - V(x)] \over \hbar^2} = {2 m [V(x) - E] \over \hbar^2} \, i^2,
\end{equation}
therefore
\begin{equation}
k =  {\sqrt{2 m [V(x) - E]} \over \hbar} \, i,
\end{equation}
and (as required)
\begin{equation}
 \varphi'(x) = k \rightarrow i \kappa = i \sqrt{2 m (V - E)} / \hbar \, .
\end{equation}
Now let 
\begin{equation}
\kappa = \sqrt{2 m (V - E)} / \hbar,
\end{equation}
and write
\begin{equation}
\varphi(x) = \varphi_{\mathrm{t} p} + i \int_{\mathrm{t} p}^{x} \kappa(y) \, \d y \, .
\end{equation}
We will now use these results to re-analyze the Shabat--Zakharov system.
\vfill
\clearpage
\noindent
Substitute the two above equations into equation (\ref{system1}), we obtain
\begin{eqnarray}
\nonumber
{\d a \over \d x} &=& +{1 \over 2 \varphi'} \bigg\{\varphi'' \, b \exp(-2 i \varphi) \bigg\},
\\
\nonumber
&=& + {\kappa' \over 2 \kappa} \, b \exp(-2 i \varphi) ,
\\
\nonumber
&=& +{\kappa' \over 2 \kappa} \, b \exp \bigg(-2 i \bigg(\varphi_{\mathrm{t} p} + i \int_{\mathrm{t} p}^{x} \kappa(y) \, \d y \bigg)\bigg) ,
\\ 
\nonumber
&=& +{\kappa' \over 2 \kappa} \, b \exp (-2 i \varphi_{\mathrm{t} p} ) \exp \bigg(+ 2  \int_{\mathrm{t} p}^{x} \kappa(y) \, \d y \bigg) ,
\\
&=& + {\kappa' \over 2 \kappa} \, b \, \exp (-2 i \varphi_{\mathrm{t} p}) \, \exp \bigg(+ 2 \int \kappa \bigg) \, .
\end{eqnarray}
Similarly to the equation (\ref{system2}),  we obtain
\begin{eqnarray}
\nonumber
{\d b \over \d x} &= &+{1 \over 2 \varphi'} \bigg\{\varphi'' \, a \exp(+2 i \varphi) \bigg\} \, ,
\\
\nonumber
&=& +{\kappa' \over 2 \kappa} \, a \exp(+2 i \varphi),
\\
\nonumber
&=& +{\kappa' \over 2 \kappa} \, a \exp \bigg(+2 i \bigg(\varphi_{\mathrm{t} p} + i \int_{\mathrm{t} p}^{x} \kappa(y) \, \d y \bigg)\bigg),
\\
\nonumber
&=& +{\kappa' \over 2 \kappa} \, a \exp (+2 i \varphi_{\mathrm{t} p} ) \exp \bigg(- 2  \int_{\mathrm{t} p}^{x} \kappa(y) \, \d y \bigg),
\\
&=& + {\kappa' \over 2 \kappa} \, a \exp(+ 2 i \varphi_{\mathrm{t} p}) \, \exp \bigg(- 2 \int \kappa \bigg) \, \, .
\end{eqnarray}
Thus we are now \emph{violating} our previous condition that $\varphi$ be real, though we still require $\varphi' \neq 0$. This is a perfectly good Shabat--Zakharov system that works in the forbidden region. But you cannot now use this to derive bounds on the transmission coefficient. The fly in the ointment resides in the fact that the formula for the probability current is modified, and that in the forbidden region the probability current is
\begin{equation}
\mathscr{J} = \mathrm{Im} \left\{\psi^* {\d \psi \over \d x} \right\} =  \mathrm{Im} \{a b^* \exp(+ 2 i \varphi_{\mathrm{t} p}) \}.
\end{equation}
We can derive the above equation by combining
\begin{eqnarray}
\mathscr{J} &=& \mathrm{Im} \left\{\psi^* {\d \psi \over \d x} \right\},
\end{eqnarray}
and
\begin{eqnarray}
\mathscr{J}  &=&  \{|a|^2 - |b|^2\}.
\end{eqnarray}
For a properly normalized flux in the allowed region $(|a|^2 - |b|^2 = 1)$, we have in the forbidden region
\begin{equation}
\mathrm{Im} \, \{a b^* \exp(+ 2 i \varphi_{\mathrm{t} p})\} = 1.
\end{equation}
While this does imply $2 |a| |b| > 1$, the inequality is unfortunately in the wrong direction to be useful for placing bounds on the transmission coefficient. It is for this reason that we have gone to the trouble of keeping track of the more general gauge condition represented by $\chi$ --- in the hope that we can use this to explore under the barrier.

A good \emph{rigorous bound} (not just a WKB approximation) on transmission under the barrier would be very useful. The best we have been able to do along these lines is presented in chapter \ref{C:consistency-10} discussing the Miller--Good transformation. 
\section{Special case 2-a}
Suppose now that $V(x)$ is continuous and monotonic increasing or decreasing, varying from $V_{-\infty} =  V(- \infty)$ to $V_{+ \infty} = V(+ \infty)$. 

Suppose $E \geq \mathrm{max} \{V_{- \infty}, V_{+ \infty}\}$ so there is no classical turning point. Then
\begin{equation}
\int_{- \infty}^{+ \infty} {|k'| \over |k|} \, \d x = \bigg|\ln \bigg({k_{+ \infty} \over k_{- \infty}}\bigg) \bigg| ,
\end{equation}
and the transmission and reflection probabilities satisfy
\begin{equation}
\label{case2a1}
T \geq {4 k_{+ \infty} k_{- \infty} \over (k_{+ \infty} + k_{- \infty})^2} ,
\end{equation}
and
\begin{equation}
\label{case2a2}
R \leq {(k_{+ \infty} - k_{- \infty})^2 \over (k_{+ \infty} + k_{- \infty})^2}. 
\end{equation}
\begin{figure}[ht] 
\centering
\includegraphics[scale=1]{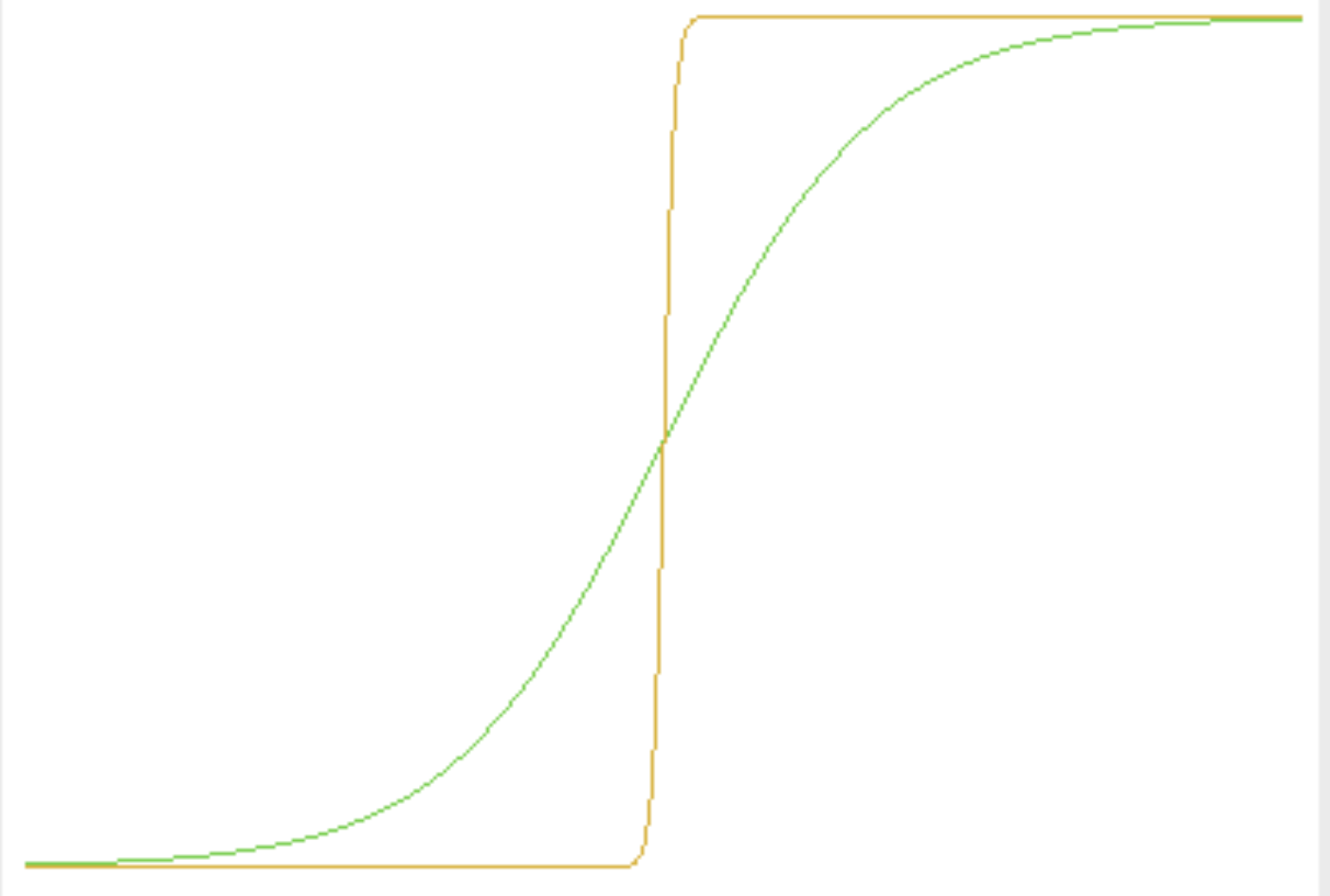}
\caption[Sharp corners maximize reflection]{\label{F:sharp-corners}Sharp corners are guaranteed to maximize reflection,  and abrupt transitions are guaranteed to maximize particle production~\cite{bounds-matt-seminar}.}
\end{figure}
\paragraph{Calculation:}
We consider
\begin{equation}
\vartheta[k(x), \varphi(x), \chi] = {1 \over 2}{|k'| \over  |k|},
\end{equation}
and the bounds become
\begin{eqnarray}
T &\geq&  \mathrm{sech^2} \bigg( \int^{+ \infty}_{- \infty}  {1 \over 2}{|k'| \over  |k|}\, \d x \bigg) =  \mathrm{sech^2} \bigg( {1 \over 2} \bigg|\ln \bigg({k_{+ \infty} \over k_{- \infty}}\bigg) \bigg|  \bigg),
\\
\nonumber
&\geq& \mathrm{sech^2} \bigg({1 \over 2} \bigg|\ln \bigg({k_{+ \infty} \over k_{- \infty}}\bigg) \bigg|  \bigg).
\end{eqnarray}
Remembering that
\begin{eqnarray}
\mathrm{sech(x)} &=& {2 \over \exp(+x) + \exp(-x)},
\end{eqnarray}
and noting
\begin{equation}
\exp \bigg({1\over 2} \bigg|\ln (k_{+ \infty}/k_{- \infty})\bigg|\bigg) = \mathrm{max} \left[\sqrt{{k_{+ \infty} \over k_{- \infty}}}, \sqrt{{k_{- \infty} \over k_{+ \infty}}}\; \right],
\end{equation}
we see that
\begin{eqnarray}
\mathrm{sech\bigg({1 \over 2}
 \bigg|\ln \bigg({k_{+ \infty} \over k_{- \infty}}\bigg) \bigg|  \bigg)} &=& {2 \sqrt{k_{+ \infty}  k_{- \infty}} \over (k_{+ \infty} + k_{- \infty})}.
\end{eqnarray}
Therefore
\begin{eqnarray}
\mathrm{sech}^2\bigg(
 {1 \over 2}\bigg|\ln \bigg({k_{+ \infty} \over k_{- \infty}}\bigg) \bigg|  \bigg) &=& {4 k_{+ \infty}  k_{- \infty} \over (k_{+ \infty} + k_{- \infty})^2},
 \end{eqnarray}
so we obtain
\begin{equation}
T \geq {4 k_{+ \infty} k_{- \infty} \over (k_{+ \infty} + k_{- \infty})^2} ,
\end{equation}
as required.
We now consider
\begin{eqnarray}
R &\leq&  \mathrm{tanh^2} \bigg({1 \over 2} \int^{+ \infty}_{- \infty}  {|k'| \over |k|}\, \d x \bigg),
\\
\nonumber
&\leq& \mathrm{tanh^2} \bigg({1 \over 2} \bigg|\ln \bigg({k_{+ \infty} \over k_{- \infty}}\bigg) \bigg|  \bigg),
\\
\nonumber
&\leq& {(k_{+ \infty} - k_{- \infty})^2 \over (k_{+ \infty} + k_{- \infty})^2},
\end{eqnarray}
as required.
These bounds are instantly understandable as the \emph{exact} analytic results for a step-function potential~\cite{Landau, Capri, Stehle, bounds1}, and the result asserts that for arbitrary smooth monotonic potentials the step function provides upper and lower bounds on the exact result. If we are interested in physical situations such as a time-dependent refractive index~\cite{Liberati, Yablonovitch}, or particle production due to the expansion of the universe~\cite{Birrell}, this technique shows that \emph{sudden} changes in refractive index or size of the universe provide a strict upper bound on particle production.
\section{Special case 2-b}
Suppose now that $V(x)$ has a single unique extremum (either a peak or a valley). Provided that $E \geq \mathrm{max} \{ V_{- \infty}, V_{\mathrm{extremum}}, V_{+ \infty}\}$
so that there is no classical turning point, then $k(x)$ moves monotonically from $k_{- \infty}$ to $k_{\mathrm{extremum}}$ and then back to $k_{+ \infty}$. Under these circumstances
we consider
\begin{eqnarray}
\int_{- \infty}^{+ \infty} {|k'| \over k} \, \d x &=& \bigg|\ln \bigg[{k_{\mathrm{extremum}} \over k_{- \infty}}\bigg]\bigg| + \bigg|\ln \bigg[{k_{\mathrm{extremum}} \over k_{+ \infty}}\bigg]\bigg|\, ,
\\
&=& \bigg|\ln \bigg[{k_{\mathrm{extremum}} \over k_{- \infty}}  \times {k_{\mathrm{extremum}} \over k_{+ \infty}}\bigg]\bigg|\, ,
\\
&=& \bigg|\ln \bigg[{k^2_{\mathrm{extremum}} \over k_{- \infty} k_{+ \infty}}\bigg]\bigg| ,
\end{eqnarray}
as required.

The bound (\ref{E:alpha}) implies
\begin{equation}
|\alpha| \leq \mathrm{cosh} \bigg|\ln \bigg[{k_{\mathrm{extremum}} \over \sqrt{k_{- \infty} k_{+ \infty}}}\bigg]\bigg|\, ,
\end{equation}
and
which yields
\begin{equation}
|\beta| \leq \mathrm{sinh} \bigg|\ln \bigg[{k_{\mathrm{extremum}} \over \sqrt{k_{- \infty} k_{+ \infty}}}\bigg]\bigg|.
\end{equation}
Numerous generalizations of these formulae are possible. For example, at the cost of a little extra notation, we also already have enough information to provide a bound on an \emph{asymmetric} barrier or \emph{asymmetric} well, as long as it has only a single extremum (maximum or minimum) we apply the previous equations to derive 
\begin{equation}
|\alpha| \leq {k^2_{\mathrm{extremum}} + k_{+ \infty} k_{- \infty} \over 2 k_{\mathrm{extremum}} \sqrt{k_{+ \infty} k_{- \infty}}},
\end{equation}
and
\begin{equation}
|\beta| \leq {|k^2_{\mathrm{extremum}} - k_{+ \infty} k_{- \infty}| \over 2 k_{\mathrm{extremum}} \sqrt{k_{+ \infty} k_{- \infty}}}.
\end{equation}
\paragraph{Calculation:} 
\begin{eqnarray}
|\alpha| &\leq& \mathrm{cosh} \bigg|\ln \bigg[{k_{\mathrm{extremum}} \over \sqrt{k_{- \infty} k_{+ \infty}}}\bigg]\bigg|, 
\nonumber
\\ 
&=& {\exp \bigg(+ \bigg|\ln \bigg[{k_{\mathrm{extremum}} \over \sqrt{k_{- \infty} k_{+ \infty}}}\bigg]\bigg|\bigg) + \exp \bigg (- \bigg|\ln \bigg[{k_{\mathrm{extremum}} \over \sqrt{k_{- \infty} k_{+ \infty}}}\bigg]\bigg|\bigg) \over 2},
\nonumber
\\
&=& {k_{\mathrm{extremum}}^2  + k_{- \infty} k_{+ \infty}  \over  2 k_{\mathrm{extremum}} \sqrt{k_{- \infty} k_{+ \infty}}},
\end{eqnarray}
as required.

In addition, we can derive
\begin{eqnarray}
|\beta| &\leq& \mathrm{sinh} \bigg|\ln \bigg[{k_{\mathrm{extremum}} \over \sqrt{k_{- \infty} k_{+ \infty}}}\bigg]\bigg|\, ,
\nonumber
\\
&=& {\exp \bigg(+ \bigg|\ln \bigg[{k_{\mathrm{extremum}} \over \sqrt{k_{- \infty} k_{+ \infty}}}\bigg]\bigg|\bigg) - \exp \bigg (- \bigg|\ln \bigg[{k_{\mathrm{extremum}} \over \sqrt{k_{- \infty} k_{+ \infty}}}\bigg]\bigg|\bigg) \over 2},
\nonumber
\\
&=& {|k_{\mathrm{extremum}}^2  - k_{- \infty} k_{+ \infty} | \over  2 k_{\mathrm{extremum}} \sqrt{k_{- \infty} k_{+ \infty}}},
\end{eqnarray}
as required.

Translated into statements about the transmission and reflection probabilities this becomes
\begin{equation}
\label{trans1}
T \geq {4 k_{+ \infty} k_{- \infty} k_{\mathrm{extremum}}^2 \over \{k_{\mathrm{extremum}}^2 + k_{+ \infty} k_{- \infty}\}^2},
\end{equation}
and
\begin{equation}
\label{trans01}
R \leq {\{k_{\mathrm{extremum}}^2- k_{+ \infty} k_{- \infty}\}^2 \over \{k_{\mathrm{extremum}}^2 + k_{+ \infty} k_{- \infty}\}^2}.
\end{equation}
\paragraph{Calculation:}
\begin{eqnarray}
T &\geq&  \mathrm{sech^2} \bigg( \int_{- \infty}^{+ \infty}{|k'| \over 2 k} \, \d x \bigg) =  \mathrm{sech^2} \bigg( {1 \over 2}\bigg|\ln \bigg[{k_{\mathrm{extremum}}^2 \over k_{- \infty} k_{+ \infty}}\bigg]\bigg|\bigg),
\\
\nonumber
&=& \mathrm{sech^2} \bigg(\bigg|\ln \bigg[{k_{\mathrm{extremum}} \over \sqrt{k_{- \infty} k_{+ \infty}}}\bigg]\bigg|\bigg),
\\
\nonumber
&=& {4 k_{+ \infty} k_{- \infty} k_{\mathrm{extremum}}^2 \over \{k_{\mathrm{extremum}}^2 + k_{+ \infty} k_{- \infty}\}^2}.
\end{eqnarray}
Also
\begin{eqnarray}
R &\leq&  \mathrm{tanh^2} \bigg( \int_{- \infty}^{+ \infty} {|k'| \over 2 k} \, \d x \bigg) =  \mathrm{tanh^2} \bigg( \bigg|\ln \bigg[{k_{\mathrm{extremum}} \over \sqrt{k_{- \infty} k_{+ \infty}}}\bigg]\bigg| \bigg).
\end{eqnarray}
But
\begin{equation}
\mathrm{tanh} (x) = {(\exp(2  x) - 1) \over (\exp(2 x) + 1)},
\end{equation}
therefore
\begin{equation}
R \leq   \mathrm{tanh^2}  \bigg|\ln \bigg[{k_{\mathrm{extremum}} \over \sqrt{k_{- \infty} k_{+ \infty}}}\bigg]\bigg|,
\end{equation}
implying
\begin{equation}
R  \leq {\{k_{\mathrm{extremum}}^2- k_{+ \infty} k_{- \infty}\}^2 \over \{k_{\mathrm{extremum}}^2 + k_{+ \infty} k_{- \infty}\}^2},
\end{equation}
as required.\hfill $\Box$

Equivalently
\begin{equation}
\label{T_equal}
T \geq {4 (E - V_{\mathrm{extremum}}) \sqrt{(E - V_{+ \infty}) (E - V_{- \infty})} \over [(E - V_{\mathrm{extremum}}) + \sqrt{(E - V_{+ \infty}) (E - V_{- \infty})}]^2},
\end{equation}
and
\begin{equation}
\label{R_equal}
R \leq {[(E - V_{\mathrm{extremum}}) - \sqrt{(E - V_{+ \infty}) (E - V_{- \infty})}]^2 \over [(E - V_{\mathrm{extremum}}) + \sqrt{(E - V_{+ \infty}) (E - V_{- \infty})}]^2}.
\end{equation}

\paragraph{Calculation:}
We consider
\begin{equation}
k_{\pm \infty} = {\sqrt{2 m \, (E-V_{\pm \infty}) \over \hbar}},
\end{equation}
and
\begin{equation}
k_{\mathrm{extremum}} = {\sqrt{2 m \, (E-V_{\mathrm{extremum}}) \over \hbar}}.
\end{equation}
we substitute the above equations into equation (\ref{trans1}) to derive equation (\ref{T_equal}). Moreover, we substitute the above equation into equation (\ref{trans01}) to derive equation (\ref{R_equal}). \hfill $\Box$

This can be compared, for example, with known analytic results for the asymmetric square well. 
\bigskip
To be more specific, if in addition $V(- \infty) = V(+ \infty) = V_{\infty}$, so that $k_{- \infty} = k_{+ \infty} = k_{\infty}$, then we have
\begin{equation}
|\alpha| \leq {k^2_{\mathrm{extremum}} + k^2_{\infty} \over 2 k_{\mathrm{extremum}} k_{\infty}},
\end{equation}
and
\begin{equation}
|\beta| \leq {|k^2_{\mathrm{extremum}} - k^2_{\infty}| \over 2 k_{\mathrm{extremum}} k_{\infty}}.
\end{equation}
Translated into statements about the transmission and reflection probabilities this becomes
\begin{equation}
\label{case2b_1}
T \geq {(E - V_{\infty}) (E - V_\mathrm{extremum}) \over (E - V_{\infty}) (E - V_\mathrm{extremum}) + {1 \over 4} (V_{\mathrm{extremum}} - V_{\infty})^2},
\end{equation}
and 
\begin{equation}
R \leq {{1 \over 4}(V_\mathrm{extremum} - V_{\infty})^2 \over (E - V_{\infty}) (E - V_\mathrm{extremum}) + {1 \over 4} (V_{\mathrm{extremum}} - V_{\infty})^2}.
\end{equation}
Equivalently
\begin{equation}
T \geq 1 - {(V_{\mathrm{extremum}} - V_{\infty})^2 \over (2E - V_{\mathrm{extremum}} - V_{\infty})^2},
\end{equation}
and
\begin{equation}
R \leq {(V_{\mathrm{extremum}} - V_{\infty})^2 \over (2E - V_{\mathrm{extremum}} - V_{\infty})^2}.
\end{equation}
For low energies, these results are weaker than the bounds derived under special case 1, (\ref{case1}, \ref{case1_1}) and (\ref{case1_2}, \ref{case1_3}), but have the advantage of requiring more selective information about the potential. For high energies,
\begin{equation}
E \gg {\hbar^2 (V_{\mathrm{extremum}} - V_{\infty})^2 \over 2 m \big( \int_{-\infty}^{+\infty} |V(x) - V_{\infty}| \, \d x \big)^2}, 
\end{equation}
the present result (when it is applicable) leads to tighter bounds on the transmission and reflection coefficients.
\section{Special case 2-c}
Suppose now that $V(x)$ has a number of extrema, (both peaks and valleys). We allow $V(+ \infty) \neq V(- \infty)$, but demand that for all extrema $E \geq \mathrm{max}\{V_{- \infty}, V_{+ \infty}, V_{\mathrm{extremum}}^i\}$ so that there is no classical turning point.

For definiteness, suppose the ordering is: $ - \infty \rightarrow$ peak $\rightarrow$ valley \dots valley $\rightarrow$ peak $\rightarrow + \infty$. Then
\begin{eqnarray}
\int_{- \infty}^{+ \infty} {|k'| \over k} \, \d x &=& \Bigg|\ln \Bigg[{k_{\mathrm{peak}}^1 \over k_{- \infty}}\Bigg]\Bigg| + \Bigg|\ln \Bigg[{k_{\mathrm{valley}}^1 \over k_{\mathrm{peak}}^1}\Bigg]\Bigg| + \dots
\nonumber
\\
&& \dots +\Bigg|\ln \Bigg[{k_{\mathrm{peak}}^n \over k_{\mathrm{valley}}^{n-1}}\Bigg]\Bigg| + \Bigg|\ln \Bigg[{k_{+ \infty} \over k_{\mathrm{peak}}^n}\Bigg]\Bigg|.
\end{eqnarray}
Defining
\begin{eqnarray}
\Pi_{p} (k) &\equiv& \prod_{\mathrm{peaks}} k^i_{\mathrm{peak}} ,
\\
\Pi_{v} (k) &\equiv& \prod_{\mathrm{valley}} k^i_{\mathrm{valley}} ,
\\
\Pi_{e} (k) &\equiv& \prod_{\mathrm{extrema}} k^i_{\mathrm{extremum}} = \Pi_{p} (k)
\, \Pi_{v} (k) ,
\end{eqnarray}
we see
\begin{equation}
\int_{- \infty}^{+ \infty} {|k'| \over k} \, \d x = \Bigg|\ln \Bigg[{\Pi_{p}^2 (k) \over k_{- \infty} k_{+ \infty} \Pi_{v}^2 (k)}\Bigg] \Bigg|.
\end{equation}
This bounds the Bogoliubov coefficients as 
\begin{equation}
|\alpha| \leq {k_{- \infty} k_{+ \infty} \Pi_{v}^2 (k) + \Pi_{p}^2 (k) \over 2 \sqrt{k_{+ \infty} k_{- \infty}} \Pi_{e} (k)} ,
\end{equation}
and
\begin{equation}
|\beta| \leq {|k_{- \infty} k_{+ \infty} \Pi_{v}^2 (k) - \Pi_{p}^2 (k)| \over 2 \sqrt{k_{+ \infty} k_{- \infty}} \Pi_{e} (k)} .
\end{equation}
Then the transmission and reflection probabilities satisfy
\begin{equation}
T \geq {4 k_{+ \infty} k_{- \infty} \Pi_{e}^2 (k) \over \big \{\Pi_{p}^2 (k) + k_{+ \infty} k_{- \infty} \Pi_{v}^2 (k) \big \}^2} ,
\end{equation}
and
\begin{equation}
R \leq {\big\{\Pi_{p}^2 (k) - k_{+ \infty} k_{- \infty} \Pi_{v}^2 (k) \big\}^2 \over \big \{\Pi_{p}^2 (k) + k_{+ \infty} k_{- \infty} \Pi_{v}^2 (k) \big\}^2}.
\end{equation}
In these formulae, peaks and valleys can be interchanged in the obvious way, and by letting the initial or final peak sink down to $V_{\pm \infty}$ as appropriate we obtain bounds for sequences such as: $- \infty \rightarrow$ valley $\rightarrow$ peak \dots valley $\rightarrow$ peak $\rightarrow + \infty$, or: $- \infty \rightarrow$ peak $\rightarrow$ valley \dots peak $\rightarrow$ valley $\rightarrow + \infty$. In the case of one or zero extrema these formulae reduce to the previously given results. [Equations (\ref{trans1})--(\ref{trans01}).]
Further modifications of these formulae are still possible, the cost is that more specific assumptions are needed to derive more specific results.
\section{Bounds: Special case 3}
In the following we will consider the bounds in special case 3. In particular, the most outstanding features of this case is: 

Let $\chi = 0$, $k_0 > 0$ and pick
\begin{equation}
\varphi' = \sqrt{\mathrm{max} \{k^2 (x), k_0^2\}},
\end{equation}
with $x_{0}^{\pm}$ defined by $k^2 (x_{0}^{\pm}) =  k_{0}^{2}$. Then we have
\begin{equation}
 \vartheta = \left\{\begin{array}{r@{\quad \quad}l}
\displaystyle{1 \over 2} {|k'| \over |k|} & k^2 > k_{0}^2; \\[15pt] 
\displaystyle{1 \over 2} {k_{0}^2 - k^2 \over k_{0}} & k^2 < k_0^2. \end{array} \right.
\end{equation}
Note that there are step function discontinuities at $x_{0}^{\pm}$, but no delta-function contribution. It now follows that
\begin{equation}
\oint \vartheta = {1 \over 2} \ln \bigg[{k_{- \infty} \over k_0}\bigg] +
 {1 \over 2 k_0} \int_{k^2 < k_0^2} [k_0^2 - k^2] \, \d x + 
 {1 \over 2} \ln \bigg[{k_{+ \infty} \over k_0} \bigg],
\end{equation}
that is
\begin{equation}
\oint \vartheta = {1 \over 2} \ln \bigg[{k_{- \infty} k_{+ \infty} \over k^2_0}\bigg] + {1 \over 2 k_0} \int_{k^2 < k_0^2} [k_0^2 - k^2] \, \d x .
\end{equation}
Consider
\begin{equation}
\int_{k^2<0} (-k^2) \, \d x = \int_{F} \kappa^2 \, \d x,
\end{equation}
so
\begin{eqnarray}
{1 \over 2 k_0} \int_{k^2 < k_0^2} [k_0^2 - k^2] \, \d x &\leq&{1 \over 2k_{0}}\bigg( \int_{F} \kappa^2 \, \d x + k_{0}^2 \, L_{k^2<k_{0}^2}\bigg) ,
\nonumber\\
&\leq& {1\over 2 k_{0}}\int \kappa^2 \, \d x + {1\over2} \, k_{0} \,  L_{k^2<k_{0}^2}.
\end{eqnarray}
So collecting terms we have 
\begin{eqnarray}
\oint \vartheta &\leq& {1 \over 2} \ln \bigg[{k_{- \infty} k_{+ \infty} \over k^2_0}\bigg] + {1\over 2 k_{0}}\int \kappa^2 \, \d x + {1\over2} \, k_{0} \,  L_{k^2<k_{0}^2}.
\end{eqnarray}
Now note
\begin{eqnarray}
T &\geq& \mathrm{sech^2} \oint \vartheta,
\nonumber\\
&\geq& \mathrm{sech^2}\bigg({1 \over 2} \ln \bigg[{k_{- \infty} k_{+ \infty} \over k^2_0}\bigg] + {1\over 2 k_{0}}\int \kappa^2 \, \d x +  {1\over2} \, k_{0} \,  L_{k^2<k_{0}^2}\bigg), 
\end{eqnarray}
so that 
\begin{equation}
T \geq{4 \over \bigg\{\displaystyle{{\sqrt{k_{-\infty}k_{+\infty}}\over k_{0}}}\exp(B(x)) + \displaystyle{{k_{0}\over \sqrt{k_{-\infty} k_{+\infty}}}} \exp(-B(x))\bigg\}^2},
\end{equation}
where
\begin{equation}
B(x) = {1\over 2k_{0}} \int \kappa^2 \, \d x +  {1\over2} \, k_{0} \, L_{k^2<k_{0}^2},
\end{equation}
when $k_{0}$ is still an adjustable parametric, though we definitely need $0<k_{0}<\mathrm{min} \{k_{+\infty}, k_{-\infty}\}$.

\section{Bounds: Special case 4}
For the fourth special case we want to derive something that looks similar to the WKB approximation, but is a strict bound instead of being an (uncontrolled) estimate.

Choose $\chi = 0$, $k_0 > 0$ and pick
\begin{equation}
\varphi' = \sqrt{\mathrm{max} \{|k^2(x)|, k_0^2\}}.
\end{equation}
Even for a potential with only a single hump there are now five regions to analyze, the two allowed regions, the forbidden region, and two transition regions enclosing the two the classical turning points. As usual we shall define our notation so that in the forbidden region:
\begin{equation}
\kappa^2 (x) = - k^2(x); \qquad \kappa > 0.
\end{equation}
In the two allowed regions
\begin{equation}
\vartheta = {1 \over 2} {|k'| \over |k|}; \qquad k^2 > k_0^2.
\end{equation}
In the two transition regions
\begin{equation}
\vartheta = {1 \over 2} {k_0^2 - k^2 \over k_0}; \qquad -k_0^2 < k^2 < k_0^2.
\end{equation}
Finally in the forbidden region
\begin{equation}
\vartheta = {1 \over 2} \sqrt{{(\kappa')^2 \over \kappa^2} + \kappa^2}; \qquad k^2 < - k_0^2.
\end{equation}
So in the forbidden region by the triangle inequality
\begin{equation}
\vartheta \leq {1 \over 2} \bigg({|\kappa'| \over \kappa} + \kappa\bigg); \qquad k^2 < -k_0^2.
\end{equation}
Collecting these
\begin{eqnarray}
\nonumber
\oint \vartheta &\leq& 
{1 \over 2} \ln \bigg[{k_{- \infty} \over k_0}\bigg] + {1 \over 2 k_0} \int_{k_0^2 > k^2 > -k_0^2}^{k^2_{\mathrm{decreasing}}} [k_0^2 - k^2] \, \d x 
\\
\nonumber
&& 
+{1 \over 2} \ln \bigg[{\kappa_{\mathrm{extremum}} \over k_0} \bigg] + {1 \over 2} \int_{\kappa^2 > k_0^2} \kappa(x)\d x + {1 \over 2} \ln \bigg[{\kappa_{\mathrm{extremum}} \over k_0}\bigg]
\\
&& 
+{1 \over 2 k_0} \int_{-k_0^2 < k^2 < k_0^2}^{k^2_{\mathrm{inclreasing}}}  [k_0^2 - k^2] \, \d x
+ {1 \over 2} \ln \bigg[{k_{+ \infty} \over k_0}\bigg].
\end{eqnarray}
Now in each of the transition regions
\begin{equation}
0 \leq k_0^2 - k^2 \leq 2 k_0^2,
\end{equation}
so 
\begin{eqnarray}
\int_{-k_0^2 < k^2 < k_0^2} [k_0^2 - k^2] \, \d x 
&\leq& 2 k_0^2 \int _{- k_0^2 < k^2 < k_0^2} 1 \, \d x ,
\\ 
&=& 
\vphantom{\Bigg|}
2k_0^2 \, L_{-k_0^2 < k^2 < k_0^2}.
\end{eqnarray}
Here $ L_{-k_0^2 < k^2 < k_0^2}$ is the combined length of the transition regions $-k_0^2 < k^2 < k_0^2$. 
That is, the integral coming from each transition region is bounded (up to a constant) by the physical width of that transition region. Furthermore in the forbidden region
\begin{equation}
\int_{\kappa^2 > k_0^2} \kappa(x) \, \d x
\leq \int_{\kappa^2 > 0} \kappa(x) \, \d x,
\end{equation}
so collecting terms we have 
\begin{equation}
\oint \vartheta \leq {1 \over 2} \ln \bigg[{k_{- \infty} k_{+ \infty} \kappa_{\mathrm{extremum}}^2 \over k_0^4}\bigg] + k_0 \, L_{-k_0^2 < k^2 <k_0^2}
+ {1 \over 2} \int_{\kappa^2 > 0} \kappa(x) \, \d x,
\end{equation}
where \emph{L} now denotes the combined total width of the two transition regions. Now note
\begin{equation}
T \geq \mathrm{sech^2} \oint \vartheta \geq \exp\bigg(-2 \oint \vartheta \bigg),
\end{equation}
so that 
\begin{equation}
T \geq {k_0^4 \over k_{- \infty} k_{+ \infty} \kappa_{\mathrm{extremum}}^2}
\, \exp [- 2 k_0 \, L(k_0)] \, \exp \bigg[ - \int_{\kappa^2 >0} \kappa(x) \d x \bigg].
\end{equation}
The bound is considerably weaker than could have been derived by making more restrictive hypotheses, but has the advantage of elegance and looking very similar to the WKB bound. Note the key requirements: We must be dealing with a single-hump potential, and $k_0$ must be in the range
\begin{equation}
0 < k_0 < \mathrm{min}\{k_{- \infty}, k_{+ \infty}, \kappa_{\mathrm{extremum}}\}.
\end{equation}
The parameter $k_0$ is otherwise arbitrary, and so can be chosen to maximize the prefactor.

\bigskip
One thing we could do is to choose a different value of $k_0$ at each transition, call them $k_1$ and $k_2$, and repeat the analysis keeping careful track of the $\kappa$ integration near the turning points. Then
\begin{eqnarray}
\nonumber
\oint \vartheta &\leq& {1 \over 2} \ln \bigg[{k_{- \infty} k_{+ \infty} \kappa_{\mathrm{extremum}}^2 \over k_1^2 \, k_2^2}\bigg]
\\
\nonumber
&& + {1 \over 2 k_1} \int_{-k_1^2 < k^2 <k_1^2} [k_1^2 - k^2] \, \d x - {1 \over 2} \int_{k_1^2 > \kappa^2 >0} \kappa(x) \, \d x
\\
\nonumber
&& + {1 \over 2 k_2} \int_{-k_2^2 < k^2 <k_2^2} [k_2^2 - k^2] \, \d x - {1 \over 2} \int_{k_2^2 > \kappa^2 >0} \kappa(x) \, \d x
\\
\nonumber
&& + {1 \over 2} \int_{\kappa^2 > 0} \kappa(x) \, \d x.
\end{eqnarray}
So far this is a rigorous bound; now we are going to adopt a linear approximation near the turning points, so that near the first turning point
\begin{equation}
k^2(x) \approx s_1 ( x - x_1),
\end{equation}
with $k(x)$ reaching the values $\pm k_1^2$ at the positions $x = x_1 \pm k_1^2/ s_1$. 
Therefore $L_1 = 2 k_1^2/ s_1$. Then
\begin{equation}
\int_{-k_1^2 < k^2 < k_1^2} [k_1^2 - k^2] \, \d x \approx \int_{x_1 - k_1^2/ s_1}^{x_1+ k_1^2 /s_1} [k_1^2 - s_1 (x -x_1)] \, \d x = {2k_1^4 \over s_1}.
\end{equation}
Similarly
\begin{equation}
\int_{k_1^2 > \kappa^2 > 0} \kappa(x) \d x \approx \int_{x_1}^{{x_1} + k_1^2/s_1} \sqrt{s_1(x-x_1)} \, \d x = {2 \over 3} {k_1^3 \over s_1},
\end{equation}
so that combining, and assuming the linear approximation is a good one
\begin{equation}
\oint \vartheta \leq {1 \over 2} \ln \bigg[{k_{- \infty} k_{+ \infty} \kappa_{\mathrm{extremum}}^2 \over k_1^2 k_2^2}\bigg] + {2 \over 3} {k_1^3 \over s_1} + {2 \over 3} {k_2^3 \over s_2} + {1 \over 2} \int_{\kappa^2 > 0} \kappa(x) \, \d x.
\end{equation}
Extremize with respect to $k_1$, then
\begin{equation}
 -{1 \over k_1} +  {2k_1^2 \over s_1} = 0; \qquad \hbox{so} \qquad k_1^3 = {1\over2} s_1,
\end{equation}
and similarly for $k_{2}$, then
\begin{equation}
\oint \vartheta \leq {1 \over 2} \ln \bigg[{k_{- \infty} k_{+ \infty} \kappa_{\mathrm{extremum}}^2 \over ((1/2) s_1)^{2/3} \, ((1/2) s_2)^{2/3}}\bigg] + {2 \over 3}
+{1 \over 2} \int_{\kappa^2 >0} \kappa(x) \, \d x,
\end{equation}
and hence
\begin{equation}
T \geq {((1/2) s_1)^{2/3} \, ((1/2) s_2)^{2/3} \over k_{- \infty} k_{+ \infty} \kappa_{\mathrm{extremum}}^2} \, \exp[-4/3] \, \exp \bigg[- \int_{\kappa^2 > 0} \kappa(x) \, \d x \bigg].
\end{equation}
Although such results are certainly a major advance in our standing of the rigorous bounds, this particular bound is not $100 \%$ rigorous due to the linear approximation. This suggests that further exploration of these ideas, with a view to obtaining a 100\% rigorous bound, might be profitable.
\section{Bounds: Future directions}
From the general definition
\begin{equation}
\vartheta[k(x), \varphi(x), \chi(x)] \equiv {\sqrt{(\varphi'' + 2 \chi \varphi')^2 + {[k^2(x) + \chi^2 + \chi' - (\varphi')^2]}^2} \over 2|\varphi'|} ,
\end{equation}
and the bound 
\begin{equation}
T \geq \mathrm{sech^2} \bigg[\oint \vartheta \, \d x \bigg],
\end{equation}
there are many other special cases you could in principle derive. The possibilities seem endless and the art is in finding something useful.
\section{Discussion}
In this chapter we dealt with some specific cases of the general bounds and 
developed a number of interesting specializations. In addition, we collected together 
a large number of results that otherwise appeared quite unrelated, including reflection above and below the barrier. The special case bounds were divided into five topics: special cases 1--4, and ``future directions''. In addition, all special cases were chosen for their directness and simplicity.

Furthermore, special case $1$, as presented in equations (\ref{case1})--(\ref{case1_1}) and (\ref{case1_2})--(\ref{case1_3}), has the advantage that it applies to both scattering over the barrier and under the barrier. On the other hand, special case $2$, as presented in equations (\ref{bound1})--(\ref{bound2}) and their specializations, applies only to scattering over the barrier but has the advantage of being much more selective in how much information is needed concerning the scattering potential.

\chapter{Parametric oscillations}
\label{C:consistency-7}
\section{Introduction}
In this chapter we shall present the basic concept of a ``parametric oscillator''. This is a simple harmonic oscillator whose parameters (its resonance frequency $\omega$ and damping time $\beta$) vary in time. The other interesting way of understanding a parametric oscillator is that it is a device that oscillates when one of its ``parameters'' (a physical entity, like capacitance) is changed~\cite{oscillator}.

We shall re-cast and represent the general bounds in terms of 
the specific mathematical structure of parametric oscillations. This time-dependent problem is closely related to the spatial properties of the time-independent Schr\"odinger equation. 

Although the discussion so far has been presented in terms of the spatial properties of the time-independent Schr\"odinger equation, the mathematical structure of parametrically excited oscillations is identical, needing only a few minor translations to be brought into the current form. For a parametrically excited oscillator we have
\begin{equation}
{\d^2 \phi \over \d t^2} = \omega(t)^2 \, \phi.
\end{equation}
Just map $t \rightarrow x, \; \omega(t) \rightarrow k(x)$, and $\phi \rightarrow \psi$. In the general analysis of equation (\ref{var_eq}) the quantity $\vartheta$ should be replaced by
\begin{equation}
\label{eqvar}
\vartheta [\varphi(t), \omega(t)] \equiv {\sqrt{(\varphi'')^2 + {[\omega^2 - (\varphi')^2]}^2} \over 2 |\varphi'|}.
\end{equation}
This is often written as
\begin{equation}
\label{eqvar2}
\vartheta [\varphi(t), \omega(t)] \equiv {\sqrt{(\ddot\varphi)^2 + {[\omega^2 - (\dot\varphi)^2]}^2} \over 2 |\dot\varphi|},
\end{equation}
but this is purely a convention, a change in notation.  (Physicists typically use dots for time derivatives and primes for space derivatives.) 
The analysis then parallels that of the Schr\"odinger equation. Some key results are given below.
\begin{figure}[!htbp] 
\hskip - 2.25 cm
\includegraphics[scale=1]{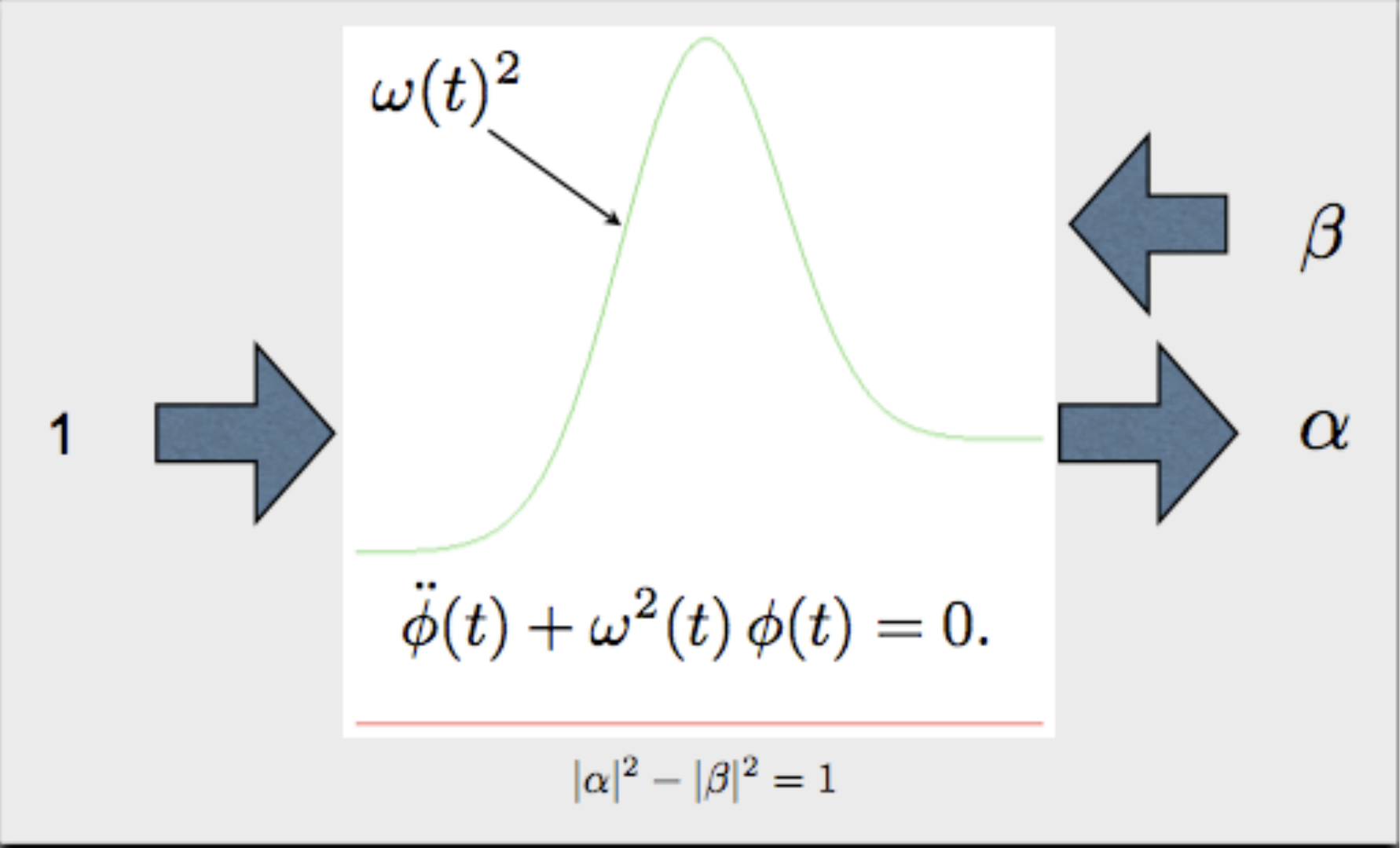}
\caption[Parameteric oscillations and Bogoliubov coefficients]{\label{F:Creation-and-destruction}For a parametric oscillator incoming ``ground state'' fluctuations from the past, (left, coefficient $1$), can in the future be amplified, (right, coefficient $\alpha$).  Quantum mechanically, the coefficient $\beta$ is related to particle excitation from the ground state~\cite{bounds-matt-seminar}.
The quantities $\alpha$ and $\beta$ are typically referred to as the Bogoliubov coefficients. }
\end{figure}
\section{Special case 1}
If $\omega(- \infty) = \omega_{0} = \omega(+ \infty) \neq 0$, then by choosing the auxiliary function to be $\varphi = \omega_0\, t$ we can use equations (\ref{E:vartheta})--(\ref{B:R0}) to deduce
\begin{equation}
|\alpha| \leq \mathrm{cosh}\bigg({1 \over 2 \omega_{0}} \int_{- \infty}^{+ \infty} |\omega^2(t) - \omega_0^2| \, \d t\bigg) ,
\end{equation}
and
\begin{equation}
|\beta| \leq \mathrm{sinh} \bigg({1 \over 2 \omega_{0}} \int_{- \infty}^{+ \infty} |\omega^2(t) - \omega_0^2| \, \d t\bigg) .
\end{equation}
\paragraph{Calculation:}
We now consider
\begin{equation}
\label{alpha_bound2}
|\alpha| \leq \mathrm{cosh} \bigg(\int_{- \infty}^{+\infty} \vartheta [\varphi(t), \omega(t)] \, \d t\bigg).
\end{equation}
We substitute equation (\ref{eqvar}) into equation (\ref{alpha_bound2}), now we derive
\begin{equation}
|\alpha| \leq \mathrm{cosh} \left(\int_{- \infty}^{+\infty} {\sqrt{(\varphi'')^2 + {[\omega^2 - (\varphi')^2]}^2} \over 2 |\varphi'|} \, \d t \right).
\end{equation}
Furthermore, under the stated assumptions we can simplify the above integral to derive
\begin{equation}
|\alpha| \leq \mathrm{cosh}\bigg({1 \over 2 \omega_{0}} \int_{- \infty}^{+ \infty} |\omega^2(t) - \omega_0^2| \, \d t\bigg) ,
\end{equation}
which automatically implies
\begin{equation}
|\beta| \leq \mathrm{sinh} \bigg({1 \over 2 \omega_{0}} \int_{- \infty}^{+ \infty} |\omega^2(t) - \omega_0^2| \, \d t\bigg),
\end{equation}
as required.
\section{Special case 2}
If $\omega(- \infty)$ and $\omega(+ \infty) \neq 0$ are both finite so that suitable asymptotic states exist, and assuming $\omega^2(t) \geq 0$ so that the frequency is always positive, then applying equations (\ref{bound1})--(\ref{bound2}) to the case of parametric resonance yields
\begin{equation}
|\alpha| \leq \mathrm{cosh} \, \bigg|\int_{- \infty}^{+ \infty}{1 \over 2}{|\omega'(t)| \over |\omega(t)|} \, \d t \bigg| \, ,
\end{equation}
and
\begin{equation}
|\beta| \leq \mathrm{sinh} \, \bigg| \int_{- \infty}^{+ \infty}{1 \over 2}{|\omega'(t)| \over |\omega(t)|} \, \d t \bigg| \, .
\end{equation}
\section{Special case 2-a}
Suppose now that $\omega^2(t)$ is positive semidefinite, continuous, and monotonic increasing or decreasing, varying from $\omega_{- \infty} = \omega(- \infty) \neq 0$ to some distinct value $\omega_{+ \infty} = \omega(+ \infty) \neq 0$. The Bogoliubov coefficients satisfy
\begin{equation}
\label{case2_a_1}
|\alpha| \leq {\omega_{- \infty} + \omega_{+ \infty} \over 2 \sqrt{\omega_{- \infty} \omega_{+ \infty}}} \; ,
\end{equation}
and
\begin{equation}
\label{case2_a_2}
|\beta| \leq {|\omega_{- \infty} - \omega_{+ \infty}| \over 2\sqrt{\omega_{- \infty} \omega_{+ \infty}}} \, .
\end{equation}
\paragraph{Calculation:}
We now consider
\begin{eqnarray}
|\alpha| &\leq& \mathrm{cosh} \, \bigg|\int_{- \infty}^{+ \infty}{1 \over 2} {|\omega'(t)| \over |\omega(t)|} \, \d t \bigg| \, ,
\\
\nonumber
&\leq& \mathrm{cosh} \,  \bigg|{1 \over 2}\ln \bigg({\omega_{+ \infty} \over \omega_{- \infty}}\bigg) \bigg| \, ,
\\
\nonumber
&\leq& {\omega_{- \infty} + \omega_{+ \infty} \over 2 \sqrt{\omega_{- \infty} \omega_{+ \infty}}} \; ,
\end{eqnarray}
as required.
Similarly 
\begin{eqnarray}
|\beta|  &\leq& \mathrm{sinh} \, \bigg|\int_{- \infty}^{+ \infty}{1 \over 2} {|\omega'(t)| \over |\omega(t)|} \, \d t \bigg| \, ,
\\
\nonumber
&\leq& \mathrm{sinh} \, \bigg|{1 \over 2}\ln \bigg({\omega_{+ \infty} \over \omega_{- \infty}}\bigg) \bigg| \, ,
\\
\nonumber
&\leq& {|\omega_{- \infty} - \omega_{+ \infty}| \over 2\sqrt{\omega_{- \infty} \omega_{+ \infty}}} \, ,
\end{eqnarray}
as required.
\section{Special case 2-b}
Under the restriction $\omega(- \infty) = \omega_{0} = \omega(+ \infty) \neq 0$, with the additional constraint that $\omega(t)$ has a single unique extremum (either a maximum or a minimum but not both), and provided that $\omega_{\mathrm{extremum}}^2 > 0$ so that we do not encounter complex frequencies (no classical turning point), the Bogoliubov coefficients satisfy
\begin{equation}
|\alpha| \leq {\omega_0^2 + \omega_{\mathrm{extremum}}^2 \over2 \omega_0 \omega_{\mathrm{extremum}}} \, ,
\end{equation}
and
\begin{equation}
|\beta| \leq {|\omega_0^2 - \omega_{\mathrm{extremum}}^2| \over2 \omega_0 \omega_{\mathrm{extremum}}} \, .
\end{equation}
\paragraph{Calculation:}
To prove the above equations, we consider
\begin{eqnarray}
\int_{- \infty}^{+ \infty} {|\omega'| \over \omega} \, \d x &=& \bigg|\ln \bigg[{\omega_{\mathrm{extremum}} \over \omega_{- \infty}}\bigg]\bigg| + \bigg|\ln \bigg[{\omega_{\mathrm{extremum}} \over \omega_{+ \infty}}\bigg]\bigg| \, ,
\\
&=& \bigg|\ln \bigg[{\omega^2_{\mathrm{extremum}} \over \omega_{- \infty} \omega_{+ \infty}}\bigg]\bigg|\, ,
\\
\nonumber
&=& \bigg|\ln \bigg[{\omega^2_{\mathrm{extremum}} \over \omega^2_0}\bigg]\bigg|.
\end{eqnarray}
This implies
\begin{eqnarray}
|\alpha| &\leq& \mathrm{cosh} \bigg|{1 \over 2}\ln \bigg[{\omega^2_{\mathrm{extremum}} \over \omega^2_0}\bigg]\bigg| \, ,
\\
\nonumber
&\leq& \mathrm{cosh} \bigg|\ln \bigg[{\omega_{\mathrm{extremum}} \over \omega_0}\bigg]\bigg| \, ,
\\
\nonumber
 &\leq& {\omega_0^2 + \omega_{\mathrm{extremum}}^2 \over2 \omega_0 \omega_{\mathrm{extremum}}}.
\end{eqnarray}
Furthermore, 
\begin{eqnarray}
|\beta| &\leq& \mathrm{sinh} \bigg|{1 \over 2}\ln \bigg[{\omega^2_{\mathrm{extremum}} \over \omega^2_0}\bigg]\bigg| \, ,
\\
\nonumber
&\leq& \mathrm{sinh} \bigg|\ln \bigg[{\omega_{\mathrm{extremum}} \over \omega_0}\bigg]\bigg| \, ,
\\
\nonumber
&\leq& {|\omega_{\mathrm{extremum}}^2 - \omega_0^2|\over2 \omega_0 \omega_{\mathrm{extremum}}}.
\end{eqnarray}

Suppose now that $\omega^2(t)$ has a single unique extremum (either a peak or a valley), but that we allow the two asymptotic frequencies to differ $\omega(+ \infty) \neq \omega(- \infty)$, and suppose further that $\omega^2(t) > 0$ so that there is no classical turning point. The Bogoliubov coefficients satisfy
\begin{equation}
|\alpha| \leq {\omega_{- \infty} \omega_{+ \infty} + \omega_{\mathrm{extremum}}^2 \over 2 \sqrt{\omega_{- \infty} \omega_{+ \infty}} \omega_{\mathrm{extremum}}} \, ,
\end{equation}
and
\begin{equation}
|\beta| \leq {|\omega_{- \infty} \omega_{+ \infty} - \omega_{\mathrm{extremum}}^2|
 \over 2 \sqrt{\omega_{- \infty} \omega_{+ \infty}} \omega_{\mathrm{extremum}}} \, .
\end{equation}
\paragraph{Calculation:}

We now show how to derive the above equations:  For the moment we shall consider
\begin{eqnarray}
|\alpha| &\leq& \mathrm{cosh} \bigg|\ln \bigg[{\omega_{\mathrm{extremum}} \over \sqrt{\omega_{- \infty}\omega_{+ \infty}}}\bigg]\bigg| \, ,
\\
\nonumber
 &\leq& {\omega_{\mathrm{extremum}}^2 + \omega_{- \infty} \omega_{+ \infty} \over2 \sqrt{\omega_{- \infty} \omega_{+ \infty}} \omega_{\mathrm{extremum}}}.
\end{eqnarray}
But this now yields
\begin{equation}
|\beta| \leq { |\omega_{\mathrm{extremum}}^2 - \omega_{- \infty} \omega_{+ \infty}| \over 2 \sqrt{\omega_{- \infty} \omega_{+ \infty}} \omega_{\mathrm{extremum}}} \, ,
\end{equation}
as required.
\section{Special case 2-c}
Suppose now that $\omega(t)$ has a number of extrema (both peaks and valleys). We allow $\omega(+ \infty) \neq \omega(- \infty)$, but demand that for all extrema $\omega_{\mathrm{extremum}}^{i} > 0$ so that there is no classical turning point.

For definiteness, suppose the ordering is: $- \infty \rightarrow$ peak $\rightarrow$ valley \dots valley $\rightarrow$ peak $\rightarrow + \infty$. Define
\begin{eqnarray}
\Pi_{p}(\omega) &\equiv& \prod_{\mathrm{peaks}} \omega_{\mathrm{peak}}^i ,
\\
\Pi_{v}(\omega) &\equiv& \prod_{\mathrm{valleys}} \omega_{\mathrm{valley}}^i ,
\\
\Pi_{e}(\omega) &\equiv& \prod_{\mathrm{extrema}} \omega_{\mathrm{extremum}}^i .
\end{eqnarray}
The Bogoliubov coefficients satisfy
\begin{equation}
|\alpha| \leq {\omega_{- \infty} \omega_{+ \infty} \Pi_{v}^2(\omega) + \Pi_{p}^2(\omega) \over \sqrt{\omega_{+ \infty}\omega_{- \infty}} \Pi_{e}(\omega)} ,
\end{equation}
and
\begin{equation}
|\beta| \leq {|\omega_{- \infty} \omega_{+ \infty} \Pi_{v}^2(\omega) - \Pi_{p}^2(\omega)| \over \sqrt{\omega_{+ \infty}\omega_{- \infty}} \Pi_{e}(\omega)} .
\end{equation}
In these formulae, peaks and valleys can be interchanged in the obvious way, and by letting the initial or final peak sink down to $\omega_{\pm \infty}$ as appropriate we obtain bounds for sequences such as: $- \infty \rightarrow$ valley $\rightarrow$ peak \dots valley $\rightarrow$ peak $\rightarrow + \infty$, or: $- \infty \rightarrow$ peak $\rightarrow$ valley \dots peak $\rightarrow$ valley $\rightarrow + \infty$. In the case of one or zero extrema these formulae reduce to the previously given results.

Again, further specializations of these formulae are still possible. As always there is a trade-off between the strength of the result and its generality.

\paragraph{Calculation:}
We see
\begin{equation}
\int_{- \infty}^{+ \infty} {|\omega'| \over \omega} \, \d t = \Bigg|\ln \Bigg[{\Pi_{p}^2 (\omega) \over \omega_{- \infty} \omega_{+ \infty} \Pi_{v}^2 (\omega)}\Bigg] \Bigg|.
\end{equation}
The Bogoliubov coefficients in this case are bounded by
\begin{equation}
|\alpha| \leq {\omega_{- \infty} \omega_{+ \infty} \Pi_{v}^2 (\omega) + \Pi_{p}^2 (\omega) \over 2 \sqrt{\omega_{+ \infty} \omega_{- \infty}} \Pi_{e} (\omega)} ,
\end{equation}
and
\begin{equation}
|\beta| \leq {|\omega_{- \infty} \omega_{+ \infty} \Pi_{v}^2 (\omega) - \Pi_{p}^2 (\omega)| \over 2 \sqrt{\omega_{+ \infty} \omega_{- \infty}} \Pi_{e} (\omega)} .
\end{equation}
\section{Bounds: Special case 3}
We let $\chi = 0$, $\omega_0 > 0$, then we can choose
\begin{equation}
\varphi' = \sqrt{\mathrm{max} \{\omega^2 (t), \omega_0^2\}},
\end{equation}
with $t_{0}^{\pm}$ defined by $\omega^2 (t_{0}^{\pm}) =  \omega_{0}^{2}$. Then we have
\begin{equation}
 \vartheta = \left\{\begin{array}{r@{\quad \quad}l}
\displaystyle{1 \over 2} {|\omega'| \over |\omega|} & \omega^2 > \omega_{0}^2; \\[15 pt]
\displaystyle{1 \over 2} {\omega_{0}^2 - \omega^2 \over \omega_{0}} & \omega^2 < \omega_0^2. \end{array} \right.
\end{equation}
Note that there are step function discontinuities at $t_{0}^{\pm}$, but no delta-function contribution. It now follows that
\begin{equation}
\oint \vartheta = {1 \over 2} \ln \bigg[{\omega_{- \infty} \over \omega_0}\bigg] + {1 \over 2 \omega_0} \int_{\omega^2 < \omega_0^2} [\omega_0^2 - \omega^2] \, \d t + {1 \over 2} \ln \bigg[{\omega_{+ \infty} \over \omega_0} \bigg].
\end{equation}
That is 
\begin{equation}
\oint \vartheta = {1 \over 2} \ln \bigg[{\omega_{- \infty} \omega_{+ \infty} \over \omega^2_0}\bigg] + {1 \over 2 \omega_0} \int_{\omega^2 < \omega_0^2} [\omega_0^2 - \omega^2] \, \d t.
\end{equation}
As usual we shall define our notation so that in the forbidden region: 
\begin{equation}
\int_{\omega^2<0} (-\omega^2) \, \d t = \int_{F} \Omega^2 \, \d t. 
\end{equation}
So
\begin{eqnarray}
{1 \over 2 \omega_0} \int_{\omega^2 < \omega_0^2} [\omega_0^2 - \omega^2] \, \d t &\leq&{1 \over 2\omega_{0}}\bigg( \int_{F} \Omega^2 \, \d t +  \omega_{0}^2 \, L_{\omega^2<\omega_{0}^2}\bigg) ,
\nonumber\\
&\leq& {1\over 2 \omega_{0}}\int \Omega^2 \, \d t + {1\over2} \omega_{0} \,  L_{\omega^2<\omega_{0}^2}.
\end{eqnarray}
Collecting terms we have 
\begin{eqnarray}
\oint \vartheta &\leq& {1 \over 2} \ln \bigg[{\omega_{- \infty} \omega_{+ \infty} \over \omega^2_0}\bigg] + {1\over 2 \omega_{0}}\int \Omega^2 \, \d t + {1\over2}\omega_{0} \,  L_{\omega^2<\omega_{0}^2},
\end{eqnarray}
and the bound on the Bogoliubov coefficients in this case become
\begin{eqnarray}
|\alpha| &\leq& \mathrm{cosh} \bigg(\oint \vartheta\bigg),
\nonumber\\
&\leq& {\sqrt{\omega_{-\infty}\omega_{+\infty}}\over 2 \omega_{0}}\exp\bigg({1\over 2\omega_{0}} \int \Omega^2 \, \d t + {1\over2}\omega_{0} \, L_{\omega^2<\omega_{0}^2}\bigg)
\nonumber\\
&&+ {\omega_{0}\over 2\sqrt{\omega_{-\infty} \omega_{+\infty}}}\exp\bigg(-{1\over 2\omega_{0}} \int \Omega^2 \, \d t  -{1\over2}\omega_{0} \, L_{\omega^2<\omega_{0}^2}\bigg),
\nonumber\\
&&
\end{eqnarray}
and
\begin{eqnarray}
|\beta| &\leq& \mathrm{sinh} \bigg(\oint \vartheta\bigg),
\nonumber\\
&\leq& \bigg|{\sqrt{\omega_{-\infty}\omega_{+\infty}}\over2\omega_{0}}\exp\bigg({1\over 2\omega_{0}} \int \Omega^2 \, \d t + {1\over2}\omega_{0} \, L_{\omega^2<\omega_{0}^2}\bigg)
\nonumber
\\
&&-{\omega_{0}\over 2\sqrt{\omega_{-\infty}\omega_{+\infty}}} \exp\bigg(-{1\over 2\omega_{0}} \int \Omega^2 \, \d t  -{1\over2}\omega_{0} \, L_{\omega^2<\omega_{0}^2}\bigg)\bigg|\, ,
\nonumber\\
&&
\end{eqnarray}
which gives us our third special case bound.
Then the transmission and reflection probabilities satisfy
\begin{equation}
T \geq {4 \over \displaystyle\left\{{\sqrt{\omega_{-\infty}\omega_{+\infty}}\over \omega_{0}} \exp(Z(t)) + {\omega_{0}\over\sqrt{\omega_{-\infty}\omega_{+\infty}}} \exp(-Z(t))\right\}^2},
\end{equation}
and
\begin{equation}
R \leq {\{\omega_{-\infty}\omega_{+\infty}\exp(Z(t)) - \omega_{0}^2 \exp(-Z(t))\}^2\over \{\omega_{-\infty}\omega_{+\infty}\exp(Z(t)) + \omega_{0}^2 \exp(-Z(t))\}^2},
\end{equation}
where
\begin{equation}
Z(t) = {1\over 2\omega_{0}} \int \Omega^2 \, \d t + {1\over2}\omega_{0} \, L_{\omega^2<\omega_{0}^2}.
\end{equation}
\section{Discussion}

Though the discussion in previous chapters has been presented in terms of the spatial properties of the time-independent Schr\"odinger equation, we have seen in this chapter that the mathematical structure of parametrically excited oscillations is essentially identical, needing only a few minor translations to be brought into the current form. 

In summary, the bounds presented in this chapter are useful in establishing qualitative analytic properties of parametric oscillators, and as such are complementary to both explicit numerical investigations and the guidance extracted from exact analytic solutions.

\chapter{Bounding the Bogoliubov coefficients}
\label{C:consistency-8}
\section{Introduction}

In this chapter we will again be considering (from a somewhat different point of view) the problem of finding \emph{approximate solutions} for wave equations in general, and quantum mechanical problems in particular. It appears that as yet relatively little work seems to have been put into the complementary problem of founding rigorous \emph{bounds} on the exact solutions. We have in mind either bounds on parametric amplification and the related quantum phenomenon of particle production (as encoded in the Bogoliubov coefficients), or bounds on transmission and reflection coefficients. 

In the last section of appendix {\bf B}, we introduce and discuss the time ordering  and give some more details of time-ordered exponentials --- these are a very convenient trick for formally solving certain matrix differential equations. 
Practising physicists and applied mathematicians will all have seen the \emph{WKB approximation} for barrier penetration probability. Unfortunately, the \emph{WKB approximation} is an example of an uncontrolled approximation, and in general we do not know if it is an over-estimate or an under-estimate.

As part of the main work, we modify and improve an approach first developed in~\cite{bounds1}. We shall examine this question by developing a \emph{formal} but \emph{exact} solution for the appropriate second-order linear ODE, in terms of a  time-ordered exponential of $2\times2$ matrices, then relating the Bogoliubov coefficients to certain invariants of this matrix. By bounding the matrix in an appropriate manner, we can thereby bound the Bogoliubov coefficients.
\section{The second-order ODE}
We would first like to present  ``the second-order ODE''  techniques developed in~\cite{bounds1}, that are applicable to numerous  physical situations; situations  which are both extremely interesting and important. Consider the ODE
\begin{equation}
\ddot \phi(t) + \omega^2(t) \,\phi(t)=0,
\label{E:time}
\end{equation}
or its equivalent in the space domain~\cite{bounds1}
\begin{equation}
 \phi''(x) + k^2(x) \,\phi(x)=0.
 \label{E:space}
\end{equation}
It is easy to see that equation (\ref{E:time}) can be viewed (in terms of the time domain) as an example of parametrically excited oscillation; it arises for instance when a wave propagates through a medium whose refractive index is externally controlled to be a function of time (though remaining spatially invariant).\footnote{For instance, situations of this type have been used to model sonoluminescence~\cite{sonoluminescence}, and more recently both quasiparticle production in analogue spacetimes~\cite{quasiparticle} and analogue signature change events~\cite{signature}.  In all these situations it is extremely useful to have rigorous and largely model-independent bounds on the amount of particle production that might reasonably be expected.}\; In contrast, the spatial version of this equation as presented in~(\ref{E:space}) arises classically in situations where the refractive index is spatially dependent (so called ``index gradient'' situations), or in a quantum physics context when considering the Schr\"odinger equation for a time-independent potential:
\begin{equation}
\label{Schrodinger-equation-chapter8}
- {\hbar^2\over2m} \, \phi''(x) + V(x) \,\phi(x)=E \,\phi(x),
\end{equation}
as long as one makes the translation
\begin{equation}
k^2(x) \leftrightarrow {2m[E-V(x)]\over \hbar^2}.
\end{equation}
However they arise, equations (\ref{E:time}) and (\ref{E:space}) are central to the study of both quantum physics and wave phenomena generally.
\begin{table}[!htb]
\begin{center}
 \setlength{\fboxsep}{0.45 cm} 
 \framebox{\parbox[t]{12.5cm}{
{\bf  Refractive index (or index of refraction)}:  This index is a measure for how much the speed of light (or other waves such as sound waves) is reduced inside a medium. The refractive index, $n$, of a medium is defined as the ratio of the phase velocity, $c$, of a wave phenomenon (for instance light or sound) in some reference medium (such as vacuum, or air at standard temperature and pressure) to the phase velocity, $v_{p}$, in the medium itself~\cite{refractive-index}:
\begin{equation}
n = {c\over v_{p}}.
\end{equation}
}}
\end{center}
\end{table}
As the result of this central significance, over the last century or more a vast body of work has gone into the question of finding \emph{approximate solutions} to equations~(\ref{E:time}) and~(\ref{E:space}). Most of these approximations are  typically based on JWKB techniques and their variants (phase integral techniques, \etc.)~\cite{approximate}.  In contrast very little work seems to have gone into the physically important question of finding \emph{explicit bounds} on the relevant Bogoliubov coefficients and/or reflection and transmission coefficients~\cite{bounds1}. 

\begin{center}
  \setlength{\fboxsep}{0.45 cm} 
   \framebox{\parbox[t]{12.5cm}{
\noindent
{\bf Index-gradient methods}: So-called index gradient optics is the branch of optics
covering optical effects produced by a gradual spatial variation of the \emph{refractive index} 
of a material~\cite{gradient-index optics}.

One can analogously speak of index gradient acoustics when the speed of sound is slowly varying as a function of position.
}}
\end{center}

In this chapter we shall modify and streamline the analysis of~\cite{bounds1}; presenting an alternative proof that is considerably more direct and focussed than that in~\cite{bounds1}.
We can make this discussion appear to be so simple and straightforward by assuming that $\omega(t)\to\omega_0$ (equivalently $k(x)\to k_0$) outside some region of compact support $[t_i,t_f]$ (equivalently $[x_i,x_f]$). That is, concentrating on the time-domain formulation of equation~(\ref{E:time}), the quantity $\omega^2(t)-\omega_0^2$ is a function of compact support.\footnote{Of course, this ``compact support'' condition is not  strictly necessary, and at the cost of a little more analysis one can straightforwardly extend the comments below to a situation where there is a finite limit $\omega(t)\to\omega_{\infty}$ as $t\to\pm\infty$~\cite{bounds1}. At the cost of somewhat more tedious additional work,  there are also useful things that can be said of the situation where $\omega(t)\to\omega_{\pm\infty}$, with $\omega_{-\infty}\neq \omega_{+\infty}$, as $t\to\pm\infty$~\cite{bounds1}.} Because of this compact support property we know that everywhere outside the region $[t_i,t_f]$ the exact solution of the wave equation (\ref{E:time}) is given by linear combinations of $\exp(\pm i \omega_0\, t)$, and that the central question to be investigated is the manner in which exact solutions on the initial domain $(-\infty,t_i)$ ``connect'' with exact solutions on the final domain $(t_f,+\infty)$. 
\begin{table}[!htb]
\begin{center}
  \setlength{\fboxsep}{0.45 cm} 
   \framebox{\parbox[t]{12.5cm}{
       {\bf Comment:}  Describing and characterizing this ``connection'' between the (known) plane wave solutions to the left and right of the barrier is exactly what the Bogoliubov coefficients are designed to do.

In order to explain these issues in more detail, in the next section we shall introduce the time-ordered exponentials, otherwise the rigorous bounds we shall derive on the transmission, reflection, and Bogoliubov coefficients will be almost impossible to understand. 
}}
\end{center}
\end{table}
Our approach will be to focus on using the above definition as a guide to the appropriate starting point. We can now systematically  develop a \emph{formal} but \emph{exact} solution for the appropriate second-order linear ODE in terms of a  time-ordered exponential of $2\times2$ matrices, then relating the Bogoliubov coefficients to certain invariants of this matrix.

We are interested in solving, \emph{exactly} but possibly \emph{formally}, the second-order ODE
\begin{equation}
\ddot \phi(t) + \omega^2(t) \,\phi(t)=0.
\end{equation}
One way of proceeding is as follows: Define a momentum
\begin{equation}
\pi  = \dot\phi,
\end{equation}
and then rewrite the second-order ODE as a system of first-order ODEs
\begin{equation}
\dot \phi = \pi;
\end{equation}
\begin{equation}
\dot\pi = -\omega^2(t)\; \phi;
\end{equation}
or in matrix notation (where we have carefully arranged all matrix elements and vector components to carry the same engineering dimensions)  
\begin{equation}
{\d\over\d t} \left[\begin{array}{c} \phi\\ \pi/\omega_0 \end{array}\right] =
\left[\begin{array}{cc} 0 & \omega_0 \\ -\omega^2/\omega_0 & 0 \end{array}\right] \; 
\left[\begin{array}{c} \phi\\ \pi/\omega_0 \end{array}\right].
\end{equation}
This matrix ODE \emph{always} has a \emph{formal} solution in terms of the so-called ``time-ordered exponential'' 
\begin{equation}
\left[\begin{array}{c} \phi\\ \pi/\omega_0 \end{array}\right]_t = 
\mathscr{T} \left\{ \exp\left( 
\int_{t_0}^t 
\left[\begin{array}{cc} 0 & \omega_0\\ -\omega^2(\bar t) /\omega_0& 0 \end{array}\right] 
\d \bar t
\right) \right\} \; \left[\begin{array}{c} \phi\\ \pi/\omega_0 \end{array}\right]_{t_0}.
\end{equation}
The meaning of the time-ordered exponential is somewhat tricky, but ultimately is just a $2\times2$ matrix specialization of the operator-valued version of the ``time-ordered exponential''  familiar from developing quantum field theoretic perturbation theory in the so-called ``interaction picture''~\cite{qft-interaction}. Specifically,  let us partition the interval $(t_0,t)$ as follows:
\begin{equation}
t_0 < t_1 < t_2 < t_3 ... < t_{n-3} < t_{n-2} < t_{n-1} < t_n=t,
\end{equation}
and define the ``mesh'' as 
\begin{equation}
M = \max_{i\in (1,n)} \{ t_i - t_{i-1} \}.
\end{equation}
Then define the time-ordered exponential as
\begin{eqnarray}
T(t) &=& \mathscr{T} \left\{ \exp\left( 
\int_{t_0}^t 
\left[\begin{array}{cc} 0 & \omega_0\\ -\omega^2(\bar t)/\omega_0 & 0 \end{array}\right] 
\d \bar t
\right) \right\} ,
\nonumber\\
&\equiv&
\lim_{M\to 0,\; (n\to \infty)} \prod_{i=0}^{n-1}  
\exp\left( 
\left[\begin{array}{cc} 0 & \omega_0\\ -\omega^2(t_{n-i})/\omega_0 & 0 \end{array}\right]  
\; (t_{n-i}-t_{n-i-1}) 
\right).
\nonumber
\\
&&
\end{eqnarray}
\begin{table}[!htb]
\begin{center}
  \setlength{\fboxsep}{0.45 cm} 
   \framebox{\parbox[t]{12.5cm}{
{\bf Comment}:   In this matrix product ``late times'' are always ordered to the left, and ``early times'' to the right.
Now we can extract \emph{all} the interesting physics by working with this time-ordered matrix. (If we work in the space domain then the equivalent matrix $T$ is ``path-ordered'', and is closely related to the so-called  ``transfer matrix''.)

\begin{itemize}
\item
Since all of the  ``complicated'' physics takes place for $t\in (t_i,t_f)$, it is also useful to define
\begin{equation}
T = \mathscr{T} \left\{ \exp\left( 
\int_{t_i}^{t_f} \left[\begin{array}{cc} 0 & \omega_0 \\ -\omega^2(\bar t)/\omega_0 & 0 \end{array}\right] \d \bar t
\right) \right\} =
 \left[\begin{array}{cc} a & b\\ c & d \end{array}\right].
 \end{equation}

\item 
We are guaranteed that $\det[T]=1$, that is $ad-bc=1$. This follows from the fact that $\det[T] = \exp\{\tr(\ln[T])\}$, and the explicit formula for $T$ above.

\item
 Another particularly nice feature is that with the current definitions the transfer matrix $T$ is manifestly \emph{real}. This is relatively rare when setting up scattering or particle production problems, so we shall make the most of it.
 \end{itemize}
}}
\end{center}
\end{table}
\section{Bogoliubov coefficients}
We have already been introduced to the concept of Bogoliubov coefficients, and also some ways to calculate them, in chapter~\ref{C:consistency-5}. Let us now extract the Bogoliubov coefficients in the present situation.  Before $t_i$, and after $t_f$, the wave-function is just linear combinations of $\exp(\pm i \omega_0 \, t)$. We can prepare things so that before $t_i$ the wavefunction is pure  $\exp(+ i \omega_0 \, t)$,
\begin{equation}
\psi(t\leq t_i) = \exp(+ i \omega_0 \, t);
\end{equation}
in which case after $t_f$ the wavefunction will be a linear combination
\begin{equation}
\psi(t\geq t_f) = \alpha \exp(+ i \omega_0 \, t) + \beta  \exp(- i \omega_0\, t) ,
\end{equation}
where the Bogoliubov coefficients $\alpha$ and $\beta$ are to be calculated. 
That is,  we have
\begin{equation}
\left[\begin{array}{c} \phi\\ \pi/\omega_0 \end{array}\right]_{t_i} =
\left[\begin{array}{c}  \exp(+ i \omega_0 \, t_i) \\  i  \exp(+ i \omega_0 \, t_i) \end{array}\right],
\end{equation}
and
\begin{equation}
\left[\begin{array}{c} \phi\\ \pi/\omega_0 \end{array}\right]_{t_f} =
\left[\begin{array}{c}  \alpha \exp(+ i \omega_0 \,t_f) + \beta  \exp(- i \omega_0 \, t_f) \\  
i \left\{ \alpha \exp(+ i \omega_0 \, t_f) - \beta  \exp(- i \omega_0 \, t_f)  \right\}\end{array}\right].
\end{equation}
But we also have
\begin{equation}
\left[\begin{array}{c} \phi\\ \pi/\omega_0 \end{array}\right]_{t_f} 
= T \left[\begin{array}{c} \phi\\ \pi/\omega_0 \end{array}\right]_{t_i},
\end{equation}
implying 
\begin{equation}
\left[\begin{array}{c}  \alpha \exp(+ i \omega_0\, t_f) + \beta  \exp(- i \omega_0 \,t_f) \\  
i  \left\{ \alpha \exp(+ i \omega_0 \,t_f) - \beta  \exp(- i \omega_0 \,t_f)  \right\}\end{array}\right]
=
\left[\begin{array}{c}  a \exp(+ i \omega_0 \, t_i) + b\, i   \exp(+ i \omega_0 \, t_i)\\  
c  \exp(+ i \omega_0 \, t_i) +d\, i  \exp(+ i \omega_0 \, t_i) \end{array}\right].
\end{equation}
Solving these simultaneous linear equations we find
\begin{equation}
\alpha = {1\over2} \left[ a+d + i\left(b-{c}\right)\right] \;\exp(-i \omega_0\,[t_f - t_i]),
\end{equation}
\begin{equation}
\beta = {1\over2} \left[ a-d + i\left(b+{c}\right)\right] \;\exp(-i \omega_0\,[t_f + t_i]),
\end{equation}
so that the Bogoliubov coefficients are simple linear combinations of elements of the matrix $T$.
Then (remember the matrix $T$ is \emph{real})
\begin{equation}
|\alpha|^2 = {1\over4} \left\{ 
 (a+d)^2 + (b-c)^2 
\right\},
\end{equation}
\begin{equation}
|\beta|^2 = {1\over4} \left\{ 
 (a-d)^2 + (b+c)^2 
\right\},
\end{equation}
and so
\begin{eqnarray}
|\alpha|^2-|\beta|^2 &=& { (a+d)^2 + (b-c)^2 - (a-d)^2  -(b+c)^2   \over4},
\\
&=&{2ad -2bc +2ad - 2bc\over 4} = {ad-bc} = 1,
\end{eqnarray}
thus verifying  that, (thanks to the unit determinant condition), the Bogoliubov coefficients are properly normalized.  Particle production is governed by the $\beta$ coefficient in the combination
\begin{eqnarray}
|\beta|^2 &=& {1\over 4} \left\{ (a-d)^2 + \left(b+{c}\right)^2 \right\},
\\
&=&  {1\over 4} \left\{ a^2 + d^2 -2 ad + b^2  + {c^2} + 2bc   \right\},
\\
&=&  {1\over 4} \left\{ a^2 + d^2+ b^2  + {c^2} - 2 \right\},
\\
&=& {1\over4} \tr\{ T \, T^T - I\}.
\end{eqnarray}
Note that the transpose $T^T$ is now \emph{time-anti-ordered}.

Similarly, we have
\begin{eqnarray}
|\alpha|^2 &=& {1\over4} \left\{  (a+d)^2 + (b-{c})^2  \right\},
\\
&=& {1 \over 4} \left\{a^2 + d^2 +2 ad + b^2 + {c^2} - 2bc\right\},
\\
&=&  {1\over 4} \left\{ a^2 + d^2+ b^2 + {c^2} + 2 \right\},
\\
&=& {1\over4} \; \tr\{ T \, T^T + I\}.
\end{eqnarray}
As a consistency check, it is now obvious that
\begin{equation}
|\alpha|^2-|\beta|^2 = {1\over2} \; \tr\{ I \} = 1.
\end{equation}
\begin{table}[!htb]
\begin{center}
  \setlength{\fboxsep}{0.45 cm} 
   \framebox{\parbox[t]{12.5cm}{
{\bf Comment}: We can always formally solve the relevant ODE, either equation~(\ref{E:time}) or its equivalent equation~(\ref{E:space}), in terms of the time-ordered exponential, and we can always formally extract the Bogoliubov coefficients in terms of traces of the form $\tr\{T\,T^T\}$. We shall now use these formal results to derive rigorous bounds on the Bogoliubov coefficients. 
}}
\end{center}
\end{table}
\section{Elementary bound:}
Now consider the quantity
\begin{eqnarray}
X(t) &=& T(t) \; T(t)^T , \\
&=& \mathscr{T} \left\{ \exp\left( 
\int_{t_i}^t 
\left[\begin{array}{cc} 0 & \omega_0\\ -\omega^2(\bar t)/\omega_0 & 0 \end{array}\right] 
\d \bar t
\right) \right\}  
\nonumber\\
&& \times 
\left[\mathscr{T} \left\{ \exp\left( 
\int_{t_i}^t 
\left[\begin{array}{cc} 0 & \omega_0\\ -\omega^2(\bar t)/\omega_0 & 0 \end{array}\right] 
\d \bar t
\right) \right\} \right]^T.
\end{eqnarray}
This object satisfies the differential equation
\begin{equation}
{\d X\over\d t} =  
\left[\begin{array}{cc} 0 & \omega_0\\ -\omega^2(t)/\omega_0 & 0 \end{array}\right]  \; X(t)  
+
X(t)\;  \left[\begin{array}{cc} 0 &-\omega^2(t)/ \omega_0\\ \omega_0 & 0 \end{array}\right],
\end{equation}
with the boundary condition
\begin{equation}
X(t_i)=I.
\end{equation}
Now note
\begin{equation}
\tr(X) = \tr\{ T \, T^T\} = a^2 +b^2 + c^2 + d^2.
\end{equation}
Furthermore 
\begin{eqnarray}
\nonumber
{\d X\over \d t} 
&=& 
 \left[\begin{array}{cc} 0 & \omega_0\\ -\omega^2/\omega_0 & 0 \end{array}\right]  
\left[\begin{array}{cc} a^2 + b^2 & ac + bd\\ ac + bd & c^2 + d^2 \end{array}\right] 
\nonumber
\\
&& 
+ \left[\begin{array}{cc} a^2 + b^2 & ac + bd\\ ac + bd & c^2 + d^2 \end{array}\right]  
\left[\begin{array}{cc} 0 &-\omega^2/ \omega_0\\ \omega_0 & 0 \end{array}\right],
\nonumber
\\
&=&\left[\begin{array}{cc} 2 \omega_0(ac + bd)  & 
\omega_0(c^2 + d^2) - (\omega^2/\omega_0) (a^2 + b^2)\\ 
\omega_0(c^2 + d^2) - (\omega^2/\omega_0) (a^2 + b^2) & 
(-2 \omega^2/\omega_0) (ac + bd)\end{array}\right],
\nonumber
\\
&&
\end{eqnarray}
and so we see
\begin{equation}
\tr \left\{ \left[\begin{array}{cc} 0 & \omega_0\\ -\omega^2/\omega_0 & 0 \end{array}\right]  X +
X  \left[\begin{array}{cc} 0 &-\omega^2/ \omega_0\\ \omega_0 & 0 \end{array}\right] \right\}
=
2(ac+bd) \left[ \omega_0-{\omega^2\over\omega_0}\right].
\end{equation}
Therefore
\begin{equation}
{\d \tr[X]\over \d t } 
=2(ac+bd) \left[ \omega_0-{\omega^2\over\omega_0}\right].
\end{equation}
Using this key result, and some very simple analysis, we shall now derive our first elementary bound on the Bogoliubov coefficients.
\begin{itemize}
\item 
For any 2 real numbers, using $(x+y)^2\geq 0$ and $(x-y)^2\geq0$, we have
 \begin{equation}
 x^2 + y^2 \geq 2|xy|.
 \end{equation}
 In particular, for any 4 real numbers  this implies
 \begin{equation}
 a^2+b^2+c^2+d^2 \geq 2 \sqrt{(a^2+b^2)(c^2+d^2)}.
 \end{equation}
 
 \item
 But we also have
 \begin{eqnarray}
\nonumber
|ac+bd|^2 + |ad-bc|^2 &=& a^2 c^2 + 2 abcd +b^2d^2 +a^2d^2 -2 abcd+ b^2 c^2, \quad
\nonumber
\\
&&
\\
& =&  (a^2+b^2)(c^2+d^2), 
\end{eqnarray}
thus, for any 4 real numbers
\begin{equation}
 a^2 +b^2 + c^2 +d^2 \geq 2 \sqrt{|ac+bd|^2+|ad-bc|^2}.
 \end{equation} 

 \item
For the particular case we are interested in we additionally have the unit determinant condition $ad-bc=1$, so the above implies
\begin{equation}
 a^2 +b^2 + c^2 +d^2 \geq 2 \sqrt{|ac+bd|^2+1},
\end{equation} 
 whence
 \begin{equation}
2  |ac+bd| \leq \sqrt{ (a^2 +b^2 + c^2 +d^2)^2 - 4}.
\end{equation} 
\end{itemize}
Then, collecting these results, we see
\begin{equation}
{\d \tr[X]\over \d t } 
=2(ac+bd) \left[ \omega_0-{\omega^2\over\omega_0}\right]
\leq 2|ac+bd| \;\left| \omega_0-{\omega^2\over\omega_0}\right|,
\end{equation}
whence
\begin{eqnarray}
{\d \tr[X]\over \d t } &\leq&  \sqrt{ (a^2 +b^2 + c^2 +d^2)^2 - 4} \;
\left| \omega_0-{\omega^2\over\omega_0}\right| ,
\\
&=& 
\sqrt{\tr[X]^2-4}  \; \left| \omega_0-{\omega^2\over\omega_0}\right|,
\end{eqnarray}
whence
\begin{equation}
{1\over \sqrt{\tr[X]^2-4}} \; {\d \tr[X]\over \d t } \leq  \left| \omega_0-{\omega^2\over\omega_0}\right|.
\end{equation}
This implies
\begin{equation}
{\d \cosh^{-1}\tr[X/2]\over\d t} \leq  \left| \omega_0-{\omega^2\over\omega_0}\right|,
\end{equation}
whence
\begin{equation}
\tr[X] \leq 2\cosh\left\{ \int _{t_i}^{t_f} \left| \omega_0-{\omega^2\over\omega_0}\right| \d t \right\}.
\end{equation}
We now have
 \begin{equation}
 |\beta|^2 =  {1\over 4} \left\{  \tr\left\{  T \;  T^T \right\} - 2 \right\} 
 =   {1\over 4} \left\{  \tr\left\{  X \right\} - 2 \right\},
 \end{equation}
so that
 \begin{eqnarray}
 |\beta|^2 
 &\leq&
 {1\over 2} \left\{ \cosh\left\{ \int _{t_i}^{t_f} \left| \omega_0-{\omega^2\over\omega_0}\right| \d t \right\} -1\right\},
\\
&=& 
\sinh^2\left\{ {1\over2} \int _{t_i}^{t_f} \left| \omega_0-{\omega^2\over\omega_0}\right| \d t \right\}.
 \end{eqnarray}
 So finally
\begin{equation}
 |\beta|^2  \leq   
 \sinh^2\left\{ {1\over2} \int _{t_i}^{t_f} \left| \omega_0-{\omega^2\over\omega_0}\right| \d t \right\},
\end{equation}
and consequently
\begin{equation}
 |\alpha|^2  \leq   
 \cosh^2\left\{ {1\over2} \int _{t_i}^{t_f} \left| \omega_0-{\omega^2\over\omega_0}\right| \d t \right\}.
\end{equation}
These bounds are quite remarkable in their generality. A version of this result was derived in~\cite{bounds1} but the present derivation is largely independent and has the virtue of being completely elementary --- in particular, the use of complex numbers has been minimized, and we have absolutely eliminated  the use of the ``auxiliary functions''  and ``gauge conditions'' that were needed for the derivation in~\cite{bounds1}.
If one translates this to the space domain, then the equivalent barrier penetration coefficient is $T_\mathrm{transmission} \leftrightarrow 1/|\alpha|^2$, and the equivalent reflection coefficient is  $R \leftrightarrow |\beta|^2/|\alpha|^2$. Making the appropriate translations
\begin{equation}
T_\mathrm{transmission} \geq 
\sech^2\left\{ {1\over2} \int _{x_i}^{x_f} \left| k_0-{k^2(x)\over k_0}\right| \d x \right\},
\end{equation}
and 
\begin{equation}
R \leq \tanh^2\left\{ {1\over2} \int _{x_i}^{x_f} \left| k_0-{k^2(x)\over k_0}\right| \d x \right\}.
\end{equation}
\begin{table}[!htb]
\begin{center}
  \setlength{\fboxsep}{0.45 cm} 
   \framebox{\parbox[t]{12.5cm}{
{\bf Comment}: For completeness we mention that reference~\cite{bounds1} provides a large number of consistency checks on these bounds by comparing them with known exact results~\cite{exact}.
}}
\end{center}
\end{table}
\section{Lower bound on $|\beta|^2$}

To obtain a lower bound on the $|\beta|$ Bogoliubov coefficient, consider any real valued parameter $\epsilon$. Then since the matrix $T$ is itself real,
\begin{equation}
\tr \left\{(T -\epsilon\, T^T)^T\;(T-\epsilon\, T^T)\right\} \geq 0,
\end{equation}
so that
\begin{equation}
(1+\epsilon^2) \, \tr (T \, T^T) - 2  \epsilon \,  \tr(T^2)  \geq 0,
\end{equation}
whence
\begin{equation}
\tr (T^T \, T) \geq {2  \epsilon\over1+\epsilon^2} \; \tr (T^2), 
\end{equation}
This bound is extremized for $\epsilon=\pm1$, whence
\begin{equation}
\tr (T^T \,T) \geq  \left|\tr (T^2)\right|, 
\end{equation}
and so
\begin{equation}
|\beta|^2 \geq {1 \over 4} \left\{  \left|\tr (T^2)\right| - 2 \right\}.
\end{equation}
This is certainly a bound, but it is not as useful as one might hope. It is useful only if $\tr[T^2]>2$. But
\begin{equation}
\tr[T^2] = a^2+d^2+2bc = a^2+d^2 +2(ad-1) = (a+d)^2-2 = (\tr[T])^2-2.
\end{equation}
So using the unit determinant condition,  $\tr[T^2]>2$ can be seen to require $|a+d|\geq2$, that is, $\tr[T]>2$. But when does this happen? For the real matrix
\begin{equation}
\left[\begin{array}{cc} a & b\\ c & d \end{array}\right] 
\end{equation}
with unit determinant the eigenvalues are
\begin{equation}
\lambda = {a+d\over2} \pm {\sqrt{ (a+d)^2 - 4}\over 2}.
\end{equation}
The condition $a+d>2$ is thus equivalent to the condition that the eigenvalues are real.  Unfortunately there seems to be no simple way to then relate this to the properties of the function $\omega(t)$.

\section{A more general upper bound}

Now let $\Omega(t)$ be an \emph{arbitrary} everywhere real and nonzero \emph{function} of $t$ with the dimensions of frequency. Then we can rewrite the Schr\"odinger ODE~(\ref{E:time}) as:
\begin{equation}
{\d\over\d t} \left[\begin{array}{c} \phi \; \sqrt{\Omega}\\ \pi/\sqrt{\Omega} \end{array}\right] =
\left[\begin{array}{cc} 
{1\over2}(\dot\Omega/\Omega) & \Omega\\
 -\omega^2(t)/\Omega & -{1\over2}(\dot\Omega/\Omega) 
 \end{array}\right] \; 
\left[\begin{array}{c} \phi\;\sqrt{\Omega}\\ \pi/\sqrt{\Omega} \end{array}\right].
\end{equation}
Again all the matrix elements have been carefully chosen to have the same engineering dimension. Again we can formally solve this in terms of the time-ordered product:
\begin{equation}
\left[\begin{array}{c} \phi \, \sqrt{\Omega}\\ \pi/\sqrt{\Omega} \end{array}\right]_t = 
\mathscr{T} \left\{ \exp\left( 
\int_{t_0}^t \left[\begin{array}{cc} 
{1\over2}(\dot\Omega/\Omega)  & \Omega\\ 
-\omega^2(\bar t)/\Omega & - {1\over2}(\dot\Omega/\Omega) 
\end{array}\right] \d \bar t
\right) \right\} \; \left[\begin{array}{c} \phi\\ \pi/\sqrt{\Omega} \end{array}\right]_{t_0}.
\end{equation}
The new $T$ matrix is
\begin{equation}
T = \mathscr{T} \left\{ \exp\left( 
\int_{t_i}^{t_f}\left[\begin{array}{cc} 
{1\over2}(\dot\Omega/\Omega)  & \Omega\\ 
-\omega^2(\bar t)/\Omega & -{1\over2}(\dot\Omega/\Omega) 
 \end{array}\right] \d \bar t
\right) \right\}.
\end{equation}
Note that the matrix $T$ is still real, and that because
\begin{equation}
\tr \left[\begin{array}{cc} 
{1\over2}(\dot\Omega/\Omega)  & \Omega\\ 
-\omega^2(\bar t)/\Omega & -{1\over2}(\dot\Omega/\Omega) 
 \end{array}\right] 
 =0.
 \end{equation}
 it still follows that $T$ has determinant unity:
 \begin{equation}
 T = \left[\begin{array}{cc} 
a & b\\ 
c & d
 \end{array}\right]; \qquad   ad-bc = 1.
\end{equation}
This means that much of the earlier computations carry through without change. In particular as long as at the initial and final times we impose $\Omega(t)\to\omega_0$ as $t\to t_f$ and $t\to t_i$, we still have
\begin{equation}
\alpha = {1\over2} \left[a+ d 
+ i\left({ b} -{ c}\right)\right] 
\exp(-i \omega_0[t_f - t_i]),
\end{equation}
\begin{equation}
\beta = {1\over2} \left[ a- d 
+ i\left({b} +{c}\right)\right] 
\exp(-i \omega_0[t_f + t_i]),
\end{equation}
\begin{equation}
|\beta|^2 =  {1\over 4} \; \tr\left\{  T \;  T^T  - I\right\},
\end{equation} 
\begin{equation}
|\alpha|^2 =  {1\over 4} \; \tr\left\{   T \;  T^T  + I\right\}.
\end{equation} 
Now consider the quantity
\begin{eqnarray}
X(t) = T(t) \; T(t)^T &=& \mathscr{T} \left\{ \exp\left( 
\int_{t_i}^t \left[\begin{array}{cc} 
{1\over2}(\dot\Omega/\Omega)  & \Omega\\ 
-\omega^2(\bar t)/\Omega & -{1\over2}(\dot\Omega/\Omega) 
 \end{array}\right] \d \bar t
\right) \right\}  
\nonumber
\\
&&\times
\left[\mathscr{T} \left\{ \exp\left( 
\int_{t_i}^t \left[\begin{array}{cc} 
{1\over2}(\dot\Omega/\Omega)  & \Omega\\ 
-\omega^2(\bar t)/\Omega & -{1\over2}(\dot\Omega/\Omega) 
 \end{array}\right] \d \bar t
\right) \right\}  \right]^T\!\!.\;\;
\nonumber
\\
&&
\end{eqnarray}
This now satisfies the differential equation
\begin{equation}
{\d X\over\d t} =  
\left[\begin{array}{cc} 
{1\over2}(\dot\Omega/\Omega)  & \Omega\\ 
-\omega^2(\bar t)/\Omega & -{1\over2}(\dot\Omega/\Omega)  
\end{array}\right]  X +
X  \left[\begin{array}{cc} 
{1\over2}(\dot\Omega/\Omega)  &-\omega^2(\bar t)/ \Omega\\ 
\Omega & -{1\over2}(\dot\Omega/\Omega)  \end{array}\right],
\end{equation}
with the boundary condition
\begin{equation}
X(t_i)=I,
\end{equation}
and
\begin{equation}
\tr[X] = a^2+b^2+c^2+d^2.
\end{equation}
A brief computation yields
\begin{eqnarray}
{\d X\over\d t} &=&  
\left[\begin{array}{cc} 
{1\over2}(\dot\Omega/\Omega)  & \Omega\\ 
-\omega^2(\bar t)/\Omega & -{1\over2}(\dot\Omega/\Omega)  
\end{array}\right]  \left[\begin{array}{cc} a^2 + b^2 & ac + bd\\ ac + bd & c^2 + d^2 \end{array}\right] 
\nonumber\\
&&
 +  \left[\begin{array}{cc} a^2 + b^2 & ac + bd\\ ac + bd & c^2 + d^2 \end{array}\right] 
 \left[\begin{array}{cc} 
{1\over2}(\dot\Omega/\Omega)  &-\omega^2(\bar t)/ \Omega\\ 
\Omega & -{1\over2}(\dot\Omega/\Omega)  \end{array}\right],
\nonumber
\\
&&
\end{eqnarray}
\begin{equation}
= \left[\begin{array}{cc} 
(\dot\Omega/\Omega)(a^2 + b^2) + 2 \Omega(ac + bd)  & \Omega (c^2 + d^2) - (\omega^2 / \Omega) (a^2 + b^2)\\ 
- (\omega^2/\Omega) (a^2 + b^2) + \Omega (c^2 + d^2) & - (2 \omega^2 / \Omega) (ac + bd) - (\dot \Omega/\Omega)(c^2 + d^2)
\end{array}\right].
\end{equation}
Then taking the trace, there is now one extra term
\begin{equation}
{\d\tr[X]\over\d t} =   
(a^2+b^2-c^2-d^2)\left[{\dot\Omega\over\Omega}\right] +   
2(ac+bd) \left[ \Omega-{\omega^2\over\Omega}\right].
\end{equation} 
Note that if $\Omega(t)\to\omega_0$ then $\dot\Omega\to 0$ and we recover the ODE of the ``elementary'' bound.  In this more general setting we now proceed by using the following facts: 
\begin{itemize}
\item 
As previously we note
 \begin{eqnarray}
|ac+bd|^2 + |ad-bc|^2 &=& a^2 c^2 + 2 abcd +b^2d^2 +a^2d^2 -2 abcd+ b^2 c^2,
\nonumber
\\
&=&  (a^2+b^2)(c^2+d^2),
\nonumber
\\
&&
\end{eqnarray}
which implies
\begin{equation}
|ac+bd| = \sqrt{ (a^2+b^2)(c^2+d^2) -1},
\end{equation}
that is
\begin{equation}
2 |ac+bd| = \sqrt{ 4 (a^2+b^2)(c^2+d^2) -4}.
\end{equation}
\item
Additionally, we use
\begin{equation}
|a^2+b^2-c^2-d^2| = \sqrt{|a^2+b^2+c^2+d^2|^2- 4 (a^2+b^2)(c^2+d^2)},
\end{equation}
implying
\begin{equation}
|a^2+b^2-c^2-d^2| ^2 + (2 |ac+bd| )^2 = |a^2+b^2+c^2+d^2|^2- 4.
\end{equation}
\end{itemize}
In particular, combining these observations,  this means that we can find an angle $\theta$ (which is in general some complicated real function of $a$, $b$, $c$, $d$) such that
\begin{equation}
2 (ac+bd) = \sqrt{ |a^2+b^2+c^2+d^2|^2- 4} \; \;\sin\theta,
\end{equation}
\begin{equation}
a^2+b^2-c^2-d^2 = \sqrt{ |a^2+b^2+c^2+d^2|^2- 4} \; \; \cos\theta,
\end{equation}
whence
\begin{equation}
{\d\tr[X]\over\d t} =   \sqrt{ |a^2+b^2+c^2+d^2|^2- 4} \; 
\left\{ \sin\theta\left[{\dot\Omega\over\Omega}\right] +   
\cos\theta \left[ \Omega-{\omega^2\over\Omega}\right] \right\}.
\end{equation} 
But for any real $\theta$ we certainly have by the Cauchy--Schwartz \emph{inequality}
\begin{equation}
\sin\theta\left[{\dot\Omega\over\Omega}\right] +   
\cos\theta \left[ \Omega-{\omega^2\over\Omega}\right] 
\leq 
\sqrt{ \left[{\dot\Omega\over\Omega}\right]^2 + \left[ \Omega-{\omega^2\over\Omega}\right]^2},
\end{equation}
implying
\begin{equation}
{\d\tr[X]\over\d t} \leq   \sqrt{ |a^2+b^2+c^2+d^2|^2- 4} \; \;
\sqrt{ \left[{\dot\Omega\over\Omega}\right]^2 + \left[ \Omega-{\omega^2\over\Omega}\right]^2}.
\end{equation}
Therefore
\begin{equation}
{\d\tr[X]\over\d t} \leq   \sqrt{ \tr[X]^2- 4} \; 
\sqrt{ \left[{\dot\Omega\over\Omega}\right]^2 + \left[ \Omega-{\omega^2\over\Omega}\right]^2},
\end{equation}
implying
\begin{equation}
{1\over \sqrt{ \tr[X]^2- 4} } {\d\tr[X]\over\d t}  \leq
\sqrt{ \left[{\dot\Omega\over\Omega}\right]^2 + \left[ \Omega-{\omega^2\over\Omega}\right]^2},
\end{equation}
whence
\begin{equation}
 {\d\cosh^{-1}(\tr[X]/2)\over\d t}  \leq
\sqrt{ \left[{\dot\Omega\over\Omega}\right]^2 + \left[ \Omega-{\omega^2\over\Omega}\right]^2},
\end{equation}
so that 
\begin{equation}
\tr[X] = \tr[T \;T^T] \leq 2 \cosh\left\{ \lint_{\!\!\!\!\!\!t_i}^{t_f} 
\sqrt{ \left[{\dot\Omega\over\Omega}\right]^2 + \left[ \Omega-{\omega^2\over\Omega}\right]^2}
\; \d t \right\}.
\end{equation}
Using the general formulae for $|\alpha|^2$ and $|\beta|^2$ in terms of $\tr\{T \, T^T\}$, and simplifying, we see
\begin{equation}
|\beta|^2 \leq  \sinh^2\left\{ {1\over 2} \int_{t_i}^{t_f} 
{1 \over |\Omega|}\sqrt{ \dot\Omega^2 + \left[ \Omega^2-{\omega^2}\right]^2}
\; \d t \right\},
\end{equation}
and
\begin{equation}
|\alpha|^2 \leq  \cosh^2\left\{ {1\over 2} \int_{t_i}^{t_f} 
{1 \over |\Omega|}\sqrt{ \dot\Omega^2 + \left[ \Omega^2-{\omega^2}\right]^2}
\; \d t \right\}.
\end{equation}
This result is completely equivalent to the corresponding result in~\cite{bounds1}; though again note that the derivation is largely independent and that it no longer requires one to introduce any ``gauge fixing'' condition, nor need we introduce any WKB-like ansatz. The current proof is much more ``direct", and at worst uses simple inequalities and  straightforward ODE theory.
If we work in the space domain instead of the time domain and make the translations $\Omega(t)\to \varphi'(x)$, $\omega(t) \to k(x)$, we see
\begin{equation}
|\alpha|^2 \leq  \cosh^2\left\{ {1\over 2} \int_{x_i}^{x_f} 
{1 \over |\varphi'|}\sqrt{ (\varphi'')^2 + \left[ (\varphi')^2-{k^2}\right]^2}
\; \d x \right\},
\end{equation}
and
\begin{equation}
|\beta|^2 \leq  \sinh^2\left\{ {1\over 2} \int_{x_i}^{x_f} 
{1 \over |\varphi'|}\sqrt{ (\varphi'')^2 + \left[ (\varphi')^2-{k^2}\right]^2}
\; \d x \right\}.
\end{equation}
This is perhaps physically more transparent in terms of the equivalent transmission and reflection coefficients
\begin{equation}
T_\mathrm{transmission} \geq  \sech^2\left\{ {1\over 2} \int_{x_i}^{x_f} 
{1 \over |\varphi'|}\sqrt{ (\varphi'')^2 + \left[ (\varphi')^2-{k^2}\right]^2}
\; \d x \right\},
\end{equation}
and
\begin{equation}
R \leq  \tanh^2\left\{ {1\over 2} \int_{x_i}^{x_f} 
{1 \over |\varphi'|}\sqrt{ (\varphi'')^2 + \left[ (\varphi')^2-{k^2}\right]^2}
\; \d x \right\}.
\end{equation}
\begin{table}[!htb]
\begin{center}
  \setlength{\fboxsep}{0.45 cm} 
   \framebox{\parbox[t]{12.5cm}{
{\bf Comment}:  For completeness we mention that reference~\cite{bounds1} provides a number of consistency checks on these more general bounds by comparing them with known exact results~\cite{exact}.
}}
\end{center}
\end{table}
\section{The ``optimal'' choice of $\Omega(t)$?}

What is the \emph{optimal} choice of $\Omega(t)$ that one can make leading to the most stringent bound on the Bogoliubov coefficients? The bound we have just derived holds for arbitrary $\Omega(t)$, subject to the two boundary conditions $\Omega(t_i) = \omega_0 = \Omega(t_f)$ and the overall constraint $\Omega(t) \neq 0$. Since both $\sinh$ and $\cosh$ are convex functions, finding the most stringent constraint on $|\beta|$ and $|\alpha|$ is thus a variational calculus problem equivalent to minimizing the action
\begin{equation}
S =  \int_{t_i}^{t_f} 
{1 \over |\Omega|}\;\sqrt{ \dot\Omega^2 + \left[ \Omega^2-{\omega^2}\right]^2}
\; \d t.
\end{equation}
The relevant Euler--Lagrange equations are quite messy, and progress (at least insofar as there is any practicable progress) is better made by using an indirect attack. The Lagrangian is 
\begin{equation}
L =  
{1 \over |\Omega|}\; \sqrt{ \dot\Omega^2 + \left[ \Omega^2-{\omega^2}\right]^2},
\end{equation}
and so the corresponding canonical momentum can be evaluated as
\begin{equation}
\pi = {\partial L\over\partial \dot\Omega} =
 {\dot \Omega \over |\Omega|\;\sqrt{ \dot\Omega^2 + \left[ \Omega^2-{\omega^2}\right]^2}}.
\end{equation}
From the boundary conditions we can deduce
\begin{equation}
\pi(t_i) = {1\over\omega_0} = \pi(t_f).
\end{equation}
The Hamiltonian is now
\begin{equation}
H = \pi\; \dot \Omega - L = 
 {\dot \Omega^2 - \left\{ \dot\Omega^2 + \left[ \Omega^2-{\omega^2}\right]^2 \right\}
 \over |\Omega|\;\sqrt{ \dot\Omega^2 + \left[ \Omega^2-{\omega^2}\right]^2}}
 =
-  {\left[ \Omega^2-{\omega^2}\right]^2
 \over |\Omega|\;\sqrt{ \dot\Omega^2 + \left[ \Omega^2-{\omega^2}\right]^2}}.
\end{equation}
\begin{table}[!htb]
\begin{center}
 \setlength{\fboxsep}{0.45 cm} 
 \framebox{\parbox[t]{12.5cm}{
{\bf Hamiltonian (classical mechanics)}:  
For a closed system the sum of the kinetic and potential energy in the system is represented by a quantity called the \emph{Hamiltonian}, which leads to a set of differential equations known as the \emph{Hamilton equations} for that system~\cite{Hamiltonian-classical-wiki}. 

The Hamilton equations are generally written as follows:
\begin{eqnarray}
\dot{p} &=& - {\partial {\cal H}\over \partial q},
\\
\dot{q} &=& {\partial {\cal H}\over \partial p},
\end{eqnarray}
the dot denotes the ordinary derivative with respect to time of the functions $p = p(t)$ (called generalized momenta) and $q = q(t)$  (called generalized coordinates), and ${\cal H}  = {\cal H}(p, q, t)$  is the so-called \emph{Hamiltonian}. Thus, a little more explicitly, one can equivalently write
\begin{eqnarray}
{\d \over \d t} p(t) &=& - {\partial \over \partial q} \, {\cal H} (p(t), q(t), t),
\\
{\d \over \d t} q(t) &=& {\partial \over \partial p} \, {\cal H} (p(t), q(t), t),
\end{eqnarray}
and specify the domain of values in which the parameter $t$ (``time'') varies.
}}
\end{center}
\end{table}
\begin{table}[!htb]
\begin{center}
 \setlength{\fboxsep}{0.45 cm} 
 \framebox{\parbox[t]{12.5cm}{
{\bf Hamiltonian (quantum mechanics)}:  The Hamiltonian $H$ is the quantum mechanical operator corresponding to the total energy of the system. It generates the time evolution of quantum states. If $| \psi(t) \rangle$ is the state of the system at time $t$, then~\cite{Hamiltonian}
\begin{equation}
H |\psi(t)\rangle = i\hbar {\partial \over \partial t} |\psi(t)\rangle,
\end{equation}
where $\hbar$ is the \emph{reduced Planck} constant. This equation is known as the (time dependent) \emph{Schr\"odinger equation}. (This is the same form as the \emph{Hamilton-Jacobi equation}, which is one of the reasons $H$ is also called the \emph{Hamiltonian}.) Given the state at some initial time $(t = 0)$, we can integrate it to derive the state at any subsequent time. If $H$ is independent of time, then
\begin{equation}
|\psi(t)\rangle = \exp \bigg(-{i H t \over \hbar}\bigg)\;  |\psi(0) \rangle.
\end{equation}

In this thesis we are sometimes using quantum Hamiltonians, and sometimes classical Hamiltonians. Whether we are in a quantum or classical situation will have to be determined from context.
}}
\end{center}
\end{table}

Unfortunately the Hamiltonian is \emph{explicitly} time-dependent [via $\omega(t)$] and so is \emph{not} conserved. The best we can say is that at the endpoints of the motion
\begin{equation}
H(t_i) = 0 = H(t_f).
\end{equation}
By solving for $\dot\Omega$ as a function of $\pi$ and $\Omega$ we can also write
\begin{equation}
\dot \Omega = {\pi \,\Omega \;\over \sqrt{1- \pi^2 \,\Omega^2}} \;  (\Omega^2-\omega^2),
\end{equation}
and
\begin{equation}
H = - {\sqrt{1-\pi^2\,\Omega^2} \;(\Omega^2-\omega^2)\over |\Omega|}.
\end{equation}
Note that $\dot\Omega$ at the endpoints \emph{cannot} in general be explicitly evaluated in terms of the boundary conditions. 

An alternative formulation which slightly simplifies the analysis is to change variables by writing
\begin{equation}
\Omega(t) = \omega_0 \; \exp[\theta(t)],
\end{equation}
where the boundary conditions are now
\begin{equation}
\theta(t_i) = 0 = \theta(t_f),
\end{equation}
and the action is now rewritten as
\begin{equation}
S =  \int_{t_i}^{t_f} 
\sqrt{ \dot\theta^2 + \omega_0^2 \left[ e^{2\theta} -{\omega^2\over\omega_0^2}\; e^{-2\theta} \right]^2}
\; \d t.
\end{equation}
Then, in terms of this new variable we have
\begin{equation}
L =  
\sqrt{ \dot\theta^2 + \omega_0^2 \left[ e^{2\theta} -{\omega^2\over\omega_0^2}\; e^{-2\theta} \right]^2},
\end{equation}
with (dimensionless) conjugate momentum 
\begin{equation}
\pi = {\partial L\over\partial \dot\theta} =
 {\dot \theta \over \sqrt{ \dot\theta^2 
 +  \omega_0^2 \left[ e^{2\theta} -{\omega^2\over\omega_0^2}\; e^{-2\theta} \right]^2}},
\end{equation}
and boundary conditions
\begin{equation}
\pi(t_i) = 1 = \pi(t_f).
\end{equation}
The (non-conserved) Hamiltonian is
\begin{equation}
H = \pi\; \dot \theta - L = 
-  {\omega_0^2 \left[ e^{2\theta} -{\omega^2\over\omega_0^2}\; e^{-2\theta} \right]^2
 \over \sqrt{ \dot\theta^2 + \omega_0^2 \left[ e^{2\theta} -{\omega^2\over\omega_0^2}\; e^{-2\theta} \right]^2}},
\end{equation}
subject to
\begin{equation}
H(t_i) = 0 = H(t_f).
\end{equation}
Inverting, we see
\begin{equation}
\dot \theta = {\pi  \over \sqrt{1- \pi^2}} \; \omega_0 \;
\left[ e^{2\theta} -{\omega^2\over\omega_0^2}\; e^{-2\theta} \right] ,
\end{equation}
and
\begin{equation}
H = - \sqrt{1-\pi^2} \;  \omega_0 \;
\left[ e^{2\theta} -{\omega^2\over\omega_0^2}\; e^{-2\theta} \right].
\end{equation}
This has given us a somewhat simpler variational problem,  unfortunately the Euler--Lagrange equations are still too messy to provide useful results.

Overall, we see that while solving the variational problem would indeed result in an optimum bound, there is no explicit general formula for such a solution. In the tradeoff between optimality and explicitness, we will have to accept the use of sub-optimal but explicit bounds.

\section{Sub-optimal but explicit bounds}

From our general bounds
\begin{equation}
|\beta|^2 \leq  \sinh^2\left\{ {1\over 2} \int_{t_i}^{t_f} 
{1 \over |\Omega|}\sqrt{ \dot\Omega^2 + \left[ \Omega^2-{\omega^2}\right]^2}
\; \d t \right\},
\end{equation}
and
\begin{equation}
|\alpha|^2 \leq  \cosh^2\left\{ {1\over 2} \int_{t_i}^{t_f} 
{1 \over |\Omega|}\sqrt{ \dot\Omega^2 + \left[ \Omega^2-{\omega^2}\right]^2}
\; \d t \right\},
\end{equation}
the following special cases are of particular interest:
\begin{description}
\item[$\Omega=\omega_0$:]  In this case we simply obtain the ``elementary'' bound considered above.

\item[$\Omega=\omega$:]  This case only makes sense if $\omega^2>0$ is always positive. (Otherwise $\omega$ and hence $\Omega$ becomes imaginary in the ``classically forbidden'' region; the matrix $T$ then becomes complex, and the entire formalism breaks down). Subject to this constraint we find
\begin{equation}
|\beta|^2 \leq  \sinh^2\left\{ {1\over 2} \int_{t_i}^{t_f} 
\left|{\dot \omega \over \omega} \right|
\; \d t \right\},
\end{equation}
and
\begin{equation}
|\alpha|^2 \leq  \cosh^2\left\{ {1\over 2} \int_{t_i}^{t_f} 
\left|{\dot \omega \over \omega} \right|
\; \d t \right\}.
\end{equation}
This case was also considered in~\cite{bounds1}.

\item[$\Omega=\omega^\epsilon \;\omega_0^{1-\epsilon}$:]  This case again only makes sense if $\omega^2>0$ is always positive. Subject to this constraint we find
\begin{equation}
|\beta|^2 \leq  \sinh^2\left\{ {1\over 2} \lint_{\!\!\!\!\!\!t_i}^{t_f} 
\sqrt{ \epsilon^2 \; {\dot\omega^2\over\omega^2}  + {
\omega^{2\epsilon} \left[ \omega_0^{2-2\epsilon}-{\omega^{2-2\epsilon}}\right]^2\over \omega_0^{2-2\epsilon}}}
\; \d t \right\},
\end{equation}
and
\begin{equation}
|\alpha|^2 \leq  \cosh^2\left\{ {1\over 2} \lint_{\!\!\!\!\!\!t_i}^{t_f} 
\sqrt{ \epsilon^2 \; {\dot\omega^2\over\omega^2}  + {
\omega^{2\epsilon} \left[ \omega_0^{2-2\epsilon}-{\omega^{2-2\epsilon}}\right]^2\over \omega_0^{2-2\epsilon}}}
\; \d t \right\}.
\end{equation}
This nicely interpolates between the two cases given above, which correspond to $\epsilon=0$ and $\epsilon=1$ respectively. 
\begin{table}[!htb]
\begin{center}
 \setlength{\fboxsep}{0.45 cm} 
 \framebox{\parbox[t]{12.5cm}{
{\bf Comment}:  The WKB approximation technique provides an approximate solution to the Schr\"odinger equation~(\ref{Schrodinger-equation-chapter8}) of a quantum mechanical system; however this technique fails at the classical turning points of the system of the potential energy function, $V(x)$~\cite{wkb-turning-point}. 
\medskip

Standard WKB theory provides a set of ``joining conditions'' for ``stepping over'' the turning points and penetrating into the classically forbidden region. Unfortunately we do not seem to have any analogue of these ``joining conditions'' in the formalism we set up in this thesis.  
}}
\end{center}
\end{table}
\item[Triangle inequality:] Since $\sqrt{x^2+y^2} \leq |x| + |y|$ we see that
\begin{equation}
|\beta|^2 \leq  \sinh^2\left\{ {1\over 2} \int_{t_i}^{t_f} 
\left|{\dot \Omega \over \Omega} \right| \; \d t
+ 
 {1\over 2} \int_{t_i}^{t_f} 
\left| \Omega-{\omega^2\over\Omega}\right|
\; \d t \right\},
\end{equation}
and
\begin{equation}
|\alpha|^2 \leq  \cosh^2\left\{ {1\over 2} \int_{t_i}^{t_f} 
\left|{\dot \Omega \over \Omega} \right| \; \d t
+ 
 {1\over 2} \int_{t_i}^{t_f} 
\left| \Omega-{\omega^2\over\Omega}\right|
\; \d t \right\}.
\end{equation}
\end{description}
These bounds, because they are explicit, are often the most useful quantities to calculate.

\vfill

\section{The ``interaction picture''}
\begin{center}
  \setlength{\fboxsep}{0.45 cm} 
   \framebox{\parbox[t]{12.5cm}{
 {\bf  Interaction picture (sometimes called the Dirac picture)}:  This is an intermediate between the Schr\"odinger picture and the Heisenberg picture. Whereas in the other two pictures either the state vector or the operators carry time dependence, in the interaction picture both carry part of the time dependence of observables. Equations that include operators acting at different times, which hold in the interaction picture, do not necessarily hold in the Schr\"odinger or the Heisenberg picture. This is because time-dependent unitary transformations relate operators in one picture to the analogous operators in the others~\cite{interaction-picture}.
 
For our purposes the interaction picture is useful because it lets us develop a perturbation theory.
}}
\end{center}
\bigskip

\noindent
If we split the function $\omega(t)^2$ into an exactly solvable piece $\omega_e(t)^2$ and a perturbation $\omega_\Delta(t)^2$ then we can develop a formal perturbation series for the transfer matrix $T$, in close analogy to the procedures for developing quantum field theoretic  perturbation theory in the interaction picture. Specifically let us write
\begin{equation}
\omega(t)^2 = \omega_e(t)^2 + \omega_\Delta(t)^2,
\end{equation}
and
\begin{equation}
{\d T(t)\over\d t} = Q(t) \; T(t) = \left[Q_e(t) + Q_\Delta(t) \right] \; T(t).
\end{equation}
Now defining
\begin{equation}
T(t) = T_e(t) \; T_\Delta(t),
\end{equation}
we shall develop a formal solution for $T_\Delta(t)$. Consider
\begin{equation}
{\d T(t)\over\d t} =  \left[Q_e(t) + Q_\Delta(t) \right] \; T_e(t) \; T_\Delta(t),
\end{equation}
and compare it with
\begin{equation}
{\d T(t)\over\d t} = {\d T_e (t)\over\d t}\; T_\Delta(t)  + T_e(t) \; {\d T_\Delta(t)\over\d t} = 
 Q_e(t) \; T_e(t) \; T_\Delta(t)  +  T_e(t) \; {\d T_\Delta(t)\over\d t}.
 \end{equation}
 Therefore
 \begin{equation}
 {\d T_\Delta(t)\over\d t} = \left\{ T_e(t)^{-1} \;  Q_\Delta(t) \; T_e(t) \right\} \; T_\Delta,
\end{equation}
whence
\begin{equation}
T_\Delta(t) = \mathscr{T} \exp\left( \int_{t_i}^t \left\{ T_e(\bar t)^{-1} \;  Q_\Delta(\bar t) \; T_e(\bar t) \right\}  \d \bar t \right).
\end{equation}
For the full transfer matrix $T$ we have
\begin{equation}
T(t) = T_e(t) \times \mathscr{T} \exp\left( \int_{t_i}^t \left\{ T_e(\bar t)^{-1} \;  Q_\Delta(\bar t) \; T_e(\bar t) \right\}  \d \bar t \right),
\end{equation}
and we have succeeded in splitting it into an exact piece $T_e(t)$ plus a distortion due to $Q_\Delta(t)$. This can now be used as the starting point for a perturbation expansion. (The analogy with quantum field theoretic perturbation theory in the interaction picture should now be completely clear.)

To develop some formal bounds on the Bogoliubov coefficients it is useful to suppress (currently) unnecessary phases by defining
\begin{equation}
\tilde \alpha = {1\over2} \left[a+ d 
+ i\left({ b} -{ c}\right)\right] ,
\end{equation}
\begin{equation}
\tilde \beta = {1\over2} \left[ a- d 
+ i\left({b} +{c}\right)\right].
\end{equation}
The virtue of these definitions is that for $T = T_e\; T_\Delta$ they satisfy a simple composition rule which can easily  be verified via matrix multiplication. From $T = T_e\; T_\Delta$ we have
\begin{equation}
\left[
\begin{array}{cc}
a  & b   \\
c  &  d 
\end{array}
\right]
=
\left[
\begin{array}{ccc}
a_e \,a_\Delta + b_e \,c_\Delta  &    a_e \,b_\Delta + b_e \,d_\Delta \\
c_e \,a_\Delta + d_e \,c_\Delta   &    c_e \,b_\Delta + d_e \,d_\Delta
\end{array}
\right].
\end{equation}
Then some simple linear algebra leads to
\begin{equation}
\tilde\beta = \tilde \alpha_e \; \tilde \beta_\Delta + \tilde \beta_e \; \tilde \alpha^*_\Delta,
\end{equation}
\begin{equation}
\tilde\alpha = \tilde \alpha_e \; \tilde \alpha_\Delta + \tilde \beta_e \; \tilde \beta^*_\Delta,
\end{equation}
But then
\begin{equation}
|\beta| = |\tilde\beta| =  \left|\tilde \alpha_e \; \tilde \beta_\Delta + \tilde \beta_e \; \tilde \alpha^*_\Delta\right|
\leq  \left|\tilde \alpha_e \; \tilde \beta_\Delta\right| + \left| \tilde \beta_e \; \tilde \alpha^*_\Delta\right|
=  \left| \alpha_e \; \beta_\Delta\right| + \left| \beta_e \;  \alpha_\Delta\right|,
\end{equation}
that is
\begin{equation}
|\beta| \leq  \left| \alpha_e \right|\; \left|\beta_\Delta\right| + \left| \beta_e \right|\;  \left|\alpha_\Delta\right|,
\end{equation}
or the equivalent
\begin{equation}
|\beta| \leq  \sqrt{1+\left| \beta_e \right|^2} \; \left|\beta_\Delta\right| + \left| \beta_e \right|\;  \sqrt{1+ \left|\beta_\Delta\right|^2}.
\end{equation}
Similarly
\begin{equation}
|\beta| = |\tilde\beta| =  \left|\tilde \alpha_e \; \tilde \beta_\Delta + \tilde \beta_e \; \tilde \alpha^*_\Delta\right|
\geq  \left| \; \left|\tilde \alpha_e \; \tilde \beta_\Delta\right| - \left| \tilde \beta_e \; \tilde \alpha^*_\Delta\right|\; \right|
=  \left| \; \left| \alpha_e \; \beta_\Delta\right| - \left| \beta_e \;  \alpha_\Delta\right| \; \right|,
\end{equation}
that is
\begin{equation}
|\beta| \geq  \left| \; \left| \alpha_e \right|\; \left|\beta_\Delta\right| - \left| \beta_e \right|\;  \left|\alpha_\Delta\right| \; \right|,
\end{equation}
or the equivalent
\begin{equation}
|\beta| \geq  \left| \; \sqrt{1+\left| \beta_e \right|^2} \; \left|\beta_\Delta\right| - \left| \beta_e \right|\;  \sqrt{1+ \left|\beta_\Delta\right|^2} \; \right|.
\end{equation}
The benefit now is that one has bounded the Bogoliubov coefficient in terms of the (assumed known) exact coefficient $\beta_e$ and the contribution from the perturbation $\beta_\Delta$. Suitably choosing the split between exact and perturbative contributions to $\omega^2$, one could in principle obtain arbitrarily accurate bounds. 

\section{Conclusion}
In this chapter we again considered rigorous bounds on transmission, reflection, and Bogoliubov coefficients. In particular, the most outstanding features of this chapter are: 
\begin{itemize}
\item
We have re-considered the general bounds on the Bogoliubov coefficients developed in~\cite{bounds1}. Additionally,  we have seen how to extend the bounds in~\cite{bounds1} in many different ways.  Moreover, we do not need to ``gauge fix'', nor do we need to appeal  to any ``WKB-like ansatz'' to get the discussion started. 
\item
We have formulated some rigorous bounds that we can place on barrier penetration probabilities, 
or equivalently on the Bogoliubov coefficients associated with a time-dependent potential.  Furthermore, we have not seen anything like these bounds anywhere else.
\item
In addition, probably there are ``optimal'' bounds still waiting to be discovered.
\item
It is apparent that the current bounds are \emph{not}  the best that can be achieved, and we strongly suspect that it may be possible to develop yet further extensions to the current formalism.
\end{itemize}


Considering the fundamental importance of the questions we are asking, it is remarkable how little work on this topic can currently be found in the literature.  

Possible extensions might include somehow relaxing the reality constraint on $\Omega(t)$ without damaging too much of the current formalism, a better understanding of the variational problem defining the ``optimal'' bound (thus hopefully leading to an explicit form  thereof), or using several ``probe functions'' [instead of the single function $\Omega(t)$] to more closely bound the Bogoliubov coefficients.

\chapter[Bounding the greybody factors]{Bounding the greybody factors for Schwarzschild black holes}
\label{C:consistency-9}
\section{Introduction}

Black hole greybody factors are important because they modify the spectrum of Hawking radiation seen at spatial infinity~\cite{Hawking}, so that it is not quite Planckian~\cite{Page}. (That is, it is no longer exactly ``blackbody radiation'', which is why [with slight abuse of language], it is called ``greybody''.)  
There is a vast scientific literature dealing with estimates of these black-hole greybody factors, using a wide variety of techniques~\cite{greybody-factor}. 

Unfortunately, most of these calculations adopt various approximations that move one away from the physically most important regions of parameter space.  Sometimes one is forced into the extremal limit, sometimes one is forced to asymptotically high or low frequencies, sometimes techniques work only away from (3+1) dimensions, sometimes the nature of the approximation is uncontrolled. As a specific example, \emph{monodromy} techniques fail for $s=1$ (photons)~\cite{monodromy}, which is observationally one of the most important cases one would wish to consider. 

Faced with these limitations, in this chapter we ask a slightly different question: Restricting attention to the physically most important situations (Schwarzschild black holes, (3+1) dimensions, intermediate frequencies, unconstrained spin and angular momentum) is it possible to at least place rigorous (and hopefully simple) analytic \emph{bounds} on the greybody factors? 

By now considering the Regge--Wheeler equation for excitations around Schwarzschild spacetime, and adapting the general analysis discussed in earlier sections of this thesis, and published in references~\cite{bounds1, bounds2}, we shall demonstrate that rigorous analytic bounds are indeed achievable. While it is certainly true that these bounds may not answer all the physical questions one might legitimately wish to ask, they are definitely a solid step in the right direction. 

Before starting the detailed calculations for this chapter, we should stress some important issues related to the greybody factors in Schwarzschild black holes:
\begin{itemize}
\item
Hawking radiation was originally derived in the geometric optics approximation where it can be shown to be  described by ideal black body radiation --- a black body is an object that absorbs all light that falls on it. No electromagnetic radiation passes through it and none is reflected. Because no light is reflected or transmitted, the object appears black when it is cold~\cite{blackbody}. This derivation led to a calculation of the temperature and entropy of a black hole. 
\item
In this chapter, we shall try to rigorously solve for bounds on the transmission probabilities for waves moving through the region of space outside of a Schwarzschild black hole --- Schwarzschild black holes are the simplest type, one that is described as a spherically symmetric body with no (electric or magnetic) charge or angular momentum (no rotation).
\item
The most important issue is to realise that  ``greybody factor'' is actually a synonym for ``transmission probability''. Indeed, the phrase ``greybody factor'' is used more in the thermodynamics and spectroscopy communities, while the phrase  ``transmission probability'' is used more in the quantum mechanics community, but these communities are referring to the same concept.
\item
The greybody factor describes the emissivity of the black hole which is not that of a perfect blackbody.
\item
As a black hole radiates energy by Hawking radiation, energy conservation implies that it will lose mass. 
\end{itemize}
\begin{table}[!htbp]
\begin{center}
  \setlength{\fboxsep}{0.45 cm} 
   \framebox{\parbox[t]{12.5cm}{
{\bf Schwarzschild black hole or static black hole}: This black hole has no charge or angular momentum. A Schwarzschild black hole is described by the Schwarzschild metric, and cannot be distinguished from any other Schwarzschild black hole except by its mass. This black hole is characterized by a spherical surface, called the event horizon, which encloses the central singularity. The event horizon is situated at the \emph{Schwarzschild radius}, often called the radius of a black hole. 

Any spherically symmetric mass distribution whose radius is smaller than the Schwarzschild radius forms a black hole~\cite{Schwarzschild-black-hole}. 
}}
\end{center}
\end{table}
\begin{table}[!htbp]
\begin{center}
  \setlength{\fboxsep}{0.45 cm} 
   \framebox{\parbox[t]{12.5cm}{
       {\bf  Evaporating black holes}: 
       There are several significant consequences that appear as a result of considering evaporating black holes:
       
(1) We first note that black hole evaporation generates a more consistent view of black hole thermodynamics, by showing how black holes interact thermally with the remainder of the universe.

(2) Secondly, the temperature of a black hole increases as it radiates away mass. In fact, the rate of temperature rises as explosively, with the most plausible of the endpoint scenarios being the complete dissolution of the black hole in a violent explosion of gamma rays. 
To make a complete description of this assumed dissolution process, one would need a model of quantum gravity, which would be needed in the final stages of the evaporation process,  when the black hole approaches Planck mass and Planck radius. (The precise details of what is going on here is the subject of much continued, and sometimes heated, debate.)

(3) Finally, the simplest models of black hole evaporation lead to the black hole information paradox.  For instance, the ``information content'' of a black hole appears to be lost when it evaporates, as under many of these models the Hawking radiation is purely random. It is felt by many (not all physicists) that the Hawking radiation must be somehow perturbed to contain the missing information, that the Hawking evaporation somehow carries the missing information to infinity. This would imply that information is \emph{not} allowed to be lost under these conditions~\cite{hawking-radiation}. (Technically, in this model the Hawking evaporation process is asserted to be ``unitary''.  The precise details of what is going on here is the subject of much continued, and sometimes heated, debate~\cite{small-dark, black-hole-in-general-relativity}.)
}}
\end{center}
\end{table}
\begin{table}[!htb]
\begin{center}
  \setlength{\fboxsep}{0.45 cm} 
   \framebox{\parbox[t]{12.5cm}{
{\bf Entropy}:
Black holes are truly thermodynamic systems with an actual and precisely calculable temperature and entropy. 
However, in terms of statistical mechanics, the entropy should be the logarithm of the number of independent states of the black hole. Understanding how to count these states would constitute a significant progress in the probe to understand quantum gravity~\cite{Ted}. (In this regard, some string-based models have led to at least partial success.)
}}
\end{center}
\end{table}
\section{Hawking radiation}
 A black body is an object that absorbs all light that falls on it. No electromagnetic radiation passes through it and none is reflected. Because no light is reflected or transmitted, the object appears black when it is cold~\cite{blackbody}. The light emitted by a black body is called \emph{black body radiation}.

Hawking radiation is an approximately thermal radiation with an approximately black body spectrum predicted to be emitted by black holes due to quantum effects. It was discovered by Stephen Hawking who provided the theoretical argument for its existence in 1974, and is closely related to work by the physicist Jacob Bekenstein who predicted that black holes should have a finite, non-zero entropy~\cite{hawking-radiation}. 
\begin{table}[!htbp]
\begin{center}
  \setlength{\fboxsep}{0.45 cm} 
   \framebox{\parbox[t]{12.5cm}{
       {\bf Comment}: 
     The most standard interpretation of the Hawking radiation theory states that implicit  ``virtual'' particle-antiparticle pairs are occasionally created outside the event horizon of a black hole. There are three things that can happen to a pair of particles just outside the event horizon~\cite{John-Chang}:
\begin{itemize}
\item
Both particles are pulled into the black hole.
\item
Both particles escape from the black hole.
\item
One particle escapes while the other is pulled into the black hole. In particular, the particle that has escaped becomes real and can consequently be observed from Earth  (or by any outside observer). The particle that was pulled into the black hole remains virtual, and to satisfy conservation of energy must have negative mass-energy. The black hole absorbs this negative mass-energy and as a result, loses mass and appears to shrink. The total rate of power emission is (to an excellent approximation) proportional to the inverse square of the black hole's mass.
\end{itemize}

This is only a visual picture, at best an aid to understanding, and many relativists would argue that the true situation can only be begun to be understood by calculating the renormalized stress-energy tensor in the vicinity of the horizon.  Such calculations agree that the key point is that \emph{negative} energy is flowing \emph{into} the black hole.

}}
\end{center}
\end{table}
\begin{table}[!htb]
\begin{center}
  \setlength{\fboxsep}{0.45 cm} 
   \framebox{\parbox[t]{12.5cm}{
{\bf Quantum field theory in curved spacetime}: Ordinary quantum field theories are defined in \emph{flat Minkowski space}, which is an outstandingly good approximation for expressing the physics of microscopic particles in weak gravitational fields like those seen on Earth. 

In contrast, one can  formulate quantum field theories in curved spacetime to describe situations in which gravity is powerful enough to influence quantum matter, while, however, it is not powerful enough to require quantization itself. Importantly, these theories depend on classical general relativity to express a curved background spacetime. A generalized quantum field theory is used to describe the behavior of quantum matter within that spacetime.  In addition, one can use this formalism to show that black holes emit a blackbody spectrum of particles known as \emph{Hawking radiation},  leading to the possibility that they evaporate over time. 
It is believed that this radiation represents a significant part of the thermodynamics of black holes~\cite{GR}.

For our purposes, we will not be dealing with curved-space quantum field theory directly. However, the general techniques we develop in this thesis can be adapted to answer specific questions in curved-space quantum field theory, such as placing limits on the transmission coefficients in black hole scattering, and bounding cosmological particle production due to the expansion of the universe. 
}}
\end{center}
\end{table}

\noindent
Greybody factors in black hole physics modify the naive Planckian spectrum that is predicted for Hawking radiation when working in the limit of geometrical optics.  We shall consider the Schwarzschild geometry in (3+1) dimensions, and analyze the Regge--Wheeler equation for arbitrary particle spin $s$ and wave-mode angular momentum $\ell$, deriving rigorous bounds on the greybody factors as a function of $s$, $\ell$, wave frequency $\omega$, and the black hole mass $m$. 
\section{Regge--Wheeler equation}
The well-known Regge--Wheeler equation describes the axial perturbation of Schwarzschild metric in the linear approximation~\cite{Heun}.
In terms of the tortoise coordinate $r_*$ the Regge--Wheeler equation ($G_N\to1$) is 
\begin{equation}
{d^2 \psi\over d r_*^2} =   [ \omega^2 - V(r) ] \psi,
\end{equation}
where for the specific case of a Schwarzschild black hole
\begin{equation}
{dr\over dr_*} = 1-{2m\over r},
\end{equation}
and the Regge--Wheeler potential is
\begin{equation}
V(r) = \left(1-{2m\over r}\right) \left[ {\ell(\ell+1)\over r^2} + {2m(1-s^2)\over r^3} \right].
\end{equation}
Here $s$ is the spin of the particle and $\ell$ is the angular momentum of the specific wave mode under consideration, with $\ell\geq s$. Thus $V(r)\geq0$ outside the horizon, where $r\in(2m,\infty)$. The greybody factors we are interested in are just the transmission probabilities for wave modes propagating through this Regge--Wheeler potential.
\begin{center}
  \setlength{\fboxsep}{0.45 cm} 
   \framebox{\parbox[t]{12.5cm}{
{\bf Tortoise coordinate:}
In a Schwarzschild spacetime the so-called ``tortoise coordinate''   is defined by~\cite{tortoise-coordinate}:
\begin{equation}
r^* = r + 2 GM \ln \bigg|{r \over 2GM} -1\bigg|.
\end{equation}

The tortoise coordinate $r^*$ approaches $-  \infty$ as  ``$r$'' approaches the Schwarzschild radius $r = 2GM$. It satisfies
\begin{equation}
{\d r^* \over \d r} = \bigg(1 - {2GM \over r}\bigg)^{-1}.
\end{equation}

If we watch an object fall towards the Schwarzschild radius, then in terms of the $r$ coordinate the object  seems to slow down and (asymptotically) stop at the horizon. In contrast,   
\begin{equation}
{\d r^* \over \d t}
\end{equation}
remains finite. In some vague analogy with the fable of the ``tortoise and the hare'', the $r^*$ coordinate has come to be known as the ``tortoise'' coordinate. 
}}
\end{center}
\bigskip

\noindent
\begin{itemize}
\item 

Despite comments often encountered in the literature, one \emph{can} explicitly solve for $r$ as a function of the tortoise coordinate $r_*$ --- in terms of the Lambert $W$ function we have the exact result
\begin{equation}
r(r_*) = 2m\left[ 1 + W( e^{[r_*-2m]/2m}) \right],
\end{equation}
whereas
\begin{equation}
r_*(r) = r + 2m\ln\left[{r-2m\over2m}\right].
\end{equation}
Unfortunately this formal result, while certainly correct and exact,  is less useful than one might suppose. (We have not been able to turn this observation into any useful calculation.) 
\item
Despite other comments often encountered in the literature, one can also explicitly solve the Regge--Wheeler equation --- now in terms of Heun functions~\cite{Heun}. Unfortunately this is again less useful than one might suppose, this time because relatively little is known about the analytical behaviour of Heun functions --- this is an area of ongoing research in mathematical analysis~\cite{Heun2}.
\end{itemize}
\begin{table}[!htb]
\begin{center}
  \setlength{\fboxsep}{0.45 cm} 
   \framebox{\parbox[t]{12.5cm}{
{\bf Heun functions}:
In mathematics, the Heun differential equation is a second-order linear ordinary differential equation of the form~\cite{Heun-wiki}
\begin{equation}
{\d^2 w \over \d z^2} + \bigg[{\gamma \over z} + {\delta \over z-1} + {\epsilon \over z-d}\bigg] \, {\d w \over \d z} + {\alpha \beta z - q \over z (z-1)(z-d)} \, w = 0.
\end{equation}
Note that the constraint $\epsilon = \alpha + \beta - \gamma - \delta +1$ is needed to ensure regularity of the singular point at $\infty$. Every second-order linear ODE in the complex plane (or on the Riemann sphere, to be more accurate) with four regular singular points can be transformed into this equation. This standardized equation has four regular singular points: $0$, $1$, $d$, and $\infty$.

Studying the properties of these functions is an area of ongoing research in mathematical analysis~\cite{Heun2}.
}}
\end{center}
\end{table}
\vfill

\section{Bounds}

The general bounds developed earlier in this thesis, and published in references~\cite{bounds1, bounds2}, can, in the current situation, be written as
\begin{equation}
T \geq \sech^2\left\{ \int_{-\infty}^{\infty}  \vartheta \; \d r_* \right\}.
\end{equation}
Here $T$ is the transmission probability (greybody factor), and $\vartheta$ is the function
\begin{equation}
\vartheta = {\sqrt{ (h')^2 + [\omega^2-V- h^2]^2}\over2 h },
\label{E:b0-1-1}
\end{equation}
where, $h$ is some positive function, $h(r_*)>0$, satisfying the limits $h(-\infty)=h(+\infty)=\omega$, which is otherwise arbitrary. Two different derivations of this general result, and numerous consistency checks, can be found in earlier chapters \ref{C:consistency-5} and \ref{C:consistency-8} of this thesis, and in  references~\cite{bounds1, bounds2}.  

(These bounds were originally developed as a technical step when studying the completely unrelated issue of sonoluminescence~\cite{sonoluminescence},  and since then have also been used to place limits on particle production in analogue spacetimes~\cite{analogue} and resonant cavities~\cite{cavity}, to investigate qubit master equations~\cite{qubit}, and to motivate further general investigations of one-dimensional scattering theory~\cite{one-dim}.)  
For current purposes, the most useful practical results are obtained by considering two special cases: 

\begin{enumerate}
\item 
If we set $h=\omega$ then
\begin{equation}
T \geq \sech^2\left\{ {1\over2\omega} \int_{-\infty}^{\infty} V(r_*) \; \d r_* \right\},
\end{equation}
whence
\begin{equation}
T \geq \sech^2\left\{ {1\over2\omega} \int_{2m}^{\infty}  
\left[ {\ell(\ell+1)\over r^2} + {2m(1-s^2)\over r^3} \right] \; \d r \right\}.
\end{equation}
Therefore, since the remaining integral is trivial, we obtain our first explicit bound:
\begin{equation}
T \geq \sech^2\left\{ {2\ell(\ell+1) +(1-s^2)\over 8\omega m} \right\}.
\label{E:b1-1-1}
\end{equation}
That is:
\begin{equation}
T \geq \sech^2\left\{ {(\ell+1)^2 +(\ell^2-s^2)\over 8\omega m} \right\}.
\label{E:b11}
\end{equation}
Note that this bound is meaningful for all frequencies. This is sufficient to tell us that at high frequencies the Regge--Wheeler  barrier is almost fully transparent, while even at arbitrarily low frequencies some nonzero fraction of the Hawking flux will tunnel through. A particularly nice feature of this first bound is that it is so easy to write down for arbitrary $s$ and $\ell$.
\item
 If we now set $h=\sqrt{\omega^2-V}$, which in this case implicitly means that we are not permitting any classically forbidden region, then
\begin{equation}
T \geq \sech^2\left\{ {1\over2} \int_{-\infty}^{\infty} \left|{h'\over h}\right| \; \d r_* \right\}.
\end{equation}
Since for arbitrary $s$ and $\ell$ the Regge--Wheeler potential is easily seen to have a unique peak at which it is a maximum, this becomes
\begin{eqnarray}
T &\geq& \sech^2\left\{ \ln\left( {h_\mathrm{peak}\over h_\infty } \right) \right\},
\\
&=& 
\sech^2\left\{ \ln\left( {\sqrt{\omega^2-V_\mathrm{peak}}\over \omega} \right) \right\},
\end{eqnarray}
which is easily seen to be monotonic decreasing as a function of $V_\mathrm{peak}$. However calculating the location of the peak, and value of the Regge--Wheeler potential at the peak is somewhat more tedious than evaluating the previous bound (\ref{E:b1-1-1}). Note that the present bound fails, and gives no useful information, once $\omega^2 < V_\mathrm{peak}$, corresponding to a classically forbidden region.  More explicitly, the bound can be rewritten as:
\begin{equation}
\label{transmission-rewritten}
T \geq {4 \omega^2 (\omega^2-V_\mathrm{peak})\over (  2\omega^2 - V_\mathrm{peak})^2} = 
1 - {V_\mathrm{peak}^2\over (  2\omega^2 - V_\mathrm{peak})^2}.
\end{equation}
\end{enumerate}
\begin{table}[!htb]
\begin{center}
  \setlength{\fboxsep}{0.45 cm} 
   \framebox{\parbox[t]{12.5cm}{
{\bf Monodromy theorem}: The monodromy theorem is a significant result regarding analytic continuation of a complex-analytic function to a larger set. The basic concept is that one can extend a complex analytic function along curves --- the process can be started in the original domain of the function and can be ended in the bigger set.  In addition, this theorem gives sufficient conditions for analytic continuation to give the same value at a given point regardless of the specific curve used to get there, so that the resulting extended analytic function would be well-defined and single-valued~\cite{monodromy-theorem}.
}}
\end{center}
\end{table}
The most interesting point of the study of black hole greybody factors~\cite{greybody-factor}, and (once one moves into the complex plane), the closely related problem of locating the quasinormal modes~\cite{monodromy,qnm,qnmv}, is a subject that has attracted a wide amount of interest. In particular, \emph{quasinormal modes} are poles of the transmission coefficient and reflect the black hole's ringdown reaction to a perturbation~\cite{monodromy}.
Unfortunately,  as a specific example, \emph{monodromy} techniques fail for $s=1$ (photons)~\cite{monodromy}, which is observationally one of the most important cases one would wish to consider. 
It is for this reason that we resort to our general bounds to extract as much information as possible.

Let us now consider various sub-cases:
\begin{itemize}
\item 
For $s=1$ (ie, photons) the situation simplifies considerably. (Remember, this is the case for which monodromy techniques fail~\cite{monodromy}.) For $s=1$ we always have $r_\mathrm{peak} = 3m$ and 
\begin{equation}
V_\mathrm{peak} = {\ell(\ell+1)\over27 m^2}.
\end{equation}
Consequently, from (\ref{transmission-rewritten})
\begin{equation}
T_{s=1} \geq {108 \omega^2 m^2 [27\omega^2 m^2-\ell(\ell+1)]\over [  54\omega^2 m^2 -\ell(\ell+1)]^2}.
\end{equation}
In almost the entire region where this bound applies ($\omega^2 > V_\mathrm{peak}$) it is in fact a better bound than (\ref{E:b1-1-1}) above.

\item
For $s=0$ (ie, scalars) and $\ell=0$ (the $s$-wave),   we have $r_\mathrm{peak} = 8m/3$ and 
\begin{equation}
V_\mathrm{peak} = {27\over1024 m^2}. 
\end{equation}
Consequently
\begin{equation}
T_{s=0,\ell=0} \geq {4096 \omega^2 m^2 [1024\omega^2 m^2-27]\over [2048\omega^2 m^2 -27]^2}.
\end{equation}
In a large fraction of the  region where this bound applies it is in fact a better bound than (\ref{E:b1-1-1}) above.

\item
For $s=0$ but $\ell\geq 1$ it is easy to see that throughout the black hole exterior, $\forall r \in(2m,\infty)$, we have 
\begin{equation}
V_{s=0,\ell\geq1}(r) < \left(1-{2m\over r}\right) \left[ {\ell^2+\ell+1\over r^2}\right],
\end{equation}
which is the $s=1$ potential with the replacement $\ell(\ell+1)\to \ell^2+\ell+1$. This bound on the potential has its maximum at $r_\mathrm{peak} = 3m$, implying
\begin{equation} 
V_{\mathrm{peak},s=0,\ell\geq1} <  {\ell^2+\ell+1\over 27 m^2}. 
\end{equation}
Therefore  the monotonicity of the bound on the greybody factor implies  
\begin{equation}
T_{s=0,\ell\geq1} >  {108 \omega^2 m^2 [27\omega^2 m^2-(\ell^2+\ell+1)]\over [  54\omega^2 m^2 -(\ell^2+\ell+1)]^2},
\end{equation}
(for $\omega$, $m$, and $\ell$ held fixed, and subject to $s\leq \ell$).

\item
For $s>1$ it is easy to see that throughout the black hole exterior, $\forall r \in(2m,\infty)$, keeping $\ell$ held fixed, we have $V_{s>1}(r) < V_{s=1}(r)$.  Therefore
\begin{equation} 
V_{\mathrm{peak},s>1} < V_{\mathrm{peak},s=1}. 
\end{equation}
Therefore  the monotonicity of the bound on the greybody factor implies  
\begin{equation}
T_{s>1} >   {108 \omega^2 m^2 [27\omega^2 m^2-\ell(\ell+1)]\over [  54\omega^2 m^2 -\ell(\ell+1)]^2},
\end{equation}
(for $\omega$, $m$, and $\ell$ held fixed, and subject to $s\leq \ell$).

\item
More generally, it is useful to define
\begin{equation}
\epsilon= {1-s^2\over \ell(\ell+1)}.
\end{equation}
Excluding the case $(s,\ell)=(0,0)$, which was explicitly dealt with above, the remainder of the physically interesting region is confined to the range $\epsilon\in(-1,+1/2]$. Then a brief computation yields
\begin{equation}
r_\mathrm{peak} = 3m \left\{ 1 - {\epsilon\over 9} + \mathcal{O}(\epsilon^2)\right\},
\end{equation}
and 
\begin{equation}
V_\mathrm{peak} = {\ell(\ell+1)\over27 m^2}  \left\{ 1 +{2\epsilon\over3} + \mathcal{O}(\epsilon^2)\right\}.
\end{equation}
In fact one can show that
\begin{equation}
V_\mathrm{peak} <  {\ell(\ell+1)\over20 m^2},
\end{equation}
over the physically interesting range. (This bound on  $V_\mathrm{peak}$ is tightest for $(s,\ell) = (0,1)$, corresponding to $\epsilon=+1/2$, where it provides a better than $1\%$ estimate, and becomes progressively weaker as one moves to $\epsilon=-1$.)  This then implies
\begin{equation}
T_{(s,\ell)\neq(0,0)} > {80 \omega^2 m^2 [20\omega^2 m^2-\ell(\ell+1)]\over [  40\omega^2 m^2 -\ell(\ell+1)]^2}.
\end{equation}
As always there is a trade-off between strength of the bound and the ease with which it can be written down.
\end{itemize}
While this second set of bounds has required a little more case by case analysis, we should in counterpoint observe that this second set of bounds provides much stronger information at very high frequencies, where in fact
\begin{equation}
T \geq 1 - \mathcal{O}[V_\mathrm{peak} \;\omega^{-4}].
\end{equation}
Unfortunately this second set of bounds is (because of details in the derivation, see earlier chapters in this thesis, and~\cite{bounds1,bounds2}) not capable of providing information once the frequency has dropped low enough for the scattering problem to develop classical turning points --- in other words a scattering problem with a classically forbidden region is not amenable to treatment using bounds of the second class considered above. For sufficently low frequencies, bounds of the form (\ref{E:b1-1-1}) are more appropriate, with
\begin{equation}
T \geq \mathcal{O}\left( \exp\{-1/\omega\}\right) . 
\end{equation}
What we have not done, at least not yet, is to use the full generality implicit in equation (\ref{E:b0-1-1}). Subject to rather mild constraints, there is a freely specifiable function $h(r_*)$ available that can potentially be used to extract tighter bounds. Work along these lines is continuing.

\section{Discussion}

The study of black hole greybody factors~\cite{greybody-factor}, and (once one moves into the complex plane), the closely related problem of locating the quasinormal modes~\cite{monodromy,qnm,qnmv}, is a subject that has attracted a wide amount of interest in both the general relativity and particle physics communities. 
In addition, we wish emphasize some specific features in this chapter:
\begin{itemize}
\item
In this chapter, we have developed a complementary set of results --- we have sought and obtained several rigorous analytic bounds that can be placed on the greybody factors.
\item
Even though these bounds are not necessarily tight bounds on the exact greybody factors they do serve to focus attention on general and robust features of these greybody factors. Moreover they provide a new method of extracting physical information. 
\item
In the current formalism, (as opposed to, for instance, monodromy techniques~\cite{monodromy}), it is obviously clear that one does not have to know anything about what is going on inside the black hole in order to obtain information regarding the greybody factors. This is as it should be, since physically the greybody factors are simply transmission coefficients relating the horizon to spatial infinity, and make no intrinsic reference to the nature of the central singularity. 
\item
Looking further afield, here should be no intrinsic difficulty in extending these results to Reissner--Nordstr\"om black holes, dilaton black holes, or to higher dimensions --- all that is really needed is an exact expression for the Regge--Wheeler potential. 
\end{itemize}
Finally, it is perhaps more interesting to see if one can significantly improve these bounds in some qualitative manner, perhaps by making a more strategic choice for the essentially free function $h(r_*)$. 

\chapter{The Miller--Good transformation}
\label{C:consistency-10}
\section{Introduction}
In this chapter, we shall consider further topics of general interest in quantum physics, such as transmission through a potential barrier --- the potential being  a region in a field of force where the force exerted on a particle is such as to oppose the passage of the particle through that region~\cite{potential-barrier-definition-miller} --- and the (formally) closely related issue of particle production from a parametric resonance. 

We have already developed a rather general bound on quantum transmission probabilities, and  in the previous chapter (chapter~\ref{C:consistency-9}) have applied it to bounding the greybody factors of a Schwarzschild black hole. In this current chapter we shall take a different tack --- we shall use the Miller--Good transformation (which maps an initial Schr\"odinger equation to a final Schr\"odinger equation for a different potential) to significantly generalize the previous bound. Moreover, we shall see that the Miller--Good transformation is an efficient method whereby to  to generalize the bound,  to make it more efficient and powerful.
\begin{table}[!htb]
\begin{center}
  \setlength{\fboxsep}{0.45 cm} 
   \framebox{\parbox[t]{12.5cm}{
{\bf Parametric resonance}:
This effect called ``parametric resonance'' is related to the phenomenon of mechanical excitation and oscillation at certain frequencies (and the associated harmonics). This effect is different from regular resonance because it often exhibits instability phenomena.

Parametric resonance occurs in a mechanical system when a system is parametrically excited and oscillates at one of its resonant frequencies. Parametric excitation differs from forcing since the action appears as a time varying modification on a system parameter. The classical example of parametric resonance is that of the vertically forced pendulum~\cite{parametric-resonance}.

For our purposes we will always view parametric resonance in terms of a time-dependent modifcation of the oscillation frequency.
}}
\end{center}
\end{table}
\section{Setting up the problem}
Consider the Schr\"odinger equation,
\begin{equation}
u(x)'' + k(x)^2 \; u(x) = 0,
\label{E:sde-10}
\end{equation}
where $k(x)^2 = 2m[E-V(x)]/\hbar^2$.  As long as $V(x)$ leads to finite (possibly different) constants $V_{\pm\infty}$ on left and right infinity, then for $E>\max\{V_{+\infty},V_{-\infty}\}$ one can set up a one-dimensional scattering problem in a completely standard manner --- see for example~\cite{Landau, Merzbacher, Capri, Messiah, branson-joachim, liboff, dicke-wittke,  shankar}.  The scattering problem is completely characterized by the transmission and reflection \emph{amplitudes} ($t$ and $r$), though the most important aspects of the physics can be extracted from the transmission and reflection \emph{probabilities} ($T= |t|^2$ and $R=|r|^2$).
Relatively little work has gone into providing general analytic bounds on the transmission probabilities, (as opposed to approximate estimates), and the only known result as far as we have been able to determine is this:
\begin{theorem}
Consider the Schr\"odinger equation (\ref{E:sde-10}). Let $h(x)>0$ be some positive but otherwise arbitrary once-differentiable function. Then the transmission probability is bounded from below by
\begin{equation}
T \geq \sech^2\left\{ \; \int_{-\infty}^{+\infty} {\sqrt{ (h')^2 + (k^2-h^2)^2} \over 2 h} \; \d x \right\}.
\label{E:bd}
\end{equation}
\end{theorem}
\noindent To obtain useful information, one should choose asymptotic conditions on the function $h(x)$ so that the integral converges --- otherwise one obtains the true but trivial result $T\geq \sech^2\infty = 0$. (There is of course a related bound in the reflection probability, $R$, and if one works with the formally equivalent problem of parametric oscillations, a bound on the resulting Bogoliubov coefficients and particle production.)

This quite remarkable bound was first derived in~\cite{bounds1}, with further discussion and an alternate proof being provided in chapter \ref{C:consistency-5} (and published in~\cite{bounds2}). These bounds were originally used as a technical step when studying a specific model for sonoluminescence~\cite{sonoluminescence},  and since then have also been used to place limits on particle production in analogue spacetimes~\cite{analogue} and resonant cavities~\cite{cavity}, to investigate qubit master equations~\cite{qubit}, and to motivate further general investigations of one-dimensional scattering theory~\cite{one-dim}. Most recently, these bounds have also been applied to the greybody factors of a Schwarzschild black hole (see previous chapter, and the related publication~\cite{greybody-factor}).

A slightly weaker, but much more tractable,  form of the bound can be obtained by applying the triangle inequality.  For $h(x)>0$:
\begin{equation}
T \geq \sech^2\left\{ \; {1\over2} \int_{-\infty}^{+\infty} 
\left[  |\ln(h)' |  + {|k^2-h^2| \over h} \right]\; \d x \right\}.
\label{E:bd0}
\end{equation}

\noindent
Five important special cases are:  
\begin{enumerate}
\item 
If we take $h=k_\infty$, where $k_\infty = \lim_{x\to\pm\infty} k(x)$,  then we have~\cite{bounds1,bounds2}
\begin{equation}
T \geq \sech^2\left\{ \; {1\over2k_\infty} \int_{-\infty}^{+\infty} |k_\infty^2-k^2| \; \d x \right\}.
\label{E:bd1}
\end{equation}
\item
If we define $k_{+\infty} = \lim_{x\to+\infty} k(x)\neq k_{-\infty} = \lim_{x\to-\infty} k(x)$,  
and take $h(x)$ to be any function that smoothly and monotonically interpolates between $k_{-\infty}$ and $k_{+\infty}$, then we have
\begin{equation}
T \geq \sech^2\left\{ \;  {1\over2} \left| \ln\left( {k_{+\infty}\over k_{-\infty}} \right) \right| 
+  {1\over2} \int_{-\infty}^{+\infty} {|k^2-h^2| \over h}   \; \d x \right\}.
\label{E:bd2}
\end{equation}
This is already more general than the most closely related  result presented in~\cite{bounds1,bounds2}.
\item
If we have a single extremum in $h(x)$ then
\begin{equation}
T \geq \sech^2\left\{ \;  {1\over2} \left| \ln\left( {k_{+\infty} k_{-\infty}\over h_\mathrm{ext}^2} \right) \right| 
+  {1\over2} \int_{-\infty}^{+\infty} {|k^2-h^2| \over h}   \; \d x \right\}.
\label{E:bd3}
\end{equation}
This is already more general than the most closely related  result presented in~\cite{bounds1,bounds2}.
\item
If we have a single minimum in $k^2(x)$, and choose $h^2=\max\{k^2,\Delta^2 \}$, assuming $k^2_\mathrm{min}\leq \Delta^2\leq k_{\pm\infty}^2$, (but still permitting $k^2_\mathrm{min} < 0$, so we are allowing for the possibility of a classically forbidden region), then 
\begin{equation}
T \geq \sech^2\left\{ \;  {1\over2} \ln\left({k_{+\infty} k_{-\infty} \over \Delta^2}\right) +  {1\over2\Delta} \int\limits_{\Delta^2>k^2} |\Delta^2 - k^2| \; \d x \right\}.
\label{E:bd4}
\end{equation}
This is already more general than the most closely related  result presented in~\cite{bounds1,bounds2}.
\item 
If $k^2(x)$ has a single minimum and  $0 < k^2_\mathrm{min} \leq k_{\pm\infty}^2$, then
\begin{equation}
T \geq \sech^2\left\{ \;  {1\over2} \ln\left( {k_{+\infty} k_{-\infty} \over k_\mathrm{min}^2} \right) \; \right\}.
\label{E:bd5}
\end{equation}
This is the limit of (\ref{E:bd4}) above as $\Delta\to k_\mathrm{min}>0$, and is one of the special cases considered in~\cite{bounds1}.
\end{enumerate}

In this chapter we shall not be seeking to \emph{apply} the general bound (\ref{E:bd}), its weakened form (\ref{E:bd0}), or any of its specializations as given in (\ref{E:bd1})--(\ref{E:bd5}) above. Instead we shall be seeking to \emph{extend} and \emph{generalize} the bound to make it more powerful. The tool we shall use to do this is the Miller--Good transformation~\cite{miller-good}.

\section{The Miller--Good transformation}
Consider the Schr\"odinger equation (\ref{E:sde-10}), and consider the substitution~\cite{miller-good}
\begin{equation}
u(x) = {1\over \sqrt{X'(x)}} \; U(X(x)).
\label{E:mg}
\end{equation}
We will want $X$ to be our ``new'' position variable, so $X(x)$ has to be an invertible function. This implies (via, for instance, the inverse function theorem) that we  need $\d X/\d x \neq 0$. In reality, since the argument will be smoothest if we arrange things so that the variables  $X$ and $x$ both agree as to which direction is left or right, we can without loss of generality  assert  $\d X/\d x > 0$, whence also $\d x/\d X > 0$.

Now compute  (using the notation $U_X= \d U/\d X$):
\begin{equation}
u'(x) = U_X(X) \,\sqrt{X'} - {1\over2} {X''\over (X')^{3/2} } \,U(X),
\end{equation}
and
\begin{equation}
u''(x) = U_{XX}(X) \; (X')^{3/2}  
- {1\over2} {X'''\over (X')^{3/2} } \,U
+{3\over4} {(X'')^2\over(X')^{5/2} } U.
\end{equation}
Insert this into the original Schr\"odinger equation, $u(x)'' + k(x)^2 u(x) = 0$, to see that
\begin{equation}
 U_{XX} + \left\{ {k^2\over (X')^{2}}  - {1\over2} {X'''\over (X')^{3} }
+{3\over4} {(X'')^2\over(X')^{4} } \right\} U = 0,
\end{equation}
which we can write as
\begin{equation}
U_{XX} + K^2 \; U = 0,
\label{E:sde2}
\end{equation}
with 
\begin{equation}
\label{K-complete}
K^2 = {1\over (X')^2} 
 \left\{ k^2  - {1\over2} {X'''\over X' }
+{3\over4} {(X'')^2\over(X')^{2} } \right\}.
\end{equation}
That is, a Schr\"odinger equation in terms of $u(x)$ and $k(x)$ has been transformed into a \emph{completely equivalent} Schr\"odinger equation in terms of $U(X)$ and $K(X)$. 
We can also rewrite this as
\begin{equation}
K^2 = {1\over (X')^2} 
 \left\{ k^2   + \sqrt{X'} \left(1\over \sqrt{X'}\right)'' \right\}.
\end{equation}
The combination
\begin{equation}
 \sqrt{X'} \left(1\over \sqrt{X'}\right)'' =  
- {1\over2} {X'''\over X' } +{3\over4} {(X'')^2\over(X')^{2}},
\label{E:Schwarzian-def}
\end{equation}
shows up in numerous \emph{a priori} unrelated branches of physics and 
is sometimes referred to as the ``Schwarzian derivative''.

\begin{table}[!htbp]
\begin{center}
  \setlength{\fboxsep}{0.45 cm} 
   \framebox{\parbox[t]{12.5cm}{
       {\bf  Definition (Schwarzian derivative)}: 
    The Schwarzian derivative is an absolute operator that is constant under all linear fractional transformations. For this reason, it commonly occurs in the theory of the complex projective line, and especially, in the theory of modular forms and hypergeometric series.
    
    The Schwarzian derivative also occurs in numerous other situations, including quantum field theory calculations related to stellar collapse and  black hole formation, moving mirror calculations related to Unruh radiation, and much more.
    
     Explicitly, the Schwarzian derivative of a function $f$ of one complex variable $x$ is described by~\cite{schwartzian-derivative}
     \begin{eqnarray}
     \nonumber
     (Sf)(x) &=& \bigg({f''(x) \over f'(x)}\bigg)' - {1 \over 2} \bigg({f''(x) \over f'(x)}\bigg)^2 ,
     \nonumber
     \\
     &=&   {f'''(x) \over f'(x)} - {3 \over 2} \bigg({f''(x) \over f'(x)}\bigg)^2.
     \end{eqnarray}
     The alternative notation $\Leftrightarrow  \{f,x\} = (S f) (x)$ is frequently used.
}}
\end{center}
\end{table}

\begin{itemize}
\item 
As previously commented,  to make sure the coordinate transformation $x\leftrightarrow X$ is well defined we want to have $X'(x)>0$, let us call this $j(x) \equiv X'(x)$ with $j(x)>0$.  We can then write
\begin{equation}
\label{K-coordinate-1}
K^2 = {1\over j^2} 
 \left\{ k^2  - {1\over2} {j''\over j }
+{3\over4} {(j')^2\over j^{2} } \right\}.
\end{equation}
Let us suppose that $\lim_{x\to\pm\infty} j(x) = j_{\pm\infty} \neq 0$; then $K_{\pm\infty} =  k_{\pm\infty}/ j_{\pm\infty}$, so if $k^2(x)$ has nice asymptotic behaviour allowing one to define a scattering problem, then so does $K^2(x)$.
\item
Another possibly more useful substitution (based on what we saw with the Schwarzian derivative) is to set   $J(x)^{-2} \equiv X'(x)$ with $J(x)>0$ in equation~(\ref{K-complete}).
When we let $J^{-2} = X'$, we  can then also write this as:
\begin{eqnarray}
\label{E:sss1}
X'' &=& -2 J' J^{-3},
\\
\label{E:sss2}
X'''&=& {-2 J'' \over J^3} + {6 (J')^2 \over J^4}.
\end{eqnarray}
We can then write
\begin{equation}
K^2 = {J^4} 
 \left\{ k^2  +{J''\over J }
\right\}.
\end{equation}
Let us suppose that $\lim_{x\to\pm\infty} J(x) = J_{\pm\infty} \neq 0$; then $K_{\pm\infty} =  k_{\pm\infty} J_{\pm\infty}^2$, so if $k^2(x)$ has nice asymptotic behaviour allowing one to define a scattering problem, so does $K^2(x)$.
\end{itemize}

\paragraph{Calculation:}
We now substitute equations (\ref{E:sss1})--(\ref{E:sss2}) above into (\ref{K-complete}), then we have
\begin{eqnarray}
\nonumber
K^2 &=& {1\over (X')^2} 
 \left\{ k^2  - {1\over2} {X'''\over X' }
+{3\over4} {(X'')^2\over(X')^{2} } \right\},
\nonumber 
\\
&=& J^4 \left\{ k^2  - {1\over2} \bigg({-2 J'' \over J^3} + {6 (J')^2 \over J^2} \bigg) 
+{3\over4}{\bigg({-2 J' \over J}\bigg)^2} \right\},
\nonumber
\\
&=& J^4 \left\{ k^2  - {1\over2} \bigg({-2 J'' \over J^3} + {6 (J')^2 \over J^4} \bigg) J^2
+{3\over4}{\bigg({-2 J' \over J}\bigg)^2} \right\},
\nonumber
\\
&=&  J^4 \left\{ k^2 + {J'' \over J} - {3 J'^2 \over J^2} +{3 J'^2 \over J^2}
\right\},
\nonumber
\\
&=&  J^4 \left\{ k^2 + {J'' \over J}
\right\},
\end{eqnarray}
as required.

\paragraph{Two theorems:}
These observations about the behaviour at spatial infinity lead immediately and naturally to the result:
\begin{theorem} 
Suppose $j_{\pm\infty} = 1$, \emph{(}equivalently, $J_{\pm\infty} = 1$\emph{)}. Then the ``potentials'' $k^2(x)$ and $K^2(X)$ have the same reflection and transmission amplitudes, and same reflection and transmission probabilities.
\end{theorem}
\noindent
This is automatic since $K_{\pm\infty} = k_{\pm\infty}$, so  equation (\ref{E:sde-10}) and the transformed equation (\ref{E:sde2}) both have the same asymptotic plane-wave solutions. Furthermore the Miller--Good transformation (\ref{E:mg}) maps any linear combination of solutions of equation (\ref{E:sde-10}) into the same linear combination of solutions of  the transformed equation (\ref{E:sde2}). QED.
\begin{theorem}
Suppose $j_{\pm\infty} \neq 1$,  \emph{(}equivalently, $J_{\pm\infty} \neq 1$\emph{)}. What is the relation between the reflection and transmission amplitudes, and reflection and transmission probabilities of the two  ``potentials'' $k^2(x)$ and $K^2(X)$? This is also trivial ---  the ``potentials'' $k^2(x)$ and $K^2(X)$ have the same reflection and transmission amplitudes, and same reflection and transmission probabilities.
\end{theorem}
\noindent
The only thing that now changes is that the properly normalized asymptotic states are distinct
\begin{equation}
{\exp(i k_\infty\, x )\over \sqrt{k_\infty} } \leftrightarrow
{\exp(i K_\infty\, x )\over \sqrt{K_\infty} },
 \end{equation}
but map into each other under the Miller--Good transformation. QED.

\section{Improved general bounds}
We already know
\begin{equation}
T \geq \sech^2\left\{ \int_{-\infty}^{+\infty}  \vartheta \; \d x \right\}.
\end{equation}
Here $T$ is the transmission probability, and $\vartheta$ is the function
\begin{equation}
\vartheta = {\sqrt{ (h')^2 + [k^2- h^2]^2}\over2 h },
\label{E:b0}
\end{equation}
with $h(x)>0$. But since the scattering problems defined by $k(x)$ and $K(X)$ have the same transmission probabilities, we also have
\begin{equation}
T \geq \sech^2\left\{ \int_{-\infty}^{+\infty}  \tilde \vartheta \; \d X \right\},
\end{equation}
with
\begin{equation}
\d X = X' \; \d x = j \; \d x,
\end{equation}
and 
\begin{eqnarray}
\tilde\vartheta 
&=& 
{\sqrt{ (h_X)^2 + [K^2- h^2]^2}\over2 h } ,
\\
&=&
 {1\over2h} \sqrt{ \left({h'\over X'}\right)^2 + \left[ {1\over j^2} 
 \left\{ k^2  - {1\over2} {j''\over j }
+{3\over4} {(j')^2\over j^{2} } \right\} - h^2\right]^2},
\\
&=& {1\over2hj } \sqrt{ (h')^2 + \left[ {1\over j}
 \left\{ k^2  - {1\over2} {j''\over j }
+{3\over4} {(j')^2\over j^{2} } \right\} - j h^2\right]^2}.
\end{eqnarray}
That is: $\forall h(x)>0,\; \forall j(x)>0$ we now have (the first form of) the improved bound
\begin{equation}
T \geq \sech^2\left\{ \int_{-\infty}^{+\infty}   {1\over2h } \sqrt{ (h')^2 + \left[ 
{1\over j}  \left\{ k^2  - {1\over2} {j''\over j }
+{3\over4} {(j')^2\over j^{2} } \right\} - j h^2\right]^2}  \; \d x \right\}.
\label{E:b1-11}
\end{equation}
Since this new bound contains \emph{two} freely specifiable functions it is definitely stronger than the result we started from,  (\ref{E:bd}). The result is perhaps a little more manageable if we work in terms of $J$ instead of $j$.
We follow the previous logic but now set
\begin{equation}
\d X = X' \; \d x = J^{-2} \; \d x,
\end{equation}
and
\begin{equation}
\tilde\vartheta 
= 
{\sqrt{ (h_X)^2 + [K^2- h^2]^2}\over2 h } 
= 
{1\over2h} \sqrt{ \left({h'\over X'}\right)^2 + \left[ J^4
 \left\{ k^2  + {J''\over J }\right\}
- h^2\right]^2}.
\end{equation}
That is: $\forall h(x)>0,\; \forall J(x)>0$ we have (the second form of) the improved bound
\begin{equation}
T \geq \sech^2\left\{ \int_{-\infty}^{+\infty}   {1\over2h  } \sqrt{ (h')^2 + \left[ 
J^2  \left\{ k^2  + {J''\over J } \right\} - {h^2\over J^2} \right]^2}  \; \d x \right\}.
\label{E:b2-11}
\end{equation}
A useful further modification is to substitute $h = H J^2$, then
$\forall H(x)>0,\; \forall J(x)>0$ we have (the third form of) the improved bound by the following argument:
When we let $h = H J^2$, then we derive
\begin{eqnarray}
h' &=& H' J^2 + 2 H J J'.
\end{eqnarray}
We now substitute above equations into (\ref{E:b2-11}), 
implying that $T$ is greater than
\begin{equation}
\sech^2\left\{ \int_{-\infty}^{+\infty}   {1\over 2 H J^2 } \sqrt{ (H' J^2 + 2 H J J')^2 + \left[ 
J^2  \left\{ k^2  + {J''\over J } \right\} - H^2 J^2 \right]^2}  \; \d x \right\}.
\end{equation}
We can simplify the above equation, and so we now also derive
\begin{equation}
T \geq \sech^2\left\{ \int_{-\infty}^{\infty}  
 {1\over2H} \sqrt{ \left[H' + 2H \; {J'\over J}\right]^2 + \left[ 
k^2  + {J''\over J } - H^2\right]^2}  \; \d x \right\}.
\label{E:b3-11}
\end{equation}
Equations (\ref{E:b1-11}), (\ref{E:b2-11}), and (\ref{E:b3-11}), are completely equivalent versions of our new bound.

\section{Some applications and special cases}

We can now use these improved general bounds,  (\ref{E:b1-11}), (\ref{E:b2-11}), and (\ref{E:b3-11}), to obtain several more specialized bounds that are applicable in more specific situations.

\subsection{Schwarzian bound}
First, take $h=(\mathrm{constant})$ in equation (\ref{E:b2-11}), then
\begin{equation}
T \geq \sech^2\left\{ {1\over2} \int_{-\infty}^{\infty}   \left|
{J^2\over h}  \left\{ k^2  + {J''\over J } \right\} - {h\over J^2} \right|  \; \d x \right\}.
\end{equation}
In order for this bound to convey nontrivial information we need the limit $\lim_{x\to\pm\infty} J^4 k^2 = h^2$, otherwise the integral diverges and the bound trivializes to $T \geq 0$.  The further specialization of this result reported in~\cite{bounds1,bounds2} and equation (\ref{E:bd1}) above corresponds to $J=(\mathrm{constant})= \sqrt{h /k_\infty}$, which clearly is a weaker bound than that reported here. In the present situation we can without loss of generality set $h\to k_\infty$ in which case 
\begin{equation}
T \geq \sech^2\left\{ {1\over2} \int_{-\infty}^{\infty}   \left|
{J^2\over k_\infty}  \left\{ k^2  + {J''\over J } \right\} - {k_\infty\over J^2} \right|  \; \d x \right\}.
\label{E:pre-Schwarzian}
\end{equation}
We now need $\lim_{x\to\pm\infty} J = 1$ in order to make the integral converge. 
If $k^2>0$, so that there is no classically forbidden region, then we can choose $J=\sqrt{k_\infty/k}$, in which case
\begin{equation}
T \geq \sech^2\left\{ {1\over2} \int_{-\infty}^{\infty}   \left|
{1\over\sqrt{k}} \left( {1\over\sqrt{k}} \right)'' \right|  \; \d x \right\}.
\label{E:Schwarzian}
\end{equation}
\begin{table}[!htbp]
\begin{center}
  \setlength{\fboxsep}{0.45 cm} 
   \framebox{\parbox[t]{12.5cm}{
{\bf Comment}:
This is a particularly elegant bound in terms of the Schwarzian derivative, [equation (\ref{E:Schwarzian-def})], which however unfortunately fails if there is a classically forbidden region. This bound is also computationally awkward to evaluate for specific potentials.
Furthermore, in the current context there does not seem to be any efficient or especially edifying way of choosing $J(x)$ in the forbidden region, and while the bound in equation (\ref{E:pre-Schwarzian}) is explicit it is not particularly useful.
}}
\end{center}
\end{table}
\subsection{Low-energy improvement}
We could alternatively set $H=(\mathrm{constant})$ in equation (\ref{E:b3-11}), to derive
\begin{equation}
T \geq \sech^2\left\{ \int_{-\infty}^{\infty}  
\sqrt{ \left[ {J'\over J}\right]^2 + {1\over4H^2} \left[ 
k^2  + {J''\over J } - H^2\right]^2}  \; \d x \right\}.
\end{equation}
In order for this bound to convey nontrivial information we need to enforce $\lim_{x\to\pm\infty}  k^2 = k_\infty^2 = H^2$,  while $\lim_{x\to\pm\infty}  J' = 0$, and  $\lim_{x\to\pm\infty}  J'' = 0$.  Otherwise the integral diverges and the bound trivializes to $T \geq 0$.  Thus
\begin{equation}
T \geq \sech^2\left\{ \int_{-\infty}^{\infty}  
\sqrt{ \left[ {J'\over J}\right]^2 + {1\over4k_\infty^2} \left[ 
k^2  + {J''\over J } - k_\infty^2\right]^2}  \; \d x \right\}.
\end{equation}
Again, the further specialization of this result reported in~\cite{bounds1,bounds2} and equation (\ref{E:bd1}) above corresponds to $J=(\mathrm{constant})$, which clearly is a weaker bound than that reported here.
To turn this into something a little more explicit, since $J(x)>0$ we can without any loss of generality write
\begin{equation}
J(x) = \exp\left[ \int \chi(x) \;\d x \right],
\end{equation}
where $\chi(x)$ is unconstrained. This permits is to write
\begin{equation}
T \geq \sech^2\left\{ \int_{-\infty}^{\infty}  
\sqrt{ \chi^2 + {1\over4k_\infty^2} \left[ 
k^2  + \chi^2-\chi' - k_\infty^2\right]^2}  \; \d x \right\}.
\end{equation}
Then by the triangle inequality
\begin{equation}
T \geq \sech^2\left\{ \int_{-\infty}^{\infty}  
\left[
 |\chi| + {1\over2k_\infty} \left| 
k^2  + \chi^2-\chi' - k_\infty^2\right|  \right] \; \d x \right\}.
\end{equation}
A further application of the triangle inequality yields
\begin{equation}
T \geq \sech^2\left\{ \int_{-\infty}^{\infty}  
\left[
 |\chi| +  {|\chi'| \over2k_\infty} +  {1\over2k_\infty} \left| 
k^2  + \chi^2- k_\infty^2\right|  \right] \; \d x \right\}.
\label{E:triangle2}
\end{equation}
Now if $k^2 \leq k_\infty^2$, (this is not that rare an occurrence, in a non-relativistic quantum scattering setting, where $k_\infty^2-k^2 = 2mV/\hbar^2$ and we have normalized to $V_\infty=0$, it corresponds to scattering from a potential that is everywhere positive), then we can choose $\chi^2= k_\infty^2-k^2$ so that
\begin{equation}
T \geq \sech^2\left\{ \int_{-\infty}^{\infty}  
\left.\left[
 |\chi| + {1\over2k_\infty} \left| 
\chi'\right|  \right] \; \d x \right\}\right|_{\chi=\sqrt{k_\infty^2-k^2}}.
\end{equation}
Assuming a unique maximum for $\chi$ (again not unreasonable, this corresponds to a single hump potential) this implies
\begin{equation}
T \geq \sech^2\left\{  
{\left.\sqrt{k_\infty^2-k^2}\right|_\mathrm{max}\over k_\infty} +
\int_{-\infty}^{\infty}  
\sqrt{k_\infty^2-k^2} \; \d x \right\}.
\label{E:nnn}
\end{equation}
This is a new and nontrivial bound, which in quantum physics language, where $k^2 = 2m(E-V)/\hbar^2$, corresponds to
\begin{equation}
T \geq \sech^2\left\{  
\sqrt{V_\mathrm{max}\over E} +
\int_{-\infty}^{\infty}  
{\sqrt{2mV}\over\hbar} \; \d x\right\}.
\label{E:nn}
\end{equation}
If under the same hypotheses we choose $\chi=0$, then the bound reported in~\cite{bounds1,bounds2} and equation (\ref{E:bd1}) above corresponds to
\begin{equation}
T \geq \sech^2\left\{  
{1\over2\sqrt{E}} \int_{-\infty}^{\infty}  
{\sqrt{2m}\,V\over\hbar} \; \d x\right\}.
\label{E:oo}
\end{equation}
Thus for sufficiently small $E$ the new bound in equation (\ref{E:nn}) is more stringent than the old bound  in equation (\ref{E:oo}) provided
\begin{equation}
\sqrt{V_\mathrm{max}} < {1\over2} \int_{-\infty}^{\infty}  
{\sqrt{2m}\,V\over\hbar} \; \d x.
\end{equation}
\begin{table}[!htb]
\begin{center}
  \setlength{\fboxsep}{0.45 cm} 
   \framebox{\parbox[t]{12.5cm}{
{\bf Comment}:
Note the long chain of inequalities leading to these results --- this suggests that these final inequalities (\ref{E:nnn}) and (\ref{E:nn}) are not optimal and that one might still be able to strengthen them considerably.
}}
\end{center}
\end{table}
\subsection{WKB-like bound}

Another option is to return to equation (\ref{E:triangle2}) and make the choice $\chi^2=\max\{0,-k^2\} = \kappa^2$, so that $\kappa=|k|$ in the classically forbidden region $k^2<0$, while $\kappa=0$ in the classically allowed region $k^2>0$. But then  equation (\ref{E:triangle2})  reduces to
\begin{equation}
T \geq \sech^2\left\{ \; \; \int\limits_{k^2<0} \kappa \; \d x + {\kappa_\mathrm{max}\over k_\infty} + {k_\infty\,L\over2} + \int\limits_{k^2>0} {|k_\infty^2-k^2|\over2 k_\infty} 
\; \d x \right\}.
\label{E:WKB-like}
\end{equation}
Key points here are the presence of $ \int\nolimits_{k^2<0} \kappa \; \d x$, the barrier penetration integral that normally shows up in the standard WKB approximation to barrier penetration, $\kappa_\mathrm{max}$ the height of the barrier, and $L$ the width of the barrier.  These is also a contribution from the classically allowed region (as in general there must be, potentials with no classically forbidden region still generically have nontrivial scattering).  Compare this with the standard WKB estimate:
\begin{equation}
T_\mathbf{WKB}  \approx 
\sech^2\left\{ \; \; \int\limits_{k^2<0} \kappa \; \d x + \ln2 \;\right\}.
\end{equation}
This form of the WKB approximation for barrier penetration is derived, for instance, in Bohm's classic textbook~\cite{bohm}, and can also be found in many other places. Under the usual conditions applying to the WKB approximation for barrier penetration we have $ \int\nolimits_{k^2<0} \kappa \; \d x \gg 1$, in which case one obtains the more well-known version
\begin{equation}
T_\mathbf{WKB}  \approx 
\exp\left\{ \; \; - 2 \int\limits_{k^2<0} \kappa \; \d x \;\right\}.
\end{equation}
The bound in equation (\ref{E:WKB-like}) is the closest we  have so far been able to get to obtaining a rigorous bound that somewhat resembles the standard WKB estimate. Again we do not expect the bound in equation (\ref{E:WKB-like}) to be optimal, and are continuing to search for improvements on this WKB-like bound.
\begin{table}[!htb]
\begin{center}
  \setlength{\fboxsep}{0.45 cm} 
   \framebox{\parbox[t]{12.5cm}{
{\bf The Classically Forbidden Region}:
In regions where the total energy is less than the potential energy, a region where in classical physics there would be zero probability of finding a particle, the amplitude of the wave function decreases. Further away from the boundary where the potential energy changes, the probability of the object being located there decreases. However, unlike classical physics, in the region where the total energy is less than the potential energy, the probability of finding the object is not always zero~\cite{Visual-Quantum-Mechanics}.
}}
\end{center}
\end{table}

\subsection{Further transforming the bound}

In an attempt to strengthen the inequalities (\ref{E:nnn}) and (\ref{E:nn}), we again  use the fact that $J(x)>0$ to (without any loss of generality) write
$J(x) = \exp\left[ \int \chi(x) \;\d x \right]$, 
where $\chi(x)$ is unconstrained. The general bound in equation (\ref{E:b3-11}) can then be transformed to: For all $H(x)>0$, for all $\chi(x)$:
\begin{equation}
\label{E:xxx}
T \geq \sech^2\left\{ \int_{-\infty}^{\infty}  
 {1\over2} \sqrt{ \left[{H'\over H} + 2 \chi \right]^2 + 
 {\left[ k^2  + \chi^2 + \chi' - H^2\right]^2\over H^2 }  }  \; \d x \right\}.
\end{equation}
This leaves us with considerable freedom. Regardless of the \emph{sign} of $k^2(x)$, we can always choose to enforce $k^2  + \chi^2 - H^2 = 0$, and so eliminate \emph{either} $\chi$ \emph{or} $H$, obtaining
\begin{equation}
T \geq \sech^2\left\{ \int_{-\infty}^{\infty}  
 {1\over2} \sqrt{ \left[{H'\over H} + 2 \sqrt{H^2- k^2} \right]^2 + 
 {\left[ (\sqrt{H^2- k^2})' \right]^2\over H^2 }  }  \; \; \d x \right\},
 \label{E:ppp1}
\end{equation}
(subject to $H(x)>0$ and $H^2(x)-k^2(x)>0$), and
\begin{equation}
T \geq \sech^2\left\{ \int_{-\infty}^{\infty}  
 {1\over2} \sqrt{ \left[{ (\sqrt{\chi^2+k^2})'\over\sqrt{\chi^2+k^2} } 
 + 2 \chi \right]^2 + 
 {  (\chi' )^2\over \chi^2+k^2 }  }  \; \; \d x \right\},
 \label{E:ppp2}
\end{equation}
(subject to $\chi^2(x)+k^2(x)>0$),
respectively.
Finding an explicit bound is now largely a matter of art rather than method. For example if we take
\begin{equation}
H^2= \max\{ k^2, \Delta^2\} \qquad \hbox{or} \qquad \chi^2 = \max\{ 0, \Delta^2-k^2\},
\end{equation}
then from either  equation (\ref{E:ppp1}) or  equation (\ref{E:ppp2}), again under the restriction that we are dealing with a single-hump positive potential,  we obtain
\begin{equation}
T \geq \sech^2\left\{ 
{1\over2} \ln\left({k_{+\infty} k_{-\infty}\over \Delta^2}\right)  
+ {(\sqrt{\Delta^2-k^2})_\mathrm{max}\over\Delta}
+ \int\limits_{\Delta^2>k^2}  \sqrt{\Delta^2-k^2} \; \d x 
\right\}.
\label{E:final}
\end{equation}
Note that $\Delta$ is a free parameter which could in principle be chosen to optimize the bound, however the resulting integral equation is too messy to be of any practical interest. This bound is somewhat similar to that reported in equations (\ref{E:bd4}) and (\ref{E:nnn}), but there are some very real differences.

\section{Summary and Discussion}
The bounds presented in this chapter are generally \emph{not} ``WKB-like'' --- apart from the one case reported in equation (\ref{E:WKB-like}) there is no need (nor does it seem useful) to separate the region of integration into classically allowed and classically forbidden regions. In fact it is far from clear how closely these bounds might ultimately be related to WKB estimates of the transmission probabilities, and this is an issue to which we hope to return in the future. 

We should mention that if one works with the formally equivalent problem of a parametric oscillator in the time domain then the relevant differential equation is
\begin{equation}
\ddot u(t) + k(t)^2 \; u(t) = 0,
\label{E:po}
\end{equation}
and instead of asking questions about transmission amplitudes and probabilities one is naturally driven to ask formally equivalent questions about Bogoliubov coefficients and particle production. The key translation step is to realize that there is an equivalence~\cite{bounds1,bounds2}:
\begin{equation}
T \leftrightarrow {1\over 1+ N};  \qquad  N\leftrightarrow {1-T\over T}.
\end{equation}
This leads to bounds on the number of particles produced that are of the form $T \geq \sech^2\{ \hbox{(some appropriate integral)} \}$, thereby  implying
\begin{equation}
N  \leq \sinh^2\{ \hbox{(some appropriate integral)} \}.  
\end{equation}

\bigskip\noindent
To be more explicit about this, our new improved bound can be written in any of three equivalent forms:
\begin{enumerate}
\item 
For all $H(x)>0$, for all $J(x) > 0$, 
\begin{equation}
\label{E:xx}
T \geq \sech^2\left\{ \int_{-\infty}^{\infty}  
 {1\over2H} \sqrt{ \left[H' + 2H \; {J'\over J}\right]^2 + \left[ 
k^2  + {J''\over J } - H^2\right]^2}  \; \d x \right\}.
\end{equation}
\item
For all $h(x)>0$,  for all $J(x) > 0$, 
\begin{equation}
T \geq \sech^2\left\{ \int_{-\infty}^{\infty}   {1\over2h  } \sqrt{ (h')^2 + \left[ 
J^2  \left\{ k^2  + {J''\over J } \right\} - {h^2\over J^2} \right]^2}  \; \d x \right\}.
\end{equation}
\item
For all $h(x)>0$,  for all $j(x) > 0$, 
\begin{equation}
T \geq \sech^2\left\{ \int_{-\infty}^{+\infty}   {1\over2h } \sqrt{ (h')^2 + \left[ 
{1\over j}  \left\{ k^2  - {1\over2} {j''\over j }
+{3\over4} {(j')^2\over j^{2} } \right\} - j h^2\right]^2}  \; \d x \right\}.
\end{equation}
\end{enumerate}
The equivalent statements about particle production are:
\begin{enumerate}
\item 
For all $H(t)>0$, for all $J(t) > 0$, 
\begin{equation}
N \leq \sinh^2\left\{ \int_{-\infty}^{\infty}  
 {1\over2H} \sqrt{ \left[H' + 2H \; {J'\over J}\right]^2 + \left[ 
k^2  + {J''\over J } - H^2\right]^2}  \; \d t \right\}.
\label{E:N1}
\end{equation}
\item
For all $h(t)>0$,  for all $J(t) > 0$, 
\begin{equation}
N \leq \sinh^2\left\{ \int_{-\infty}^{\infty}   {1\over2h  } \sqrt{ (h')^2 + \left[ 
J^2  \left\{ k^2  + {J''\over J } \right\} - {h^2\over J^2} \right]^2}  \; \d t \right\}.
\label{E:N2}
\end{equation}
\item
For all $h(t)>0$,  for all $j(t) > 0$, 
\begin{equation}
N \leq \sinh^2\left\{ \int_{-\infty}^{+\infty}   {1\over2h } \sqrt{ (h')^2 + \left[ 
{1\over j}  \left\{ k^2  - {1\over2} {j''\over j }
+{3\over4} {(j')^2\over j^{2} } \right\} - j h^2\right]^2}  \; \d t \right\}.
\label{E:N3}
\end{equation}
\end{enumerate}
In closing, we reiterate that these general bounds reported in equations (\ref{E:b1-11}), (\ref{E:b2-11}), and (\ref{E:b3-11}), their specializations in equations (\ref{E:pre-Schwarzian}), (\ref{E:Schwarzian}), (\ref{E:nnn}), (\ref{E:nn}), (\ref{E:WKB-like}),  and (\ref{E:final}), and the equivalent particle production bounds in equations (\ref{E:N1})--(\ref{E:N3}), are all general purpose tools that are applicable to a wide variety of physical situations~\cite{analogue, cavity, qubit, one-dim, greybody-factor, sonoluminescence}. Furthermore we strongly suspect that further generalizations of these bounds are still possible.

\chapter{Analytic bounds on transmission probabilities}
\label{C:consistency-11}
\section{Introduction}
In this chapter, we shall develop some additional and novel analytic bounds on transmission probabilities (and the related reflection probabilities and Bogoliubov coefficients) for generic one-dimensional scattering problems.  We shall review the basic concepts underlying this fascinating topic by rewriting the Schr\"odinger equation for some complicated potential whose properties we are trying to investigate in terms of some simpler potential whose properties are assumed known plus a (possibly large) ``shift'' in the potential. Doing so permits us to extract considerable useful information without having to exactly solve the full scattering problem.

\enlargethispage{20pt}

In earlier chapters of this thesis, and in several  published papers~\cite{bounds1, bounds2, greybody-factor, Miller-good-transformation-articles}, we  have derived a number of rigorous bounds on transmission probabilities (and reflection probabilities, and Bogoliubov coefficients) for one-dimensional scattering problems. The derivation of these bounds generally proceeds by rewriting the Schr\"odinger equation in terms of some equivalent system of first-order equations, and then analytically bounding the growth of certain quantities related to the net flux of particles as one sweeps across the potential.

In this chapter we shall obtain significantly different results, of both theoretical and practical interest. While a vast amount of effort has gone into studying the Schr\"odinger equation and its scattering properties~\cite{Landau, Merzbacher, Capri, Messiah, branson-joachim, liboff, bohm, dicke-wittke, shankar}, it appears that relatively little work has gone into providing general analytic bounds on the transmission probabilities, (as opposed to approximate estimates). The only known results as far as we have been able to determine are presented in~\cite{bounds1}, in the earlier chapters of this thesis, and the related publications~\cite{bounds2, greybody-factor, Miller-good-transformation-articles} based on work reported in this thesis. Several quite remarkable bounds were first derived in~\cite{bounds1}, with further discussion and an alternate proof being provided in~\cite{bounds2}. 

These bounds were originally used as a technical step when studying a specific model for sonoluminescence~\cite{sonoluminescence},  and since then have also been used to place limits on particle production in analogue spacetimes~\cite{analogue} and resonant cavities~\cite{cavity}, to investigate qubit master equations~\cite{qubit}, and to motivate further general investigations of one-dimensional scattering theory~\cite{one-dim}. Recently, these bounds have also been applied to the greybody factors of a Schwarzschild black hole~\cite{greybody-factor}. Most recently, significant extensions of the original bounds have been developed by adapting the Miller--Good transformations~\cite{Miller-good-transformation-articles}.

In this chapter we shall return to this problem, developing a new set of techniques that are more amenable to the development of both \emph{upper} and \emph{lower} bounds. For technical reasons the new techniques are also more amenable to investigating behavior ``under the barrier''.  The basic idea is to re-cast the Schr\"odinger equation for some complicated potential whose properties we are trying to investigate in terms of some simpler potential whose properties are assumed known, plus a ``shift'' in the potential.

\vfill

\section{From Schr\"odinger equation to system of ODEs}

We are interested in the scattering properties of the Schr\"odinger equation,
\begin{equation}
\psi''(x) + k(x)^2 \; \psi(x) = 0,
\label{E:sde}
\end{equation}
where $k(x)^2 = 2m[E-V(x)]/\hbar^2$.  As long as $V(x)$ tends to finite (possibly distinct) constants $V_{\pm\infty}$ on left and right infinity, then for $E>\max\{V_{+\infty},V_{-\infty}\}$ one can set up a one-dimensional scattering problem in a completely standard manner --- see, for example, standard references such as~\cite{Landau, Merzbacher, Capri, Messiah, branson-joachim, liboff, bohm, dicke-wittke, shankar}, and the background discussion presented in earlier chapters of this thesis.  The scattering problem is completely characterized by the transmission and reflection \emph{amplitudes} (denoted $t$ and $r$), although the most important aspects of the physics can be extracted from the transmission and reflection \emph{probabilities} ($T= |t|^2$ and $R=|r|^2$).

\subsection{Ansatz}

The idea is to try to say things about exact solutions to the ODE
\begin{equation}
\psi''(x) + k^2(x)\; \psi(x) = 0,
\end{equation}
by comparing this ODE to some ``simpler'' one
\begin{equation}
\psi_0''(x) + k_0^2(x)\; \psi_0(x) = 0,
\end{equation}
for which we are assumed to the know exact solutions $\psi_0(x)$.
In a manner similar to the analysis in references~\cite{bounds1, bounds2}, we will start by introducing the ansatz
\begin{equation}
\label{E:representation}
\psi(x) = 
a(x) \;\psi_0(x)+ b(x) \;\psi_0^*(x).
\end{equation}
This representation is
of course extremely highly redundant, since \emph{one} complex number $\psi(x)$ has
been traded for \emph{two} complex numbers $a(x)$ and $b(x)$.  This redundancy allows us, without any loss of generality, to enforce one auxiliary constraint connecting $a(x)$ and $b(x)$. We find it particularly useful to enforce the auxiliary condition
\begin{equation}
\label{E:gauge}
{\d a\over \d x} \; \psi_0+ 
{\d b\over \d x} \; \psi_0^*
= 0.
\end{equation}
Subject to this auxiliary constraint on the derivatives of $a(x)$ and $b(x)$, the derivative of $\psi(x)$ takes on the especially simple form
\begin{equation}
\label{E:gradient}
{\d\psi\over \d x} = a \; \psi_0' + b \; \psi_0^*{}'.
\end{equation}
(This ansatz is largely inspired by the techniques of references~\cite{bounds1, bounds2}, where JWKB estimates for the wave function were similarly used as a ``basis'' for formally writing down the exact solutions.)

\subsection{Probability density and probability current}

For the probability density we have:
\begin{eqnarray}
\rho  &=& \psi^* \psi ,
\\
&=& \big|  a(x) \psi_0
+  b(x) \psi_0^* \big|^2 ,
\\
&=& \{ |a|^2 + |b|^2| \} |\psi_0|^2 + 2 \mathrm{Re} \, \{a b^* \psi_0^2  \} ,
\\
&=& \{ |a|^2 + |b|^2| \} \rho_0 + 2 \mathrm{Re} \, \{a b^* \psi_0^2  \}.
\end{eqnarray}
Furthermore, for the probability current:
\begin{eqnarray}
\mathscr{J} &=& \mathrm{Im} \, \bigg\{ \psi^* {\d \psi \over \d x}\bigg\} ,
\\
&=& \mathrm{Im} \, \bigg\{ \left[ a^* \psi_0^* + b^* \psi_0 \right] \; 
\left[a \psi_0' + b \psi_0^*{}'  \right] \bigg\} , \qquad\;
\\
&=&\mathrm{Im} \, \Bigg\{ |a|^2 \psi_0^* \psi_0' + |b|^2 \psi_0 \psi_0^*{}' + ab^* \psi_0 \psi_0' + a^*b \psi_0^*\psi_0^*{}'
 \Bigg\} ,
 \\
&=& \{ |a|^2- |b|^2 \} \; \mathrm{Im} \, \{ \psi_0^* \; \psi_0'\} ,
\\
\label{probability-current-for-analytic-function}
&=&   \{ |a|^2- |b|^2 \}  \; \mathscr{J}_0.
\end{eqnarray}
Under the conditions we are interested in, (which correspond to a time-independent solution of the Schr\"odinger equation), we have $\dot\rho=0$, and so $\partial_x \mathscr{J}=0$. (And similarly  $\dot\rho_0=0$, so $\partial_x \mathscr{J}_0=0$.) That is, $\mathscr{J}$ and   $\mathscr{J}_0$ are position-independent constants,  an observation which then puts a constraint on the amplitudes $|a|$ and $|b|$.  Applying an appropriate boundary condition, which we can take to be $a(-\infty)=1$, $b(-\infty)=0$, we then see
\begin{equation}
 |a|^2 - |b|^2 = 1.
\end{equation}
This observation justifies interpreting $a(x)$ and $b(x)$ as ``position-dependent Bogoliubov coefficients''. Furthermore without any loss in generality we can choose the normalizations on $\psi$ and $\psi_0$ so as to set the net fluxes to unity: $\mathscr{J}=\mathscr{J}_0=1$.

\subsection{Second derivatives of the wavefunction}
%
We shall now re-write the Schr\"odinger equation in terms of two coupled
first-order differential equations for these position-dependent
Bogoliubov coefficients $a(x)$ and $b(x)$. To do this, evaluate $\d^2\psi/ \d x^2$ making
repeated use of the auxiliary condition (\ref{E:gauge})
\begin{eqnarray}
\label{E:double-gradient}
{\d^2\psi\over \d x^2} 
&=& 
{\d\over \d x} 
\left( a \,\psi_0' + b \, \psi_0^*{}' \right) ,
\\
&=&
a'  \, \psi_0' + b'  \, \psi_0^*{}' +  a  \, \psi_0'' + b \, \psi_0^*{}'' ,
\\
&=& 
a'  \, \psi_0' - a'  \,  {\psi_0\over\psi_0^*}  \,  \psi_0^*{}' -  a  \, k_0^2  \, \psi_0 - b  \, k_0^2  \,  \psi_0^* ,
\\
&=& {a'\over\psi_0^*}  \, \{ \psi_0^* \psi_0' - \psi_0 \psi_0^*{}' \} 
-  
k_0^2  \, [ a  \psi_0 + b \psi_0^*] ,
\\
&=& {2i\J_0 a'\over \psi_0^*} 
-  
k_0^2  \,  [ a  \psi_0 + b \psi_0^*]  ,
\\
&=& {2i a'\over \psi_0^*} 
-  
k_0^2  \,  [ a  \psi_0 + b \psi_0^*].
\end{eqnarray}
Where in the last line we have finally used our normalization choice $\J_0=1$.
This is one of the two relations we wish to establish. 
Now use the gauge condition to eliminate $\d a/\d x$ in favour of $\d b/\d x$ to obtain a second relation for  $\d^2\psi/ \d x^2$.  This now permits us to write $\d^2\psi/ \d x^2$ in
either of the two equivalent forms
\begin{eqnarray}
\label{E:double-gradient2}
{\d^2\psi\over \d x^2} 
&=& {2i a'\over \psi_0^*} 
-  
k_0^2 \,  [ a  \psi_0 + b \psi_0^*];
\\
&=& - {2i b'\over \psi_0} 
-  
k_0^2 \,  [ a  \psi_0 + b \psi_0^*].
\end{eqnarray}
%

\subsection{SDE as a first-order system}

Now insert these formulae for the second derivative of the wavefunction into the Schr\"odinger equation written in the
form
\begin{equation}
{\d^2\psi\over \d x^2} + k(x)^2 \; \psi  = 0,
\end{equation}
to deduce the \emph{pair} of first-order ODEs:
\begin{eqnarray}
\label{E:system-a}
{\d a\over \d x} &=& +
{i\over2} [k^2-k_0^2]\;
\{ a\; |\psi_0|^2 + b\; \psi_0^*{}^2 
\};
\\
\label{E:system-b}
{\d b\over \d x} &=& -
{i\over2} [k^2-k_0^2]\;
\{ a \;\psi_0^2 + b \; |\psi_0|^2 \}.
\end{eqnarray}
It is easy to verify that this first-order system is compatible with
the auxiliary condition (\ref{E:gauge}), and that by iterating the
system twice (subject to this auxiliary condition) one recovers exactly
the original Schr\"odinger equation. 
We can re-write this 1st-order system of ODEs in matrix form as
\begin{equation}
{\d\over \d x} \left[\begin{matrix} a \\ b\end{matrix}\right] = 
{i[k^2-k_0^2]\over2} 
\left[
\begin{matrix}
|\psi_0|^2 & \psi_0^*{}^2\\
-\psi_0^2 & - |\psi_0|^2
\end{matrix}
\right]
\left[ \begin{matrix} a \cr b\end{matrix}\right].
\end{equation}
\begin{table}[!htb]
\begin{center}
  \setlength{\fboxsep}{0.45 cm} 
   \framebox{\parbox[t]{12.5cm}{
{\bf Comment}:
Matrix ODEs of this general form are often referred to as Shabhat--Zakharov  or Zakharov--Shabat systems~\cite{bounds1}. This matrix ODE can be used to write down a formal solution to the SDE in terms of ``path-ordered exponentials'' as in references~\cite{bounds1, bounds2}.  We choose not to adopt this route here, instead opting for a more direct computation in terms of the magnitudes and phases of $a$ and $b$.
}}
\end{center}
\end{table}
\begin{table}[!htbp]
\begin{center}
  \setlength{\fboxsep}{0.45 cm} 
   \framebox{\parbox[t]{12.5cm}{
{\bf The path-ordered exponentials (or the ordered exponential)}: In a general formal context, this mathematical object is defined in arbitrary non-commutative algebras, and it is the closest equivalent possible to the exponential function of the integral in the commutative algebras. 

In practice the values lie in matrix and operator algebras.  For the element $A(t)$ from the algebra $(g, ^*)$ (the set $g$ with the non-commutative product $^*$), where $t$ is the ``time parameter'' the ordered exponential~\cite{ordered-exponential}.
\begin{equation}
OE[A](t):= \exp\bigg(\int_{0}^{t} \, \d t' A(t')\bigg).
\end{equation}
of $A$ can be defined via one of several equivalent approaches:
\begin{itemize}
\item
As the limit of the ordered product of the infinitesimal exponentials:
\begin{equation}
\!\!\!\!
OE[A](t) = \lim_{N \rightarrow \infty} \bigg\{\exp(\epsilon A(t_{N})) \times \exp(\epsilon A(t_{N-1}))  \times \dots \times \exp(\epsilon A(t_{0}))\bigg\}. \qquad
\end{equation}
where the time moments $\{t_{0},t_{1},\dots, t_{N}\}$ are defined as $t_{j} = j \times \epsilon$ for $j = 0, \dots N$, and $\epsilon = t / N$.
\item
Via the initial value problem, where the $OE[A](t)$ is the unique solution of the system of equations:
\begin{equation}
{\partial OE[A](t) \over \partial t} = A(t) \times OE[A](t), 
\end{equation}
where $OE[A](0) =1$.
\item
Via an integral equation:
\begin{eqnarray}
OE[A](t) &=&  1 + \int_{0}^{t} \d t_{1} A(t_{1}) + \int_{0}^{t} \d t_{1} \, \int_{0}^{t_{1}} \d t_{2} A(t_{1}) \times A(t_{2}) 
\nonumber
\\
&+& \int_{0}^{t} \d t_{1} \int_{0}^{t_{1}} \d t_{2} \int_{0}^{t_{2}} \d t_{3} A(t_{1}) \times A(t_{2}) \times A(t_{3}) + \dots
\nonumber
\\
&&
\end{eqnarray} 
\end{itemize}
}}
\end{center}
\end{table}

\subsection{Formal (partial) solution}

Define magnitudes and  phases by
\begin{equation}
a = |a|\; e^{i\phi_a}; \qquad b = |b|\; e^{i\phi_b}; \qquad \psi_0 = |\psi_0|\; e^{i\phi_0}.
\end{equation}
Calculate
\begin{equation}
a' =  |a|' \, e^{i\phi_a} +  i |a| \, e^{i\phi_a} \, \phi_a' = e^{i\phi_a} \left\{ |a|' + i |a| \, \phi_a'\right\},
\end{equation}
whence
\begin{equation}
|a|' + i |a| \, \phi_a' = {i\over2} [k^2-k_0^2]\;  |\psi_0|^2 \; \{ |a|\; + |b|\; e^{-i(\phi_a-\phi_b+2\phi_0)} \}.
\label{E:a}
\end{equation}
Similarly we also have
\begin{equation}
|b|' + i |b| \, \phi_b' = -{i\over2} [k^2-k_0^2]\;  |\psi_0|^2 \; \{ |b|\; + |a|\; e^{-i(\phi_b-\phi_a-2\phi_0)} \}.
\label{E:b}
\end{equation}
Now take the real part of both these equations, 
whence
\begin{equation}
|a|' = +{1\over2} [k^2-k_0^2] \;  |b| \; |\psi_0|^2  \sin(\phi_a-\phi_b+2\phi_0);
\end{equation}
\begin{equation}
|b|' = +{1\over2} [k^2-k_0^2] \;  |a| \; |\psi_0|^2  \sin(\phi_a-\phi_b+2\phi_0).
\end{equation}
Therefore
\begin{equation}
|a|' =   {1\over2} [k^2-k_0^2] \; |\psi_0|^2  \sin(\phi_a-\phi_b+2\phi_0)  \;   \sqrt{|a|^2-1}.
\end{equation}
That is
\begin{equation}
{|a|' \over \sqrt{|a|^2-1}}  =  {1\over2} [k^2-k_0^2] \; |\psi_0|^2  \sin(\phi_a-\phi_b+2\phi_0),
\end{equation}
whence
\begin{equation}
\left\{ \cosh^{-1} |a| \right\}_{x_1}^{x_2}  = {1\over2}
 \int _{x_1}^{x_2} [k^2-k_0^2] \; |\psi_0|^2  \sin(\phi_a-\phi_b+2\phi_0) \; \d x.
\end{equation}
Now apply the boundary conditions: At $x=-\infty$ we have both $a(-\infty)=1$, and $b(-\infty)=0$. Therefore
\begin{equation}
\cosh^{-1} |a(x)|  = {1\over2} 
 \int _{-\infty}^{x}  [k^2-k_0^2] \; |\psi_0|^2  \sin(\phi_a-\phi_b+2\phi_0) \; \d x,
\end{equation}
and so
\begin{equation}
\label{E:pre-Theta}
|a(x)|  = \cosh\left\{ {1\over2} 
 \int _{-\infty}^{x} [k^2-k_0^2] \; |\psi_0|^2  \sin(\phi_a-\phi_b+2\phi_0) \; \d x \right\}.
\end{equation}
In particular
\begin{equation}
\cosh^{-1} |a(\infty)|  = {1\over2} 
 \int _{-\infty}^{+\infty}  [k^2-k_0^2] \; |\psi_0|^2  \sin(\phi_a-\phi_b+2\phi_0) \; \d x,
\end{equation}
or equivalently
\begin{equation}
|a(\infty)|  = \cosh\left\{ {1\over2} 
 \int _{-\infty}^{+\infty} [k^2-k_0^2] \; |\psi_0|^2  \sin(\phi_a-\phi_b+2\phi_0) \; \d x \right\}.
\end{equation}
Of course this is only a \emph{formal} solution since $\phi_a(x)$ and $\phi_b(x)$ are, (at least at this stage), ``unknown''. But we shall argue that this formula still contains useful information. In particular, in view of the normalization conditions relating $a$ and $b$, and the parity properties of $\cosh$ and $\sinh$,  we can also write
\begin{equation}
\label{E:a1}
|a(\infty)|  = \cosh\left| {1\over2} 
 \int _{-\infty}^{+\infty} [k^2-k_0^2] \; |\psi_0|^2  \sin(\phi_a-\phi_b+2\phi_0) \; \d x \right|;
\end{equation}
\begin{equation}
\label{E:b1}
|b(\infty)|  = \sinh\left| {1\over2} 
 \int _{-\infty}^{+\infty} [k^2-k_0^2] \; |\psi_0|^2  \sin(\phi_a-\phi_b+2\phi_0) \; \d x \right|.
\end{equation}

\subsection{First set of bounds}

To determine the first elementary set of bounds on $a$ and $b$ is now trivial. We just note that
\begin{equation}
|\sin(\phi_a-\phi_b+2\phi_0)| \leq 1.
\end{equation}
Therefore
\begin{equation}
|a(\infty)|  \leq \cosh\left\{ {1\over2} 
 \int _{-\infty}^{+\infty} {|k^2-k_0^2| \;|\psi_0|^2} \, \d x \right\};
\end{equation}
\begin{equation}
|b(\infty)|  \leq \sinh\left\{ {1\over2} 
 \int _{-\infty}^{+\infty} {|k^2-k_0^2| \;|\psi_0|^2} \, \d x \right\}.
\end{equation}
What does this now tell us about the Bogoliubov coefficients?
\subsection{Bogoliubov coefficients}
The slightly unusual thing, (compared to our earlier work in this thesis and in references~\cite{bounds1, bounds2, Miller-good-transformation-articles}),  is that now the ``known'' function $\psi_0$ may also have its own Bogoliubov coefficients. Let us assume we have set our boundary conditions so that for the ``known'' situation 
\begin{equation}
\psi_0(x\approx - \infty) \sim \exp\{ik(-\infty) x\},
\end{equation}
and
\begin{equation}
\psi_0(x\approx + \infty) \sim \alpha_0 \exp\{ik(+\infty) x\} 
+ \beta_0  \exp\{-ik(+\infty) x\}.
\end{equation}
Then the way we have set things up, for the ``full'' problem we still have
\begin{equation}
\psi(x\approx - \infty) \sim \exp\{ik(-\infty) x\},
\end{equation}
whereas
\begin{eqnarray}
\psi(x\approx + \infty) &\sim& a(\infty) \, \psi_0(x) + b(\infty) \, \psi_0^*(x) , \\
&\sim& 
[\alpha_0 \,  a(\infty) + \beta_0^* \, b(\infty) ] \,\exp\{ik(+\infty) x\}
\nonumber\\
&&
+ [\beta_0 \, a(\infty) + \alpha_0^* \, b(\infty) ] \, \exp\{-ik(+\infty) x\}.
\end{eqnarray}
That is, the overall Bogoliubov coefficients satisfy
\begin{equation}
\alpha = \alpha_0 \, a(\infty) + \beta_0^* \, b(\infty);
\end{equation}
\begin{equation}
\beta = \beta_0 \, a(\infty) + \alpha_0^* \, b(\infty).
\end{equation}
These equations relate the Bogoliubov coefficients of the ``full'' problem $\{\psi(x),\,k(x)\}$ to those of the simpler ``known'' problem $\{\psi_0(x),\,k_0(x)\}$, plus the evolution of the $a(x)$ and $b(x)$ coefficients.
Now observe that
\begin{equation}
|\alpha| \leq |\alpha_0| \; |a(\infty)| + |\beta_0| \; |b(\infty)|.
\end{equation}
But we can define
\begin{equation}
|\alpha_0| = \cosh \Theta_0; \quad |\beta_0| = \sinh \Theta_0; 
\qquad 
 |a(\infty)| = \cosh \Theta; \quad  |b(\infty)| = \sinh \Theta; 
\end{equation}
in terms of which 
\begin{equation}
|\alpha| \leq \cosh \Theta_0 \cosh \Theta +\sinh \Theta_0 \sinh \Theta
= \cosh\left(\Theta_0 + \Theta\right).
\end{equation}
That is:  Since we know 
\begin{equation}
\Theta \leq \Theta_\mathrm{bound} 
\equiv {1\over2}   \int _{-\infty}^{+\infty} {|k^2-k_0^2| \;|\psi_0|^2} \d x,
\end{equation}
we can deduce
\begin{equation}
|\alpha|  \leq \cosh\left\{ \cosh^{-1}|\alpha_0| + {1\over2} 
 \int _{-\infty}^{+\infty} {|k^2-k_0^2| \;|\psi_0|^2} \d x \right\};
\end{equation}
\begin{equation}
|\beta|  \leq \sinh\left\{ \sinh^{-1}|\beta_0| + {1\over2} 
 \int _{-\infty}^{+\infty} {|k^2-k_0^2| \;|\psi_0|^2} \d x \right\}.
\end{equation}
\subsection{Second set of bounds}
A considerably trickier inequality, now leading to a \emph{lower bound} on the Bogoliubov coefficients, is obtained by considering what the phases would have to be to achieve as much destructive interference as possible. That implies
\begin{equation}
|\alpha| \geq |\alpha_0| \; |a(\infty)| - |\beta_0| \; |b(\infty)|,
\end{equation}
whence
\begin{equation}
|\alpha| \geq 
\cosh\left|\Theta_0 - \Theta\right|.
\end{equation}
Therefore, using $\Theta \leq \Theta_\mathrm{bound}$, it follows that \emph{as long as} $ \Theta_\mathrm{bound} < \Theta_0$, one can deduce 
\begin{equation}
|\alpha| \geq 
\cosh\left\{\Theta_0 - \Theta_\mathrm{bound}\right\}.
\end{equation}
(If on the other hand $\Theta_\mathrm{bound} \geq \Theta_0$, then one only obtains the trivial bound $|\alpha|\geq 1$.) Another way of writing these bounds is as follows
\begin{equation}
|\alpha|  \geq \cosh\left\{ \cosh^{-1}|\alpha_0| - {1\over2} 
 \int _{-\infty}^{+\infty} {|k^2-k_0^2| \;|\psi_0|^2} \d x \right\};
\end{equation}
\begin{equation}
|\beta|  \geq \sinh\left\{ \sinh^{-1}|\beta_0| - {1\over2} 
 \int _{-\infty}^{+\infty} {|k^2-k_0^2| \;|\psi_0|^2} \d x \right\};
\end{equation}
with the tacit understanding that the bound remains valid only so long as argument of the hyperbolic function is positive.
\subsection{Transmission probabilities}

As usual, the transmission probability (barrier penetration probability) is related to the Bogoliubov coefficient by
\begin{equation}
T = {1\over|\alpha|^2},
\end{equation}
whence
\begin{equation}
T  \geq \sech^2\left\{ \cosh^{-1}|\alpha_0| + {1\over2} 
 \int _{-\infty}^{+\infty} {|k^2-k_0^2| \;|\psi_0|^2} \d x \right\}.
\end{equation}
That is
\begin{equation}
T  \geq \sech^2\left\{ \cosh^{-1}(T_0^{-1/2}) + {1\over2} 
 \int _{-\infty}^{+\infty} {|k^2-k_0^2| \;|\psi_0|^2} \d x \right\},
\end{equation}
or even
\begin{equation}
T  \geq \sech^2\left\{ \sech^{-1}(T_0^{1/2}) + {1\over2} 
 \int _{-\infty}^{+\infty} {|k^2-k_0^2| \;|\psi_0|^2} \d x \right\}.
\end{equation}
Furthermore, as long as the argument of the $\sech$ is positive, we also have the upper bound
\begin{equation}
T  \leq \sech^2\left\{ \sech^{-1}(T_0^{1/2}) - {1\over2} 
 \int _{-\infty}^{+\infty} {|k^2-k_0^2| \;|\psi_0|^2} \d x \right\}.
\end{equation}
If one wishes to make the algebraic dependence on $T_0$ clearer, by expanding the hyperbolic functions these formulae may be recast as the statement that $T$ is \emph{greater} than
\begin{equation}
{ T_0 \over 
\left[ \cosh\left\{ {1\over2} \int _{-\infty}^{+\infty} {|k^2-k_0^2| \;|\psi_0|^2} \d x \right\} 
+ \sqrt{1-T_0}  \sinh\left\{ {1\over2} \int _{-\infty}^{+\infty} {|k^2-k_0^2| \;|\psi_0|^2} \d x \right\} \right]^2},
\end{equation}
and (as long as the numerator is positive before squaring), that $T$ is \emph{less} than
\begin{equation}
{ T_0 \over 
\left[ \cosh\left\{ {1\over2} \int _{-\infty}^{+\infty} {|k^2-k_0^2| \;|\psi_0|^2} \d x \right\} 
- \sqrt{1-T_0}  \sinh\left\{ {1\over2} \int _{-\infty}^{+\infty} {|k^2-k_0^2| \;|\psi_0|^2} \d x \right\} \right]^2}.
\end{equation}
\section{Consistency check}
There is one special case in which we can easily compare with the previous results of chapters \ref{C:consistency-5} and \ref{C:consistency-8} of this thesis, and references~\cite{bounds1, bounds2}. Take $k_0 = k(\pm\infty)$ to be independent of position, so that our comparison problem is a free particle. In that case 
\begin{equation}
\psi_0 = {\exp(ik_0 x)\over\sqrt{k_0}}; \qquad
|\psi_0|^2 = {1\over k_0} ; \qquad \J_0 = 1; \quad \alpha_0 = 1; \qquad \beta_0 = 0.
\end{equation}
Then the bounds derived above simplify to
\begin{equation}
|\alpha|  \leq \cosh\left\{  {1\over2k_0} 
 \int _{-\infty}^{+\infty} |k^2-k_0^2| \;\d x \right\},
\end{equation}
\begin{equation}
|\beta|  \leq \sinh\left\{ {1\over2k_0} 
 \int _{-\infty}^{+\infty} |k^2-k_0^2| \;\d x \right\}.
\end{equation}
This is  ``Case I'' of reference~\cite{bounds1}, and the ``elementary bound'' of reference~\cite{bounds2}, which demonstrates consistency whenever the formalisms overlap. 
(Note that it is not possible to obtain ``Case II'' of reference~\cite{bounds1} or the ``general bound'' of reference~\cite{bounds1, bounds2}  from the present analysis --- this is not a problem, it is just an indication that this new bound really is a \emph{different} bound that only partially overlaps with the previous results of references~\cite{bounds1, bounds2, Miller-good-transformation-articles}.)

A second (elementary) check is to see what happens if we set $\psi(x)\to\psi_0(x)$, effectively assuming that the full problem is analytically solvable. In that case $T\to T_0$,  (and similarly both $\alpha\to\alpha_0$ and $\beta\to\beta_0$), as indeed they should.

\section{Keeping the phases?}
We can extract a little more information by taking the imaginary parts of equations (\ref{E:a}) and (\ref{E:b}) to obtain:
\begin{equation}
\phi_a' = {1\over2} [k^2-k_0^2]\;  |\psi_0|^2 \; \left\{ 1\; + {|b|\over |a|} \; \cos(\phi_a-\phi_b+2\phi_0) \right\};
\label{E:aa}
\end{equation}
\begin{equation}
\phi_b' = -{1\over2} [k^2-k_0^2]\;  |\psi_0|^2 \; \left\{ 1 + {|a|\over |b|} \; \cos(\phi_b-\phi_a-2\phi_0) \right\}.
\label{E:bb}
\end{equation}
Subtracting
\begin{equation}
(\phi_a-\phi_b)' = [k^2-k_0^2]\;  |\psi_0|^2 \; \left\{ 1\; + {1\over 2 |a| \, |b|} \; \cos(\phi_a-\phi_b+2\phi_0) \right\}.
\label{E:ab}
\end{equation}
This is now a differential equation that only depends on the difference in the phases --- the overall average phase $(\phi_a+\phi_b)/2$ has completely decoupled. (Moreover, in determining the transmission and reflection probabilities, this average phase also neatly decouples). 
To see how far we can push this observation, let us now define a ``nett'' phase
\begin{equation}
\Delta  = \phi_a-\phi_b+2\phi_0.
\end{equation}
Furthermore,  as per the previous subsections, we retain the definitions
\begin{equation}
|a| = \cosh\Theta; \qquad |b| = \sinh\Theta.
\end{equation}
Then equation (\ref{E:pre-Theta}) becomes
\begin{equation}
\label{E:Theta}
\Theta(x)  = \left\{ {1\over2} 
 \int _{-\infty}^{x} [k^2-k_0^2] \; |\psi_0|^2  \sin(\Delta(x)) \; \d x \right\}.
\end{equation}
while the ``nett'' phase satisfies
\begin{equation}
\Delta(x)' = \left\{ [k^2-k_0^2]\;  |\psi_0|^2 \; + 2\phi_0' \right\} + { [k^2-k_0^2]\;  |\psi_0|^2 \over \sinh[2\Theta(x)]} \; \cos[\Delta(x)] .
\label{E:ab2}
\end{equation}
We can even substitute for $\Theta(x)$ and thus rewrite this as a single \emph{integro-differential} equation for $\Delta(x)$:
\begin{eqnarray}
\Delta(x)' &=& \left\{ [k^2-k_0^2]\;  |\psi_0|^2 \; + 2\phi_0' \right\} 
\nonumber\\
&&+ 
{ [k^2-k_0^2]\;  |\psi_0|^2 \over 
\sinh\left(  \int _{-\infty}^{x} [k^2-k_0^2] \; |\psi_0|^2  \sin[\Delta(x)] \; \d x  \right)} \; \cos[\Delta(x)].
\qquad\qquad
\label{E:ab3}
\end{eqnarray}
This equation is completely equivalent to the original Schr\"odinger equation we started from. Unfortunately further manipulations seem intractable, and it does not appear practicable to push these observations any further.
\section{Application: Small shift in the potential}

Let us now consider the situation
\begin{equation}
V(x) = V_0(x) + \epsilon\; \delta V(x),
\end{equation}
for $\epsilon$ ``sufficiently small''.  

\subsection{First-order changes}
To be consistent with previous notation (\ref{introduce-the-notation}) let us define 
\begin{equation}
k^2 = k_0^2 + \epsilon \; \left\{ {2m \; \delta V\over\hbar^2}\right\} \equiv k_0^2 + \epsilon\; \delta v.
\end{equation}
Using equation (\ref{E:pre-Theta}) we obtain the preliminary estimates
\begin{equation}
|a(x)|  = 1+ \mathcal{O}(\epsilon^2),
\end{equation}
and similarly 
\begin{equation}
|b(x)|  = \mathcal{O}(\epsilon).
\end{equation}
It is now useful to change variables by introducing some explicit phases so as to define
\begin{equation}
a = \tilde a \; \exp\left(+{i\over2} \int [k^2-k_0^2]\, |\psi_0^2|\, \d x\right); 
\end{equation}
\begin{equation}
b = \tilde b \; \exp\left(-{i\over2} \int [k^2-k_0^2]\, |\psi_0^2|\, \d x\right).
\end{equation}
Doing so modifies the system of differential equations (\ref{E:system-a}, \ref{E:system-b}) so that it becomes
\begin{eqnarray}
\label{E:system-aa}
{\d \tilde a\over \d x} &=& 
+{i\over2} [k^2-k_0^2] \; \tilde b\; \psi_0^*{}^2 \; \exp\left(-i \int [k^2-k_0^2] \, |\psi_0^2|\, \d x\right);
\\
\label{E:system-bb}
{\d \tilde b\over \d x} &=& 
- {i\over2} [k^2-k_0^2]\; \tilde a \;\psi_0^2 \; \exp\left(+i \int [k^2-k_0^2] \, |\psi_0^2|\, \d x\right).
\end{eqnarray}
The advantage of doing this is that in the current situation we can now estimate
\begin{eqnarray}
\label{E:system-aaa }
{\d \tilde a\over \d x} &=&  \mathcal{O}(\epsilon^2),
\\
\label{E:system-bbb}
{\d \tilde b\over \d x} &=& 
-  {i\epsilon\over2} \; \delta v(x) \;\psi_0^2(x) \; \exp\left(+i \int \epsilon \; \delta v\; |\psi_0^2| \; \d x\right) +  \mathcal{O}(\epsilon^3). \qquad
\end{eqnarray}
Integrating
\begin{equation}
\label{E:b1b}
\tilde b(\infty)  =  -  {i\epsilon\over2} 
 \int _{-\infty}^{+\infty} \delta v(x) \; \psi_0^2(x)  \; \exp\left(+i \int \epsilon \; \delta v\; |\psi_0^2| \; \d x\right)  \; \d x  + \mathcal{O}(\epsilon^3).
\end{equation}
This is not the standard Born approximation, though it can be viewed as an instance of the so-called ``distorted wave Born approximation''~\cite{born-approximation}.  In terms of the absolute values we definitely have
\begin{equation}
|\tilde b(\infty)| =  |b(\infty)| \leq  \; {\epsilon\over2} \;
 \int _{-\infty}^{+\infty} |\delta v(x)| \; |\psi_0^2(x)|  \; \d x  + \mathcal{O}(\epsilon^3).
\end{equation}
\subsection{Particle production}

When it comes to considering particle production we note that
\begin{equation}
\beta =\beta_0 \; a(\infty)+ \alpha_0^* \; b(\infty)  = \beta_0 + \alpha_0^* \; b(\infty)  + \mathcal{O}(\epsilon^2),
\end{equation}
so, since $N=|\beta|^2$, the \emph{change} in the number of particles produced is
\begin{equation}
\delta N = \delta|\beta^2| = \Re\left\{2 \beta^*\, \delta \beta\right\}= \Re\left\{ 2 \alpha_0^* \, \beta_0\; b(\infty) \right\} +  \mathcal{O}(\epsilon^2).
\end{equation}
In particular
\begin{equation}
\left| \,\delta N \,\right| \leq  \epsilon \, |\alpha_0| \, |\beta_0|\;  \int _{-\infty}^{+\infty} |\delta v(x)| \; |\psi_0^2(x)| \;  \d x  +  \mathcal{O}(\epsilon^2).
\end{equation}
Since 
\begin{equation}
|\alpha_{0}| |\beta_{0}| = \sqrt{1+|\beta_{0}|^2} \, |\beta_{0}| = \sqrt{N_{0} +1} \, \sqrt{N_{0}} = \sqrt{N_{0}(N_{0}+1)},
\end{equation}
can also write as
\begin{equation}
\left| \,\delta N \,\right|\leq  \epsilon \;\sqrt{N_0(N_0+1)} \;  \int _{-\infty}^{+\infty} |\delta v(x)| \; |\psi_0^2(x)| \;  \d x  +  \mathcal{O}(\epsilon^2).
\end{equation}
Note that one will only get an order $\epsilon$ change in the particle production if the ``known'' problem $\{\psi_0,\, k_0\}$ already results in nonzero particle production.  

\subsection{Transmission probability}

To see how a small shift in the potential affects the transmission probability we note
\begin{equation}
T = {1\over|\alpha|^2} = {1\over| \alpha_0 \, a(\infty) + \beta_0^* \, b(\infty)|^2} =  {1\over| \alpha_0  + \beta_0^* \, b(\infty) + \mathcal{O}(\epsilon^2)|^2}.
\end{equation}
But then
\begin{equation}
T =  {1\over| \alpha_0 |^2 \; \left|1 + \{\beta_0^* \, b(\infty)/\alpha_0\} + \mathcal{O}(\epsilon^2)\right|^2},
\end{equation}
implying
\begin{equation}
T =   T_0 \;  \left\{1  - 2 \Re\left\{{\beta_0^* \, b(\infty)\over\alpha_0}\right\} + \mathcal{O}(\epsilon^2) \right\}.
\end{equation}
So the \emph{change} in the transmission probability is
\begin{equation}
\delta T =   - T_0 \;  \left\{2 \Re\left\{{\beta_0^* \; b(\infty)\over\alpha_0}\right\} + \mathcal{O}(\epsilon^2) \right\}.
\end{equation}
Taking absolute values one obtains
\begin{equation}
|\delta T| \leq  \epsilon \; T_0 \; \sqrt{1-T_0} \, \int _{-\infty}^{+\infty} |\delta v(x)| \; |\psi_0^2(x)| \;  \d x+ \mathcal{O}(\epsilon^2).
\end{equation}
Note that one will only get an order $\epsilon$ change in the transmission probability if the ``known'' problem $\{\psi_0,\, k_0\}$ already results in nonzero transmission (and nonzero reflection).  
\section{Discussion}
We wish emphasize the advantages of the particular bounds derived in this chapter:
\begin{itemize}
\item 
They are very simple to derive --- the algebra is a lot less complicated than some of the other approaches that have been developed in earlier chapters of this thesis, and published in several papers~\cite{bounds1, bounds2, greybody-factor, Miller-good-transformation-articles}. (And a lot less complicated than some of the blind alleys we have explored.)
\item 
Under suitable circumstances the procedure of this chapter yields both upper and lower bounds. Obtaining both upper and lower bounds is in general very difficult to do --- see in particular the attempts in~\cite{bounds2}.
\item 
All of the other bounds we have developed in earlier chapters of this thesis, and published in several papers~\cite{bounds1, bounds2, greybody-factor, Miller-good-transformation-articles}  needed some condition on the phase of the wave-function, (some condition similar to $\varphi'\neq 0$), which had the ultimate effect of making it difficult to make statements about tunnelling ``under the barrier''.  There is no such requirement in the present analysis. (The closest analogue is that we need $\J_0\neq0$, which we normalize without loss of generality to $\J_0=1$.) In particular this means that there should be no particular difficulty in applying the bound in the classically forbidden region --- the ``art'' will lie in finding a suitable form for $\psi_0$ which is simple enough to carry out exact computations while still providing useful information.
\end{itemize}
In closing, we reiterate the fact that generic one-dimensional scattering problems, which have been extensively studied for close to a century,  nevertheless still lead to interesting features and novel results.

\chapter{Discussion}
\label{C:consistency-12}
  \section{What we have achieved}
In this chapter we shall discuss the overall concept and achievements of this thesis. To a large extent much of the work has already has summarised at the end of each section. This thesis has been written with the goal of making it fully accessible to people with a basic background in non-relativistic quantum physics, especially in the physics of transmission, reflection, and Bogoliubov coefficients. Mathematically, the key feature is an analytic study of the properties of certain second-order linear differential equations, and the derivation of analytic bounds on the growth of solutions of these equations (as a function of position and/or time).
In this thesis we divided our efforts into analyzing four separate problems relating to rigorous bounds on transmission, reflection,  and Bogoliubov coefficients --- these are considered under four separate themes: 
\begin{enumerate}
\item
 Bounding the Bogoliubov coefficients, 
 \item
Bounding the greybody factors for Schwarzschild black holes, 
\item
Transmission probabilities and the Miller--Good transformation, and
 \item
 Analytic bounds on transmission probabilities.
\end{enumerate}
In addition, all four of these separate themes which are reported in this thesis, are also the seeds for the various published journal articles (plus one submitted article) that have already arisen from this thesis.

This thesis is divided into twelve main chapters. In the following, we shall summarise and anaylse the main work we derived in each chapter.
\section[The main analysis: Structure of the thesis]{The main analysis: 
\\ Structure of the thesis}

We first provided sufficient context for the reader  to appreciate the role played by the various topics to be discussed in this thesis, and to place them into a wider perspective. Firstly, we introduced the Schr\"odinger equation --- a specific partial differential equation used in the development of the ``new'' (1925) quantum theory. Secondly, we provided the basic theory underlying the \emph{WKB approximation}, and the concept of the time-independent  Schr\"odinger equation --- both are foundations for all our subsequent analyses. We have presented a very general introduction to these concepts first --- so that the bounds we derived on transmission, reflection, and Bogoliubov coefficients were easier to understand. 

We mainly considered the scattering theory in one space dimension --- because it  is mathematically simple and physically transparent.

In particular, it is interesting to show how to derive  the basic ideas of transmission and reflection directly by using scattering theory. 
In addition, we have just seen an important connection between reflection and transmission amplitudes. 
We called the probability that a given incident particle is reflected as the ``reflection coefficient''. While the probability that it is transmitted is called the ``transmission coefficient''.

In chapter \ref{C:consistency-3}, we collected many known analytic results in a form amenable to comparison with the general results we subsequently derived. In addition, we also introduced the concept of \emph{quasinormal modes} [QNM]. We used these tools for comparing the bounds with known analytic results. Moreover, we reproduced  some of the analytically known results, and showed (or at least sketched) how to derive their scattering amplitudes, and so calculate quantities such as the tunnelling probabilities and quasinormal modes. We did this explicitly for the delta--function potential, double--delta--function potential, square potential barrier, tanh potential, sech$^2$ potential, asymmetric square-well potential, the Poeschl--Teller potential and its variants, and finally the Eckart--Rosen--Morse--Poeschl--Teller potential.

We also obtained a number of significant bounds, considerably stronger than those in~\cite{bounds1}, of both theoretical and practical interest. Even though the calculations we have presented are sometimes somewhat tedious, we feel however, they are more than worth the effort --- since there is a fundamental lesson to be learnt from them. Technically, we demonstrated that the Schr\"odinger equation can be written as a Shabat--Zakharov or a Zakharov--Shabat system, which  can then be re-written in $2\times2$ matrix form. 

In chapter~\ref{C:consistency-5}, we have again moved our attention back to a  Shabat--Zakharov system of ODEs by re-casting and describing the first derivation of scattering bounds as presented by Visser in reference~\cite{bounds1}. The formalism as developed here works in terms of one free function $\varphi(x)$. In other parts of this thesis we have established generalized bounds; some in terms of \emph{two} arbitrary functions $\varphi(x)$ and $\chi(x)$, and some in terms of \emph{three} arbitrary functions  $\varphi(x)$, $\Delta(x)$,  and $\chi(x)$. The derivation of this chapter is noteworthy because of its brevity and simplicity. 

All of the above techniques from chapter~\ref{C:consistency-5} are important to develop a number of interesting bounds in chapter~\ref{C:consistency-6}. We dealt with some specific cases of these bounds and develop a number of interesting specializations.  We have collected together a large number of results that otherwise appear quite unrelated, including reflection above and below the barrier. In addition, we have divided the special case bounds we considered into five special cases: special cases 1--4, and ``future directions''. Finally, we took further specific cases of these bounds and related results to reproduce many analytically known results.

Consequently, we have re-cast and represented these bounds (from chapter~\ref{C:consistency-6}) in terms of the mathematical structure of parametric oscillations. This time-dependent problem is closely related to the spatial properties of the time-independent Schr\"odinger equation.

In chapter~\ref{C:consistency-8}, we re-assessed the general bounds on the Bogoliubov coefficients developed in earlier chapters of this thesis, and published in reference~\cite{bounds1}, providing a new and largely independent derivation of the key results, one that short-circuits much of the technical discussion in chapter~\ref{C:consistency-5}, and published in reference~\cite{bounds1}.

After this investigation about bounding the Bogoliubov coefficients and their techniques we have moved to study the greybody factors in Schwarzs\-child black hole.  The ``greybody factor'' is actually a synonym for ``transmission probability''. Indeed, the phrase ``greybody factor'' is used more in the thermodynamics and spectroscopy communities, while the phrase  ``transmission probability'' is used more in the quantum mechanics community, but they are referring to the same concept. In this thesis, we developed a complementary set of results --- we derived several rigorous analytic bounds that can be placed on the greybody factors.  Even though these bounds are not necessarily tight bounds on the 
exact greybody factors, they do serve to focus attention on general and robust features of these greybody factors. Moreover they provide a new method of extracting physical information. Furthermore, we considered the greybody factors in black hole physics, which modify the naive Planckian spectrum that is predicted for Hawking radiation when working in the limit of geometrical optics.

We used the Miller--Good transformation (which maps an initial 
Schr\"od\-in\-ger equation to a final Schr\"od\-in\-ger equation for a different potential) to significantly generalize the previous bound. Moreover, we shall see that the Miller--Good transformation 
is an efficient process to generalize the bound, to make it more efficient and powerful.

Finally, we have again shifted our attention back to the analytic bounds and transmission probabilities context.  We developed a new set of techniques that are more amenable to the development of \emph{both} upper and lower bounds. Moreover, we derived significantly different results (a number of rigorous bounds on transmission probabilities for one dimensional scattering problems), of both theoretical and practical interest. 
\begin{table}[!ht]
\centerline{Several ways to derive bounds for arbitrary wave phenomena} 
\centerline{(including Schwarzchild black hole greybody factors)}
\bigskip
\hskip - 1.5 cm 
\begin{tabular}{| l | l |}
\hline
\hline
Method & Characteristics and properties of bounds \\
\hline
\hline
(0). Standard WKB & -- uncontrolled approximation.
\\
                                       & -- do not know if the approximation
\\                                         
                                       & \quad is high or low.
\\                                      
\hline
(1). WKB ``basis''   & -- rigorous bounds.
\\
                                       & -- use ``gauge fix'' method.
\\                                       
\hline
(2). Bogoliubov coefficients & -- rigorous bounds.
\\
                                          & -- do not use ``gauge fix''.
\\
                                         & -- short-circuits the technical details of method (1).
\\
\hline
(3). general bounds on the & -- obtained several rigorous analytic bounds
\\
\quad  \quad  greybody factors &  \quad  that can be placed on the greybody factors.
\\
\hline
(4). the Miller--Good transformation  &  -- generally not ``WKB--like''.
\\
                                                                     & -- no need to separate the region into
\\                                                                    
                                                                     & \quad  classically allowed and forbidden regions.
\\
\hline
(5). the analytic bounds  &   --  very simple to derive the bounds.
\\
                                                                     & --  procedure yields both upper and lower bounds.
\\                                                                     
\hline
\end{tabular}
\caption[Several ways to derive bounds for arbitrary wave phenomena]{This table shows several ways to derive bounds for arbitrary wave phenomena (including the greybody factors of Schwarzchild black holes), and also the key properties of these bounds. (See chapters \ref{C:consistency-8}, \ref{C:consistency-9}, \ref{C:consistency-10}, and \ref{C:consistency-11} for more details.)}
\end{table}
Instead of explaining the details of the analysis  yet again, 

we would like to stress a few points  that we believe are useful to understand the overall concept of the thesis:
\begin{itemize}
\item
The Schr\"odinger equation  describes the space --and time-- dependence of quantum mechanical systems, and its application to the wave function are the basic idea that describes the wavelike properties of a subatomic system.
\item
The probability current express the reflection and transmission coefficients. The probability current is based on the assumption that the intensity of a beam is the product of the speed of its 
particles and their linear number density. It is then a mathematical theorem that this probability current is conserved. 
\item
The WKB approximation is generally applicable to problems of wave propagation in which the frequency of the wave is very high or equivalently, the wavelength of the wave is very short. 
\item
The ideas of reflection and transmission of waves in both unbound and bound 
states are important. By considering reflection and transmission of waves in unbound states, 
we have seen that in principle they are completely specified by the potential 
function $V (x)$. 
\item
The \emph{quasinormal modes} [QNM] are the modes of energy dissipation of a perturbed object or field. In particular, the most outstanding and well-known example is the perturbation of a wine glass with a knife: the glass begins to ring, it rings with a set, or superposition, of its natural frequencies -- its modes of sonic energy dissipation. As previously explained, when the glass went on ringing forever, we can call these modes normal. For instance, here the amplitude of oscillation decays in time, so we call its modes \emph{quasi-normal}~\cite{quasinormal-mode}. 
\item
The Schr\"odinger equation can be written as a Shabat--Zakharov system, which  can then be re-written in $2\times2$ matrix form. We rearranged this formation in terms of a generalized position-dependent ``transfer matrix''  involving  the symbol $\mathcal{P}$ which denotes ``path ordering''. 
\item
A ``parametric oscillator'' is a simple harmonic oscillator whose parameters (its resonance frequency $\omega$ and damping time $\beta$) vary in time. The other interesting way of understanding a parametric oscillator is that it is a device that oscillates when one of its ``parameters'' (a physical entity, like capacitance) is changed.
\end{itemize}
\section{Further interesting issues}
There are some interesting ways this thesis could be extended in future work.
We would like to wrap-up by providing a list of things that we believe are interesting to continue to analyze:
\begin{itemize}
\item
This present research certainly deserves more work on how to extend the bounds in many different ways. While we have already established several powerful techniques to derive rigorous bounds on transmission, reflection, and Bogoliubov coefficients, we feel,  however, that there are probably  ``optimal'' bounds still waiting to be discovered. 
\item
 In particular, it is apparent that the current bounds in chapter~\ref{C:consistency-8} are not the best that can be 
achieved, and we strongly suspect that it may be possible to develop yet further extensions to the current formalism. It is in fact possible that the ``more general''  bounds are close to being discovered and will have further development in the near future. 
 \item
 The bounds presented in chapter~\ref{C:consistency-10} are generally \emph{not} ``WKB-like'' --- apart from the one case reported in equation~(\ref{E:WKB-like}) there is no need (nor does it seem useful) to separate the region of integration into classically allowed and classically forbidden regions. In fact it is far from clear how closely these bounds might ultimately be related to WKB estimates of the transmission probabilities, and this is an issue to which we hope to return in the future. 
\item
Finally, we have seen that even though the topic considered in this thesis  is ultimately a quantum mechanics subject, dating back to 1925, this does not mean that everything has already been done. It is conceivable that the new techniques in this project help us to derive more rigorous and tighter bounds for the barrier penetration probability.
\end{itemize}

All the above suggestions would be interesting and feasible, although some of them would be more tedious to work on than others. 

In summary, this thesis provides a platform for better understanding the rigorous bounds that one can place on the Bogoliubov coefficients associated with a time-dependent potential, and the several rigorous analytic bounds that can be placed on the greybody factors. This thesis developed a way of looking for nice and accurate bounds. Furthermore, one primary goal of this thesis was to explore the best way of finding barrier penetration probability. 
In conclusion, we can say that this project will be another step to improving our understanding of quantum mechanics, in particular, non-relativistic quantum physics, especially in regard to transmission, reflection, and Bogoliubov coefficients. 

\addcontentsline{toc}{chapter}{Bibliography}




\begin{thebibliography}{99}
\bibitem{Transfer}
Transfer matrix techniques are discussed, at varying levels of detail, by Merzbacher \cite{Merzbacher}, Singh \cite{Singh}, and Mathews \cite{Mathews}.
\bibitem{Transfer-matrix} 
Transfer matrix techniques are discussed, at
varying levels of detail, by Merzbacher~\cite{Merzbacher},
Singh~\cite{Singh}, and Mathews~\cite{Mathews}.
\end{thebibliography}
\end{document}